\documentclass[a4wide,11pt]{article}

\usepackage{amssymb}
\usepackage{amsfonts}
\usepackage{epsfig}

\newcommand{\vb}[1]{{\mathbf{#1}}}
\newcommand{\lb}[1]{\label{#1}}

\newcommand{\bc}{\begin{center}}
\newcommand{\ec}{\end{center}}
\newcommand{\be}{\begin{equation}}
\newcommand{\ee}{\end{equation}}
\newcommand{\bea}{\begin{eqnarray}}
\newcommand{\eea}{\end{eqnarray}}
\newcommand{\ba}[1]{\begin{array}{#1}}
\newcommand{\ea}{\end{array}}

\newcommand{\bt}[1]{\begin{table}[ht]\centering\begin{tabular}{#1}}
\newcommand{\et}[1]{\end{tabular}\caption{\it\small#1}\end{table}}
\newcommand{\fig}[4]{\begin{figure}[htb]\epsfxsize=#2\bigskip\centerline{\epsfbox{#1}}\caption{\small\it #3 \label{#4}}\bigskip\end{figure}}

\begin{document}

\thispagestyle{empty}

\title{A Locally Anisotropic Metric for Matter in an Expanding Universe:\\[2mm] I. The Ansatz and the Modified Newton Law\\[.5cm]}

\author{P. Castelo Ferreira\\[2mm]\small \tt pedro.castelo.ferreira@ist.utl.pt\\[2mm]\small CENTRA, IST, Av. Rovisco Pais 1, 1100-001 Lisboa, Portugal\\[.4cm]}

\date{}

\maketitle

\begin{abstract}

\noindent It is suggested a metric ansatz describing matter in an expanding universe,
hence interpolating between the Schwarzschild metric near central bodies
of mass $M$ and the Friedman-Lema\'{\i}tre-Robertson-Walker metric
for large radial coordinate, given by\\[-4mm]
\begin{center}$
\displaystyle ds^2=Z\,c^2dt^2-\frac{1}{Z}\left(dr_1-\frac{H\,r_1}{c}\,Z^{\frac{\alpha}{2}+\frac{1}{2}}\,cdt\right)^2-r_1^2\,d\Omega\ ,
$\end{center}
where $Z=1-2GM/(c^2r_1)$, $G$ is the Newton constant, $c$ is the speed of light, $H=H(t)$ is the
time-dependent Hubble rate, $d\Omega=d\theta^2+\sin^2\theta\,d\varphi^2$ is the solid angle element
and we are employing Schwarzschild expanding coordinates $r_1$ (also known as physical coordinates
for expanding space-time).
For constant exponent $\alpha=0$ it is retrieved the isotropic McVittie
metric and for $\alpha=1$ it is retrieved the locally anisotropic Cosmological-Schwarzschild metric,
both already discussed in the literature. It is shown that, for constant exponent $\alpha$, the event
horizon at the Schwarzschild
radius $r_1=2GM/c^2$ is only singularity free for $\alpha\ge 3$ and space-time is asymptotically
flat for $\alpha>5$ which excludes these known cases. Also it is shown that, to strictly maintain the
Schwarzschild mass pole at the origin $r_1=0$ without the presence of more severe singularities,
hence describing a complete space-time with finite total mass-energy within a shell of finite radius,
it is required a radial coordinate dependent exponent $\alpha(r_1)=\alpha_0+\alpha_1\,2GM/(c^2\,r_1)$
with a negative coefficient $\alpha_1<0$ such that at the event horizon,
for $\alpha_0-|\alpha_1|\ge 3$ space-time is singularity free and for $\alpha_0-|\alpha_1|>5$ space
time is asymptotically flat. This metric may solve the long standing puzzle of describing local
matter distributions in an expanding universe firstly addressed by McVittie.\\

\noindent The curvature, curvature invariant and stress-energy tensor are analyzed in detail being
derived the allowed bounds for the parameters $\alpha_0$ and $\alpha_1$ that allow simultaneously
space-time to be singularity free (except for the Schwarzschild mass pole at the origin) and
the mass-energy density to be positive definite outside the event horizon. It is shown that,
although space-time is locally anisotropic near the mass $M$, isotropy at spatial infinity is
maintained. This characteristic is qualitatively consistent both with the experimental evidence
of local anisotropy due to matter structures and global spatial isotropy. The modified Newton law for
this metric is derived being shown that for planetary scales the usual General Relativity Newton law is
approximately maintained for the full allowed range of the parameter $\alpha_0$ while for galaxy
scales and large values of the parameter $\alpha_0>c^2/\sqrt{q_0(GM\,H_0)^2}$
there is a significant deviation from this law which may contribute, for instance, for the flattening
of galaxy velocity curves, hence allowing, at least partially, to describe dark matter effects, here
interpreted as due to the universe expansion. Are derived solutions for planetary orbits both
in the circular orbit and perturbative static elliptic orbits approximations and estimated the
orbital precession, period corrections and time variation of the orbital radius which are well
within the existing experimental bounds for the solar system.

\end{abstract}

\newpage

\tableofcontents

\newpage

\section{Introduction\lb{sec.introd}}

In 1929 by analyzing the red-shift for radiation received from nearby galaxies
Hubble~\cite{Hubble} found experimental evidence that the universe was expanding
at a rate given by the Hubble constant $H=\dot{a}/a$, where $a$ is the scale factor
of the universe. Later, in 70's, it was realized that the expansion was not constant,
at that time it was believed to be decelerating at a rate given by
$-q_0H_0^2$~\cite{Sandage}, where $q_0$ is the deceleration parameter.
More accurate measurements have been accomplished and today it is known that the universe
is in accelerated expansion, hence $q_0<0$. These observations raised several theoretical
(and phenomenological) puzzles both for small and large spatial scales.

In particular, for small spatial scales, the definition of a metric that describes both an
expanding background (the universe) and local matter distributions is today an open
problem. As far as the author is aware there
are only two distinct metrics that have been proposed. Not long after Hubble's work,
in the 30's, McVittie derived a metric in isotropic coordinates~\cite{McVittie}
that preserves locally spatial isotropy interpolating between the Schwarzschild (SC)
metric~\cite{Schwarzschild} which describes point-like massive bodies for small spatial scales
and the Friedman-Lema\'{\i}tre-Robertson-Walker (FLRW) metric~\cite{RW} which describes the expanding
universe for large spatial scales. It is known to be the only metric that maintains local spatial
isotropy. Allowing for local spatial anisotropy (still maintaining spatial isotropy for large scales)
has been recently proposed a Cosmological-Schwarzschild metric~\cite{cosm_SC}.
As has been discussed in~\cite{sing} for the McVittie metric and will be analyze in detail
in this work, both these metrics have singularities at the event horizon (the Schwarzschild radius).
In this way the respective space-time is not complete and the metrics do not converge asymptotically
to the SC metric near the point-like mass. For further discussions in this topic see
also~\cite{review_expanding_1,review_expanding_2, Bolen, Nolan,Lambda}.
Our main objective in the present work is to address this problem and try to solve it by deriving
(or building) a new metric ansatz. We will take a phenomenological approach not having a fundamental
theoretical basis for the results derived, the derivation of the new metric is only based in the
assumptions that a metric describing matter in an expanding background (the universe) must interpolate
between the SC and the FLRW metric maintaining space-time singularity free except for the SC mass pole
at the center of mass. We will allow for local anisotropy as long as isotropy is recover for large
spatial scales (specifically that the FLRW metric is recovered at spatial infinity). We note that this
characteristic is not unwelcome, there is experimental evidence that local matter distribution does
generate local anisotropy~\cite{anisotropy,WMAP}, hence the imposition by McVittie for a fully
isotropic metric seems excessive.

With respect to large spatial scales, the same observations concerning the expansion of the universe
led to today's cosmological inflationary models~\cite{Inflation}. The deceleration
is commonly attributed to the existence of dark energy which constitute 72.6\% of the
total mass-energy of the universe. The most common description of dark energy is through the cosmological
constant~\cite{L} which describes a constant background (vacuum) energy with negative equation
of state. Alternative approaches do exist that conciliate
inflationary theories with the existence of a non-null vacuum energy. Such examples are
quintenssence~\cite{5essence}, phantom energy~\cite{phantom}, modified theories
of gravity~\cite{modgrav} and string based/inspired theories~\cite{strings}.
For a review in these topics see~\cite{review}. Although not originally related to expansion
there is also for large and/or very massive astrophysical systems experimental evidence
of non-radiating matter~\cite{Zwicki} moving at non-relativistic speed, usually named
cold dark matter (CDM), which is responsible for significant deviations with respect
to the usual General Relativity predictions, namely causing the flattening of galaxy
rotation curves~\cite{galaxies_DM}, increase of the gravitational
weak lensing effects by astrophysical large structures~\cite{lensing_DM} (as is the case of
the bullet cluster~\cite{bc_DM}) and the modulation of the weak lensing effects near heavy
astrophysical bodies~\cite{modulation_DM}. The most direct interpretation of these effects
is the existence of physical non-radiating matter that as not been accounted for, hence named
dark matter. Alternatives to this interpretation are based in modifications to the usual
General Relativity Newton law, specifically there are two distinct approaches, either known
as MOND~\cite{MOND} for which the General Relativity Newton law becomes proportional to the inverse of the radial
distance for galactic scales and the Scalar-Tensor-Vector theory~\cite{STV} which considers extra
fields interacting gravitationally. For reviews in these topics see~\cite{DM}. It is interesting
that the metric we derive in this work, although based in very simple assumptions, also allows
at galactic scales for a significant deviation from the usual General Relativity Newton law,
hence in our framework these effects may be interpreted as a consequence of the universe expansion.

It is today well accepted that both dark matter and dark energy contribute to the universe
evolution. In particular this relation led to the most
successful model for today's universe, the $\Lambda CDM$ model~\cite{LCDM} which
accounts for the contributions of these forms of matter and energy to the total
universe matter-energy density and pressure.
Based in this model and the combination of distant supernova of type Ia red-shifts (SN) measurements~\cite{SN}, the baryon acoustic oscillations (BAO) measurements~\cite{BAO} and
the five-year Wilkinson microwave anisotropy probe (WMAP) measurements~\cite{WMAP}
the parameters for the expansion of today's universe can be estimated. According to these
estimative the universe is flat and we are taking the following experimental values for the Hubble rate
\be
H=\frac{\dot{a}}{a}\ \ ,\ \ H_0=\left.\frac{\dot{a}}{a}\right|_{t=t_0}=h\ Mpc^{-1}\,Km\,s^{-1}=2.28\pm 0.04\,\times 10^{-18}\ s^{-1}\ ,
\lb{H}
\ee
and the deceleration parameter
\be
q=-\frac{\ddot{a}}{a}\,\left(\frac{\dot{a}}{a}\right)^{-2}\ \ ,\ \ q_0=\left.-\frac{\ddot{a}}{a}\frac{1}{H^2}\right|_{t=t_0}=-\frac{\ddot{a}_0}{a_0\,H_0^2}\approx -0.589\ .
\lb{H_q0}
\ee
This last value is computed from the equation of state for the universe and we will
address the details of this calculation later on. Similarly the third derivative of
the scale factor can be defined in terms of a parameter $s$
\be
s=\frac{\dot{\ddot{a}}}{a}\,\left(\frac{\dot{a}}{a}\right)^{-1}\,\left(\frac{\ddot{a}}{a}\right)^{-1}\ \ ,\ \ s_0=-\left.\frac{\dot{\ddot{a}}}{a}\frac{1}{q\,H^3}\right|_{t=t_0}=-\frac{\dot{\ddot{a}}_0}{a_0}\frac{1}{q_0\,H_0^3}\ ,
\lb{H_s0}
\ee
which encodes the variation of the deceleration of the universe. As far as the author
is aware there are no experimental estimative for the current value of $s_0$. We also refer
that the characteristic lengths of the expansion effects are the Hubble time
\be
t_H=\frac{1}{H}\ \ ,\ \ t_{H_0}=\left.\frac{1}{H_0}\right|_{t=t_0}=\frac{1}{H_0}\approx 4.39\times 10^{17}\ s\ ,
\lb{H_t}
\ee
and the Hubble length
\be
l_H=\frac{c}{H}\ \ ,\ \ l_{H_0}=\left.\frac{c}{H}\right|_{t=t_0}=\frac{c}{H_0}\approx 1.32\times 10^{26}\ m\ .
\lb{H_length}
\ee
Depending on the model being considered to describe inflation these correspond approximately (or at least
give an order of magnitude) to the age of the universe and the distance to the cosmological horizon.

The information from beyond a distance of the value of the Hubble length cannot reach the observer
within the life-time of the universe,
hence it is causally disconnect from the observer. In cosmology this is known as the horizon problem
which, together with the flatness problem (why the universe is flat today?) led to theories where
the fundamental constants vary with time~\cite{VSL,Dilaton,Uzan}. In particular theories of varying
Newton constant $G$ may be relevant for planetary system physics. For General Relativity in flat
backgrounds, the variation of this constant is equivalent to a time varying orbital radius,
hence experimental measurements of either of this variations are equivalent. When considering planetary
orbits in an expanding background this is no longer the case and the relation between a varying $G$ and
a varying orbital radius is distinct. In particular, as we will show, for an expanding background the
orbital radius varies with time for a fixed $G$. We note that this discussion is relevant when
interpreting experimental data or setting bounds on these variations.

We have organized this work as follows:

In section~\ref{sec.review} we review the description of an expanding universe in terms of the
FLRW metric. In particular we discuss in detail the relation between the choice of coordinates
and the physical observables defining the several coordinate choices used in the remaining of
the work. We also discuss other relevant issues as the derivation of the stress-energy tensor 
for the FLRW metric and the derivation of the deceleration parameter from the cosmological parameters.

In section~\ref{sec.matter} we review how matter is usually described in terms of a metric both
for flat and expanding backgrounds. We start by analyzing the Schwarzschild metric deriving
the most relevant quantities for this work. In particular we review how to derive the General
Relativistic corrections to the precession and period of planetary orbits. We also introduce
the McVittie metric and the Cosmological-Schwarzschild metric for expanding backgrounds discussing
their properties.

In section~\ref{sec.metric_generic} we build a locally anisotropic metric ansatz
interpolating between the SC and FLRW metric. We start by a one parameter metric and analyze
the properties of the respective space-time. In order to regularize the essential singularities
at the origin it will be required to refine the ansatz by introducing a second parameter.
Once this is accomplished we derive the stress-energy tensor explicitly analyzing its
properties, in particular we discuss the local anisotropy and define the allowed relative
bounds of the parameters that ensure mass-energy positiveness outside the event horizon.

In section~\ref{sec.Newton} we derive and analyze the modified Newton law for
this new metric. First we study the radial acceleration component due to a central mass
and how the transition between the General Relativity Newton law and the Newton law for
the expanding background (obtained from the FLRW metric) is affected by the metric
parameter values. We further analyze the orbital motion for the
new metric both for circular orbits and static elliptical orbits approximation estimating
the time variation of the orbital radius and the corrections to the orbital precession and
orbital period comparing these with available experimental data.

In section~\ref{sec.conclusions} we resume the results obtained in this work and discuss
further directions of research in this topic, namely the possible relevance of our results
for dark matter effects.

We have gathered the technical details of the main calculations in the appendixes.
In appendix~\ref{A.defs} we list the conventions and definitions considered in this work.
In appendix~\ref{A.FRW} we present the connections, curvature and Einstein tensor for
the FLRW metric and derive the lower order series expansion for the time evolution
of the Hubble rate. In appendix~\ref{A.McV} we present the McVittie metric for several
coordinate choices. In appendix~\ref{A.generic} are listed the connections, curvature,
Einstein tensor and curvature invariant for the new metric, both for the one parameter
and two parameter ansatze. Are also derived the differential equations for stationary
orbital motion.

\section{Reviewing the Background Cosmological Metric\lb{sec.review}}
\setcounter{equation}{0}

In this section we review the theoretical effects of space expansion
for free travelling massive particles and radiation interacting only with the cosmological
background. These effects are well known and can easily be found in the
literature~\cite{Peebles,Gravitation,Kenyon}. We try to approach this subject in a
pedagogical fashion from a non-expert perspective.
We discuss distinct coordinate systems by rescaling the spatial coordinates
by the universe scale factor $a$, in particular we consider expanding coordinates
(for which spatial coordinates expand over time) and non-expanding coordinates (for which the
spatial coordinates do not expand over time) as defined in the appendix~\ref{A.defs}.
This discussion is relevant whenever we want to compare theoretical results with physical
measurements. Also, for technical simplification
of the calculations, different coordinate choices are often considered independently of the
physical measurable and respective interpretation. For these reasons it is relevant to
discuss what coordinate and unit choice is more adequate when studying a particular physical
system and relating theoretical and experimental results.

It is known that the physical laws do not depend on frame nor coordinate choice in the sense
that the relations between different quantities remain the same, however the way we measure
those quantities depend on the units and coordinates and consequently our interpretation
of physical phenomena, as well as the definition and values of the fundamental constants, does
depend on our measurement procedure~\cite{Uzan}. As simple examples, considering to
use natural units $c=\hbar=1$ is not suitable when trying to measure in the laboratory the
speed of light and radiative decays with a clock that measures time in seconds. Also if we
have a rod that measures lengths $\tilde{r}=1/r$ (being $r$ the usual distance for which
the area of the sphere is $4\pi\,r^2$ while for $\tilde{r}$ it is $4\pi/\tilde{r}^2$)
the classical Newton law of gravitation would be stated as that the gravitational acceleration
has direction along growing $\tilde{r}$ being proportional to the cubic distance plus the velocity over
distance. Specifically we would obtain
\be
\ddot{r}=-\frac{GM}{r^2}\ \Leftrightarrow\ \ddot{\tilde{r}}=+GM\tilde{r}^3+\frac{\dot{\tilde{r}}}{\tilde{r}}\ .
\lb{Newton_inverse}
\ee
Hence the physical interpretation does depend on the coordinate and unit choices
which reflect the way we interact with reality and how we perform our measurements
of the physical quantities. Nevertheless the laws of physics are invariant under
a change of coordinates, in the sense that once we obtain a law, it is still valid under
any transformation of coordinates applied directly to
the specific physical relation, as we have just exemplified in the above equation for the Newton law.
If necessary, for technical simplification or any other reason, we can use any coordinate
system we wish for calculation purposes, as long as we recall to transform the final
quantities (or expressions) back to the coordinate system that is directly related
to our reality. Of particular relevance to the present work is
the definition of measurement for space and time lengths which dictates how we
perceive our universe and the coordinate choice(s) which \textit{translate}
between our theoretical constructions and reality (or the other way around).

In General Relativity the specific expressions for the metric components depend on the
coordinate and unit choice. In particular when we are living in an expanding space it
is relevant to know if our measurement rods are also expanding or not. A material rod
is surely expanding with space, hence we will not be able to measure any
expansion with it. However radiation is sensitive to space expansion, it was
through the red-shift of radiation emitted by distant objects that the Hubble
rate~(\ref{H}) was originally measured~\cite{Hubble}.

\subsection{Working with Non-Expanding Coordinates\lb{sec.review_nexp}}

We will take the global geometry of space-time to be null $\kappa=0$ which, based in
experimental data, is today the most probable case~\cite{WMAP}.
The usual metric taken to describe a flat homogenous and isotropic universe
for an observer at rest at the origin of the coordinate frame is the
Friedmann-Lema\'{\i}tre-Robertson-Walker type metric
\be
\displaystyle ds^2=c^2dt^2-a^2dr^2-a^2r^2\left(d\theta^2+\sin^2\theta d\varphi^2\right)= c^2dt^2-a^2\delta_{ij}dx^idx^j\ ,
\lb{g_FRW}
\ee
We will henceforth refer to it as the FLRW metric (note that, for $\kappa=0$, this
metric is often referred only as Robertson-Walker metric~\cite{RW}), here expressed 
both in spherical coordinates and Cartesian coordinates. For this coordinate choice we
have the geometrical identification with a time-dependent area of the sphere $A(t)=4\pi\,a^2\,r^2$
(with commoving radius $ar$). We refer to this coordinate choice as \textit{non-expanding coordinates},
although the three-dimensional space lengths do expand, as will be shown next,
the coordinates (as given by the coordinate equations of motion) for massive particles do not. Here $a=a(t)$ stands for the time dependent scale factor of the universe as appearing
in the definition of the Hubble rate~(\ref{H}).
$a$ is dimensionless and we assume that at some reference time $t_0$ corresponding to the
most recent estimative for the size of the universe its value is normalize to $a(t_0)=a_0=1$.
This is achieved either by rescaling the spatial coordinates by this constant or, alternatively,
to consider a ratio $a/a_0$ (instead of $a$) in the original metric. However, unless otherwise stated,
we will keep it in the equations to keep track of the several technical steps and identify how the
scale factor $a=a(t)$ affects the results discussed in this work.
Also, as much as possible, we will derive model independent results such that $a$ is not a metric
degree of freedom being some generic time-dependent function describing the background. The technical details for this metric
including connections and the Einstein equations can be found in
appendix~\ref{A.FRW}.

Let us start by defining the three-dimensional geometrical lengths corresponding to this metric.
This length is defined through the internal product corresponding to the three-dimensional
(spatial) hyper-surface for some fixed time $t$. Specifically the three-dimensional
intrinsic metric for the FLRW metric~(\ref{g_FRW}) is $^{(3)}g_{ij}=-g_{ij}=a^2\delta_{ij}$
and the coordinates in this spatial hyper-surface coincide with the four-dimensional spatial
coordinates, $^{(3)}x^i=x^i$. We use the suffix '$^{(3)}$' to distinguish between the
four-dimensional and three-dimensional metrics and coordinates. Then, in the three-dimensional
hyper-surface, the square of the geometric length $l_{\mathrm{geom}}$ is
\be
\delta_{ij}l_{\mathrm{geom}}^il_{\mathrm{geom}}^j=\,^{(3)}g_{ij}\,^{(3)}x^i\,^{(3)}x^k=a^2\delta_{ij}x^ix^j\ .
\ee
As for time lengths, neglecting the relativity effects ($\gamma=dt/d\tau\approx 1$),
the proper time and the coordinate time coincide. Then, given this approximation,
any measurement of coordinate time coincides with measurements of proper time.
Hence we obtain the following definitions for geometrical lengths, geometrical velocities
and geometrical accelerations
\be
\ba{rcl}
l_{\mathrm{geom}}^i&=&a\,x^i\ ,\\[6mm]
v_{\mathrm{geom}}^i&=&a\,\dot{x}^i\ ,\\[6mm]
a_{\mathrm{geom}}^i&=&a\,\ddot{x}^i\ .
\ea
\lb{l_geom_FRW}
\ee
The geometrical length is the physical observable length for the FLRW metric~(\ref{g_FRW}). We do not
measure coordinates lengths, once a metric is considered we measure these geometrical lengths.
However, as we will analyze in detail, neither the geometrical velocity, neither the geometrical
acceleration coincide with the physical measurable velocities and accelerations. Instead these will be given, as expect in the framework of General Relativity, by the geometrical covariant velocities and
acceleration.

To derive the classical limit, for which an Euclidean metric $\delta_{ij}$
can be used, the definition of three-dimensional coordinates (in the spatial hyper-surface
for each fixed $t$) is not enough. We need to consider the above definitions of geometrical spatial
lengths that play the role of the classical local Euclidean coordinates in the Newtonian limit.
Here local means that this construction is only valid in the neighborhood of the
space-time event we are considering. Specifically we deal only with
time-independent metric components, hence our results will be valid in a neighborhood of
some time $t_0$. More generally when the spatial curvature is significant we can,
at most, define a Euclidean metric in the neighborhood of some spatial point $x_0^i$
(the most common example is the earth surface, being a sphere, only locally it is approximately flat). 
In the following we use the notation $l^i$ for the exact expressions and $\bar{l}^i$
for the respective classical quantity (an approximation to $l_{\mathrm{mass}}^i$) that represents
the Newtonian limit.

To further proceed without specifying a cosmological model that would fix the expression
for the scale factor $a$, let us consider a series expansion of $a^p$ in
the neighborhood of the reference time $t_0$ 
\be
\ba{rcl}
a^p&=&\displaystyle a_0^p\left(1+p\frac{\dot{a}_0}{a_0}(t-t_0)+\frac{p}{2}\left(\frac{\ddot{a}_0}{a_0}+(p-1)\left(\frac{\dot{a}_0}{a_0}\right)^2\right)(t-t_0)^2\right)+ O\left(t^3\right)\\[5mm]
&\approx&\displaystyle a_0^p\left(1+p H_0(t-t_0)-\frac{p}{2}\left(q_0-p+1\right)\,H_0^2(t-t_0)^2\right)+ O\left((H_0\,(t-t_0))^3\right)\ .
\ea
\lb{a_exp}
\ee
Here $a_0=a(t_0)$, $\dot{a}_0=\dot{a}(t_0)$ and $\ddot{a}_0=\ddot{a}(t_0)$ are the scale factor 
and its time derivatives evaluated at the time $t=t_0$. In the last approximation we have replaced these
derivatives by the Hubble rate $H_0=\dot{a}_0/a_0$ and deceleration
parameter $q_0=-\ddot{a}_0/(a_0\,H_0^2)$ also evaluated at $t=t_0$. Taking a phenomenological approach
we assume that the values for these parameters are known at time $t_0$ and it is not necessary to
actually specify a cosmological evolutionary model to study and estimate the effects of the
expansion today. This is a valid approach as long as we are dealing with short time scales
$t\ll 1/H\sim 10^{17}\ s\sim 10^{10}\ year$ and spatial scales $x^i\ll l_{H_0}=c/H_0\sim 10^{26}\ m$.

We can now compute the time evolution for a coordinate point $x_0^i$ with null coordinate
velocity $\dot{x}^i$. The evolution of the respective geometrical spatial
lengths $l_{\mathrm{0}}^i$ in the neighborhood of $t=t_0$ is given by
\be
l_{\mathrm{0}}^i(t)=a(t)x_0^i\approx \bar{l}^i_{\mathrm{0}}(t)=l_{\mathrm{0}(0)}^i+l_{\mathrm{0}(0)}^iH_0(t-t_0)-\frac{1}{2}q_0l_{\mathrm{0}(0)}^iH_0^2(t-t_0)^2+O(t^3)\ ,
\lb{l_i_geom_FRW}
\ee
where $l_{\mathrm{0}(0)}^i=a_0x_0^i$ is the spatial length at $t=t_0$, we have considered the series
expansions~(\ref{a_exp}) for $p=1$ and explicitly wrote the time dependence of the several
quantities. The underscore indexes '$0(0)$' represent respectively the fixed coordinate index ($x_0^i$)
and the evaluation time $t=t_0$. Then, as expected, the length will increase with a \textit{velocity}
given by the product of Hubble rate $H_0$ by the initial length $l_{\mathrm{0}(0)}^i$ and an
\textit{acceleration} given by the product of the deceleration rate $-q_0H_0^2$ by the same initial
length $l_{\mathrm{0}(0)}^i(t_0)$.

To compute the geodesic path for massive particles let us first compute and solve the
equations of motion for a particle travelling with some given initial velocity $\dot{x}^i_0$
as observed by a static observer at the origin of the coordinate frame,
$x^i_{\mathrm{obs}}=(0,0,0)$. We assume that the only physical interactions are due the
expanding background (given by the metric) and static is meant with respect to that background
such that the observers coordinate velocity is $\dot{x}^i_{\mathrm{obs}}=(0,0,0)$.
In this section we use Cartesian coordinates. The connections are given in
equation~(\ref{A.FRW.conn_cart}) and the generic expression for the relativistic
factor $\gamma=dt/d\tau$ is given by equation~(\ref{A.gamma}), which for the metric~(\ref{g_FRW})
is
\be
\gamma=\frac{dt}{d\tau}=\left(1-\frac{a^2\delta_{ij}\dot{x}^i\dot{x}^j}{c^2}\right)^{-\frac{1}{2}}\ .
\lb{gamma_FRW}
\ee
The equations of motion for the spatial coordinates are given by the usual geodesic equation
as expressed in~(\ref{A.geo}). For the FLRW metric~(\ref{g_FRW}) we obtain
\be
\ddot{x}^i=-2c\Gamma^{i}_{\ 0j}\dot{x}^j-\gamma^{-1}\dot{\gamma}\,\dot{x}^i\approx-2\frac{\dot{a}}{a}\,\dot{x}^i\approx -2\dot{x}^i\,H_0\ .
\lb{ddx_FRW}
\ee
Here we have approximated the time derivatives of the scale factor by
the Hubble rate evaluated at the reference time $t=t_0$, $H_0=\dot{a}_0/a_0$~(\ref{H}), and we have taken the non-relativistic limit considering $\gamma\approx 1$ and $\dot{\gamma}\approx 0$. This limit
corresponds to non-relativistic speeds $\dot{x}^i\ll c$ and short-scale spatial distances
such that the spatial coordinates are well bellow the characteristic length scale of the system $x^i\ll l_H$~(\ref{H_length}). An estimative for the relativistic corrections is
\be
\ba{rcl}
\ddot{x}^i_{\mathrm{rel}}&=&\displaystyle-\gamma^{-1}\dot{\gamma}\,\dot{x}^i=+\left(\frac{\dot{a}a\delta_{jk}\,\dot{x}^j\dot{x}^k}{c^2}+\frac{a^2\delta_{jk}\,\ddot{x}^j\dot{x}^k}{c^2}\right)\gamma^{2}\,\dot{x}^i\\[5mm]
                         &=&\displaystyle- \frac{\dot{a}}{a}\,\frac{\delta_{jk}\,\dot{x}^j\dot{x}^k}{c^2}\,\gamma^{2}\,a^2\,\dot{x}^i
\approx-\left(1+\frac{a^2\dot{x}^2}{c^2}\right)\frac{a^2\dot{x}^2}{c^2}\,H\,\dot{x}^i+O\left(\frac{\dot{x}^6H}{c^6}\dot{x}^i\right)\ ,
\ea
\lb{ddx_rel_FRW}
\ee
where in the last line we have replaced the second time derivative $\ddot{x}^i$ by the equations of
motion~(\ref{ddx_FRW}), expanded the relativistic factor in first order in $\dot{x}^2/c^2$ and
used the short hand notation $x^2=\delta_{ij}x^ix^j$. For non-relativistic coordinate velocities ($\dot{x}^2\ll c^2$) and small-scale spatial coordinates ($x^i\ll l_{H_0}$) this correction is negligible
with respect to the above expression~(\ref{ddx_FRW}).

Integrating the equations of motion~(\ref{ddx_FRW}) we obtain
\be
\frac{\ddot{x}^i}{x^i}=-2\frac{\dot{a}}{a}\ \Rightarrow\ \dot{x}^i=\left(\frac{a_0}{a}\right)^2\,\dot{x}_0^i\ .
\lb{1st_int_FRW}
\ee
Here we have set the integration constant to $a_0^2\dot{x}_0^i=a(t_0)\dot{x}^i(t_0)$ which
corresponds to the equation evaluated at the reference time $t=t_0$. Given the
expansion~(\ref{a_exp}) with $p=-2$ we can further integrate this equation
in from the reference time $t_0$ to some generic time $t$ obtaining the coordinate path
for massive particles
\be
\ba{rcl}
\displaystyle \int_{x_0}^x dx'&=&\displaystyle \dot{x}^i_0a_0^2\int_{t_0}^t \frac{dt'}{a^2}\approx\dot{x}^i_0 \int_{t_0}^t dt'\left[1-2H_0(t'-t_0)+\left(q_0+3\right)\,H^2(t'-t_0)^2\right]\ ,\\[6mm]
\displaystyle x^i_{\mathrm{mass}}(t)&\approx&\displaystyle x^i_0+\dot{x}^i_0\left((t-t_0)-H_0(t-t_0)^2+\frac{1}{3}\left(q_0+3\right)\,H^2_0(t-t_0)^3\right)+O\left(H^3_0t^4\right)\ .
\ea
\lb{x_FRW}
\ee
We use the notation $x'$ and $t'$ to distinguish between the integration variable and integration
limits $x$ and $t$. We note that the coordinate acceleration~(\ref{ddx_FRW}) cannot be directly
interpreted as the Newtonian acceleration and, as we have already discussed, similarly
to the coordinates and coordinate velocities, cannot be directly measured, we measure the
geometrical acceleration~(\ref{l_geom_FRW}). We will return to this discussion later.

Finally we can define the geometrical lengths $l_{\mathrm{mass}}^i$ for a massive particle
describing its trajectory. In order to do so we take the solutions for the coordinate equations
of motion~(\ref{x_FRW}) and the series expansion for the scale factor~(\ref{a_exp})
obtaining
\be
\ba{rcl}
\displaystyle l_{\mathrm{mass}}^i&=&\displaystyle a\,x^i_{\mathrm{mass}}\\[5mm]
\bar{l}_{\mathrm{mass}}^i&=&\displaystyle \bar{l}_{0(0)}^i+\bar{v}_c^i\,(t-t_0)+\frac{1}{2}\bar{a}_c(t-t_0)^2+O\left(t^3\right)\ ,\\[6mm]
\bar{l}_{0(0)}^i&=&a_0x_0^i=l^i_{\mathrm{geom}}(t_0)\ ,\\[5mm]
\bar{v}_c^i&=& a_0\,\left(\dot{x}_0^i+x_0^i\,H_0\right)=v^i_{\mathrm{geom}}(t_0)+l^i_{\mathrm{geom}}(t_0)\,H_0 ,\\[5mm]
\bar{a}_c^i&=&-q_0\,l^i_{\mathrm{geom}}(t_0)\,H_0^2\ .\\[5mm]
\ea
\lb{v_a_classic_FRW}
\ee
We interpret $\bar{v}_c^i=\dot{\bar{l}^i}_{\mathrm{mass}}$ as the classical velocity and
$\bar{a}_c^i=\ddot{\bar{l}^i}_{\mathrm{mass}}$ as the classical acceleration in the
neighborhood of $t_0$. Only the spatial coordinate lengths at the origin $\bar{l}_{0(0)}^i$
coincide with the geometrical spatial lengths evaluated at the initial time $t_0$. Neither the $\bar{v}_c^i$
nor $\bar{a}_c^i$ coincide with the respective geometrical quantities, instead are given by the sum
of the background velocity and acceleration~(\ref{l_i_geom_FRW}) with the particle geometrical quantities~(\ref{l_geom_FRW}). For the example just presented
only the geometrical velocity $v^i_{\mathrm{geom}}(t_0)=a_0\dot{x}^i_0$ is present, more generally
we can consider as well some external acceleration such that $\bar{a}_c^i=a^i_{\mathrm{geom}}(t_0)-q_0\,l_{0(0)}\,H_0^2$ with $a^i_{\mathrm{geom}}(t_0)=a_0\ddot{x}_{\mathrm{ext}}^i(t_0)$.

This result was actually expected, we recall that in General Relativity the physical observable velocity
and acceleration are given (at all times) by the projection of the respective covariant quantities.
Hence let us show how these quantities are related with the classical quantities~(\ref{v_a_classic_FRW})
computed above. The covariant velocity and acceleration are computed as usual by considering the parallel
transport derivatives
\be
\ba{rclcl}
\displaystyle \mathrm{v}^i&=&\displaystyle\frac{D x^i}{D\tau}=\gamma\frac{D x^i}{Dt}=\gamma\left(\dot{x}^i+c\,\Gamma^{i}_{\ 0j}x^j\right)\\[6mm]
&=&\displaystyle\gamma\left(\dot{x}^i+\frac{\dot{a}}{a}x^i\right)=\gamma\left(\left(\frac{a_0}{a}\right)^2\,\dot{x}_0^i+\frac{\dot{a}}{a}x^i\right)&\approx&\displaystyle\left(\frac{a_0}{a}\right)^2\,\dot{x}_0^i+\frac{\dot{a}}{a}\,x^i\ ,\\[6mm]
\displaystyle \mathrm{a}^i&=&\displaystyle\frac{D^2 x^i}{D\tau^2}=\gamma\frac{D \mathrm{v}^i}{Dt}=\gamma\frac{d\mathrm{v}^i}{dt}+c\,\gamma\,\Gamma^{i}_{\ 0j}\mathrm{v}^j=\\[6mm]
&=&\displaystyle\gamma^2\left(\ddot{x}^i+2\frac{\dot{a}}{a}\dot{x}^i+\frac{\ddot{a}}{a}x^i+\gamma^{-1}\dot{\gamma}\left(\dot{x}^i+\frac{\dot{a}}{a}x^i\right)\right)\\[6mm]
&=&\displaystyle\gamma^2\left(\frac{\ddot{a}}{a}x^i+\gamma^{-1}\dot{\gamma}\frac{\dot{a}}{a}x^i\right)&\approx&\displaystyle\frac{\ddot{a}}{a}\,x^i\ .
\ea
\lb{va_FRW}
\ee
In the final expressions we used the equations of motion for $\ddot{x}^i$~(\ref{ddx_FRW}) and
$\dot{x}^i$~(\ref{1st_int_FRW}) and in the final approximations we have neglect the
relativistic corrections setting $\gamma\approx 1$ and $\dot{\gamma}\approx 0$.
Considering the projection for these vectors into the spatial hyper-surface at fixed time $t_0$,
we obtain the expressions for the respective classical quantities valid in a neighborhood
of $t_0$
\be
\ba{rclcrcl}
^{(3)}g_{ij}\,^{(3)}\mathrm{v}^i\,^{(3)}\mathrm{v}^i&=&a^2\delta_ {ij}\mathrm{v}^i\mathrm{v}^j&\Rightarrow&\bar{v}_c^i&=&a_0\,\mathrm{v}^i(t_0)\ ,\\[5mm]
^{(3)}g_{ij}\,^{(3)}\mathrm{v}^i\,^{(3)}\mathrm{v}^i&=&a^2\delta_ {ij}\mathrm{a}^i\mathrm{a}^j&\Rightarrow&\bar{a}_c^i&=&a_0\,\mathrm{a}^i(t_0)\ .
\ea
\lb{va_v_a_FRW}
\ee
In this way we confirm that at the reference time $t=t_0$ the classical spatial
Euclidean coordinates (the spatial lengths) $\bar{l}^i_0$, velocities $\bar{v}_c^i$ and
accelerations $\bar{a}_c^i$ do correspond to the spatial coordinates $x^i$, covariant velocity
$\mathrm{v}^i$ and covariant acceleration $\mathrm{a}^i$ projected to the spatial hyper-surface
at $t=t_0$. Taking in consideration the corrections due to the expanding background, the usual laws
of classical mechanics are valid in a neighborhood of $t_0$ as expressed by~(\ref{v_a_classic_FRW}). However it is important to stress that these values are
only valid as an approximation. More generally neither of these quantities are constant over large
time scales ($t\sim 1/H_0$) neither large spatial scales ($x\sim l_{H_0}$) for which it is necessary to properly take in consideration the dynamics
of cosmological evolution of the universe~\cite{Inflation,review}. From the equations of
motion~(\ref{ddx_FRW}) we obtain the expression for the coordinate
velocity to be $\dot{x}^i=a_0^2\,\dot{x}^i_0/a^2$ and, as we have just discussed,
due to the internal product being given in terms of the intrinsic metric, we obtain
the usual expression for the coordinate peculiar velocities
$v_{\mathrm{geom}}^i=a\dot{x}^i=a_0\,(a_0\dot{x}^i_0)/a=a_0\,v_{\mathrm{geom}}(t_0)^i/a$ observed
experimentally between astrophysical objects~\cite{Peebles}. These are
due to the expansion of space being exact within General Relativity and have no classical analogy.
In addition, either assuming that we can estimate and subtract the coordinate velocity $a\dot{x}^i$,
or in the limit of relatively large distances and small coordinate velocities $\dot{x}^i\ll \bar{l}_{\mathrm{geom}}^i$ or simply by having a large enough statistical sample for which the
coordinate velocities average to zero, from the expression for the covariant velocity~(\ref{va_FRW}),
we retrieved the Hubble law for lengths (or distances). The relation between the length $l_{\mathrm{geom}}^i=ax^i$ and its time variation $\delta l_{\mathrm{geom}}^i/\delta t$ is
\be
\frac{\delta l_{\mathrm{geom}}^i}{\delta t}=a\,\frac{Dx^i}{Dt}=a\mathrm{v}^i\approx\frac{\dot{a}}{a}\,l_{\mathrm{geom}}^i\ .
\lb{H_law_FRW}
\ee
The approximation in the last equality is only due to have neglecting the peculiar velocities
and the variation $\delta$, as stated in the first equality, considers both the parallel
transport of spatial coordinate and the projection to three dimensions. We stress that
this relation is exact within the framework of General Relativity being valid at all times,
only for short time scales and spatial scales the rate of the scale factor can be approximated by the constant
Hubble rate $H_0=\dot{a}_0/a_0$. Taken the classical limit,
as given in~(\ref{v_a_classic_FRW}), this relation is (exactly) valid only at the time $t=t_0$,
for which $\dot{\bar{l}^i}(t=t_0)=\bar{v}^i_c\approx H(t=t_0) \bar{l}_0$. 
We remark that this relation is coordinate dependent in the sense that, for a different
coordinate choice, the respective spatial lengths will be expanding at different rates. 
We will discuss this issue later on.

So far we have not discussed any measurement procedure. Let us discuss two common methods
for measuring distances. Specifically we address two kind of light measurements,
range measurements where the travelling time is measured (usually the proper time
using atomic clocks, see for instance section 16.4 of~\cite{Gravitation} for further details) and the distance
computed by considering the speed of light, and red-shift measurements which allow
to directly measure the variation of the wave length due to the expansion of the light
path from the emission point to the reception point. The later was the original technique used when
setting the Hubble law.

Let us start by defining the line element for light travelling in the background metric~(\ref{g_FRW})
along some coordinate direction $x$. A light-like trajectory corresponds to the shortest possible path
$ds^2=0$, then from~(\ref{g_FRW}) we obtain the infinitesimal relation and respective differential equation
\be
c\,dt=a\,\delta_{ij}e^idx^j\ \Rightarrow \frac{dx^i}{dt}=\frac{c}{a}\,e^i\ ,
\lb{eq_diff_light_FRW}
\ee
where $e^i$ stands for a unity vector with respect to the Euclidean metric $\delta_{ij}e^ie^j=1$
and is introduced to define the propagation direction of radiation. This equation can be reduce
to one single direction $x$, for which the internal product $x.e$ reduces to an overal $\pm$ sign $\delta_{ij}e^ix^j=\pm |x|$. Again, to keep the derivation model independent, let us consider
the expansion for the scale factor~(\ref{a_exp}) with $p=-1$ valid in the neighborhood
of $t_0$ and integrate the expression from the reference time $t_0$ to some genric time $t$
obtaining the light coordinate at time $t$ 
\be
\ba{rcl}
x_{\mathrm{light}}^i(t)&\approx&\displaystyle x_0^i+\frac{c\,e^i}{a_0}\,\int_{t_0}^t\,dt'\,\left[1-H_0\,(t'-t_0)+\frac{1}{2}\left(q_0-2\right)H_0^2\,(t'-t_0)^2+O\left((H_0^3(t-t_0)^3\right)\right]\\[6mm]
&=&\displaystyle x_0^i+\frac{c\,e^i}{a_0}\,\left((t-t_0)-\frac{1}{2}H_0\,(t-t_0)^2+\frac{1}{6}\left(q_0-2\right)H_0^2\,(t-t_0)^3+O\left(H_0^4(t-t_0)^4\right)\right)\ .
\ea
\lb{x_light_FRW}
\ee
We recall that we are considering an observer at the origin of the coordinate frame and our aim
is to measure the geometrical distance between the point $x_0^i$ (at $t=t_0$) and the observer
for which the coordinate is null (at some time $t$). Hence the coordinate distance for light is
\be
\Delta x_{\mathrm{light}}^i=x_0^i-x_{\mathrm{light}}^i(t)\ .
\ee
From this coordinate length we can infer the respective light length corresponding
to the point $x_0$ from where the light signal was emitted
\be
\bar{l}_{\mathrm{light}}^i=a\Delta x_{\mathrm{light}}^i= -\left(c\,e^i\right)(t-t_0)-\frac{1}{2}\left(c\,e^iH_0\right)(t-t_0)^2+O(t^3)\ .
\lb{l_0_FRW}
\ee
We note that due to light being travelling to the origin of the coordinate frame,
$e^i$ has the opposite sign of the value of the coordinate $x^i_{0}$ which is correctly
offset by the overall minus sign.

The above light length does not correspond to the geometrical spatial
lengths~(\ref{l_i_geom_FRW}), this is easily explained by noting
that the spatial expansion affects the geometrical lengths by a factor $+xH_0$~(\ref{l_i_geom_FRW})
proportional to the coordinate $x$, hence for light travelling from $x^i=x_0^i$ to $x^i=0$
we obtain an average effect such that $H_0\,(x_0^i+0)/2$.
Also we note that the geometrical lengths are time dependent such that
they do not match at the time of emission and time of reception of the light signal,
respectively are $\bar{l}_{0}^i(t=t_0)\bar{l}_{0(0)}^i=a_0x^i_0$ and $\bar{l}_{0}^i(t)=a(t)x^i_0$.
The above equation estimates the lengths at the time the light is received ($t$ in the
previous equation) since it is already corrected (projected) by the scale factor $a$.

Therefore when computing distances by considering the light travel time $\Delta t$
there will be a correction to the usual classical light length $\bar{l}_{0(0)}=a_0x^i_0=c\Delta t$
both for the geometrical length corresponding to the light signal emission~(\ref{l_0_FRW}) as
well as for the estimative of the geometrical length distances to the massive
particle~(\ref{v_a_classic_FRW}). Specifically, to first order in $H_0$, we obtain
\be
\ba{rcl}
\displaystyle \bar{l}_{0(0)}&=&\displaystyle c\Delta t\ ,\\[6mm]
\displaystyle \bar{l}_{\mathrm{light}}&=&\displaystyle\sqrt{\delta_{ij}\bar{l}_{\mathrm{light}}^i\bar{l}_{\mathrm{light}}^j}\\[6mm]
&=&\displaystyle \bar{l}_{0(0)}+\frac{1}{2}\,H_0\,\frac{\bar{l}_{0(0)}^2}{c}+O\left(H_0^2\,\bar{l}_{(0)}^2/c\right)\ ,\\[6mm]
\displaystyle \bar{l}_{\mathrm{mass}}&=&\displaystyle\sqrt{\delta_{ij}\bar{l}_{\mathrm{mass}}^i\bar{l}_{\mathrm{mass}}^j}\\[6mm]
&=&\displaystyle +\bar{l}_{0(0)}+H_0\,\frac{\bar{l}_{0(0)}^2}{c}+O\left(H_0^2\,\bar{l}_{(0)}^2/c\right)\ .
\ea
\lb{Range_corr_FRW}
\ee
Due to the very small value of the ratio $H_0/c=1/l_{H_0}\approx 2.7\times 10^{-27}\ m^{-1}$,
these corrections are for most purposes negligible being usually below experimental accuracy.
In figure~\ref{fig.1} are represented the evolution of the several spatial lengths
during a light range measurement.
\fig{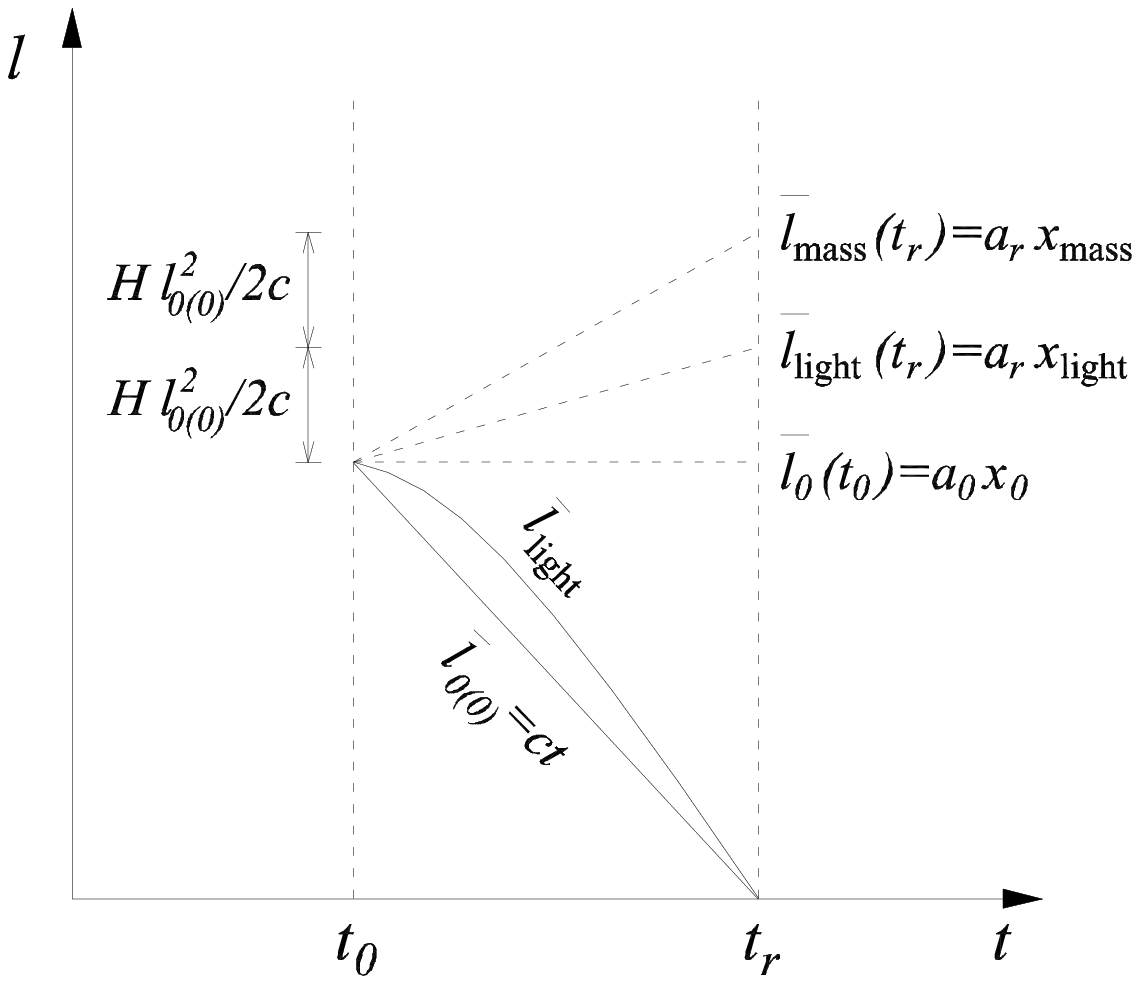}{80mm}{Evolution of spatial geometrical lengths $l$ in a range measurement
corresponding to the spatial coordinates $x_0^i$ (fixed at time $t_0$), light coordinates $x^i_{\mathrm{light}}$ and
massive particle coordinates $x^i_{\mathrm{mass}}$ from the emission time $t_e$ to the reception
time $t_r$ as given by equation~(\ref{Range_corr_FRW}).}{fig.1}

For the case for which the signal is emitted by the observer,
reflected by the massive particle and received back at the origin, the problem
can be symmetrized with respect to the time at which the reflection takes place,
both the geometrical lengths given in equations~(\ref{v_a_classic_FRW}) and~(\ref{l_0_FRW}) are symmetric
with respect to time inversion. Then, taking the reflection time
to be $t_0$, we have that the emission time $t_e$ and reception time $t_r$ are spaced apart from $t_0$
by the same time length, $t_0-t_e=t_r-t_0$. Hence, for fixed Hubble rate $H_0=H(t_0)$, the measured
distance is the same in both directions and we get twice the travelling time corresponding to twice
the light length $\bar{l}_0^i$.

We proceed to briefly compute the cosmological red-shift. The most straight forward way to do
so is to consider the 4-momentum for travelling radiation with frequency $\omega/2\pi$ and
wave number $k^i$, $k^\mu=(\omega/c,k^i)$. Its square is null $g_{\mu\nu}k^\mu k^\nu=\omega^2/c^2-a^2k^2=0$ (light is massless),
such that by energy-momentum conservation we obtain the following optical dispersion
relation and respective solution for the wave number
\be
\omega=c\,a(t)\sqrt{\delta_{ij}k^ik^j}\ \Rightarrow\ k^i=\frac{k_0^i}{a}\ .
\ee
The direction of propagation for the light is given by the wave vector $k_0^i$
which corresponds to the evaluation of $k^i$ at some reference time $t_0$.
Alternatively, with out specifying a reference time it can be set to
$k_0^i=\omega\,e^i/c $ where $e^i$ is a unit vector ($\delta_{ij}e^ie^j=1$)
which defines the direction of light propagation. By further noting that the wave
vector is inversely proportional to the wave length $\lambda$
$k^i=2\pi\, e^i/\lambda$, we obtain the expression
\be
\lambda(t)=\frac{a(t)\,\omega}{c}=\frac{a(t)}{a_0}\lambda_0\ ,
\lb{lambda}
\ee
where $\lambda_0$ and $a_0$ stand for the wave length and scale factor values at some time $t_0$. 
Then the usual cosmological red-shift $z$ is straight forwardly defined as
\be
z=\frac{\lambda_r-\lambda_e}{\lambda_e}=\frac{\frac{a_r}{a_e}\lambda_e-\lambda_e}{\lambda_e}=\frac{a_r\lambda_e-a_e\lambda_e}{a_e\lambda_e}=\frac{a_r}{a_e}-1\ ,
\ee
where the indexes $e$ and $r$ stand for emission and reception of the radiation.
Considering the series expansion of the right-hand side of this equation we obtain,
to first order, the red-shift Hubble law
\be
\ba{rcl}
z&\approx&\displaystyle H_0 \Delta t-\frac{1}{2}q_0H_0^2\, \Delta t^2+O(H_0^3\Delta t^3)\\[6mm]
&=&\displaystyle H_0\, \frac{\bar{l}_{0(0)}}{c}-\frac{1}{2}q_0H_0^2\, \frac{ \bar{l}_{0(0)}^2}{c^2}+O\left(H_0^3\,\bar{l}^3/c^3\right)\ .
\ea
\lb{z_FRW}
\ee
In the last equality we have replaced the time interval $\Delta t=t_r-t_e$ by the classical
length interval $\bar{l}_{0(0)}=c \Delta t$~(\ref{Range_corr_FRW}). The corrections to the classical
lengths are mostly relevant only for observation of very far objects and, otherwise, are usually
below the experimental accuracy.

Although these deductions are helpful, both as a reference and to introduce several techniques,
we note that the historical path has been the reversed. The Hubble law was set
experimentally~\cite{Hubble} implying that the universe was expanding which led to the derivation
of the FLRW metric~(\ref{g_FRW})~\cite{RW}.

\subsection{Working with Expanding Coordinates\lb{sec.review_exp}}

We have just shown that for non-expanding coordinates $x$ corresponding to metric~(\ref{g_FRW}),
the three-dimensional coordinates $x^i$ do not expand, for a fixed coordinate point $\dot{x}_0^i=0$
the coordinate velocity~(\ref{1st_int_FRW}) and acceleration~(\ref{ddx_rel_FRW}) are null. The
physical measurable quantities correspond to the respective geometrical quantities and do expand
with the background, the geometrical velocity is corrected by the Hubble velocity due to expansion,
$+H\,x^i$, and the geometrical acceleration by the acceleration of the background,
$-q\,H^2\,x^i$~(\ref{va_v_a_FRW}). Classically it is only possible to define local corrections
that match these background effects at some fixed time~(\ref{v_a_classic_FRW}). These results raise
the question whether new four-dimensional coordinates can be defined that coincide with these
geometrical lengths which would have a more intuitive physical meaning. We also expect this
new coordinate choice to simplify the technical details and interpretation of the several results.

The new set of three-dimensional spatial coordinates $x_1$ that we are considering in this section is
\be
x^i=\frac{x_1^i}{a}\ \ \ ,\ \ \ r=\frac{r_1}{a}\ .
\lb{coord_transf}
\ee
We name these expanding coordinates due to their value being changing over
time with the scale factor $a$. Here the underscore index '$1$' stands
for a new coordinate set following the convention describe in appendix~\ref{A.defs}.
For spherical coordinates this coordinate transformation is equivalent to
transformation~(\ref{A.r1.r}). We note that the transformation~(\ref{coord_transf}) corresponds to
the (simple) three-dimensional coordinate transformation. The respective generalized coordinate
transformation $\Lambda^\mu_{\ \nu}$ for four-vectors and other tensorial contravariant quantities
is computed from the linear transformation for the infinitesimal forms $dx^0$ and $dx^i$
\be
dx^0=dx_1^0\ \ ,\ \ dx^i=\frac{dx_1^i}{a}-\frac{\dot{a}}{a}\,\frac{x_1^i}{a}\,\frac{dx_1^0}{c}\ ,
\lb{form_transf}
\ee
such that we obtain
\be
\ba{rclcrcl}
\Lambda^0_{\ 0}&=&\displaystyle\frac{\partial x^0}{\partial x_1^0}=1&,&\Lambda^0_{\ i}&=&\displaystyle\frac{\partial x^0}{\partial x_1^i}=0\\[6mm]
\Lambda^i_{\ 0}&=&\displaystyle\frac{\partial x^i}{\partial x_1^0}=-\frac{\dot{a}}{a}\,\frac{x_1^i}{a}\,\frac{1}{c}&,&\Lambda^i_{\ j}&=&\displaystyle\frac{\partial x^i}{\partial x_1^j}=-\delta^i_{\ j}\,\frac{\dot{a}}{a}\,\frac{x_1^j}{a}\,\frac{1}{c}\ ,
\ea
\lb{gen_coord_transf}
\ee
where, in the last component, no summation over the repeated indexes $j$ is implied. 
As usual the transformation $\tilde{\Lambda}$ for covariant tensors is obtained by inverting the transformation matrix $\Lambda$.

In particular, the metric for these new coordinates, is obtained either from the infinitesimal
forms transformation~(\ref{form_transf}) applied to the infinitesimal length square $ds^2$ or
the transformation $\tilde{\Lambda}$ applied to the covariant metric components $g_{\mu\nu}$.
Specifically it is
\be
\ba{rcl}
\displaystyle ds^2&=&\displaystyle c^2\left(1-\left(\frac{\dot{a}\,r_1}{a\,c}\right)^2\right)dt^2+2c\left(\frac{\dot{a}\,r}{a\,c}\right)\,dr\,dt-dr_1^2-r_1^2\left(d\theta^2+\sin^2\theta d\varphi^2\right)\ ,\\[6mm]
&=&\displaystyle c^2\left(1-\left(\frac{\dot{a}}{a\,c}\right)^2\delta_{ij}x^i_1x^j_1\right)dt^2+2c\left(\frac{\dot{a}}{a\,c}\delta_{ij}x_1^j\right)\,dx_1^i\,dt-\delta_{ij}dx_1^idx_1^j\ .
\ea
\lb{g_FRW_1}
\ee
Where we are omitting the index '$1$' on the $x^0$ coordinates, the coordinate transformation
is trivial for the time components ($x^0=x_1^0$).

When dealing with this metric, the main technical difference with respect to the previous subsection,
is due to the non-null metric component $g_{0i}$. In particular when considering a projection to
the three-dimensional hyper-surface it is necessary to decompose the four-dimensional
infinitesimal square length into a time length minus
a \textit{proper} spatial length, specifically $ds^2=c^2dt^2-(dx_1-\dot{a}/a\,dt)^2$.
It is this spatial length that corresponds to the three dimensional measurable length
\be
l_{\mathrm{geom}.1}^2=-\left(g_{ij}x_1^ix_1^j-g_{0j}x_1^0x_1^j\right)=\delta_{ij}x_1^ix_1^j\left(1-\frac{\dot{a}}{a}t\right)^2\ .
\ee
However this result is not enough to proceed. It is desirable to have a projection
definition that can be applied to generic vectors, including velocity and acceleration
as was considered in the previous section. There are several ways to define this projection
(see for example section~21.4 of~\cite{Gravitation}). We use the construction originally
suggested in~\cite{Israel} such that the definitions for the intrinsic metric $^{(3)}g_{ij}$
and the three-dimensional spatial projection for any vector is given in
equation~(\ref{Israel_proj}) of the appendix~\ref{A.defs}.
For the specific case of the spatial coordinates we obtain
\be
^{(3)}g_{ij}=\delta_{ij}\ \ ,\ \ ^{(3)}x_1^i= x_1^i\left(1-\frac{\dot{a}}{a}t\right)\ .
\lb{proj_FRW_1}
\ee
We also note that the off diagonal term $g_{0i}$ usually
represents a frame velocity $v^i_{\mathrm{frame}}$ and, when this velocity is constant,
a diagonal metric can be considered by shifting the space coordinates
$\hat{x}^i_1=x^i_1-v^i_{\mathrm{frame}}t$. In this case this velocity is both time and space
dependent, nevertheless we can expand this term recursively and, by successive shifts of
the spatial coordinates $x_1^i$, to obtain the same definition for the three-dimensional
coordinates to any desired order (in powers of $H_0$).

For this coordinate choice, neglecting Special Relativity effects,
the proper time coincides with the coordinate time, $\tau=t$. Then,
given the above projection~(\ref{proj_FRW_1}), the geometrical lengths,
geometrical velocities and geometrical accelerations are defined as
\be
\ba{rcl}
l_{\mathrm{geom}.1}^i&=&\displaystyle x_1^i\left(1-\frac{\dot{a}}{a}t\right)\ ,\\[6mm]
v_{\mathrm{geom}.1}^i&=&\displaystyle \dot{x}_1^i-\frac{\dot{a}}{a}\frac{x_1^i}{c}\dot{x}_1^0=\dot{x}_1^i-\frac{\dot{a}}{a}x_1^i\ ,\\[6mm]
a_{\mathrm{geom}.1}^i&=&\displaystyle \ddot{x}_1^i-\frac{\dot{a}}{a}\frac{x_1^i}{c}\ddot{x}_1^0=\ddot{x}_1^i\ .
\ea
\lb{l_geom_FRW_1}
\ee
In the last equalities of the last line we explicitly expressed the derivatives of the
covariant coordinate $x_1^0=ct$ by the respective time coordinate expressions $\dot{x}_1^0=1$
and $\ddot{x}_1^0=0$. 
As before, in order to obtain the classical limit, it is necessary to expand the projection
factor $1-\dot{a}/a\,t$. However we note that in a neighborhood of a generic time $t_0$,
it will contain factors depending linearly in this value, specifically we obtain
\be
\ba{rcl}
\displaystyle 1-\frac{\dot{a}}{a}\,t&=&\displaystyle\left(1-\frac{\dot{a}_0}{a_0}t_0\right)+\left(-\frac{\dot{a}_0}{a_0}-\frac{\ddot{a}_0}{a_0}t_0+\left(\frac{\dot{a}_0}{a_0}\right)^2t_0\right)(t-t_0)\\[6mm]
& &\displaystyle+\left(-\frac{\ddot{a}_0}{a_0}+\left(\frac{\dot{a}_0}{a_0}\right)^2-\frac{\dot{\ddot{a}}}{2}\,\frac{t_0}{2}+\frac{\ddot{a}_0}{a_0}\,\frac{\dot{a}_0}{a_0}\,\frac{3t_0}{2}-\left(\frac{\dot{a}_0}{a_0}\right)^3t_0\right)(t-t_0)^2+O(t^3) \ .
\ea
\ee
The dependence on $t_0$ is not desirable since this value is, in principle, arbitrary.
Depending on the physical system we are considering it can be either the time we start a
specific experiment or one century ago. There are several ways to deal with this issue,
the simpler is to shift the time coordinate by $t_0$ and perform the expansion in the
neighborhood of $t=0$. We note that none of the metric components change under this coordinate
transformation due to the time dependence being encoded in the dimensionless scale
factor $a$. Then the expansion in the neighborhood of $t=0$ is
\be
1-\frac{\dot{a}}{a}\,t\approx 1-H_0t+(q_0+1)H_0^2t^2+O((H_0t)^3)\ .
\lb{proj_exp_FRW_1}
\ee
Then the geometrical length $l^i_{1.0}$ for a spatial coordinate point $x^i_{1.0}$
(fixed to the background) is
\be
\ba{rcl}
l_{1.0}^i(t)&=&\displaystyle x_{1.0}^i\left(1-\frac{\dot{a}}{a}\,t\right)\approx\bar{l}^i_{1.\mathrm{0}}(t)\\[6mm]
&=&\displaystyle x_{1.0(0)}^i-x_{1.0(0)}^iH_0\,t+\left(q_0+1\right)x_{1.0(0)}^iH_0^2\,t^2+O((H_0t)^3)\\[6mm] &=&\displaystyle l_{1.0(0)}^i+v_{1.0(0)}^iH_0\,t+\left(q_0+1\right)l_{\mathrm{0}(0)}^iH_0^2\,t^2+O((H_0t)^3)\ .
\ea
\lb{l_i_geom_FRW_1}
\ee
The two underscore indexes '$_{1.0}$' in the coordinate stands for the coordinate
label ($x_1$ in this case) and the evaluation time ($t_0$), $x_{1.0}=x_{1}(t=t_0)$,
while in the geometrical lengths we use the notation '$_{1.0(0)}$' to specify
the coordinate choice, the type of length (as in the previous section '$0$' for a
coordinate fixed to the background, 'mass' for a massive particle and 'light' for
radiation) and the evaluation time (inside the brackets). In the last expression we have replaced
the coordinate quantities by the respective geometrical quantities as defined in~(\ref{l_geom_FRW_1})
with $\dot{x}_{1.0(0)}^i=0$, the length at the initial time coincides with the geometrical
length evaluated at $t=0$, $l_{1.0(0)}^i=l_{\mathrm{geom}.1}^i(t=0)=x^i_{1.0}$
and the velocity corresponds to the geometrical velocity also evaluated at
$t=0$, $v_{1.0(0)}^i=v_{\mathrm{geom}.1}^i(t=0)=-x^i_{1.0}\,H_0$.
As for the geometrical acceleration it corresponds to the spatial expansion
acceleration for this coordinate choice.

The remaining of the calculations closely follow the ones of the previous section.
However we repeat them here. The main reason to do so is to clearly identify
the relation between the several results derived in both expanding coordinates $x$ and
non-expanding coordinates $x_1$ allowing to identify its relation with our world
physical measurable as well as to identify the technical advantages of each choice.

The relativistic factor $\gamma$ and its derivative with respect to the
time coordinate $\dot{\gamma}$ are
\be
\ba{rcl}
\displaystyle\gamma=\frac{dt}{d\tau}&=&\displaystyle\left(1-\left(\frac{\dot{a}}{a}\right)^2\frac{\delta_{ij}x^i_1x^j_1}{c^2}+2\left(\frac{\dot{a}}{a}\right)\frac{\delta_{ij}x_1^i\dot{x_1}^j}{c^2}-\frac{\delta_{ij}\dot{x}_1^i\dot{x_1}^j}{c^2}\right)^{-\frac{1}{2}}\\[6mm]
 &=&\displaystyle \left(1-\frac{1}{c^2}\left(\dot{x}_1-\frac{\dot{a}}{a}x_1\right)^2\right)^{-\frac{1}{2}}\\[6mm]
 &\approx&\displaystyle 1+\frac{1}{2c^2}\left(\dot{x}_1-\frac{\dot{a}}{a}x_1\right)^2\ ,\\[6mm]
\dot{\gamma}&=&\displaystyle\frac{\gamma^2}{c^2}\left[\ddot{x}_1-\left(\frac{\ddot{a}}{a}-\left(\frac{\dot{a}}{a}\right)^2\right)x_1-\frac{\dot{a}}{a}\dot{x}_1\right].\left[\dot{x}_1-\frac{\dot{a}}{a}x_1\right]\ .
\ea
\lb{gamma_FRW_1}
\ee
The generic spatial coordinate equations of motion are given by the geodesic equation~(\ref{A.geo})
of appendix~\ref{A.defs} and the connections for this metric in Cartesian coordinates are given in
equation~(\ref{A.FRW_1.conn_cart}) appendix~\ref{A.FRW}. Hence we obtain the following equations
of motion
\be
\ba{rcl}
\ddot{x}^i_1&=&\displaystyle-c^2\Gamma^i_{\ 00}-2c\Gamma^i_{\ 0j}\dot{x}^j-\Gamma^i_{\ jk}\dot{x}^j\dot{x}^k-\gamma^{-1}\dot{\gamma}\dot{x}^i\\[6mm]
&=&\displaystyle+\frac{\ddot{a}}{a}x_1^i-\frac{x_1^i}{c^2}\left(\frac{\dot{a}}{a}\right)^2\left(\dot{x}_1-\frac{\dot{a}}{a}x_1\right)^2-\gamma^{-1}\dot{\gamma}\dot{x}^i\\[6mm]
&\approx&\displaystyle+\frac{\ddot{a}}{a}x_1^i \approx -q_0\,H_0^2\,x_1^i\ .
\ea
\lb{ddx_FRW_1}
\ee
In the last equality of the last line we replaced the second derivative of the scale factor
by the deceleration parameter value at the reference time $t_0$, $q_0$~(\ref{H_q0}) and from the
second to the third line we took the approximation for non-relativistic limit, coordinate velocities
much smaller than the speed of light, $\dot{x}^i\ll c$, and spatial distances much lower than
the characteristic length (Hubble length) $x^i\ll c/H_0=l_{H_0}$~(\ref{H_length})
such that $\gamma\approx 1$, $\dot{\gamma}\approx 0$ and $(\dot{x}-H_0x)^2/c^2\ll q_0$.
The relativistic corrections corresponding to the second and third terms of the second line
can be evaluated by replacing the expression for $\ddot{x}$ in the time derivative for
$\dot{\gamma}$~(\ref{gamma_FRW_1})
\be
\ba{rcl}
\ddot{x}^i_{1.\mathrm{rel}}&=&\displaystyle\frac{1}{c^2}\frac{\dot{a}}{a}\left(\dot{x}_1-\frac{\dot{a}}{a}x_1\right)^2\left(\dot{x}_1^i-\frac{\dot{a}}{a}x_1^i\right)+\frac{\dot{x}^i}{2c^4}\frac{\dot{a}}{a}\left(\dot{x}_1-\frac{\dot{a}}{a}x_1\right)^4\ ,\\[6mm]
&\approx&\displaystyle H_0\left(\frac{\dot{x}_1}{c}-\frac{x_1}{l_{H_0}}\right)^2\left(\dot{x}_1^i-H_0x_1^i\right)+\frac{\dot{x}^i}{2}H_0\left(\frac{\dot{x}_1}{c}-\frac{x_1}{l_{H_0}}\right)^4+O\left(H_0^5\,t\right)\ .
\ea
\lb{ddx_rel_FRW_1}
\ee
Again, in the last line, we have replaced the time derivatives of the scale factor by the
Hubble rate $H_0$ and length $l_{H_0}$ at the reference time $t_0=0$.

The equations of motion in the non-relativistic limit~(\ref{ddx_FRW_1}) can be integrated directly such that we obtain
\be
\dot{x}_1^ia-x_1^i\dot{a}=\dot{x}^i_{1.0}a_0-x^i_{1.0}\dot{a}_0\ ,
\lb{1st_int_FRW_1}
\ee
where the integration constant was set to the value of the functions evaluated
at the reference time $t_0=0$.
As expected, under a coordinate transformation $x_1^i=ax$, this equation coincides with the
differential equation~(\ref{1st_int_FRW}) of the previous subsection. Hence taking in consideration
the solution for $x$~(\ref{x_FRW}) and the series expansion in the neighborhood of $t_0$
for the scale factor~(\ref{a_exp}) with $p=1$, we obtain the
geometrical length~(\ref{v_a_classic_FRW}) for the coordinates $x$ which coincides with the
coordinate $x_1^i$ solutions for the path of massive particles
\be
x_{1.\mathrm{mass}}^i=a\,x^i_{\mathrm{mass}}=\bar{l}^i_{\mathrm{mass}}=x^i_{1.0}+\dot{x}^i_{1.0}\,t-\frac{1}{2}q_0x^i_{1.0}H_0^2\,t^2+O(t^3)\ .
\lb{x_FRW_1}
\ee
In the last equality we have used the coordinate transformation to define the following relations
$x^i_{1.0}=a_0x^i_0$ and $\dot{x}^i_{1.0}=a_0(\dot{x}^i_{0}+x^i_{0}H_0)$.

Considering the expansion of the projection factor~(\ref{proj_exp_FRW_1}) and the solution
for $x_1$~(\ref{x_FRW_1}) we obtain the geometrical length for massive particles
\be
\ba{rcl}
\bar{l}_{\mathrm{mass}.1}^i&=&\displaystyle \bar{l}_{1.0}^i+\bar{v}_{1.c}^i\,t+\frac{1}{2}\bar{a}_{1.c}t^2+O\left(t^3\right)\ ,\\[6mm]
\bar{l}_{1.0}^i&=&\displaystyle x_{1.0}^i=l^i_{\mathrm{geom}.1}(0)\ ,\\[5mm]
\bar{v}_{1.c}^i&=&\displaystyle \dot{x}_{1.0}^i-x_{1.0}^i\,H_0=v^i_{\mathrm{geom}.1}(0) ,\\[6mm]
\bar{a}_{1.c}^i&=&\displaystyle q_0\,x_{1.0}^i\,H_0^2=q_0\,l^i_{\mathrm{geom}.1}(0)\,H_0^2\ .
\ea
\lb{v_a_classic_FRW_1}
\ee
We note that both the velocity and acceleration corrections due to the time dependence of the
background have the opposite sign of the ones obtained for non-expanding
coordinates~(\ref{v_a_classic_FRW}). This means that for this coordinate choice, maintaining
the same experimental values of the Hubble rate $H_0$ and deceleration parameter $q_0$, space would be
in decelerated deflation.

Again we note that these classical quantities correspond to the respective
covariant quantities evaluated at the initial time $t=0$.
The covariant velocity components are
\be
\ba{rcl}
\mathrm{v}_1^i&=&\displaystyle\frac{Dx_1^i}{D\tau}=\gamma\left(\dot{x}_1^i+c^2\Gamma^i_{\ 00}t+c\Gamma^i_{\ 0j}x_1^j\right)\\[6mm]
&=&\displaystyle \gamma\left(\dot{x}_1^i-\frac{\ddot{a}}{a}x_1^it-x_1^i\,\frac{x_1^2}{c^2}\left(\frac{\dot{a}}{a}\right)^3\left(1-\frac{\dot{a}}{a}t\right)\right)\\[6mm]
&=&\displaystyle \gamma\left(\frac{\dot{a}}{a}x_1^i+\frac{\dot{x}_{1.0}^ia_0-x_{1.0}^i\dot{a}_0}{a}-\frac{\ddot{a}}{a}x^it-x_1^i\,\frac{x_1^2}{c^2}\left(\frac{\dot{a}}{a}\right)^3\left(1-\frac{\dot{a}}{a}t\right)\right)\\[5mm]
&\approx&\displaystyle \frac{\dot{a}}{a}x_1^i+\frac{\dot{x}_{1.0}^ia_0-x_{1.0}^i\dot{a}_0}{a}\ ,\\[6mm]
\mathrm{v}^0&=&\displaystyle\frac{Dx^0}{D\tau}=c\gamma\left(1+c^2\Gamma^0_{\ 00}t+c\Gamma^0_{\ 0i}x_1^i\right)\\[6mm]
&=&\displaystyle c\gamma\left(1-\frac{x_1^2}{c^2}\left(\frac{\dot{a}}{a}\right)^2\left(1-\frac{\dot{a}}{a}t\right)\right)\\[6mm]
&\approx&c\ ,
\ea
\lb{v_cov_FRW_1}
\ee
and the covariant acceleration is
\be
\ba{rcl}
\mathrm{a}_1^i&=&\displaystyle\frac{D^2x_1^i}{D\tau^2}=\gamma\left(\mathrm{\dot{v}}^i+c\Gamma^i_ {\ 0j}\,\mathrm{v}^j+c\Gamma^i_{\ 00}\,\mathrm{v}^0\right)\\[6mm]
&\approx&\displaystyle-\frac{\ddot{a}}{a}x^i\ ,\\[6mm]
\mathrm{a}^0&=&\displaystyle\frac{D^2x_1^0}{D\tau^2}=\gamma\left(\mathrm{\dot{v}}^0+c\Gamma^0_ {\ 0i}\,\mathrm{v}^i+c\Gamma^0_{\ 00}\,\mathrm{v}^0\right)\\[6mm]
&\approx& 0\ ,
\ea
\lb{a_cov_FRW_1}
\ee
where we have used the equations of motion for $\ddot{x}_1^i$~(\ref{ddx_FRW_1}) and $\dot{x}^i$~(\ref{1st_int_FRW_1}) to replace the time derivatives in these equations
and, in the last approximations, we have considered the limit of non-relativistic velocities and
distances keeping only the lower order terms in the Hubble rate. By considering the projection of these quantities into the three-dimensional
hyper-surface at some fixed space-time point $(t_0,x_{1.0})$ it is straight forward to obtain
the respective classical quantities~(\ref{v_a_classic_FRW_1}).

The Hubble law corresponding to the geometrical lengths
$l_{\mathrm{geom}.1}^i=x_1^i(1-(\dot{a}/a)t)$~(\ref{l_geom_FRW_1})
is directly derived from the expression for the covariant velocity~(\ref{v_cov_FRW_1})
by projecting it into the three-dimensional hyper-surface
\be
\ba{rcl}
\displaystyle\frac{\delta l_{\mathrm{geom}.1}^i}{\delta t}&=&\displaystyle\left(\frac{Dx_1^i}{d\tau}-\frac{\dot{a}}{a}\,\frac{x^i}{c}\,\frac{Dx_1^0}{d\tau}\right)=\left(\mathrm{v}_1^i-\frac{\dot{a}}{a}\,\frac{x^i}{c}\,\mathrm{v}^0\right)\\[6mm]
&\approx&\displaystyle-\frac{a_0}{a}\dot{x}_{1.0}-\frac{\dot{a}_0}{a}\,x_{1.0}-\frac{\ddot{a}}{a}\,x_1\,t\approx \frac{a_0}{a}v_{\mathrm{geom}.1}^i(0)+q\,H^2\,l_{\mathrm{geom}.1}^i\,t+O(H^3t^2)\ .
\ea
\lb{H_law_FRW_1}
\ee
In the last approximation we have use the definition of the geometrical velocity and length
to replace the $\dot{x}_{1.0}^i$ and $x_1^i$ such that we obtain an extra term of order $H^3$.
The first term correspond to the peculiar velocities discussed in the previous
section, a particle moving in the absence of any other interactions will decrease its
geometrical velocity, $v_{\mathrm{geom}.1}=v_{\mathrm{geom}.1(0)}(a_0/a)$.
However, for expanding coordinates $x_1$, the linear term in the Hubble rate $H=\dot{a}/a$
is absent, instead we have a second order term proportional to the deceleration parameter.

Let us complete our analysis by computing the light path lengths and radiation frequency-shift.
From the metric~(\ref{g_FRW_1}) we define the respective line element corresponding to $ds^2=0$
\be
c^2dt^2=\delta_{ij}\left(dx_1^i-\frac{\dot{a}}{a}\,x^i_1\,d\,t\right)\left(dx_1^j-\frac{\dot{a}}{a}\,x^j_1\,d\,t\right)\ \Rightarrow\ \frac{dx^i_1}{dt}=\left(ce^i+\frac{\dot{a}}{a\,c}x^i\right)\ ,
\ee
where $e^i$ is a unit vector defining the propagation direction for radiation. Under a transformation
of coordinates $x_1=ax$ this equation coincides with the respective equation for non-expanding
coordinates $x$ derived in the previous section~(\ref{eq_diff_light_FRW}). Then, multiplying the solution~(\ref{x_light_FRW}), by the scale factor $a$ we obtain the solution for the above equation
in expanding coordinates $x_1^i$
\be
x^i_{\mathrm{light}.1}=x^i_{1.0}+\left(c e^i+x^i_{1.0}H_0\right)t+\frac{1}{2}\left(c e^iH_0-q_0x^i_{1.0}H_0^2\right)t^2+O(t^3)\ .
\lb{x_light_FRW_1}
\ee
Where we have also transformed the integration constants taking in consideration the
coordinate transformation $x_1=ax$. We note that for light travelling from $x^i_{1.0}$
to the origin of the coordinate frame, similirarly to the case of the previous section~(\ref{l_0_FRW}),
we obtain the coordinate distance $\Delta x^i_{\mathrm{light}.1}(t)=x^i_{1.0}-x^i_{\mathrm{light}.1}(t)$,
such that considering the projection into the three-dimensional hyper-surface~(\ref{proj_exp_FRW_1})
we obtain the geometrical light length
\be
\bar{l}^i_{\mathrm{light}.1}=\Delta x^i_{\mathrm{light}.1}\left(1-\frac{\dot{a}}{a}t\right)\approx-ce^i\,t+\frac{1}{2}ce^i\,H_0\,t^2+O(H_0^3t^3)\ .
\ee

Then, given some travelling time $\Delta t$ for radiation emitted from $x_{1.0}$ 
and received by the observer at the origin of the reference frame, we obtain the following
estimative for the light length $\bar{l}_{\mathrm{light}.1}$ and the geometrical length
$\bar{l}_{\mathrm{geom}.1}$ expressed in terms of the fixed background classical length
estimative $\bar{l}_{0(0)}$
\be
\ba{rcl}
\displaystyle \bar{l}_{0(0)}&=&\displaystyle c\Delta t\ ,\\[6mm]
\displaystyle \bar{l}_{\mathrm{light}.1}&=&\displaystyle\sqrt{\delta_{ij}\bar{l}_{\mathrm{light}.1}^i\bar{l}_{\mathrm{light}.1}^j}\\[6mm]
&=&\displaystyle \bar{l}_{0(0)}-\frac{1}{2}\,H_0\,\frac{\bar{l}_{0(0)}^2}{c}+O\left(H_0^2\,\bar{l}_{(0)}^2/c\right)\ ,\\[6mm]
\displaystyle \bar{l}_{\mathrm{geom}.1}&=&\displaystyle\sqrt{\delta_{ij}\bar{l}_{\mathrm{geom}}^i\bar{l}_{\mathrm{geom}}^j}\\[6mm]
&=&\displaystyle +\bar{l}_{0(0)}-H_0\,\frac{\bar{l}_{0(0)}^2}{c}+O\left(H_0^2\,\bar{l}_{(0)}^2/c\right)\ .
\ea
\lb{Range_corr_FRW_1}
\ee
We note that the corrections obtained have the inverse sign of the ones obtained for expanding
coordinates $x$ in the previous section.

In order to compute the frequency-shift let us consider the square of four-momentum vector $k^\nu=(\omega/c,k^i)$,
$g_{\mu\nu}k_1^\mu k_1^\nu=\omega^2/c^2-(k_1-(\dot{a}/a)x_1\omega/c^2)^2=0$, obtaining
the following dispersion relation and respective solution for the wave number vector $k^i$
\be
\omega^2=\delta_{ij}\left(ck_1^i-\frac{\dot{a}}{a}\frac{x_1^i\,\omega}{c}\right)\left(ck_1^j-\frac{\dot{a}}{a}\frac{x_1^j\,\omega}{c}\right)\ \Rightarrow\ k_1^i=\frac{\omega}{c}\left(e^i-\frac{\dot{a}}{a}\,x_1^i\right)\ .
\ee
When deriving the solution it is relevant to note that, for light travelling to the origin
of the coordinate frame, $k^i_1$ and $x_1^i$ have opposite signs such that $x_1.e=-|x_1|$ and $\omega=c|k_1|+(\dot{a}/a)|x_1|\omega/c$. Given an observed wavelength $\lambda_{1.r}$ at the
reception (the origin $x_1=0$), the same radiation wavelength at some generic point $x_1$ is
\be
\lambda_1=\frac{2\pi}{|k_1|}=\frac{\lambda_{1.r}}{1-\frac{\dot{a}}{a}\frac{|x_1|}{c}}\ .
\lb{lambda_1}
\ee
Then, considering radiation from a source at a coordinate length $x_{1.e}$ corresponding to
a received wavelength $\lambda_{1.r}=\lambda_{1.e}(1-(\dot{a}/a)|x_{1.e}|/c)$ we obtain
the following frequency-shift
\be
z_1=\frac{\lambda_{1.r}-\lambda_{1.e}}{\lambda_{1.e}}=\frac{\lambda_{1.r}}{\lambda_{1.e}}-1=-\frac{\dot{a}}{a}\frac{|x_{1.e}|}{c}\approx-H_0\frac{\bar{l}_{0(0)}}{c}+\left(q_0+1\right)H_0^2\frac{\bar{l}^2_{0(0)}}{c^2}\ .
\lb{z_FRW_1}
\ee
Hence, maintaining the numerical values of $H_0$~(\ref{H}) and $q_0$~(\ref{H_q0}) we obtain a blue-shift.

The results obtained in this section are consistent among themselves, when working with expanding
coordinates $x_1$ (and assuming the experimentally numerical values for $H_0$~(\ref{H}) and $q_0$~(\ref{H_q0})) we obtain that space is deflating at a decelerated rate~(\ref{l_i_geom_FRW_1})
such that light trajectories are decreased with respect to the fixed background approximation~(\ref{Range_corr_FRW_1}) and radiation is blue-shifted~(\ref{z_FRW_1}).
These results are the opposite to the results obtained for non-expanding coordinates $x$
in the previous section and do not correspond to our physical measurable quantities, instead are
it is due to the coordinate choices which must be properly accounted for each coordinate system.
Next we discuss this topic comparing the results obtained for both coordinates choices $x$ and
$x_1$ and its relation to physical measurable as well as the technical advantage of using
expanding coordinates $x_1$.

\subsection{Coordinate Choice(s) and Physical Measurable\lb{sec.review_map}}

From the derivations in the two previous sections we have just concluded that the specific
form of the Hubble law depends on the coordinate choice. As the Newton law does~(\ref{Newton_inverse}),
or any other physical law. Furthermore, in the framework of General Relativity, instead of the
usual spatial coordinates, velocity and acceleration, the measurable quantities are the respective
geometrical quantities (projected to the spatial three-dimensional hyper-surface) which we derived
for two distinct coordinate choices (non-expanding
coordinates $x$ in section~\ref{sec.review_nexp} and expanding coordinates in
section~\ref{sec.review_exp}). As expected, for a given value of the Hubble rate $H$ and
deceleration parameter $q$, in each coordinate system we obtain distinct time evolutions for
these quantities. Recalling that the coordinate choice
is a representation of our physical measurements, we can safely conclude that the
coordinate system for which the (General Relativity) geometrical quantities directly
corresponds to our world are non-expanding coordinates $x$.
The Hubble law, both for distances~(\ref{H_law_FRW}) as for the red-shift~(\ref{z_FRW})
has been verified experimentally and is directly described only by the coordinates $x$
with metric given by~(\ref{g_FRW}). This conclusion was actually expected since,
as already mentioned, the metric~(\ref{g_FRW}) was originally derived to describe
the experimental Hubble law~\cite{RW}. 

Employing any other coordinate system predicts different expansion rates as we have
just exemplified, for expanding coordinates $x_1$ radiation travelling towards an
observer is blue-shifted~(\ref{z_FRW_1}) which implies that the light path expressed in
this coordinate system is shrinking. This result is actually consistent, as long as we
recall that we have to correctly map the results obtained back to the coordinate system
which corresponds to our measurable quantities. Depending of the quantities being discussed
this mapping is achieved either by inverting the simple coordinate transformation~(\ref{coord_transf})
(for scalar and non-covariant spatial coordinate quantities), inverting the generalized coordinate
transformation~(\ref{gen_coord_transf}) -for covariant tensors- or account for the
three-dimensional projections (for three-dimensional projected quantities). In particular
we note that the Special Relativity corrections obtained through the factor $\gamma$ (a scalar)
are directly mapped by applying the coordinate transformation~(\ref{coord_transf}) to the respective
expressions. It is also important
to remark that this discussion is equivalent to a frame choice.
In General Relativity there is no preferred frame and one of the cornerstones of the
the theory is to be frame independent. There is no inconsistency between our discussion
and this statement, simply a given observer lives at a specific frame and,
although the physical laws can be inferred for any other frame, the physical measurements
are necessarily obtained at the observer frame.

Depending in the specific problem being solved, it is common to employ coordinate systems
that do not correspond to our physical measurable. In particular working with expanding
coordinates $x_1$ has a clear technical advantage. These coordinates exactly match
the measurable three-dimensional projections for the non-expanding coordinates, $x_1=ax=l$, hence
they correspond to the physical lengths. Consistently the equations of motion for $x_1$~(\ref{ddx_FRW_1})
(the Newton law for an expanding background) match the respective equations for the
geometrical quantities for the coordinates $x$. This result can be explicitly checked by directly
considering the non-covariant coordinate transformation $x=x_1/a$~(\ref{coord_transf}) in the
respective equation of motion~(\ref{ddx_FRW}). Consistently also the classical solutions for the $x$
coordinates for the massive particle lengths $l^i_{\mathrm{mass}}$~(\ref{v_a_classic_FRW}) coincides
with the classical solutions for the massive particle coordinates $x_{1.\mathrm{mass}}^i$~(\ref{x_FRW_1}),
to show it is enough to consider the coordinate map $x=x_1/a$~(\ref{coord_transf}) including the
the mapping of the integration constants such that $x_{1.0}^i=a_0x_0^i$ and
$\dot{x}_{1.0}^i=a_0\dot{x}_0^i+\dot{a}_0x_0^i$. Consistently with out discussion, to compute the physical distances squared
from the $x_1$ coordinates (as well as for velocities and accelerations) we must employ the
Euclidean metric $\delta_{ij}$, we recall that for non-expanding coordinates the distance is
defined as $g_{ij}x^ix^j=a^2\delta_{ij}x^ix^j=\delta_{ij}x^i_1x^j_1$.
We note that the time coordinate is the same for
both coordinate systems, this ensures that quantities with time-derivatives (such as velocities
and accelerations) as well as the Special Relativity corrections can be directly mapped by
the spatial coordinate map.

Hence the conclusion is that we can perform most of our calculations
using expanding coordinates $x_1$ such that the physical measurable results 
are given directly by the coordinate expressions. However it is important to note
that this mapping is only valid for physical relations (laws) that are expressed in terms of
the coordinates $x_1$ and its derivatives. For non three-dimensional coordinate quantities
it is necessary to account for the specific map between them and the coordinate
(or geometrical quantities) transformations in each distinct coordinate system. Specifically
the red-shift $z$~(\ref{z_FRW}) for non-expanding coordinates $x$ and the blue-shift
$z_1$~(\ref{z_FRW_1}) for expanding coordinates $x_1$ are written in terms of geometrical
wavelengths $\lambda$~(\ref{lambda}) and $\lambda_1$~(\ref{lambda_1}), not of coordinate wavelengths.
Hence to derive the map between these quantities it is necessary to consider a transformation
that accounts for the geometrical wavelengths definitions, hence $\lambda=(\lambda_1/a)(1-(\dot{a}/a)(|x_1|/c))$. As for covariant quantities such as
the covariant velocity and acceleration the mapping is given by the generalized coordinate
transformations~(\ref{gen_coord_transf}).

In the literature the expanding coordinates $x_1$ are often referred to as
\textit{physical coordinates}, this is understood as that the coordinates $x_1$ correspond
to the physical observable lengths in our specific frame. Given our previous discussion
we should not forget that the physical metric corresponds to non-expanding coordinates
$x$~(\ref{g_FRW}). The generic map between the physical quantities corresponding to the coordinates $x^i_1$, coordinate velocities $\dot{x}_1^i$ and coordinate accelerations $\ddot{x}^i_1$ and the
respective measurable geometrical quantities for the $x$ coordinates as given in~(\ref{l_geom_FRW}),
the geometrical lengths $l_{\mathrm{geom}}^i$, geometrical velocities $v_{\mathrm{geom}}^i$ and
geometrical accelerations $a_{\mathrm{geom}}^i$ is
\be
\left\{\ba{rcrclcl}
l_{\mathrm{phys}}&=&x_1^i&=&ax^i&=&l_{\mathrm{geom}}\\[5mm]
v_{\mathrm{phys}}&=&\dot{x}_1^i&=&\displaystyle a\dot{x}^i+\frac{\dot{a}}{a}(ax^i)&=&\displaystyle v_{\mathrm{geom}}+\frac{\dot{a}}{a}l_{\mathrm{geom}}\\[5mm]
a_{\mathrm{phys}}&=&\ddot{x}_1^i&=&\displaystyle a\ddot{x}^i+\frac{\ddot{a}}{a}(ax^i)+2\frac{\dot{a}}{a}(a\dot{x}^i)&=&\displaystyle a_{\mathrm{geom}}+\frac{\ddot{a}}{a}l_{\mathrm{geom}}+2\frac{\dot{a}}{a}v_{\mathrm{geom}}
\ea\right.
\lb{lengths_x1_phys}
\ee

Once we have settled our \textit{world} coordinate system (meaning the observer frame) and concluded
what the measurable physical quantities are, we can compute the estimative for the expansion effects in
some physical systems for which the spatial scale is much lower than the Hubble length $x^i\ll l_{H_0}$~(\ref{H_length}). Based in the results for the $x^i$ coordinates obtained for light range
measurements~(\ref{Range_corr_FRW}) and the red-shift measurements~(\ref{z_FRW}) we present
some examples in table~\ref{table.range_corrections} which are clearly below experimental accuracy
(the accuracy of Cassini spacecraft measurements~\cite{Cassini} are still one order of magnitude
below these effects).
\begin{table}[ht]
\tiny\begin{center}
\begin{tabular}{ccccccc}
System & Av. Distance & Range Correction& $z$ & Method Used & Accuracy & Reference \\
 & ($meter$) & $\Delta\bar{l}_{(1)}$~(\ref{Range_corr_FRW}) ($meter$) & $z$~(\ref{z_FRW}) &  &  &  \\
\hline\hline\\[-2mm]
Earth-Sun & $1.50\times 10^{11}$ &$+1.71\times 10^{-4}$ & $1.14\times 10^{-15}$ & -- & -- & --\\[2mm]
 Earth-Moon & $3.85\times 10^{8}$ &$+1.13\times 10^{-9}$ &  $2.93\times 10^{-18}$ & Laser Range & $\Delta l\approx 2.00\times 10^{-2}\, m$&\cite{Lunar_01}\\[2mm]
Pioneer 10-Earth & $>2.85\times 10^{12}$ &$>6.17\times 10^{-2}$ & $>2.17\times 10^{-14}$  & Doppler shift &$\Delta\nu/\nu_0\approx 10^{-12}$&\cite{Pioneer}\\[2mm]
 & $<6.60\times 10^{12}$ &$<3.31\times 10^{-1}$ & $<5.02\times 10^{-14}$  & & & \\[2mm]
Cassini-Earth & $1.11\times 10^{11}$ &$9.44\times 10^{-3}$ & $8.47\times 10^{-15}$  & Doppler shift &$\Delta\nu/\nu_0\approx 10^{-14}$&\cite{Cassini}\\[2mm]\hline
\end{tabular}
\caption{\it \small Range corrections and cosmological red-shifts for Earth-Sun and some current experiments.
Not enough accuracy is today achieved to detect expansion effects in these range and Doppler shift measurements. \lb{table.range_corrections}}
\end{center}
\end{table}

In the introduction we mentioned the existence of the Hubble horizon at the Hubble lenght $l_H$
(when considering the observer at the origin of the coordinate frame). A specific proof for the
existence of an horizon can be obtained from the infinitesimal proper distance for light,
$ds^2=0=c^2-(\dot{r}_1-H\,r_1)^2$. From this expression we obtain the following
expression for the speed of light in an expanding background
\be
\dot{r}_1=c-H\,r_1\ \ \Leftrightarrow\ \ \dot{r}_1=0\ \mathrm{for}\ r_1=\frac{c}{H}=l_H\ .
\ee
Hence, as expected from an horizon, travelling radiation freezes at $r_1=l_H$ such that
information cannot be exchange between the two causally disconnected regions, $r_1<l_H$ and $r_1>l_H$.
So far we have not discussed the large spatial scale Newton law for the expanding background.
The higher order corrections on the Hubble constant to the classical acceleration that we have neglected
for the short spatial scale approximation~(\ref{ddx_FRW_1}), will become relevant near the
cosmological horizon $r_1\sim l_H$. For spherical expanding coordinates $r_1$ we obtain
\be
\ddot{r}_1\approx-c^2\Gamma^1_{\ 00}\approx-qH^2\,r_1-\frac{H^4}{c^2}\,r_1^3\ .
\lb{F_Newton_FRW_1}
\ee
Assuming $q<0$, the specific radial distance for which the net effect is null is
\be
r_1=\sqrt{-q}\,l_H\ .
\lb{r_FRW_1}
\ee
Taking in consideration the estimative for the deceleration parameter $q_0=-0.589$~(\ref{H_q0})
we obtain that for small distances the first term is dominant and expansion has a
repulsive effect (massive bodies will increase its relative distances) while
for large distances the second term is dominant and expansion has an attractive effect
(massive bodies will decrease its relative distances). Also we can obtain a rough
estimative for the above value of the radial coordinate in today's universe to be
$r_1\approx 0.77\,l_H$. However we note that an observer at the present time located
at the origin of the coordinate frame is observing distant events ($r_1\sim l_H$) that
occurred a long time ago $t\sim 1/H$. This is due to the time that information takes to
travel from the event location to the observer location.
Let us note that we have considered for most purposes the
reference time $t_0$ for which the Hubble rate $H_0$ and the deceleration parameter $q_0$
are evaluated to coincide with the initial time at which we measure the initial distances and speeds. Usually the expansion
parameters $H$ and $q$ are known exactly only for the time $t_{\mathrm{exp}}$ corresponding to their
experimental measurement. Taking a phenomenological approach we can consider a time series expansion
for these parameters such that their value at the time $t_0$ is related to their value at the time
$t_{\mathrm{exp}}$ by the relations expressed in equation~(\ref{A.H_0_exp}) of appendix~\ref{A.cosm_H}.
For relatively small scales these corrections are well within the measurement error. As an example let
us take the galaxy NGC~U2885 which, for an observer at the solar system, is at a distance of
$r_1\approx 118\ Mpc=3.63\times 10^{24}\ m$ (about $r_1\sim l_H/10$). Then any event observed
will have happen at a time $t_0$ in the past such that
$t_0-t_{\mathrm{exp}}=-r_1/c=-1.21\times 10^{16}\ s$.
Taking this negative time span we obtain the following relations
\be
H_0=H_{\mathrm{exp}}\left(1+0.011\right)\ \ ,\ \ q_0=q_{\mathrm{exp}}\left(1+0.027\times(s_0+0.178)\right)\ .
\ee
These correction are not negligible however they are within the experimental error. For the current
value of the Hubble rate~(\ref{H}) the error is $\pm 0.04$. As for the deceleration parameter an
estimative for the error is $\pm 0.15$ (we will briefly address how to derive this value in the next
section~\ref{sec.T_qo}) and the author is unaware of any estimative for the value
of the parameter $s$, here we assume that it is of the same or lower order of magnitude
than $q$. Hence in the remaining of this work, when computing quantitative estimative,
we will assume today's values for the Hubble rate $H_0$ and deceleration parameter $q_0$
given in~(\ref{H}) and~(\ref{H_q0}). We note that although the universe is in accelerated
expansion ($H_0>0$ and $q_0<0$) the Hubble rate $H=\dot{a}/a$ decreases with time. This is not
inconsistent, it means that the derivative of the scale factor $\dot{a}$ increases slower
with time than the scale factor $a$. As for the time evolution of the deceleration parameter
$q$ it decreases or increases depending on the value of the parameter $s_0$. Today's value
for $q_0$ is negative, hence for $s_0>2q_0+1$ it increases
(its absolute value decreases), for $s_0=2q_0+1$ is constant and for $s_0<2q_0+1$ it
decreases (its absolute value increases). A more rigorous definition of the time evolution
of these parameters for large spatial scales requires to consider a cosmological evolutionary
model to account for the large time scale evolution of these quantities.

Next we will discuss the stress-energy tensor and shortly address how the parameter $q$
can be estimated from the equation of state for today's universe.

\subsection{The Stress-Energy Tensor and the Deceleration Parameter $q$\lb{sec.T_qo}}

In cosmology it is usually assumed that the background corresponds to a homogeneous and isotropic
perfect fluid. Hence the diagonal components of the stress-energy tensor corresponding to the
gravitational pressure are equal to each other and the mass-energy density is
isotropic. This tensor is related to the Einstein tensor by the usual relation
$T_{\mu\nu}=c^4/(8\pi\,G)\, G_{\mu\nu}$ which, for the FLRW metric in expanding
coordinates~(\ref{g_FRW_1}), is given in equation~(\ref{A.FRW_1.EEQ}) of appendix~\ref{A.FRW}.
Spatial isotropy necessarily implies spherical symmetry (the opposite statement is not valid,
spherical symmetry does not imply spatial isotropy), hence for technical simplification we
will use spherical coordinates in the following discussion. To work in a orthogonal locally
Lorentz frame (the observer locally flat frame) let us consider a tetrad
$e_{\mu}$ relating the coordinate metric with the Minkowski metric,
$g_{\mu\nu}=\eta_{\hat{\mu}\hat{\nu}}e^{\hat{\mu}}_{\ \mu}e^{\hat{\nu}}_{\ \nu}$,
where the hatted indexes represent the indexes of the flat space-time.
Specifically for the FLRW metric the non-null tetrad components are
\be
e^{\hat{0}}_{\ 0}=e^{\hat{1}}_{\ 1}=1\ \ ,\ \ e^{\hat{1}}_{\ 0}=-\frac{\dot{a}\,r_1}{a\,c}\ \ ,\ \ e^{\hat{2}}_{\ 2}=r\ \ ,\ \ e^{\hat{3}}_{\ 3}=r\sin\theta\ .
\ee 
The stress-energy tensor components in the Cartan frame and in the coordinate frame
are related by this tetrad as
$T_{\hat{\mu}\hat{\nu}}=e^{\ \mu}_{\hat{\mu}}e^{\ \nu}_{\hat{\nu}}T_{\mu\nu}$,
where the inverse components of the tetrad $e^{\ \mu}_{\hat{\mu}}$
are straight forwardly computed by raising and lowering the indexes with the respective
Minkowski and coordinate metrics. Considering the stress-energy tensor for a background
perfect fluid in the commoving frame as given in equation~(\ref{A.EEQ}) of appendix~\ref{A.defs}
with fluid velocity $u_\mu=(c,0,0,0)$ and the Einstein tensor for the FLRW
metric~(\ref{A.FRW_1.EEQ}), we obtain the following density and pressure
\be
\ba{rcl}
\rho_{H}&=&\displaystyle T_{\hat{0}\hat{0}}\,=\,\frac{3}{8\pi\,G}\left(\frac{\dot{a}}{a}\right)^2=\frac{3H^2}{8\pi\,G}\ ,\\[6mm]
p_{H}&=&\displaystyle T_{\hat{i}\hat{i}}=\,-\frac{c^2}{8\pi\,G}\left(\,\left(\frac{\dot{a}}{a}\right)^2+2\left(\frac{\ddot{a}}{a}\right)\right)=-\frac{c^2(1-2q)H^2}{8\pi\,G}\ ,
\ea
\lb{p_rho_FRW_1}
\ee 
where no summation over the repeated indexes $\hat{i}$ is implied. We also note that the
non-null components $T_{01}=T_{10}$ do not correspond to a measurable physical
stress nor momentum flux (shear), their value is due to the coordinate choice, in the
Cartan frame $T_{\hat{0}\hat{1}}=0$. Consistently, for non-expanding coordinates $r$ there
is no non-null off-diagonal terms in the coordinate stress-energy tensor (the Einstein tensor for non-expanding coordinates given in equation~(\ref{A.FRW.EEQ}) of appendix~\ref{A.FRW} is diagonal).

These equations allow us to estimate today's deceleration factor $q_0=-(\ddot{a}_0/a_0)H_0^2$
as given in equation~(\ref{H_q0}). The combined WMAP+BAO+SN data~\cite{WMAP}
gives a cosmological constant (dark energy) relative density of $\Omega_\Lambda=0.726\pm 0.015$
with an equation of state $p_\Lambda=c^2\,\omega_\Lambda \rho_\Lambda$, where $\omega_\Lambda=-1^{+0.12}_{-0.14}$. Assuming that the
background is a perfect fluid, that the remaining matter is pressureless
($\omega_m=0\ \Rightarrow\ p_m=c^2\,\omega_m\rho_m=0$) and neglecting the radiation contributions ($p_r\approx 0$ and $\rho_r\approx 0$) we obtain the universe equation of state
\be
p_{H}=c^2\,\omega_{H}\rho_{H}\ \ ,\ \ \omega_{H}\approx\omega_\Lambda\frac{\Omega_\Lambda}{\Omega_{H}}=-0.726\ ,
\ee
where the total relative density is unity by definition, $\Omega_{H}=1$.
Then, taking the expressions for the FLRW pressure and density~(\ref{p_rho_FRW_1}),
and solving the equation of state for $q_0$ we obtain
\be
q_0\approx\frac{1}{2}+\frac{3}{2}\,\omega_T=-0.589\ .
\ee
This derivation can be found, for instance, in section~29 of~\cite{Gravitation}.
The error associated to this estimative can be inferred directly from the values for $\Omega_\Lambda$
and $\omega_\Lambda$ being the upper error bar $+0.167$ and the lower error bar $-0.146$. However these 
values have no statistical significance, the value for $\omega_\Lambda$ is quoted to $95\%$ confidence level while $\Omega_\Lambda$ is quoted to $68\%$ confidence level. Also we note that this value
for $q_0$ was computed assuming that the FLRW metric~(\ref{g_FRW_1}) does describe
the universe today and within the $\Lambda$CDM model which takes into account baryonic matter (the
usual heavy matter), dark energy (the cosmological constant) and cold dark matter (missing
non-interacting or weakly interacting matter).

\subsection{Observer not at the Origin of the Coordinate Frame\lb{sec.review_obs}}

So far we have always considered the observer at the origin of the coordinate frame.
For this case the geodesic equation for the observer is trivial, $\vb{r_1}=0$ at all times.
The origin of the coordinate frame is the only spatial point not affect by expansion.
Generally we can consider a commoving observer at some generic spatial point
$\vb{r_{1.\mathrm{obs}}}\neq 0$. The equations of motion for such an observer
are given by the geodesic equations which have the solutions~(\ref{l_i_geom_FRW_1}).
All space is expanding and the observer necessarily is attached to it (see for instance
the reviews~\cite{review_expanding_1,review_expanding_2} for a discussion in
this topic). Hence a commoving observer at a generic spatial observing an event at some other
spatial point $\vb{r_1}$ will not perceive the geodesic path of the event, instead will be
measuring the differences between its own geodesic path and the one of the event. Then, from the
perspective of a commoving observer, instead of the geodesic equations it would be the geodesic
deviation equations that describe the evolution of the event
\be
\frac{d^2(x_1^\delta-x^\delta_{1.\mathrm{obs}})}{d\tau^2}=\displaystyle-R^\delta_{\ \mu \rho\nu}(x_1^\rho-x^\rho_{1.\mathrm{obs}})\,\frac{d(x_1^\mu-x^\mu_{1.\mathrm{obs}})}{d\tau}\frac{d(x_1^\nu-x^\nu_{1.\mathrm{obs}})}{d\tau}\ .
\ee
These equations are valid when the geodesic separation $\varepsilon^\mu=(x_1^\mu-x^\mu_{1.\mathrm{obs}})$
is much lower than the characteristic lengths of the physical system being studied. For the expansion effects these are the Hubble time $t_{H_0}\sim 10^{17}\ s$~(\ref{H_t}) and the Hubble length
$l_{H_0}\sim 10^{26}\,m$~(\ref{H_length}). Hence for most astrophysical observations the geodesic deviation equations, in the absence of any other interactions, are valid. To lowest order in $H$ in the
non-relativistic limit ($\dot{r}_1\ll c$) and  assuming only radial motion
($\dot{\theta}=\dot{\varphi}=0$) these equations are
\be
t\approx t_{\mathrm{obs}}\ \ ,\ \ \ddot{r}_1\approx -qH^2(r_1-r_{1.\mathrm{obs}})\ ,
\ee
hence matching the lowest order equations of motion for the expanding coordinates~(\ref{ddx_FRW_1})
which correspond to the Newtonian limit of General Relativity.
The relativistic corrections are substantially distinct from the expressions~(\ref{ddx_rel_FRW_1})
obtained in the previous sections depending both in the geodesic separation $\varepsilon^\mu$
and the observer coordinates $x^\mu_{1.\mathrm{obs}}$ such that an observer at $\vb{r_1}$ will
have to perform simultaneous measurement of its own position with respect to the origin of the
coordinate frame, $\vb{r_{1.\mathrm{obs}}}$, as well as of the distance to the event $\vb{\varepsilon}$.

We will not develop in full detail these calculations here, it is relevant to mention this
issue to alert for the need to correct the observables due to the geodesic of the observer
as well as to recall that only for geodesic separation much lower than the characteristic
length of the physical system it is a valid approach to consider the geodesic deviation equation.
A possible alternative to this construction, as has been carried out in the previous sections,
is to consider a fixed observer at the origin of the coordinate frame which has a trivial geodesic
path such that the relative motion of the event being observed is directly given by its geodesic
path. When considering matter effects, depending on the physical measurements we want
to estimate, this issue may be relevant. However for the main results presented in this work
this discussion plays no role.

Next we review how matter in flat an expanding backgrounds is usually described.

\section{Reviewing Matter in an Expanding Background\lb{sec.matter}}
\setcounter{equation}{0}

In this section we review how the gravitational interactions of matter are described at the level
of the metric. We start by resuming the relevant characteristics
of the Schwarzschild metric~\cite{Schwarzschild} for this work which, within the framework of
General Relativity, is considered to be an exact description for space-time in the neighborhood
of a point-like mass (central mass) in a flat background (Minkowski space-time). We also resume
the existing metrics and the main characteristics of the respective space-times for
matter in an expanding background given by the FLRW metric~(\ref{g_FRW_1}),
namely the McVittie metric~\cite{McVittie} and the
Cosmological-Schwarzschild metric~\cite{cosm_SC}.

\subsection{Schwarzschild Metric}

The gravitational potential for a central mass (a point-like massive body) has spherical symmetry with
respect to the center of mass, hence both for technical simplification and easier physical
interpretation, it is often assume that the mass is centered at the origin of the
coordinate frame and employed spherical coordinates. Specifically we will use 
the following definition for the dimensionless Schwarzschild gravitational
potential $U_{\mathrm{SC}}$
\be
U_{\mathrm{SC}}=-\frac{2GM}{c^2\,r_1}\ .
\lb{U_SC}
\ee
Here $M$ is the gravitational mass of the massive body, $G$ is the Newton gravitational constant
and $c$ is the speed of light. Semi-classically it is often considered the dimensionfull
gravitational potential $\phi_{\mathrm{SC}}=c^2 U_{\mathrm{SC}}/2$. 
The Schwarzschild (SC) metric~\cite{Schwarzschild} for such a point-like mass
is written in terms of the gravitational potential $U_{\mathrm{SC}}$ as
\be
\ba{rcl}
ds^2&=&\displaystyle c^2\left(1-\frac{2GM}{c^2\,r_1}\right)dt^2-\frac{dr_1^2}{1-\frac{2GM}{c^2\,r_1}}-r_1^2\left(d\theta^2+\sin^2\theta d\varphi^2\right)\\[6mm]
&=&\displaystyle c^2\left(1-\frac{2GM}{c^2\sqrt{\delta_{ij}x_1^ix_1^j}}\right)\,dt^2-\left(\delta_{ij}+\frac{2GM\delta_{ik}\delta_{jl}x_1^ix_1^l}{c^2\left(\delta_{kl}x_1^kx_i^l\right)^\frac{3}{2}\left(1-\frac{2GM}{c^2\sqrt{\delta_{kl}x_1^kx_1^l}}\right)}\right)dx_1^idx_1^j ,
\ea
\lb{g_SC}
\ee
where we wrote the infinitesimal length square element both for spherical coordinates
(for which the metric is explicitly spherically symmetric) and Cartesian coordinates.
No effects for space expansion are yet considered, however let us consider expanding
coordinates $r_1$ which correspond to the physical measurable distances as discussed in
section~\ref{sec.review_map}. The Ricci scalar (scalar curvature) and the curvature invariant
are
\be
R_{\mathrm{SC}}=-8\pi\,\frac{GM}{c^2}\delta^{(3)}(\vb{r_1})\ \ ,\ \ {\mathcal{R}}_{\mathrm{SC}}=R_{\alpha\beta\rho\delta}R^{\alpha\beta\rho\delta}=\frac{48(GM)^2}{c^4\,r_1^6}\ ,
\lb{R_SC}
\ee
where $\delta^{(3)}(\bf{r_1})$ stands for the three-dimensional Dirac-delta. For Cartesian coordinates
it is the product of three one-dimensional Dirac-delta $\delta^{(3)}((x,y,z))=\delta(x)\delta(y)\delta(z)$,
while for spherical coordinates it is given only by a one-dimensional Dirac-delta on the radial coordinate
divided by the area of the sphere such that its volume integral is normalized to unity
\be
\delta^{(3)}(\vb{r_1)}=\frac{\delta(r_1)}{4\pi r_1^2}\ \ \Leftrightarrow\ \ \int_{-\pi}^\pi d\theta\int_{0}^{2\pi} d\varphi\int_0^{+\infty} dr_1\sqrt{-g}\delta^{(3)}(\vb{r_1)}=1\ .
\lb{Dirac_delta}
\ee

Therefore the SC metric has a singularity at $r_1=0$ which corresponds to the mass pole
at the center of mass, the origin of the coordinate frame.
The respective mass density can be computed by considering the Einstein tensor components written
in the Cartan frame. For the SC metric~(\ref{g_SC}) the Cartan Tetrad is $e^{\hat{\mu}}_{\ \mu}=\sqrt{|g_{\mu\mu}|}$ (without summation over repeated indexes implied) and the Ricci tensor components are $R_{\hat{\mu}\hat{\nu}}=R_{\mu\nu}/g_{\mu\nu}=4\pi(GM/c^2)\delta^{(3)}(\vb{r_1})$ (also without
summation over repeated indexes implied). Assuming the stress-energy tensor for a perfect fluid
we obtain the mass density from the $T_{\hat{0}\hat{0}}$ component~(\ref{A.EEQ})
\be
T^{\mathrm{SC}}_{\hat{0}\hat{0}}=\frac{c^4}{8\pi\,G}\,G^{\mathrm{SC}}_{\hat{0}\hat{0}}=c^2\,M\,\delta^{(3)}(\vb{r_1})\ \ \Leftrightarrow\ \ \rho_{\mathrm{SC}}=M\delta^{(3)}(\vb{r_1}).
\lb{SC_T}
\ee
Hence, for the SC metric, $M$ is both the gravitational mass~(\ref{U_SC}), the mass pole value
and the total mass of the space-time which is obtained by integrating the mass-energy density
$\rho_{\mathrm{SC}}$~(\ref{SC_T}).

There is a relevant remark to the remaining of this work, to have a mass pole at the
center of mass (here at the origin) the curvature invariant $\mathcal{R}$~(\ref{R_SC})
must be at most divergent by the inverse square of the volume element,
i.e $\mathcal{R}\sim V^{-2}\sim r_1^{-6}$. This requirement implies that the '$00$' component
of the stress-energy tensor is at most divergent by the volume element,
$T_{\hat{0}\hat{0}}\sim r_1^{-3}$, such that the pole value is finite and can be obtained
by considering the volume integral of the mass-energy density inside a shell of fixed
radius $r_1$ and taking the limit $r_1\to 0$. For more severe divergences, $\mathcal{R}\sim r^{-n}$
with $n>6$, we obtain a essential singularity which no longer can be interpreted as a mass pole.
The volume integral of the mass-energy density inside a shell of radius $r_1$ is divergent
(as well as its limit $r_1\to 0$).

With respect to the metric components for the Schwarzschild metric~(\ref{g_SC}) are singular
at the Schwarzschild radius, the well known coordinate singularity,
\be
r_{1.\mathrm{SC}}=\frac{2GM}{c^2}\ ,
\lb{r_SC}
\ee
which is interpreted as an event horizon for an external observer.

For astrophysical systems, in order to work perturbatively in the gravitational field,
it is often considered the (gravitational) weak field approximation
\be
U_{\mathrm{SC}}=\frac{2GM}{c^2\,r_1}\ll 1\ .
\lb{weak_field}
\ee
This inequality is obeyed for distances much larger than the Schwarzschild radius~(\ref{r_SC}).
Taking the solar system as an example and considering the gravitational field at the
surface of the sun, corresponding to the mass $M_{\mathrm{Sun}}\approx 1.98\times 10^{30}\ Kg$ and radius
$r_{\mathrm{Sun}}\approx 6.96\times 10^{8}\ m$, we obtain $U\sim 10^{-6}\ll 1$ (for further example
see box 19.2 of~\cite{Gravitation}).

With respect to the proper time $\tau_{\mathrm{SC}}$ and proper lengths $l_{1.\mathrm{SC}}$,
they do not generally coincide with the coordinate time $t$ and coordinate distance $r_1$.
However for non-relativistic velocities $\dot{r}_1,\dot{\theta},\dot{\varphi}\ll c$ and
weak gravitational fields~(\ref{weak_field}) they are approximately the same.
Considering an observer at $r_{1.\mathrm{obs}}$, far enough of the mass $M$,
observing events near the mass at $r_1\approx 0$ such that $r_{1.\mathrm{obs}}\gg r_1$,
we obtain
\be
\tau=\sqrt{1-\frac{2\,GM}{|r_{1.\mathrm{obs}}-r_1|}}\,t\approx t\ \ ,\ \ r_1=\frac{1}{\sqrt{1-\frac{2\,GM}{|r_{1.\mathrm{obs}}-r_1|}}}\,r_1\approx r_1\ .
\lb{SC_lengths}
\ee
Again, this is a valid approximation for most astrophysical measurements.

As for the classical Newton law for a test particle of mass $m$ and the respective radial acceleration
are as usual obtained directly from the gravitational potential
\be
\vb{F_{\mathrm{Newton}}}=m\,\frac{c^2}{2}\,\vb{\nabla} U\ \ \Leftrightarrow\ \ \ddot{r}_1=-\frac{GM}{r^2}\ .
\lb{F_Newton_classic}
\ee
In the Newtonian limit the General Relativity corrections to the Newton law can be obtained
directly from the geodesic equations. Neglecting the corrections due to the velocities of the test
mass the relevant equation is the radial component of the geodesic equations, specifically we obtain
\be
\ddot{r}_1\approx-c^2\Gamma^1_{\ 00}=-\frac{GM}{r_1^2}+\frac{2(GM)^2}{c^2\,r_1^3}\ .
\lb{F_Newton_GR}
\ee
The first term coincides with the classical Newton acceleration~(\ref{F_Newton_classic}).
The second term is due to General Relativity correction and, outside the Schwarzschild radius
$r_{1.\mathrm{SC}}=2GM/c^2$~(\ref{r_SC}), is never dominant being its value always below
the first term. At the Schwarzschild radius (the event horizon) both terms have the same
value such that the net gravitational acceleration is null. Either when the test mass is
relatively close to the central mass or when a higher accuracy for theoretical estimative
is required the geodesic deviation equations are employed instead of the geodesic equations.

Being relevant to the present work, we next discuss what is understood by local
anisotropy in the presence of a massive body. Light travelling in a gravitational
potential generated by a massive object is either blue-shifted or red-shifted depending on
weather it is travelling away or to the mass, respectively, while if travelling
(approximately) tangentially to the radial direction its frequency is not affected. This is
only an apparent spatial anisotropy, although the speed of light is not constant along all
directions, space-time is still isotropic. Specifically from the SC metric~(\ref{g_SC})
we obtain directly from the infinitesimal line element for travelling radiation
the following expressions for the speed of light
\be
\ba{rcccl}
\displaystyle c_r&=&\displaystyle \frac{dr_1}{dt}&=&\displaystyle c\left(1-\frac{2\,GM}{r_1\,c^2}\right)\ ,\\[6mm]
\displaystyle c_\theta=c_\varphi&=&\displaystyle r_1\frac{d\theta}{dt}&=&\displaystyle r_1\sin\theta\frac{d\varphi}{dt}=c\sqrt{1-\frac{2\,GM}{r_1\,c^2}}\ .
\ea
\ee
Nevertheless there is a specific coordinate choice $r_2$ for which the speed of light is isotropic.
These coordinates are usually called isotropic coordinates and the coordinate transformation
from the coordinates $r_1$ to the coordinates $r_2$ is given in equation~(\ref{A.r2.r1})
of appendix~\ref{A.defs}. For the new radial coordinate $r_2$ we obtain the metric
\be
\displaystyle ds^2=c^2\frac{\left(1-\frac{GM}{2\,r_2\,c^2}\right)^2}{\left(1+\frac{GM}{2\,r_2\,c^2}\right)^2}dt^2-\left(1+\frac{GM}{2\,r_2\,c^2}\right)^4\left(dr_2^2+r_2^2d\theta^2-r_2^2\sin^2\varphi d\varphi^2\right)\ ,
\lb{g_SC_iso}
\ee
for which the speed of light is constant along all spatial directions
\be
c_r=c_\theta=c_\varphi=\frac{dr_2}{dt}= c\,\frac{1-\frac{2\,GM}{|r_2-r_{2.0}|\,c^2}}{\left(1+\frac{2\,GM}{|r_2-r_{2.0}|\,c^2}\right)^3}\ .
\ee
Although not directly corresponding to the physical coordinates (the Schwarzschild coordinates)
these are widely employed in astrophysics, in particular in the PPN formalism~\cite{PPN}.

A general definition of spatial isotropy is with respect to mass-energy and
momentum fluxes which are encoded in the energy-momentum tensor $T_{\mu\nu}$. Isotropy of space
implies that, in the Lorentz frame with Minkowski metric $\eta_{\hat{\mu}\hat{\nu}}$ related to
the coordinate metric by $g_{\mu\nu}=e^{\hat{\mu}}_{\ \mu}e^{\hat{\nu}}_{\ \nu}\eta_{\hat{\mu}\hat{\nu}}$
through a tetrad $e_\mu$ the mass-energy flux is null, $T_{\hat{0}\hat{i}}=0$, and the momentum fluxes are
the same (isotropic) along the spatial dimensions being identified with the matter pressure $p$, $T_{\hat{i}\hat{j}}=\delta_{\hat{i}\hat{j}}p$. For the case of the SC metric~(\ref{g_SC}) the energy-momentum tensor is identically null except for the mass pole at the origin~(\ref{SC_T}),
hence the matter density and pressure are null everywhere else, $p=\rho=0$,
which is consistent with the assumption of space being empty of any sort of matter.

\subsection{Planetary Orbits\lb{sec.rev_orbits}}

Here we shortly review how to derive the orbit solutions for planetary motion
around a central mass and the theoretical orbital properties both for Keplerian and
General Relativistic orbits, namely we discuss orbital precessions and periods.

We will take the usual approach for two body orbital systems by considering a
variational functional given by the Lagrangian
\be
\frac{{\mathcal{L}}}{m}=g_{\mu\nu}\frac{dx^\mu}{d\tau}\frac{dx^\mu}{d\tau}=c^2\ ,
\lb{L_orbits}
\ee
where $m$ is the orbiting mass which factors out from the Lagrangian (see for instance section~8
of~\cite{Kenyon} or section~25.1 of~\cite{Gravitation}). For the Schwarzschild
metric~(\ref{g_SC}), considering an orbit lying in the plane of constant coordinate
$\theta=\pi/2$ such that $d\theta=0$ and $\sin\theta=1$, we obtain the following equation
\be
\frac{{\mathcal{L}}_{\mathrm{SC}}}{m}=\left(1-U_{\mathrm{SC}}\right)\,\left(c\,\frac{dt}{d\tau}\right)^2-\frac{1}{1-U_{\mathrm{SC}}}\,\left(\frac{dr_1}{d\tau}\right)^2-r_1^2\left(\frac{d\varphi}{d\tau}\right)^2=c^2\ .
\lb{L_orbit_SC}
\ee
The Lagrangian does not depend explicitly either in the time coordinate $t$ nor on the
angular coordinate $\varphi$,
hence there are two constant of motions. Respectively $E$ and $J$ defined as
\be
\ba{rclcl}
\displaystyle\frac{2E}{mc}&=&\displaystyle\frac{1}{m}\frac{\delta{\mathcal{L}}}{\delta(c\,dt/d\tau)}&=&\displaystyle 2\left(1-U_{\mathrm{SC}}\right)\left(c\,\frac{dt}{d\tau}\right)\ ,\\[6mm]
2J&=&\displaystyle\frac{1}{m}\frac{\delta{\mathcal{L}}}{\delta(d\varphi/d\tau)}&=&\displaystyle-2r_1^2\left(\frac{d\varphi}{d\tau}\right)\ .
\ea
\lb{orbital_constants}
\ee
We note that these constants of motion correspond to angular momentum conservation ($J$)
and energy conservation ($E$). Replacing both of them in the Lagrangian~(\ref{L_orbit_SC}),
multiplying by a factor of $(1-U_{\mathrm{SC}})$, gathering the constant terms on the left-hand
side of the equation, considering the derivative transformation from proper time $\tau$ to the
angular coordinate $\varphi$ and redefining the radial coordinate to $u=1/r_1$ such that
\be
\left\{\ba{rcl}
\displaystyle\frac{dr_1(\tau)}{d\tau}&=&\displaystyle\frac{dr_1(\varphi)}{d\varphi}\,\frac{d\varphi(\tau)}{d\tau}=-J\,\frac{r_1'(\varphi)}{r_1^2(\varphi)}\\[5mm]
u(\varphi)&=&\displaystyle\frac{1}{r_1(\varphi)}
\ea\right. \ \Rightarrow\ \ u'(\varphi)=\frac{d}{d\varphi}\frac{1}{r_1(\varphi)}=-\frac{r_1'(\varphi)}{r_1^2(\varphi)}\ ,
\lb{orb_var_transf}
\ee  
we obtain the following differential equation on the function $u=u(\varphi)$
\be
c^2\left(1-\left(\frac{E}{mc^2}\right)^2\right)=-J^2\,(u')^2-J^2\,u^2+2GM\,u+J^2\,\frac{2GM}{c^2}\,u^3\ .
\lb{orb_deq_1st}
\ee
In the variable and function transformations~(\ref{orb_var_transf}) we have explicitly written the dependence on proper time $\tau$ and $\varphi$ while in the differential equation~(\ref{orb_deq_1st})
we have omitted the variable dependence. The first three terms in the left-hand side of this equation
correspond to the Keplerian orbit equation while the last term is the General Relativity correction.

The differential equation~(\ref{orb_deq_1st}) is non-linear, the usual linear second order orbital
equation is obtained by differentiating with respect to the variable $\varphi$ and factoring out
an overall factor of $2u'$ such that we obtain
\be
u''+u=\frac{GM}{J^2}+\frac{3GM}{c^2}\,u^2\ .
\lb{eqd_u}
\ee
Again, the first term on the right-hand side of the equation match the classical Keplerian orbits,
while the second term is the General Relativity correction which has as effect a small perihelion
advance with respect to classical solutions (which is equivalent to an orbital preccession effect).

The classical (Keplerian) orbital ellipse solution is
\be
u_0''+u_0=\frac{GM}{J^2}\ ,
\lb{eqd_uo}
\ee
with solution given by
\be
u_0=\frac{1+e\cos\varphi}{d}\ ,
\lb{uo}
\ee
which corresponds to an ellipse with the mass $M$ on one of its foci where $e\in[0,1[$ is the orbit eccentricity. As for the parameter $d$ and the constant of
motion $J^2$ are related to the semi-major orbit axis $r_{1.\mathrm{orb}}$ as
\be
d=r_{1.\mathrm{orb}}(1-e^2)\ \ \Leftrightarrow\ \ J^2=r_{1.\mathrm{orb}}\,GM\,(1-e^2)\ .
\lb{d_a_J}
\ee
We note that both $e$ and $r_{1.\mathrm{orb}}$ must be set as initial boundary conditions for each orbit
and are not derivable theoretically, however the constants of motion $J$ and $E$ are not
 independent. $J$ is set from the relation between the differential
equation~(\ref{eqd_uo}) and the physical (geometrical) interpretation of the parameter $d$~(\ref{d_a_J}).
As for the constant of motion $E$
can be approximated for a classical orbit by replacing the solution $u_0$~(\ref{uo}) in the non-linear differential equation~(\ref{orb_deq_1st}) such that neglecting the last term on the
right-hand side, we obtain
\be
E^2\approx E_0^2=(m\,c^2)^2\,\left(1-\frac{GM}{r_{1.\mathrm{orb}}\,c^2}\right)\ .
\lb{E_E0}
\ee
The period $T_0$ of this classical solution is known to be independent of the orbit
eccentricity $e$ being given by Kepler's third law
\be
T_0=\frac{2\pi\,r_{1.\mathrm{orb}}^\frac{3}{2}}{\sqrt{GM}}\ .
\lb{T_Kepler}
\ee

The General Relativity corrections are obtained by considering the differential
equation~(\ref{eqd_u}) with the last term evaluated with the classical solution $u_0$~(\ref{uo})
\be
\ba{rcl}
\displaystyle u&=&\displaystyle u_0+u_{\mathrm{GR}}\ ,\\[5mm]
\displaystyle u_{\mathrm{GR}}''+u_{\mathrm{GR}}&=&\displaystyle \frac{3GM}{c^2}\,\left(\frac{1+e\,\cos\varphi}{d}\right)^2\ .
\ea
\lb{eqd_GR}
\ee
We note that this is a valid approximation due to the GR correction term being much smaller than
the classical term (assuming large radial coordinate, hence $r\gg 1$ and $u=1/r\ll 1$).
The solution of the differential equation~(\ref{eqd_GR}) is
\be
u_{\mathrm{GR}}=\frac{3GM}{c^2\,d^2}\,\left(\left(1+\frac{e^2}{2}\right)-\frac{e^2}{6}\cos 2\varphi+e\,\varphi\,\sin\varphi\right)\ ,
\lb{uGR}
\ee
where we have set the integration constants so that no simple oscillating terms are present
containing either $\cos\varphi$ or $\sin\varphi$. The reason to do so is simply to maintain the
same values for the eccentricity $e$ and parameter $d$ than in the unperturbed solution~(\ref{uo}).
Specifically we could have set the integration constants such that a term $e\,k_{\mathrm{GR}}\,\cos\varphi$
with $k_{\mathrm{GR}}=3GM/(c^2\,d^2)(1+e^2/2)$ is present which maintains the same value for the orbit eccentricity. Then together with the first constant terms in the above solution~(\ref{uGR}),
it could be included in the unperturbed solution $u_0$ by re-defining the parameter $d$ by a small shift
\be
\ba{rcl}
u&=&\displaystyle u_0+\bar{u}_{\mathrm{GR}}=\frac{1+k_{\mathrm{GR}}}{d}\left(1+e\cos\varphi\right)+\frac{3GM}{c^2\,d^2}\,\left(-\frac{e^2}{6}\cos 2\varphi+e\,\varphi\,\sin\varphi\right)\\[6mm]
\displaystyle\frac{1}{\tilde{d}}&=&\displaystyle\frac{1+k_{\mathrm{GR}}}{d}\ \ \Rightarrow\ \tilde{d}\approx d(1-k_{\mathrm{GR}})\ .
\ea
\lb{integration_constants}
\ee
Here we have used the notation $\bar{u}_{\mathrm{GR}}$ to distinguish this solution from
$u_{\mathrm{GR}}$~(\ref{uGR}). For this example this corresponds to a redefinition of the semi-major axis
$r_{1.\mathrm{orb}}$ by the amount $k_{\mathrm{GR}}$. A similar construction can be carried for which either the eccentricity $e$ or both $e$ and $r_{1.\mathrm{orb}}$ are corrected. However we note that
these correction, besides being negligible when compared with the remaining terms, have no particular physical meaning, we recall that both the orbital eccentricity $e$ and semi-major axis $r_{1.\mathrm{orb}}$
are set as boundary conditions (or initial value) for the differential equations, hence we are simply
re-defining our equation parameters and its relation to the physical (geometrical) quantities of the
physical orbit.

Hence, with respect to the several terms in the correction~(\ref{uGR}), we have that the first term
is a constant that, following the previous discussion, can be neglected, the second term is a periodic
deformation of the ellipse (the orbit) that slightly increases the orbital path contributing
a small deviation to the period with respect to the classical one, $T_0$~(\ref{T_Kepler}), and the last
term increases steadily over time (here with the angular coordinate) and is responsible for the well
known orbital precession (the ellipse rotates steadily around the center of the mass $M$).
To analyze the effect of these terms let us define the correction parameter
\be
\alpha_{GR}=\frac{3GM}{c^2\,d}\ ,
\lb{alpha_GR}
\ee
and consider the following approximation to the sum of the term $\cos\varphi$ on the
solution $u_0$~(\ref{uo}) and the term $\alpha_{\mathrm{GR}}\,\varphi\sin\varphi$ in the solution
$u_{\mathrm{GR}}$~(\ref{uGR}) by considering the respective lower order series expansion
\be
\ba{rcl}
\cos\varphi+\alpha_{GR}\,\varphi\sin\varphi&\approx&\displaystyle 1-\frac{\varphi^2}{2}+2\alpha_{GR}\,\frac{\varphi^2}{2}\ \approx\  1-(1-2\alpha_{GR})\,\frac{\varphi^2}{2}\\[6mm]
&\approx&\displaystyle \cos(\sqrt{1-2\alpha_{GR}}\,\varphi)\ \approx\ \cos((1-\alpha_{GR})\,\varphi).
\ea
\lb{cos_exp}
\ee
From the first to the second line we have approximate the series expansion by the cosine function
and in the last equality we have expanded the square root to lowest order in the correction parameter
$\alpha_{\mathrm{GR}}$. Hence we obtain the following approximate solution of the full
differential equation~(\ref{eqd_u})
\be
\ba{rcl}
u&\approx&\displaystyle \frac{1}{d}\left(1+e\,\cos\left((1-\alpha_{GR})\varphi\right)\right)+u_{\mathrm{osc.GR}}\ ,\\[6mm]
u_{\mathrm{osc.GR}}&=&\displaystyle -\frac{\alpha_{GR}}{6d}\,e^2\cos(2\varphi)\ .
\ea
\lb{u_GR_exp}
\ee
The precession amount per turn of the orbit is directly taken from the argument of the cosine
\be
\frac{\Delta\varphi_{\mathrm{GR}}}{2\pi}=\alpha_{GR}\ ,
\lb{precession_GR}
\ee
where we are using radians to measure angles. As for the period correction
it can be evaluated from the term $u_{\mathrm{osc.GR}}$ by noting that the infinitesimal time
displacement $dt$ is given in terms of the tangential velocity to the orbital path, $v_\perp=dx_\perp/dt$, as
\be
\ba{rcl}
dt&=&\displaystyle\frac{dx_\perp}{v_\perp}=\frac{1}{u\,v_\perp}\,d\varphi\ ,\\[6mm]
v_\perp&=&\displaystyle r_1\frac{d\varphi}{dt}=\frac{1}{u}\frac{d\varphi}{d\tau}\,\left(\frac{dt}{d\tau}\right)^{-1}\approx J\,u\,\sqrt{1-\frac{2GM}{c^2}\,u}\ ,
\ea
\lb{vt_GR}
\ee
where $v_\perp$, $r_1=1/u$ and $u$ are functions of the angular coordinate $\varphi$.
In the first line we have used the relation between
the infinitesimal spatial displacement on the orbit path $dx_\perp$ and the infinitesimal
angular displacement $d\varphi$, $dx_\perp=r_1\,d\varphi=d\varphi/u$. In the last equality we have used the definitions of the
constant of motion $J$~(\ref{orbital_constants}) and considered the limit
of non-relativistic velocities $\dot{x}^\mu\ll c$ such that
$(dt/d\tau)^{-1}=\gamma^{-1}\approx\sqrt{1-2GM\,u/c^2}$. Hence, considering the integration
of the infinitesimal time displacement $dt$~(\ref{vt_GR}) over one turn of the orbit, we obtain, to lowest order,
the period correction due to the term $u_{\mathrm{osc.GR}}$
\be
\ba{rcl}
\displaystyle\int_0^T\,dt&=&T\ =\ T_0+\Delta T_{\mathrm{GR}}\ ,\\[6mm]
\Delta T_{\mathrm{GR}}&\approx&\displaystyle -\frac{2}{|J|}\,\int_0^{2\pi}d\varphi\,\frac{u_{\mathrm{osc.GR}}}{u_0^3}\,\left(1-\frac{GM}{c^2}\,u_0\right)\ .
\ea
\lb{dT_GR}
\ee
To derive the final expression for $\Delta T_{\mathrm{GR}}$ we have expanded the expression for
$dt$~(\ref{vt_GR}) with $u\approx u_0+u_{\mathrm{osc.GR}}$ to first order in $u_{\mathrm{osc.GR}}$
and integrated the term containing this oscillatory correction. The term containing only the solution
$u_0$ corresponds to the period $T_0$.
Later on, when discussing both the precession and period corrections due to the
expanding background, for comparative purposes of the magnitude of the several effects
we will give numerical estimative for equations~(\ref{precession_GR}) and~(\ref{dT_GR}).

The remaining correction to the orbital motion are due to solar oblateness (Sun quadrupole moment $J_2$)
which contributes significantly to the orbit precession (see for instance section 40.5,
box 40.3, of~\cite{Gravitation}) and the interactions between the several celestial bodies.
It is only possible to estimate the ephemerides of the planets by the use of extensive
numerical calculations in the PPN formalism~\cite{PPN} including most of the known bodies in the solar system, see for instance~\cite{Pitjeva} for further details. We will not give further details on these
topics and will compare the estimative obtained when considering an expanding background with the
General Relativity corrections for flat backgrounds discussed previously. We note that the
precession effects due to General Relativity are
usually significant and are taken in consideration in theoretical calculations, however the
orbital period corrections are usually negligible when compared with the other corrections due, mostly,
to the interactions with other bodies, hence dealt only numerically. We considered these corrections
here for comparison purposes only.

Next we briefly resume the existing metrics to describe matter in an expanding background.

\subsection{McVittie Metric}

The McVittie metric~\cite{McVittie} was derived originally in non-expanding isotropic coordinates
$r_3$ demanding both spherical symmetry and spatial isotropy such that the stress-energy tensor
is shear free. For this coordinate choice this metric is given in equation~(\ref{A.McV.g_r3}) of appendix~\ref{A.McV}.
For expanding coordinates $r_1$~(\ref{A.McV.g_r1}) it reads
\be
\ba{rcl}
ds^2&=&\displaystyle \left(1-\frac{2\,GM}{c^2\,r_1}\right)c^2\,dt^2-r_1^2\left(d\theta^2+\sin^2\theta d\varphi^2\right)\\[6mm]
& &\displaystyle-\frac{1}{1-\frac{2\,GM}{c^2\,r_1}}\left(dr_1-\frac{\dot{a}\,r_1}{a\,c}\,\left(1-\frac{2\,GM}{r_1\,c^2}\right)^\frac{1}{2}c\,dt\right)^2\ ,
\ea
\lb{g_McV}
\ee
where we explicitly factorize the length square infinitesimal element into proper-time and
proper-distance. This metric, for expanding isotropic coordinates $r_2$ and non-expanding coordinates $r$,
is given respectively in equations~(\ref{A.McV.g_r2}) and~(\ref{A.McV.g_r}) of appendix~\ref{A.McV}.

The McVittie metric is a sobreposition of the Schwarzschild metric~(\ref{g_SC}) describing
small spatial scales with the FLRW metric~(\ref{g_FRW}) describing large spatial scales.
As can directly be inferred from the metric expression~(\ref{g_McV}) it has the following properties:
\begin{enumerate}
\item{} asymptotically, it coincides at spatial infinity ($r_1\to\infty$) and in
massless limit ($M\to 0$) with the cosmological FLRW metric~(\ref{g_FRW_1});
\item{} in the static limit ($a\to 1$) it coincides with the
Schwarzschild metric~(\ref{g_SC});
\item{} for expanding coordinates $r_1$ it has Lorentzian measure which corresponds to the
spherical coordinate measure to be $\sqrt{-g}=r_1^2\sin\theta$.
\end{enumerate}

In addition it is isotropic as well as stress and shear free.
The non-null spatial components of the stress-energy tensor in the orthonormal Lorentz frame
(Cartan frame) are $T_{\hat{i}\hat{i}}=T_{ii}/g_{ii}=T_{01}/g_{01}=p$ (for $i=1,2,3$), where no
summation over repeated indexes is implied. Later we will write the explicit expressions
for this tensor as a particular case of a more generic metric. The non-null off-diagonal
component $T_{01}$ is not a measurable physical stress, it is due to the coordinate choice as we have
discussed earlier in section~\ref{sec.review_obs} for the FLRW metric~(\ref{g_FRW_1}) in expanding coordinates $r_1$, specifically in the Lorentz frame we have that $T_{\hat{0}\hat{1}}=0$.

If we insist in that space isotropy must be maintained it is accepted
that there is no other possible choice~\cite{review_expanding_1,review_expanding_2}.

At the origin ($r_1=0$) the McVittie metric has the same pole of the Schwarzschild metric,
the dominant singular term of the curvature invariant in the neighborhood of the origin is
\be
{\mathcal{R}}_{\mathrm{McV}}(r_1\sim 0)\sim\frac{48(GM)^2}{c^4\,r_1^6}\ .
\ee
In addition it has also a singularity at the Schwarzschild radius
$r_1=r_{\mathrm{SC}}=2GM/c^2$~\cite{sing}, in its neighborhood the dominant divergent term
of the curvature invariant is
\be
{\mathcal{R}}_{\mathrm{McV}}(r_1\sim r_{1.\mathrm{SC}})\sim \frac{12(1-q)^2H^4}{1-\frac{2GM}{c^2\,r_1}}\ .
\ee
This is an extended singularity corresponding to a two sphere. 
Hence, although in the static
limit $a\to 1$ the McVittie metric converges to the Schwarzschild metric,
in the limit $r_1\to 2GM/c^2$ (near the Schwarzschild radius) the McVittie
metric~(\ref{g_McV}) does not converge asymptotically to the SC
metric~(\ref{g_SC}) having a clearly distinct behavior due to the singularity at the horizon.
We note that it is reasonable to expect
that a metric describing both expansion and matter effects asymptotically converges
to the SC metric at the SC horizon, in this limit the matter
effects should be dominant with respect to the background fluid effects
describing expansion. Also within the framework of General Relativity, the SC
metric has been extensively and successfully employed for different coordinate choices
being in agreement with experimental data for most astrophysical systems (see, for
instance, section~38 of~\cite{Gravitation}). From a more theoretical perspective this metric
describes a non-complete space-time and the total mass inside a shell of finite radius
$r_1>r_{1.\mathrm{SC}}$ is divergent due to the divergence at the SC radius $r_{1.\mathrm{SC}}$.
These are clearly unwelcome properties.

We recall that the original derivation of this metric~\cite{McVittie} has been carried
for isotropic coordinates which are employed for large radial coordinate (usually
in astrophysical systems), in particular the map between these coordinates and the usual
expanding Schwarzschild coordinates~(\ref{A.r2.r1}) is only defined outside the Schwarzschild
horizon ($r_1\ge 2GM/c^2$ and $r_2\ge GM/(2c^2)$), hence no analysis was carried for
the asymptotic limit near this horizon.

\subsection{The Cosmological-Schwarzschild Anisotropic Metric}

Global space isotropy is commonly accepted as a fact,
both due to theoretical reasoning, Poincar\'e invariance
is only strictly maintained for isotropic space-times, as well
as due to large-scale observations of our universe, the
background radiation in our universe is globally isotropic~\cite{WMAP}.
However there is experimental evidence for local anisotropy corroborated
both by local deviations of the Hubble flow in nearby astrophysical systems~\cite{anisotropy},
as well as due to the local anisotropies of the background radiation~\cite{WMAP}.

Hence it is not physically unconceivable that a description of matter in an expanding
background may generate a local spatial anisotropy. Assuming that this
may be the case there is one metric that has already been considered
by several authors employing distinct technical approaches~\cite{cosm_SC}.
It can be justified by noting that we have been measuring spatial lengths
without taking in account spatial expansion. Then, being the classical Newton law
our starting point, for non-expanding coordinates $r$, we can rewrite it taking in
consideration the spatial expansion as
\be
a\ddot{r}=-\frac{GM}{a^2r^2}\ .
\lb{Newton_alpha_1}
\ee
The extra factor of $a$ multiplying $\ddot{r}$ is due to the projection
to the three-dimensional hyper-surface as discussed in section~\ref{sec.review},
also the acceleration \textit{expands} with space.

The metric corresponding to the above Newton law~(\ref{Newton_alpha_1}),
both for non-expanding coordinates $r$ and for expanding coordinates $r_1$, is
\be
\ba{rcl}
ds^2&=&\displaystyle \left(1-\frac{2GM}{a\,c^2\,r}\right)c^2\,dt^2-a^2\,r^2\left(d\theta^2+\sin^2\theta d\varphi^2\right)-\frac{a^2\,dr^2}{1-\frac{2GM}{a\,c^2\,r}}\\[5mm]
&=&\displaystyle \left(1-\frac{2GM}{c^2\,r_1}\right)c^2\,dt^2-r_1^2\left(d\theta^2+\sin^2\theta d\varphi^2\right)\\[6mm]
& &\displaystyle-\frac{1}{1-\frac{2GM}{c^2\,r_1}}\left(dr-\frac{\dot{a}\,r_1}{a\,c}\left(1-\frac{2GM}{c^2\,r_1}\right)c\,dt\right)^2\ .
\ea
\lb{g_cosm_SC}
\ee
Where we explicitly factorize the infinitesimal length square into proper-time and proper-distance.
We remark that by considering the Schwarzschild metric expression~(\ref{g_SC}) to be written in
non-expanding coordinates $r$, this metric is obtained by a direct replacing of the non-expanding
radial coordinate $r$ with an expanding radial coordinate $r\to a\,r$ and correcting the radial component
of the metric accordingly $dr\to a\,dr$. Due to this relation we refer to this metric as
Cosmological-Schwarzschild metric (CSC).

It has the same three properties listed for the McVittie metric maintaining
spherical symmetry, however space-time is locally anisotropic, the components of the
stress-energy tensor in the Cartan frame obey the following relations
$T_{\hat{1}\hat{1}}=T_{11}/g_{11}=T_{01}/g_{01}\neq T_{\hat{2}\hat{2}}=T_{\hat{3}\hat{3}}=T_{22}/g_{22}=T_{33}/g_{33}$,
hence the radial component and angular components have distinct values which
explicitly shows the existence of spatial anisotropy.

As the McVittie metric, also this metric describe space-time with two singularities, at the origin of
the coordinate frame $r_1=0$ and at SC radius $r_1=2GM/c^2$.
In the neighborhood of the origin the curvature invariant dominant term coincides
with the Schwarzschild curvature invariant~(\ref{R_SC})
\be
{\mathcal{R}}_{\mathrm{S}}(r_1\sim 0)\sim\frac{48(GM)^2}{c^4\, r_1^6}\ .
\ee
As for the neighborhood of the Schwarzschild radius $r_{1.\mathrm{SC}}=2GM/c^2$~(\ref{r_SC})
the dominant term in the curvature invariant is
\be
{\mathcal{R}}_{\mathrm{S}}(r_1\sim r_{1.\mathrm{SC}})\sim\frac{4(GM)^2}{c^4\, r_1^2}\frac{H^4(1+q)^2}{\left(1-\frac{2GM}{c^2\,r_1}\right)^2}\ .
\ee
Hence it has the same unwelcome properties of the McVittie metric, does not converge
asymptotically to the SC metric near the central mass ($r_1\sim r_{1.\mathrm{SC}}$),
space-time is not complete and the mass inside a shell of finite fixed radius is divergent
due to the singularity at the SC radius.

Next we will consider a more generic metric ansatz(e) that has, as particular cases, both the McVittie
metric~(\ref{g_McV}) and the CSC metric~(\ref{g_cosm_SC}). We will also obtain a parameter range for
which only the mass pole at the origin is present such that the SC horizon is singularity free
and the SC metric~(\ref{g_SC}) is asymptotically obtained at the SC
radius.

\section{A Locally Anisotropic Metric for Matter in Expanding Space-Time\lb{sec.metric_generic}}
\setcounter{equation}{0}

Once we consider a background fluid, as for the case of the FLRW metric~(\ref{g_FRW}), are generally
present a non-null background gravitational density and pressure. It is a very conservative and
well accepted assumption to expected that, by considering a point-like massive object in such
a background, we obtain a spherically symmetric pattern deformation asymptotically vanishing
at spatial infinity. We also remark that for a central mass, the direction of the gravitational
interaction is radial, hence there is a preferred spatial direction. Hence it is not unconceivable,
even being physically intuitive, to consider the fluid deformation to be anisotropic with respect
to the radial and angular directions. Also, as already discussed, local anisotropy is consistent
with experimental observations~\cite{anisotropy, WMAP}, as long as global spatial isotropy is
preserved.

In this section we will build a metric ansatz describing matter in an expanding background
that generalizes the McV metric~(\ref{g_McV}) and the CSC metric~(\ref{g_cosm_SC}) and interpolates between
the FLRW metric~(\ref{g_FRW_1}) and the Schwarzschild metric~(\ref{g_SC}) maintaining space-time
free of singularities except for the Schwarzschild mass-pole at the origin. Although, for the
metric ansatz, space-time is locally anisotropic, spatial isotropy is recovered at spatial
infinity. We start by the simpler case of a one parameter metric analyzing the singularities
for the several values of the parameter such that the space-time is regular
at the Schwarzschild radius, hence obtaining a complete space-time asymptotically flat at
the SC horizon. However the singularity at the origin is more severe than the SC mass-pole such that
the mass inside a shell of finite radius is divergent. To strictly maintain the SC mass-pole
at the origin we refine the ansatz by considering one further regularization parameter obtaining
a space-time with finite total mass inside a shell of finite radius.
We further analyze the stress-energy tensor defining the range of the parameters for which
the mass-energy is positive outside the event horizon.

\subsection{The Ansatz I: A First Approach\lb{sec.ansatz_I}}

In the following we consider that the expansion of space is a global effect due to the
background matter and energy spread across all universe such that cannot be locally
eliminated. This is not a widely accepted assumption, except for the cosmological constant
effect (dark energy) which accounts for $72.6\%$ of all background gravitational effects~\cite{WMAP}
and, due to be a constant, cannot be physically excluded from any system.
A common argument concerning the remaining $27.4\%$ of matter and energy is that a stationary low density dust (matter background) in the neighborhood of a
massive stellar object is attracted and rapidly aggregated by that massive object. Nevertheless
we note that, generally, also stars and other massive objects are responsible for matter and radiation
emissions (hence also contributing for the background matter and energy density), moreover expansion
is a global effect mostly due to long range gravitational interactions, the matter and
energy density far from a massive object also contribute to the local expansion and, in the same
fashion are also affected both by the gravitational field of the several local massive objects
and their long range interactions. Hence our assumption is that there is no natural mechanism to set a local cutoff for which expansion is locally eliminated.

We next proceed to generalize both the McV metric~(\ref{g_McV}) and
the SCS metric~(\ref{g_cosm_SC}). We consider expanding spherical coordinates $r_1$
and will build an ansatz intended to describe local matter in an expanding background
interpolating between the FLRW metric~(\ref{g_FRW_1}) which describes the cosmological expanding
background and the Schwarzschild metric~(\ref{g_SC}) which describes local matter
in a flat background. With respect to the metric properties, when setting up this ansatz,
we consider the following assumptions:
\begin{enumerate}
\item{} in the massless limit $M\to 0$ and at
spatial infinity converges asymptotically to the FLRW metric~(\ref{g_FRW_1});
\item{} in the static limit $a\to 1$ ($\dot{a}\to 0$) coincide with the Schwarzschild metric~(\ref{g_SC});
\item{} has Lorentzian measure $\sqrt{-g}=1$ for Cartesian coordinates $x_1$ which corresponds
to the spherical coordinate measure $\sqrt{-g}=r_1^2\sin\theta$ such that the area of the sphere is $A=4\pi\,r_1^2$.
\end{enumerate}
These characteristics are shared by both the McV metric~(\ref{g_McV}) and the CSC metric~(\ref{g_cosm_SC})
discussed in the previous section. The first two assumptions are common to both these metrics,
we wish to interpolate between two distinct metrics (the FLRW and SC metrics) and these assumptions
ensure that in the limiting cases we retrieve the two original metrics.
The second assumption can become stronger, we could demand the ansatz
to converge asymptotically to the SC metric at the origin. This
case is actually relevant to strictly maintain the SC mass pole at the center of mass
and will be addressed in the next section. In addition we will also obtain bounds
for the metric parameter for which the ansatz converges asymptotically also to the SC metric at
the event horizon (the SC radius), hence for which space-time is asymptotically flat near the
point-like mass.
In order to both maintain some anchorage with already existing results, namely the McVittie metric~(\ref{g_McV}) and CSC metric~(\ref{g_cosm_SC}), and to keep track of the several
steps in the ansatz(e) building and the respective physical interpretations,
we will proceed with this milder requirement.

As for the last assumption it may, generally, be lifted and it is clearly coordinate dependent.
Nevertheless for expanding coordinates $r_1$ (which correspond to the physical measurable lengths as
discussed in section~\ref{sec.review_map}) it is expected that we obtain the Lorentz measure which is shared by both the FLRW metric~(\ref{g_FRW_1}) and the Schwarzschild metric~(\ref{g_SC}). This is not
by chance and it reflects the way we perceive our world, in particular with the way we perform
measurements and with Local Lorentz invariance. Specifically Euclidean geometry is the basis of
all our spatial measurements such that our measurable spatial sphere has area of
$A=4\pi\,r_1^2$ and the geometrical generalizations to four-dimensional space-time manifolds are
necessarily Locally Lorentz, at least in a patch including the physical system being studied.
In simple terms this means that a space-like vector remains space-like and a time-like vector
remains time-like being each orthogonal to each other such that an orthonormal Lorentz basis
(corresponding to the Cartan frame with Minkowski metric) can be considered locally at each space-time point (see, for instance section~13 of~\cite{Gravitation} for further details).

We also remark that, when working with metrics having Lorentz measure, we can
directly infer from the splitting of the infinitesimal length square into proper time
and proper spatial length whether space-time is locally Lorentz or not, considering an ADM
parameterization of the metric we obtain that $d\tau=Ndt^2$ must be time-like and
$dl^2=-g_{ij}(dx^i+N^idt)(dx^j+N^jdt)$ must be space-like ($N$ is the lapse function and $N^i$
are the shift functions). With respect to the specific metrics discussed here, we have that
the SC metric~(\ref{g_SC}) describes local Lorentz space-time outside the event horizon $r_1>r_{1.\mathrm{SC}}$ and the FLRW metric~(\ref{g_FRW_1}) describes local Lorentz space-time
within the cosmological horizon $r_1<l_H$. Also considering this splitting allows us to build
an ansatz for matter in an expanding background following a very simple procedure. We note that
for expanding coordinates $r_1$ the deformation of the Minkowski metric due to spatial expansion
which corresponds to the FLRW metric~(\ref{g_FRW_1}) is given by an additive radial shift
function
\be
N^1_{\mathrm{FLRW}}=\frac{\dot{a}\,r_1}{a\,c^2}=\frac{r_1}{l_H}\ ,
\lb{N_1_FRW_1}
\ee
such that the radial infinitesimal element is shifted to $dr_1\to dr_1-N^1_{\mathrm{FLRW}}\, cdt$,
where $l_H=c/H$ is the time-dependent Hubble length~(\ref{H_length}). To show it explicitly let
us rewrite the FLRW metric~(\ref{g_FRW_1}) factorizing the infinitesimal length square into
proper-time and proper-length obtaining
\be
ds^2=c^2\,dt^2-\left(dr_1-\frac{\dot{a}\,r_1}{a\,c}\,c\,dt\right)^2-r_1^2\left(d\theta^2+\sin^2\theta d\varphi^2\right)\ .
\lb{g_FRW_1_fact}
\ee
The Lorentzian measure corresponding to the spherical coordinates measure 
$\sqrt{-g}=r^2\sin\theta$ is maintained whenever the corrections are encoded in an additive
radial shift function. Hence, considering a deformation of the Schwarzschild metric~(\ref{g_SC})
by a generic shift function dependent on the radial coordinate and time $N^1(r,t)$ such that
the infinitesimal radial element is shifted to $dr_1\to dr_1-N^1 dt$, ensures
that both spherical symmetry and the metric measure are maintained. To describe
expansion effects $N^1$ has to necessarily contain the expansion factor
$N^1_{\mathrm{FLRW}}=r_1/l_H$~(\ref{N_1_FRW_1}) which ensures that at spatial infinity
we retrieve asymptotically the FLRW metric~(\ref{g_FRW_1_fact}).
As for the SC metric, it converges asymptotically to the Minkowski metric due to its dependence
on the factor $(1-U_{\mathrm{SC}})=1-2GM/(c^2\,r_1)$ which converges asymptotically to is unity, $\lim_{r_1\to\infty}(1-U_{\mathrm{SC}})=1$.
Furthermore, at the Schwarzschild radius this factor is null $\lim_{r_1\to r_{1.\mathrm{SC}}}(1-U_{\mathrm{SC}})=0$.
Then, in addition, considering the shift function $N^1$ to depend also on a multiplicative
positive power of the factor $(1-U_{\mathrm{SC}})<1$ simultaneously ensures that near the
massive object the expansion effects decrease being exactly null at the SC radius and, at spatial
infinity, the FLRW shift function $N^1_{\mathrm{FLRW}}$~(\ref{N_1_FRW_1}) is recovered. In this
way the value of the exponent of the factor $(1-U_{\mathrm{SC}})$ fine-tunes the intensity of
the expansion effects felt near the mass. Specifically increasing the positive exponent will
decrease the value of the shift function for relatively small radial coordinate $r_1$, still
maintaining the convergence to the asymptotic FLRW metric for large values of the radial coordinate.
It is also sensitive to note that both the isotropic McVittie metric~(\ref{g_McV}) and the
anisotropic CSC metric~(\ref{g_cosm_SC}) discussed in the previous section are obtained from the
SC metric by considering a deformation by such a shift function, respectively with powers of the
factor $(1-U_{\mathrm{SC}})$ of exponents $1/2$ and $1$.

Given the previous discussion we are considering the deformation to the Schwarzschild metric, due to
expanding background, to be given by a radial shift function of the form
\be
N^1_\alpha=\frac{\dot{a}\,r_1}{a\,c}\,\left(1-\frac{2GM}{c^2\,r_1}\right)^{\frac{\alpha}{2}+\frac{1}{2}}
\lb{N_1_alpha_1}
\ee
such that we obtain the metric ansatz
\be
\ba{rcl}
ds^2&=&\displaystyle\left(1-\frac{2GM}{c^2\,r_1}\right)c^2\,dt^2-r_1^2\left(d\theta^2+\sin^2\theta d\varphi^2\right)\\[6mm]
& &\displaystyle-\frac{1}{1-\frac{2GM}{c^2\,r_1}}\left(dr_1-\frac{\dot{a}\,r_1}{a\,c}\left(1-\frac{2GM}{c^2\,r_1}\right)^{\frac{\alpha}{2}+\frac{1}{2}}c\,dt\right)^2\ .
\ea
\lb{g_generic}
\ee
Here we are considering $\alpha$ to be a (real) constant. This metric ansatz is asymptotically
isotropic at spatial infinity, being only globaly isotropic for $\alpha=0$ which corresponds
to the McVittie metric~(\ref{g_McV}). As for the CSC metric~(\ref{g_cosm_SC}) it corresponds to
$\alpha=1$. By direct inspection this metric has the desired properties, either in the massless
limit and at spatial infinity converges asymptotically to the FLRW metric~(\ref{g_FRW_1}), in the
static limit $(\dot{a}\to 0)$ coincides with the Schwarzschild metric~(\ref{g_SC}) and it has
Lorentzian measure $\sqrt{-g}=r^2\sin\theta$ coinciding, for all space-time, with the measure of both
SC metric and FLRW metric. Hence this metric describes local Lorentz space-time in between
the Schwarzschild horizon $r_1=r_{1.\mathrm{SC}}=2GM/c^2$ and the corrected cosmological
horizon corresponding to the solution of the equation $l_H/r_1=(1-2GM/(c^2\,r_1))^{(\alpha+1)/2}$.
Assuming the weak field approximation~(\ref{weak_field}) and taking the solution for the
cosmological horizon to first order in the gravitational field $U_{\mathrm{SC}}$~(\ref{U_SC}),
we obtain that space time is locally Lorentz in the range
\be
r_1\in\left]\frac{2GM}{c^2},l_H+\frac{(\alpha+1)GM}{c^2}\right[\ .
\lb{LL}
\ee

There is one more property of this metric that has relevant physical implications.
We note that by taking the constant parameter $\alpha$ to infinity we recover the Schwarzschild
metric. As already pointed out this feature allows to fine-tune the expansion effects in local
systems, outside the Schwarzschild radius the factor $(1-U_{\mathrm{SC}})$ increases
monotonically with $r_1$ up to spatial infinity being strictly less than
unity (being null at the SC radius and unity at spatial infinity). Hence for relatively
large positive values of $\alpha$ the expansion effects are highly suppressed near the massive object
being still relevant for large values of the radial coordinate $r_1$, at spatial infinity the metric
still converges to the FLRW metric~(\ref{g_FRW_1}). It is therefore physically intuitive to
expect relatively high values for this parameter, expansion effects are for small scale
astrophysical systems (for instance the solar system) negligible within the experimental accuracy
and the SC metric~(\ref{g_SC}) describe to a very high accuracy these systems.

Next we analyze the singularities, as well as the asymptotic behavior of the scalar curvature $R_\alpha$
and the curvature invariant ${\mathcal{R}}_\alpha$ at the Schwarzschild horizon $r_1=r_{1.\mathrm{SC}}$
and at the center of mass $r_1=0$, computing the allowed ranges of the parameter $\alpha$ for
which the SC horizon is singularity free, space-time is asymptotically flat at this horizon
(such that the ansatz converges asymptotically to the SC metric) and the SC mass pole
is maintained at the origin (such that the mass inside a shell of finite radius is finite).

\subsection{Singularities and Curvature at the Schwarzschild Radius: Lower $\alpha$ Bounds\lb{sec.alpha_r_sc}}

Depending on the value of the
parameter $\alpha$, the locally anisotropic metric~(\ref{g_generic}) has a space-time singularity
at the Schwarzschild radius $r_1=2GM/c^2$, this is the case for $\alpha=0$
and $\alpha=1$ corresponding to the metrics~(\ref{g_McV}) and~(\ref{g_cosm_SC}) already discussed
in section~\ref{sec.matter}. Although in the previous section we have assumed a positive exponent
of the shift function, for completeness of our analysis, we are proceeding with the analysis for
the full possible range of the parameter $\alpha\in]-\infty,+\infty[$.
The Ricci scalar $R_\alpha$ (scalar curvature) and the curvature invariant ${\mathcal{R}}_\alpha$ are
given, respectively, in equations~(\ref{A.generic_R}) and~(\ref{A.generic_RR}) of
appendix~\ref{A.generic}. By direct inspection of these expressions, we obtain
the regularity and asymptotic leading expressions of these quantities at the Schwarzschild radius,
$r_1\to r_{1.\mathrm{SC}}=2GM/c^2$, as listed in table~\ref{table.R_generic} and
table~\ref{table.RR_generic}, respectively, for the several distinct ranges of the parameter $\alpha$.
\begin{table}[ht]
\begin{center}
\begin{tabular}{lcc}
$\alpha$& $R_\alpha$ regularity & $R_\alpha$ leading asymptotic expressions\\
 &for $r_1\to r_{1.\mathrm{SC}}$ &for $r_1\sim r_{1.\mathrm{SC}}$\\\hline\hline\\[-2mm]
$\displaystyle\alpha\in\left]+3,+\infty\right[$&finite&$R_\alpha(r_{1.\mathrm{SC}})=R_{\mathrm{SC}}(r_{1.\mathrm{SC}})=0$\\[6mm]
$\displaystyle\alpha=+3$&finite&$ R_\alpha(r_{1.\mathrm{SC}})=3\left(\left(\frac{\dot{a}}{a\,c}\right)^2-\frac{\ddot{a}}{a\,c^2}\right)$\\[6mm]
$\displaystyle\alpha\in[-1,3[$&divergent&$\displaystyle \sim\left(1-\frac{2GM}{r_1c^2}\right)^{\frac{\alpha}{2}-\frac{3}{2}}$\\[6mm]
$\displaystyle\alpha\in]-\infty,-1[$&divergent&$\displaystyle \sim\left(1-\frac{2GM}{r_1c^2}\right)^{\alpha-1}$\\
\\\hline
\end{tabular}
\caption{\it \small Regularity and asymptotic values of the Ricci scalar $R_\alpha$~(\ref{A.generic_R}) (the scalar curvature) for expanding coordinates near the Schwarzschild radius $r_1\sim 2GM/c^2$.\lb{table.R_generic}}
\end{center}
\end{table}
\begin{table}[ht]
\begin{center}
\begin{tabular}{lcc}
$\alpha$& ${\mathcal{R}}_\alpha$ behaviour & ${\mathcal{R}}_\alpha$ leading assymptotic expressions\\
 &for $r_1\to r_{1.\mathrm{SC}}$ &for $r_1\sim r_{1.\mathrm{SC}}$\\\hline\hline\\[-2mm]
$\displaystyle\alpha\in\left]+3,+\infty\right[$&finite&${\mathcal{R}}_\alpha(r_{1.\mathrm{SC}})={\mathcal{R}}_{\mathrm{SC}}(r_{1.\mathrm{SC}})=0$\\[6mm]
$\displaystyle\alpha=+3$&finite&${\mathcal{R}}_\alpha(r_{1.\mathrm{SC}})=12\left(\frac{c^2}{2GM}\right)^4$\\[4mm]
 & & $-12\left(\frac{c^2}{2GM}\right)^2\left(\left(\frac{\dot{a}}{a\,c}\right)^2-\frac{\ddot{a}}{a\,c^2}\right)+9\left(\left(\frac{\dot{a}}{a\,c}\right)^2-\frac{\ddot{a}}{a\,c^2}\right)^2$\\[6mm]
$\displaystyle\alpha\in[+1,3[$&divergent&$\displaystyle \sim\left(1-\frac{2GM}{r_1c^2}\right)^{\alpha-3}$\\[6mm]
$\displaystyle\alpha\in]-\infty,1[/\{0\}$&divergent&$\displaystyle \sim\left(1-\frac{2GM}{r_1c^2}\right)^{2\alpha-4}$\\[6mm]
$\displaystyle\alpha=0$&divergent&$\displaystyle \sim\left(1-\frac{2GM}{r_1c^2}\right)^{-1}$\\
\\\hline
\end{tabular}
\caption{\it \small Regularity and asymptotic leading expressions for the curvature invariant ${\mathcal{R}}_\alpha$~(\ref{A.generic_RR})
for expanding coordinates near the Schwarzschild radius $r_1\sim 2GM/c^2$.\lb{table.RR_generic}}
\end{center}
\end{table}

For $\alpha<3$ the Ricci scalar $R$ (the scalar curvature) has a singularity at the Schwarzschild radius,
while for $\alpha=3$ it is finite and for $\alpha>3$ it is null. It is explicitly checked from
the curvature invariant $\mathcal{R}$ that for $\alpha<3$ the Schwarzschild radius is singular. 
Based in these results we conclude that only for $\alpha\ge 3$ space-time is free of singularities
at the event horizon. Neither the McVittie metric~(\ref{g_McV}) nor
the CSC metric~(\ref{g_cosm_SC}) discussed in the previous section~\ref{sec.matter} obey this bound.
A space-time singularity at the Schwarzschild radius is clearly a significant deviation from the Schwarzschild metric meaning that, near the massive object, the expansion effects are dominant with
respect to the matter effects described by the original Schwarzschild metric. Such result is clearly
in disagreement with the existing experimental evidence, the Schwarzschild
metric~(\ref{g_SC}) has been widely applied in small scale astrophysical systems being in very close
agreement with experimental data as well as with most tests of the General Relativity Newton law,
in particular in the solar system and earth based experiments. Hence it is expected that near the
massive objects we should obtain, at most, small corrections to the Schwarzschild metric, a
divergence is clearly not a small
correction. Also we expect that, to exist, any space-time singularity to lie at the origin (as it does
for the SC metric). Given this discussion we conclude that the bound $\alpha\ge 3$ must be obeyed.

This bound can also be justified at the level of the metric by noting that in the
limit $r_1\to r_{1.\mathrm{SC}}$ the asymptotic leading expressions for the metric components are
\be
\ba{rcl}
g_{00}&=&\displaystyle \left(1-\frac{2GM}{r_1\,c^2}\right)^1-\frac{\dot{a}^2\,r_1^2}{a^2\,c^2}\left(1-\frac{2GM}{r_1\,c^2}\right)^\alpha\sim 0^1+0^\alpha\\[6mm]
g_{0r}&=&\displaystyle -\frac{\dot{a}\,r_1}{a\,c}\left(1-\frac{2GM}{r_1\,c^2}\right)^{\frac{\alpha}{2}-\frac{1}{2}}\sim 0^{\frac{\alpha}{2}-\frac{1}{2}}\\[6mm]
g_{rr}&=&\displaystyle -\frac{1}{1-\frac{2GM}{r_1\,c^2}}\sim \frac{1}{0^{1}}\ .
\ea
\ee
A necessary condition for the metric to converge to the SC metric at the Schwarzschild radius is to demand that, in this limit, all terms containing corrections due to expanding background vanish at least as fast
as the terms corresponding to the original metric components of the Schwarzschild metric~(\ref{g_SC}).
From the component $g_{0r}$ we obtain the lower bound $\alpha/2-1/2\ge 1$ which coincides with the
above bound, $\alpha\ge 3$.

For all values of $\alpha$, at spatial infinity ($r_1\to+\infty$), the curvature
converges asymptotically to the FLRW curvature
\be
\lim_{r_1\to\infty} R_\alpha=-6\left(\left(\frac{\dot{a}}{a\,c}\right)^2+\frac{\ddot{a}}{a\,c^2}\right)=R_{\mathrm{FLRW}}=6(q-1)H^2\ ,
\ee
The expression for $R_{\mathrm{FLRW}}$ is given in equation~(\ref{A.FRW_R}) of appendix~\ref{A.FRW}
and we have replaced the time derivatives of the scale factor by the time-dependent Hubble rate $H$
and deceleration parameter $q$. This result is expected, by construction the locally anisotropic
metric~(\ref{g_generic}) converges asymptotically to the FLRW metric~(\ref{g_FRW_1}). The asymptotic value of the curvature at spatial infinity will be positive for $q>1$, null for $q=1$ and negative for $q<1$.

Related to the previous discussions, there is one more relevant issue that we want to address, the sign
of the curvature near the Schwarzschild radius and the value of its spatial derivative. This analysis
allows to determine the asymptotic behavior of the curvature near massive bodies which is related
to the gravitational interactions for small spatial scales (small astrophysical scales as planetary
systems). We will only consider the case for $\alpha\ge 3$ for which the SC radius is non-singular
as we have shown. Close to the massive body, there are several distinct asymptotic behaviors
which depend on the value of the deceleration parameter $q$. To identify these distinct asymptotic
regimes let us consider the leading term for the spatial derivative of the curvature near the SC radius
\be
\left.\frac{\partial R_\alpha}{\partial r_1}\right|_{r_1\sim r_{1.\mathrm{SC}}}\sim\frac{3(1+q_0)\,H^2\,(GM)^2}{c^6\,r_1^2}\,\left(\frac{\alpha}{2}-\frac{3}{2}\right)\,\left(1-\frac{(6-\alpha)\,GM}{6c^2\,r_1}\right)\left(1-\frac{2GM}{c^2\,r_1}\right)^{\frac{\alpha}{3}-\frac{5}{2}}\ .
\lb{dR_generic}
\ee
The regularity and asymptotic values for this expression at the event horizon, for $\alpha\ge 3$,
are listed in table~\ref{table.dR_generic}.
\begin{table}[ht]
\begin{center}
\begin{tabular}{lcc}
$\alpha$& $R'_\alpha$ regularity & $R'_\alpha$ asymptotic value\\
 &for $r_1\to r_{1.\mathrm{SC}}$ &for $r_1\to r_{1.\mathrm{SC}}$\\\hline\hline\\[-2mm]
$\displaystyle\alpha\in\left]+5,+\infty\right[$&finite&$R'_\alpha(r_{1.\mathrm{SC}})=0$\\[6mm]
$\displaystyle\alpha=+5$&finite&$ R'_\alpha(r_{1.\mathrm{SC}})=5\left(\left(\frac{\dot{a}}{a\,c}\right)^2-\frac{\ddot{a}}{a\,c^2}\right)$\\[4mm]
$\displaystyle\alpha\in]3,5[$&divergent&$\displaystyle R'_\alpha(r_{1.\mathrm{SC}})=-\infty$ for $q>-1$\\[6mm]
 &finite&$\displaystyle R'_\alpha(r_{1.\mathrm{SC}})=0$ for $q=-1$\\[6mm]
 &divergent&$\displaystyle R'_\alpha(r_{1.\mathrm{SC}})=+\infty$ for $q<-1$\\[6mm]
$\displaystyle\alpha=3$&finite&$ R'_\alpha(r_{1.\mathrm{SC}})=-3\left(\left(\frac{\dot{a}}{a\,c}\right)^2+\frac{\ddot{a}}{a\,c^2}\right)$\\
\\\hline
\end{tabular}
\caption{\it \small Regularity and asymptotic values for the spatial derivative of the curvature
$R'_\alpha$~(\ref{dR_generic}) for expanding coordinates near the Schwarzschild radius
$r_1\sim 2GM/c^2$.\lb{table.dR_generic}}
\end{center}
\end{table}
From these asymptotic values we conclude that space-time is asymptotically flat near
the Schwarzschild radius ($R_\alpha(r_{1.\mathrm{SC}})=R'_\alpha(r_{1.\mathrm{SC}})=0$)
for $\alpha>5$ and all values of $q$ or for $\alpha> 3$ and $q=-1$ (for this case the curvature is negative up to spatial infinity).

With respect to the curvature behavior in the radial coordinate
range $r_1\in]r_{1.\mathrm{SC}},+\infty[$ we obtain that: for $\alpha>5$, depending on the
value of $q$, we will have distinct behaviors for the curvature , for $q\le -1$ the curvature
is strictly negative,
for $q\ge+1$ it is strictly positive and for $q\in]-1,+1[$ it is positive near the SC radius $r_1=r_{1.\mathrm{SC}}$, with growing $r_1$ it grows to a maximum and then decreases becoming
negative and converging to the FLRW value at spatial infinity;
for $\alpha\in]3,5]$ the curvature is null at the SC radius, being its derivative negative or null
for $q<-1$ or $q=-1$, respectively (for which cases the curvature is negative up to spatial infinity),
or positive for $q>1$ (for which case the curvature is positive until spatial infinity); for the
particular case of $\alpha=3$, at the SC radius $r_{1.\mathrm{SC}}$, the curvature is negative, null or positive for $q<-1$, $q=-1$ and $q>-1$, respectively, and its derivative is negative, null or positive for $q<1$, $q=1$ and $q>1$, hence for the deceleration parameter range $q\in]-1,1[$ the curvature will be
positive near the SC radius and negative for large values of the radial coordinate (up to spatial infinity).

Specifically in the range of the deceleration parameter $q\in]-1,1[$, which contains as a particular
case the estimative for today's value of this parameter $q_0=-0.589$, and for $\alpha\ge 3$ the curvature
will always be positive near the massive object becoming negative with growing radial coordinate.
The value of the radial coordinate for which the curvature changes sign is, approximately to first
order in the gravitational field $U_{\mathrm{SC}}$ and the parameter $\alpha$,
\be
R_\alpha=0\ \Rightarrow\ r_1\approx\frac{2GM}{c^2}\,\frac{2\alpha(2-q)+3(1+q)}{6(1-q)}\ \mathrm{for}\ q\in]-1,+1[\ ,\ \alpha\ge 3\ .
\lb{r_R_0}
\ee
This expression is derived considering an expansion of the curvature in the weak field
approximation being valid only for $\alpha U_{\mathrm{SC}}=2\alpha GM/(c^2\,r_1)< 1$. We note that
for relatively high values of the parameter $\alpha$ for which $\alpha>1/U_{\mathrm{SC}}$ it
is required to compute numericaly the solution of the equation $R_\alpha=0$ by considering the
exact expressions for the curvature.

Resuming, from the analysis of the space-time singularities at the Schwarzschild radius carried
in this section, we have obtained the following lower bounds for the parameter $\alpha$
\be
\ba{rcll}
\alpha&\ge& 3&\Rightarrow\ \ \mathrm{no\ space-time\ singularities\ at}\ r_1=r_{1.\mathrm{SC}}\ ,\\[6mm]
\alpha&>& 5&\Rightarrow\ \ \mathrm{space-time\ asymptotically\ flat\ at}\ r_1=r_{1.\mathrm{SC}}\ .
\ea
\lb{alpha_bounds}
\ee
Some examples for several values of the parameter $\alpha$ and the deceleration parameter evaluated
at the reference time $t_0=0$, $q_0=q(0)$ are presented in figure~\ref{fig.R_alpha}. 
\fig{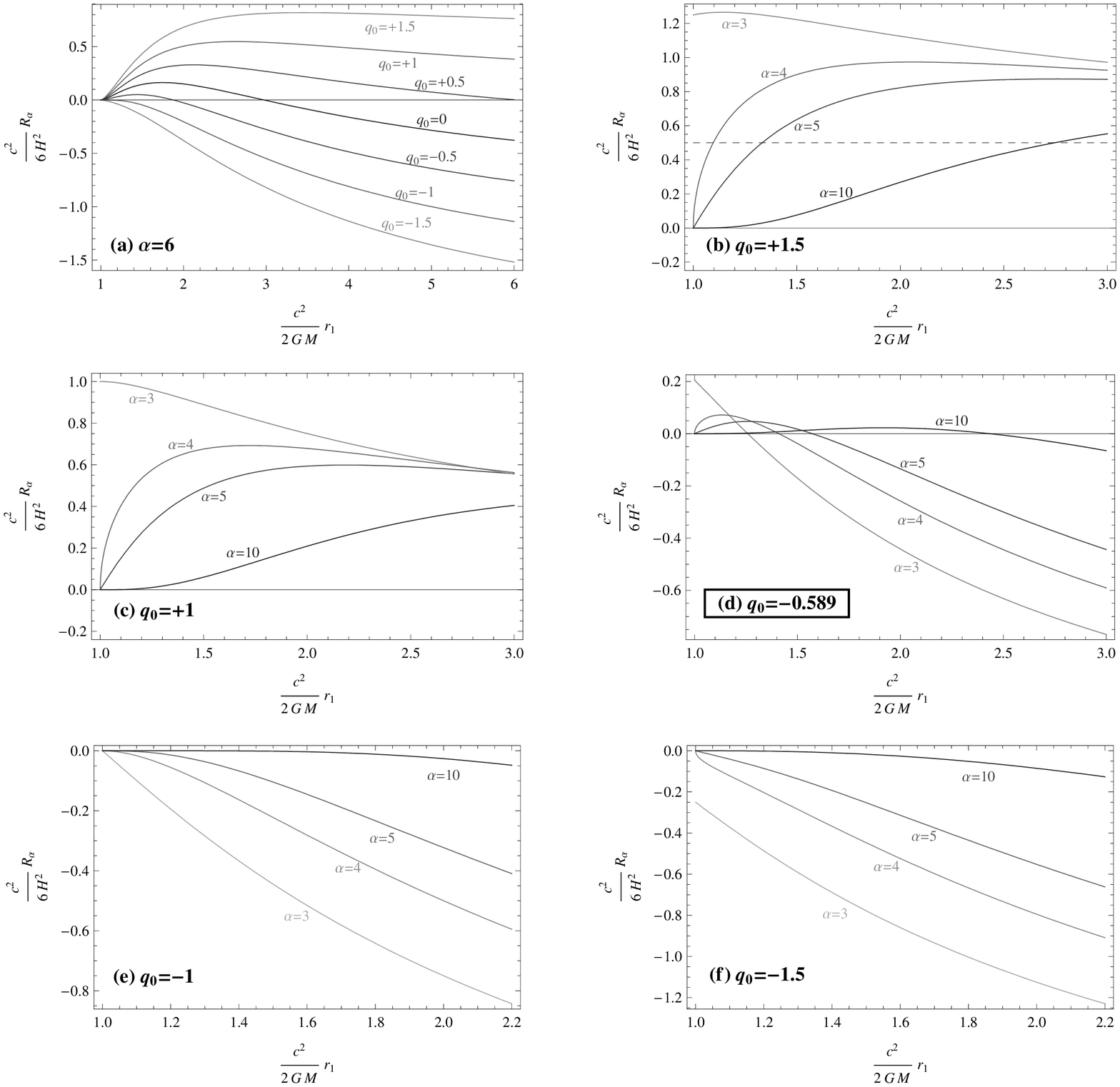}{150mm}{Curvature $R$~(\ref{A.generic_R}) as a function of $r_1$ for several
values of the deceleration parameter $q_0$ evaluated at the reference time
$t_0$ and parameter $\alpha$ (see table~\ref{table.R_generic} and table~\ref{table.dR_generic}
for comparison):\hfill\break
\ {\bf(a)} example for $\alpha=6$ and several values of $q_0=-1.5,-1,-0.5,0,+0.5,+1,+1.5$, at the
SC radius $R=R'=0$ for all plotted lines and, at spatial infinity, all converge to $q_0-1$;\hfill\break
\ {\bf(b)} an example for $q_0>1$ for $\alpha=3,4,5,10$; at the SC radius only for $\alpha=3$,
$R>0$, $R'>0$ and only for $\alpha>5$, $R=R'=0$, the plotted lines converge at spatial infinity to $q_0-1=0.5$ represented by a dotted horizontal line and the curvature is positive for $r_1\in]r_{1.\mathrm{SC}},+\infty[$;\hfill\break
{\bf(c)} an example for the particular case $q_0=+1$, at the SC radius only for $\alpha=3$,
$R>0$, $R'=0$ and only for $\alpha>5$, $R=R'=0$, at spatial infinity the plotted lines converge to $q_0-1=0$ and the curvature is positive for $r_1\in]r_{1.\mathrm{SC}},+\infty[$;\hfill\break
{\bf(d)} estimative for today's value of the deceleration parameter $q_0=-0.589$~(\ref{H_q0}) for $\alpha=3,4,5,10$ which is an example of $q\in]-1,+1[$, at the SC radius only for $\alpha=3$,
$R>0$, $R'=0$ and only for $\alpha>5$, $R=R'=0$, at spatial infinity the plotted lines converge to $q_0-1=-1.589$, the
curvature is positive near the massive body and negative at spatial infinity, changing sign at a
finite value of $r_1$~(\ref{r_R_0});\hfill\break
{\bf(e)} the particular case of $q_0=-1$ for $\alpha=3,4,5,10$, at the SC radius only
for $\alpha=3$, $R=0$, $R'< 0$ and for all other values of $\alpha> 3$, $R=R'=0$, at spatial infinity the
plotted lines converge to $q_0-1=-2$, the curvature is negative
in the range $r_1\in]r_{1.\mathrm{SC}},+\infty[$;\hfill\break
{\bf (f)} an example for $q_0<-1$ case for $\alpha=3,4,5,10$, at the SC radius
only for $\alpha=3$, $R<0$, $R'<0$ and only for $\alpha>5$, $R=R'=0$, at spatial infinity the
plotted lines converge to $q_0-1=-2.5$, the curvature is always negative in the range $r_1\in]r_{1.\mathrm{SC}},+\infty[$.}{fig.R_alpha}
\clearpage

These results are physically appealing, in particular for today's value of the
deceleration parameter $q_0=-0.589$ and for $\alpha>5$, near the massive bodies space-time is
asymptotically flat. Hence the usual gravitation laws for flat backgrounds obtained either
from General Relativity or the Newtonian limit approximation should be asymptotically (or at least
approximately retrieved to a very good accuracy). We recall that these laws are experimentally
well established being tested to a very high precision for small scale spatial scales systems
such as the solar system. Far away from the massive objects, the expansion effects dominate and the
curvature converges asymptotically to the FLRW curvature which is also experimentally well established
result from large scale observations (for instance the Hubble law). In between these two
asymptotic regions there is a transition region for which the gravitational interactions
will be modified such that an interpolation between the two asymptotic limiting cases is obtained.
The parameter $\alpha$ fine-tunes the transition between these distinct regimes.

We will study these three regimes in detail later on, for now let us complete our
analysis of the space-time singularities by analyzing the curvature $R_\alpha$ and curvature
invariant ${\mathcal{R}}_\alpha$ at the center of mass, the origin of the coordinate frame $r_1=0$.

\subsection{Singularities at the Center of Mass: Mass Divergences and Upper $\alpha$ Bounds\lb{sec.alpha_r_0}}

Having already discussed the space-time singularities at the Schwarzschild radius, it is
still necessary to analyze the singularities at the center of mass $r_1=0$. Independently of
the value of the parameter $\alpha$, the mass pole divergence for the Schwarzschild
metric~(\ref{R_SC}), is also present for the locally anisotropic metric~(\ref{g_generic}).
However, depending on the value of $\alpha$, there will exist other contributions for the
singularity at the origin. We list the regularity and the asymptotic
leading terms at the origin for the Ricci scalar $R_\alpha$ (curvature) and the curvature invariant
${\mathcal{R}}_\alpha$ in table~\ref{table.R_generic_0} and table~\ref{table.RR_generic_0}.
\begin{table}[ht]
\begin{center}
\begin{tabular}{lcc}
$\alpha$& $R_\alpha$ regularity & $R_\alpha$ leading assymptotic expression\\
 &for $r_1\to 0$ &for $r_1\sim 0$\\\hline\hline\\[-2mm]
$\displaystyle\alpha\in\left]0,+\infty\right[$&divergent&$\displaystyle\sim\frac{1}{r_1^\alpha}$\\[6mm]
$\displaystyle\alpha\in]-\infty,0]$&mass pole&$\displaystyle R_\alpha(0)=R_{\mathrm{SC}}(0)=-8\pi\frac{GM}{c^2}\delta^{(3)}(0)$\\
\\\hline
\end{tabular}
\caption{\it \small Regularity and asymptotic leading expressions for the Ricci scalar $R_\alpha$~(\ref{A.generic_R}) (curvature) for expanding coordinates near the center of mass $r_1\sim 0$.\lb{table.R_generic_0}}
\end{center}
\end{table}
\begin{table}[ht]
\begin{center}
\begin{tabular}{lcc}
$\alpha$& ${\mathcal{R}}_\alpha$ behavior & ${\mathcal{R}}_\alpha$ leading asymptotic expression\\
 &for $r_1\to 0$ &for $r_1\sim 0$\\\hline\hline\\[-2mm]
$\displaystyle\alpha\in\left]+3,+\infty\right[$&divergent&$\displaystyle\sim\frac{1}{r_1^{2\alpha}}$\\[6mm]
$\displaystyle\alpha\in\left[-9,+3\right]$&divergent&$\displaystyle\sim\frac{1}{r_1^6}$\\[6mm]
$\displaystyle\alpha\in\left]-\infty,-9\right[$&divergent&$\displaystyle{\mathcal{R}}_\alpha(r_1\sim 0)={\mathcal{R}}_{\mathrm{SC}}(r_1\sim 0)\sim\frac{1}{r_1^6}$\\
\\\hline
\end{tabular}
\caption{\it \small Regularity and asymptotic leading expressions of the curvature invariant
${\mathcal{R}}_\alpha$~(\ref{A.generic_R}) for non-expanding coordinates
near the center of mass $r_1\sim 0$.\lb{table.RR_generic_0}}
\end{center}
\end{table}

At the center of mass $r_1=0$, for $\alpha\leq 0$, the Ricci scalar $R_\alpha$ coincides with the
Schwarzschild curvature~(\ref{R_SC}), while for $\alpha> 0$ it is divergent by
the power of the radial coordinate $r_1^{-\alpha}$. As for the curvature
invariant ${\mathcal{R}}_\alpha$, for $\alpha<-9$, it asymptotically coincides with
the Schwarzschild curvature invariant ${\mathcal{R}}_{\mathrm{SC}}\sim r_1^{-6}$~(\ref{R_SC}),
while for $\alpha\in[-9,+3]$ the leading divergent term in the neighbourhood of $r_1=0$
is still the SC curvature invariant $\sim r_1^{-6}$, however the divergence at the center
of mass has several other contributions of lower $1/r_1$ powers. For $\alpha> 3$ the curvature invariant leading divergent term is $r_1^{-2\alpha}$, hence a more severe divergence than the SC mass pole
divergence such that for this range of the parameter the total mass at the origin diverges.

This discussion is not complete without explicitly computing the mass-energy density
and the mass content of the space-time described by the locally anisotropic
metric~(\ref{g_generic}). In the commoving frame of
the background cosmological fluid the mass-energy density is, generally, $\rho=T_{\hat{0}\hat{0}}/c^2$,
given in terms of the component $T_{\hat{0}\hat{0}}$ of the stress-energy tensor in the Lorentz
frame (Cartan frame). We give the details of this computation in section~\ref{sec.gen_stress},
for now let us quote the expression for the total mass-energy density
\be
\ba{rcl}
\rho_{\mathrm{tot}(\alpha)}(r_1)&=&\displaystyle \rho_{\mathrm{SC}}+\rho_\alpha\ ,\\[6mm]
\rho_{\mathrm{SC}}&=&\displaystyle M\,\delta^{(3)}(\vb{r_1})\ ,\\[6mm]
\rho_\alpha&=&\displaystyle+\frac{c^2}{8\pi\,G}\,\left(\frac{\dot{a}}{a\,c}\right)^2\,\left(3+\frac{2(\alpha-3)GM}{c^2\,r_1}\right)\,\left(1-\frac{2GM}{c^2\,r_1}\right)^{\alpha-1}\ ,
\ea
\lb{rho_tot}
\ee
where we are explicitly considering the Schwarzschild mass pole contribution~(\ref{SC_T}) to the
mass-energy density such that this expression has both a contribution from the standard
point like mass, the SC mass pole, a localized distribution at $r_1=0$, and a contribution from the
background, hence an extended mass-energy distribution $\rho_\alpha$. We recall
that the background mass-energy density in the absence of the local mass $M$ is
$\rho_H=3(\dot{a}/a)/(8\pi G)$ as given in equation~(\ref{p_rho_FRW_1}) and note that for
all values of $\alpha$, $\rho_\alpha$ asymptotically converges to this mass-energy density
at spatial infinity, $\lim_{r_1\to+\infty}\rho_\alpha=\rho_H$. Hence the deformation of the
background mass-energy density is encoded in $\rho_\alpha$, such that to evaluate
the deformation (or correction) $\rho_{\mathrm{def}}$ to the background mass-energy density due
the presence of the mass $M$ is necessary to subtract the background mass-energy density $\rho_H$
\be
\rho_{\mathrm{def}(\alpha)}=\rho_{\mathrm{SC}}+\rho_{\alpha}-\rho_H\ .
\lb{rho_grav}
\ee
We note that in this discussion we are interpreting the locally anisotropic metric~(\ref{g_generic})
as a deformation (or correction) to the cosmological expanding background given by the FLRW
metric~(\ref{g_FRW_1}) due to the presence of a local central mass $M$. When deriving the ansatz
for this metric we have considered the opposite interpretation, that it is a deformation (or correction)
to the Schwarzschild metric~(\ref{g_SC}). Both these interpretations are correct being compatible with
each other, we recall that the locally anisotropic metric~(\ref{g_generic}) is interpolating between
both the FLRW and the SC metrics, hence it is necessarily a deformation (or correction) to both of them.

Generally the mass (or equivalently, the energy) within a shell of fixed radius is obtained by
integrating the mass-energy density over the respective volume.
Specifically the total space-time mass and the mass of the deformation due to the presence of
a point-like mass $M$ within a shell of fixed radius $r_1$, are given by
\be
\ba{rcl}
M_{\mathrm{tot}(\alpha)}(r_1)&=&\displaystyle M+\Delta M_{\mathrm{tot}(\alpha)}(r_1)\ ,\\[6mm]
M_{\mathrm{def}(\alpha)}(r_1)&=&\displaystyle M+\Delta M_{\mathrm{tot}(\alpha)}(r_1)-M_H(r_1)\ ,\\[6mm]
\Delta M_{\mathrm{tot}(\alpha)}(r_1)&=&\displaystyle \tilde{M}_\alpha(r_1)-\tilde{M}_\alpha(0)\ ,
\ea
\lb{M_tot}
\ee
where $M$ is the usual Schwarzschild gravitational mass contribution from the density
$\rho_{\mathrm{SC}}$, $\Delta M_{\mathrm{tot}(\alpha)}$ is the contribution due to
the expanding background when the mass $M$ is present from the density $\rho_\alpha$
and $M_H$ is the background mass in the absence of the mass $M$ corresponding to the
density $\rho_H$. Specifically we have that $\tilde{M}_\alpha(r_1)$ is given by the indefinite
volume integral of $\rho_\alpha$
\be
\tilde{M}_\alpha(r_1)=\int_{-\pi}^{\pi}d\theta\int_0^{2\pi}d\varphi\int dr_1\sqrt{-g}\,\rho_\alpha=\left(\frac{\dot{a}}{a}\right)^2\,\frac{r_1^3}{2G}\left(1-\frac{2GM}{c^2\,r_1}\right)^\alpha\ ,
\lb{MM}
\ee
and $M_H$ by the volume integral of $\rho_H$
\be
M_H(r_1)=\int_{-\pi}^\pi d\theta\int_{0}^{2\pi} d\varphi \int_0^{r_1}dr'_1\sqrt{-g}\,\rho_H=\frac{r_1^3}{2G}\left(\frac{\dot{a}}{a\,c}\right)^2\ .
\lb{M_H}
\ee
The measure is given, as usual, by $\sqrt{-g}=r_1^2\sin\theta$.

At the origin $\tilde{M}_\alpha(0)$ has three distinct behaviors which depend
on the value of $\alpha$ being greater, equal or smaller than $3$,
\be
\tilde{M}_\alpha(0)=
\left\{\ba{lcl}
\displaystyle(-1)^\alpha\infty&,&\alpha>3\ ,\\[6mm]
\displaystyle-M\,\left(\frac{\dot{a}}{a\,c}\right)^2\left(\frac{GM}{c^2}\right)^2&,&\alpha=3\ ,\\[6mm]
0&,&\alpha<3\ ,
\ea\right.
\lb{MM_0}
\ee
therefore the total mass for a shell at fixed radial coordinate $r_1$, including both the SC
mass pole contribution $M$ and the contribution of the extended deformation due to the expanding
background $\Delta M_{\mathrm{tot}(\alpha)}(r_1)$ is
\be
M_{\mathrm{tot}(\alpha)}(r_1)=
\left\{\ba{lcl}
\displaystyle\pm\infty&,&\alpha>3\ ,\\[6mm]
\displaystyle M+\left(\frac{\dot{a}}{a}\right)^2\left(\frac{M}{c^2}\left(\frac{GM}{c^2}\right)^2+\frac{1}{2G}\left(r_1-\frac{2GM}{c^2}\right)^3\right)&,&\alpha=3\ ,\\[6mm]
\displaystyle M+\frac{r_1^3}{2G}\left(\frac{\dot{a}}{a}\right)^2\left(1-\frac{2GM}{c^2\,r_1}\right)^\alpha&,&\alpha<3\ .
\ea\right.
\lb{d_M}
\ee
Hence, for $\alpha>3$, the total mass inside a shell of finite radius $r_1$ is divergent
while for $\alpha\le 3$ it is finite. This result is consistent with the analysis of the curvature invariant singularities given in table~\ref{table.RR_generic_0}, for $\alpha\le 3$ the leading
divergence of ${\mathcal{R}}_\alpha$ is a power of the radial coordinate coinciding with the
inverse volume squared $\sim r_1^{-6}$, while for $\alpha>3$ it has a more severe divergence
by a power of $\sim r_1^{-2\alpha}$. We also note that for $\alpha\le 3$ the mass pole
value at the origin is maintained,
$\lim_{r_1\to 0}M_{\mathrm{tot}(\alpha\le 3)}(r_1)=M$.

We recall that at spatial infinity the density $\rho_\alpha$ converges asymptotically to the
background density $\rho_H$ however, depending on the value of the parameter $\alpha$,
the respective mass expressions do not match this behavior. This is due to the value of
the indefinite integral $\tilde{M}_\alpha$ at the origin. Hence taking the limit of large radius
for the volume integration shells we obtain
\be
M_{\mathrm{tot}(\alpha)}(r_1\sim+\infty)=
\left\{\ba{lcl}
\displaystyle\pm\infty&,&\alpha>3\ ,\\[6mm]
\displaystyle M+M_H+M\left(\frac{\dot{a}}{a\,c}\right)^2\left(\frac{GM}{c^2}\right)^2&,&\alpha=3\ ,\\[6mm]
\displaystyle M+M_H&,&\alpha<3\ .
\ea\right.
\lb{d_M_infty}
\ee
For $\alpha>3$ the total mass is divergent, for $\alpha=3$ it corresponds to the background mass
$M_H$ plus the SC gravitational mass $M$ of the massive body plus a small correction proportional
to $H^2$ and for $\alpha<3$ it is exactly given by the background mass $M_H$ plus the SC gravitational
mass $M$.

Resuming the results obtained in this section, from the analysis of the singularities at the center
of mass, we have set the following bounds on the parameter $\alpha$
\be
\ba{rclll}
\alpha&>& 3&\Rightarrow&\mathrm{space-time\ singularities\ at}\ r_1=0\ \mathrm{diverge\ with\ the\ power}\ \sim r_1^{-2\alpha}\ ,\\[5mm]
& & & &\mathrm{total\ mass\ inside\ shell\ of\ radius}\ r_1\ \mathrm{is\ \bf divergent}\ ,\\[6mm]
\alpha&=& 3&\Rightarrow&\mathrm{space-time\ singularities\ at}\ r_1=0\ \mathrm{diverge\ with\ the\ power}\ \sim r_1^{-6}\ ,\\[5mm]
& & & &\mathrm{total\ mass\ inside\ shell\ of\ radius}\ r_1\ \mathrm{is\ \bf finite}\\[5mm]
& & & &\mathrm{being\ above}\ M+M_H\ \mathrm{for}\ r_1\sim+\infty\ ,\\[6mm]
\alpha&<& 3&\Rightarrow&\mathrm{space-time\ singularities\ at}\ r_1=0\ \mathrm{diverge\ with\ the\ power}\ \sim r_1^{-6}\ ,\\[5mm]
& & & &\mathrm{total\ mass\ inside\ shell\ of\ radius}\ r_1\ \mathrm{is\ \bf finite}\\[5mm]
& & & &\mathrm{being}\ M+M_H\ \mathrm{for}\ r_1\sim+\infty\ .
\ea
\lb{alpha_bounds_0}
\ee
Then we conclude that for the locally anisotropic metric~(\ref{g_generic}) only for $\alpha\le 3$
the origin is free of essential singularities such that the total mass within a shell
of finite radius is finite. Hence, in order to avoid mass divergences, this bound should
be imposed. In addition we note that only for $\alpha<3$ at spatial infinity we recover
that the total mass is the sum of the background mass $M_H$ with the SC gravitational mass $M$
while for $\alpha=3$ there is a small positive correction, $H^2G^2M^3/c^3$. If we
require to exactly maintain the relation between total mass and gravitational mass in the
universe the stronger bound $\alpha<3$ should be considered, however we note that this correction
is for most purposes negligible. These bounds are the opposite of the ones obtained
previously when analysing the space-time singularities at the Schwarzschild 
radius, $r_1=r_{1.\mathrm{SC}}$~(\ref{alpha_bounds}).
In the next section we discuss and deal with this (in)compatibility. 

As a final remark let us note that within the Schwarzschild radius the
metric~(\ref{g_generic}) becomes, generally,
complex, except for odd integer values of the parameter $\alpha$ for which a signature flip is
obtained. The locally anisotropic metric~(\ref{g_generic}) does not describe a local Lorentz
space-time inside the event horizon~(\ref{LL}), this characteristic is inherited from the Schwarzschild
metric and it is attributed to the coordinate choice, the Schwarzschild coordinates only
describe the physical space-time outside of the event-horizon. In order to fully
describe the horizon inner region and properly compute geodesic paths it is necessary to consider
other coordinate choice such as Novikov coordinates or Kruskal-Szekeres coordinates~\cite{KS}.
We are not further discussing this topic here, for further details see for instance section~31.4 of~\cite{Gravitation}. We also note that the above expressions for the total mass $M_{\mathrm{tot}(\alpha)}$~(\ref{M_tot}), $\tilde{M}_\alpha$~(\ref{MM}) and $\Delta M_{\mathrm{tot}(\alpha)}$~(\ref{d_M})
are well defined as long as we consider the integration shell of fixed radius $r_1$ outside the SC
event horizon, i.e. $r_1>r_{1.\mathrm{SC}}=2GM/(c^2\,r_1)$, and that for $\alpha\ge 3$ the limit
$\lim_{r_1\to 0}\tilde{M}_{\mathrm{tot}(\alpha)}=\tilde{M}_{\mathrm{tot}(\alpha)}(0)$~(\ref{MM_0})
is also well defined being a real value which is enough for mass computation purposes and evaluation of
singularities.

\subsection{The Ansatz II: Removing Essential Singularities at the Center of Mass\lb{sec.ansatz_II}}

The results of the two previous sections raise a problem that must be solved.
In section~\ref{sec.alpha_r_sc} we have concluded that only for $\alpha\ge 3$, at the Schwarzschild
radius $r_1=r_{\mathrm{SC}}$, space-time is singularity free~(\ref{alpha_bounds}) while in
section~\ref{sec.alpha_r_0} we concluded, from the analysis of the singularities at the origin,
that only for $\alpha\le 3$ the total mass inside a shell of finite radius is
finite~(\ref{alpha_bounds_0}). Compatibility between these
two bounds leave us with the only possible value for this parameter to be $\alpha=3$.
This value is theoretically consistent however, based in the argument that, when compared to
experimental data, the Schwarzschild metric describes astrophysical gravitational systems to a
very high precision, we expect that the deformation of this metric, as encoded in the
ansatz~(\ref{g_generic}), has negligible corrections for short
spatial scales. Namely, in agreement with the General Relativity and the Newtonian gravitational laws,
we may expect that near massive bodies, space-time is asymptotically flat. This requirement
corresponds to the lower bound $\alpha>5$~(\ref{alpha_bounds}) which are clearly incompatible
with the upper bound $\alpha\leq 3$~(\ref{alpha_bounds_0}).

We will next solve this incompatibility by modifying our metric ansatz~(\ref{g_generic}).
So far we have considered the parameter $\alpha$ to be a constant, generally we may assume
it to be a space-time dependent function. We recall that in the physical system being addressed,
matter in an expanding background, there are only two dimensionless quantities, the FLRW expansion
shift function $N^1_{\mathrm{FLRW}}=r_1/l_H$~(\ref{N_1_FRW_1}) and the SC gravitational potential
$U_{\mathrm{SC}}=2GM/(c^2\,r_1)$~(\ref{U_SC}). As discussed in the previous
section~\ref{sec.alpha_r_sc} and section~\ref{sec.alpha_r_0} we further note that the
parameter $\alpha$ fine-tunes the expansion effects relatively close to the massive objects, hence
its value is mostly relevant for relatively small radial scales (specifically for $r_1\ll l_H$)
for which the matter effects are dominant. As for large radial scales (specifically for $r_1\sim l_H$) the expansion effects are dominant, the gravitational potential $U_{\mathrm{SC}}$ decreases with the radial coordinate ($\lim_{r_1\to+\infty} U_{\mathrm{SC}}=0$)
and the expansion shift function increases with the radial coordinate ($N^1_{\mathrm{FLRW}}\sim r_1$)
such that the factor $(1-U_{\mathrm{SC}})^\alpha$ converges asymptotically to unity at spatial
infinity independently of the value of $\alpha$ (as long as $\alpha$ is finite). It is therefore
consistent to expect a space-time dependent parameter $\alpha$ to depend only on the dimensionless gravitational field. Hence let us assume that its dependent on $U_{\mathrm{SC}}$ is linear,
specificaly
\be
\alpha(r_1)=\alpha_0+\alpha_1\,U_{\mathrm{SC}}(r_1)=\alpha_0+\alpha_1\,\frac{2GM}{c^2\,r_1}\ .
\lb{alpha_r_1}
\ee
This exponent maintains spherical symmetry of the metric~(\ref{g_generic}) converging
asymptotically to the constant coefficient $\alpha_0$ at spatial infinity. For relatively
small values of the radial coordinate $r_1$, near the massive body, the effects of its
dependence on the gravitational potential $U_{\mathrm{SC}}$ become relevant. Hence the
coefficient $\alpha_1$ fine-tunes the exponent dependence on the gravitational field
near the massive object.

Next we discuss the effects of this new term in the exponent.
The connections, curvature and curvature invariant for a space-dependent exponent $\alpha$ are
given in equations~(\ref{A.gen_connections_r}),~(\ref{A.gen_R_r}) and~(\ref{A.generic_RR_r}) of
Appendix~\ref{A.generic}. With respect to the singularities at the Schwarzschild radius we note
that, for this value of the radial coordinate, the exponent is
$\alpha(r_1=r_{1.\mathrm{SC}})=\alpha_0+\alpha_1$ such that the same bounds on $\alpha$
as expressed in~(\ref{alpha_bounds}) are valid for the exponent evaluated at the event horizon, \be
\ba{rcll}
\alpha(r_{1.\mathrm{SC}})=\alpha_0+\alpha_1&\ge& 3&\Rightarrow\ \mathrm{no\ space-time\ singularities\ at}\ r_1=r_{1.\mathrm{SC}}\ ,\\[6mm]
\alpha(r_{1.\mathrm{SC}})=\alpha_0+\alpha_1&>&5&\Rightarrow\ \mathrm{space-time\ is\ asymptotically\ flat\ at}\ r_1=r_{1.\mathrm{SC}}\ .
\ea
\lb{alpha_bounds_0_1}
\ee
The proof of these bounds is straight forwardly obtained by noting that $\lim_{r_1\to r_{1.\mathrm{SC}}} (1-U_{\mathrm{SC}})^p\log(1-U_{\mathrm{SC}})=0$ for $p>0$ such that all terms in the curvature
and curvature invariance containing the derivative of the exponent $\alpha'$ vanish at the event
horizon.

As for the singularities at the center of mass, we retrieve a constant exponent
for $\alpha_1=0$ corresponding to the case analyzed in the previous section with
$\alpha=\alpha_0$, for $\alpha_1>0$ the divergences at the origin, independently of the value of
$\alpha_0$, are more severe than for the case of constant $\alpha$ such that
we always obtain a divergence by a positive infinite power of $1/r_1$
\be
{\mathcal{R}}_{(\alpha_0,\alpha_1>0)}(r_1\sim 0)\sim \left(\frac{1}{r_1}\right)^{+\frac{|\alpha_1|}{r_1}}\ .
\lb{RR_alpha_0_1_+}
\ee
Finally, for $\alpha_1<0$, the exponent $\alpha(r)$ diverges at the center of mass, the origin of the
coordinate frame, to $\lim_{r_1\to 0}\alpha(r_1)=-\infty$. Hence by direct inspection of the
curvature invariant ${\mathcal{R}}_\alpha$ given in equation~(\ref{A.generic_RR_r})
of appendix~\ref{A.generic} we conclude that all the terms which dependent on the scale factor $a$
vanish at $r_1=0$. This result is established by noting that, for all $p$, $q$ and for $\alpha_1<0$,
$\lim_{r_1\to 0}r^p(1-U_{\mathrm{SC}})^{-|\alpha_1|/r_1}=0$ and
$\lim_{r_1\to 0}r^p(1-U_{\mathrm{SC}})^{-|\alpha_1|/r_1}\log(1-U_{\mathrm{SC}})^q=0$
such that, for all values of $\alpha_0$, the only contribution to the space-time singularity
is exactly given by the Schwarzschild curvature invariant~(\ref{R_SC})
\be
{\mathcal{R}}_{(\alpha_0,\alpha_1<0)}(r_1\sim 0)={\mathcal{R}}_{\mathrm{SC}}=\frac{48(GM)^2}{c^4\,r_1^6}\ .
\lb{RR_alpha_0_1_-}
\ee

The total mass-energy density for a radial coordinate dependent exponent $\alpha(r_1)$
is given, similarly to~(\ref{rho_tot}), by
$\rho_{\mathrm{tot}(\alpha(r_1))}=\rho_{\mathrm{SC}}+\rho_{\alpha(r_1)}$,
where the extended distribution $\rho_{\alpha(r_1)}$ is
\be
\ba{rcl}
\rho_{\alpha(r_1)}&=&\displaystyle\frac{1}{c^2}\,T^{(\alpha(r_1))}_{\hat{0}\hat{0}}=\frac{c^2}{8\pi\,G}\,G^{(\alpha(r_1))}_{\hat{0}\hat{0}}\\[6mm]
&=&\displaystyle \frac{1}{8\pi\,G}\left(\frac{\dot{a}}{a}\right)^2\left(\left(3+\frac{2(\alpha-3)GM}{c^2\,r_1}\right)\left(1-\frac{2GM}{c^2\,r_1}\right)^{\alpha-1}\ \ \ \right.\\[6mm]
&&\displaystyle\hfill\left.+\left(1-\frac{2GM}{c^2\,r_1}\right)^{\alpha}\,r_1\log\left(1-\frac{2GM}{c^2\,r_1}\right){\alpha}'\right)\ .
\ea
\lb{rho_alpha_r}
\ee
In this expression $G^{(\alpha(r_1))}_{\hat{0}\hat{0}}$ and $T^{(\alpha(r_1))}_{{\hat{0}\hat{0}}}$ are
the Einstein tensor and stress-energy tensor for the locally anisotropic metric~(\ref{g_generic})
in the Cartan-frame (excluding the SC mass pole contribution). The details of the
computation of these tensors are discussed in the next section.

The total mass inside a shell of constant radius $r_1$ is obtained, as usual, by considering
the respective volume integral. We follow the same definitions for the several mass
quantities of the previous section~\ref{sec.alpha_r_0}. Similarly to the definition of
the quantity $\tilde{M}$~(\ref{MM}), for the specific exponent
$\alpha(r_1)=\alpha_0+2GM/(c^2\,r_1)$~(\ref{alpha_r_1}) and considering the indefinite
integral of the above density~(\ref{rho_alpha_r}) we obtain that
\be
\tilde{M}_{(\alpha_0,\alpha_1)}(r_1)=\left(\frac{\dot{a}}{a}\right)^2\frac{r_1^3}{2G}\left(1-\frac{2GM}{c^2\,r_1}\right)^{\alpha_0+\alpha_1\frac{2GM}{c^2\,r_1}}\ .
\ee
At the origin, this quantity is divergent for $\alpha_1>0$ and null for $\alpha_1<0$
\be
\tilde{M}_{(\alpha_0,\alpha_1)}(0)=\left\{
\ba{rcl}
e^{i\delta}\infty&,&\alpha_1>0\ ,\\[6mm]
0&,&\alpha_1<0\ ,
\ea\right.
\ee
while for $\alpha_1=0$ we retrieve the case of constant exponent $\alpha=\alpha_0$ analyzed in
the previous section~\ref{sec.alpha_r_0}. In the above expression $e^{i\delta}$ is a generic
complex phase. Hence consistently with the singularities analysis of the curvature invariant ${\mathcal{R}}_{(\alpha_0,\alpha_1)}$ as expressed in equations~(\ref{RR_alpha_0_1_+})
and~(\ref{RR_alpha_0_1_-}), independently of the value of $\alpha_0$, the total mass
is divergent for $\alpha_1>0$ and is finite for $\alpha_1<0$.
Furthermore, for $\alpha_1<0$ and all values of $\alpha_0$, at the origin the singularity exactly matches the SC mass pole singularity such that the mass coincides with the SC mass pole value $M$.
As for the total mass within a shell of large radius ($r_1\sim+\infty$), for $\alpha_1<0$ and all values
of $\alpha_0$, is given by the sum of the SC gravitational mass $M$ with the background mass
(in the absence of the local mass $M$) $M_H$~(\ref{M_H}),
$M_{\mathrm{tot}(\alpha_0,\alpha_1<0)}(r_1\sim+\infty)=M+M_H$.

Resuming the results obtained in this section from the analysis of the singularities
at the center of mass, we have concluded that when considering a radial coordinate dependent
exponent $\alpha(r_1)=\alpha_0+\alpha_1\,U_{\mathrm{SC}}$~(\ref{alpha_r_1}), independently
of the value of the parameter $\alpha_0$, we obtain the following bounds on the parameter $\alpha_1$
\be
\ba{rclcl}
\alpha_1&>&0&\Rightarrow&\mathrm{space-time\ singularity\ at}\ r_1=0\ \mathrm{diverges\ by}\ \sim r_1^{+\infty}\ ,\\[5mm]
& & & &\mathrm{total\ mass\ inside\ shell\ of\ radius}\ r_1\ \mathrm{is\ \bf divergent}\ ,\\[6mm]
\alpha_1&<&0&\Rightarrow&\mathrm{space-time\ singularity\ at}\ r_1=0\ \mathrm{coincides\ with\ SC\ singularity}\\
& & & &\hfill \left({\mathcal{R}}_{\alpha_0,\alpha_1>0}={\mathcal{R}}_{\mathrm{SC}}\sim r_1^{-6}\right)\ ,\\[5mm]
& & & &\mathrm{total\ mass\ inside\ shell\ of\ radius}\ r_1\ \mathrm{is\ \bf finite}\ .
\ea
\lb{bounds_alpha_r_1}
\ee
Then we have accomplished our main objective for this section, by considering a radial coordinate
dependent exponent $\alpha(r_1)=\alpha_0+\alpha_1\,U_{\mathrm{SC}}$~(\ref{alpha_r_1}) with
$\alpha_1<0$ we have removed the essential singularities at the origin (except for the SC mass pole)
maintaining the main properties of the locally anisotropic metric~(\ref{g_generic}). Furthermore for this
exponent choice it strictly converges asymptotically at the origin to the Schwarzschild metric.

Also we note that assuming the above bound $\alpha_1<0$~(\ref{bounds_alpha_r_1}) and either of the bounds~(\ref{alpha_bounds_0_1}) we obtain the bounds $\alpha_0\ge|\alpha_1|+3$ and $\alpha_0>|\alpha_1|+5$
such that the following inequality applies
\be
\alpha_0>|\alpha_1|\ .
\lb{alpha_0_alpha_1}
\ee
As a consequence of this relation the corrections to the several quantities outside
the SC event horizon ($r_1>r_{1.\mathrm{SC}}=2GM/c^2$) due to the coefficient $\alpha_1$ are
smaller than the ones due to the coefficient $\alpha_0$, $\alpha_0>\alpha_1\,U_{\mathrm{SC}}(r_1>r_{\mathrm{SC}})$ such that the results derived
in section~\ref{sec.alpha_r_sc} with respect to the curvature behavior
close to the Schwarzschild radius are qualitatively maintained (see figure~\ref{fig.R_alpha}).
In particular the bounds for $\alpha$ at the event horizon~(\ref{alpha_bounds}) are valid
for the exponent value evaluate at the SC radius $\alpha(r_{1.\mathrm{SC}})=\alpha_0-|\alpha_1|$
such that for $\alpha_0>5+|\alpha_1|$ space-time is asymptotically flat at the event horizon.

Next we compute the stress-energy tensor in the Lorentz frame (Cartan frame) and
both the mass-energy density and pressures of the background fluid for the
locally anisotropic metric~(\ref{g_generic}) analyzing their properties.

\subsection{The Stress-Energy Tensor with Anisotropic Pressures and\\ Positive Definite Mass-Energy Density\lb{sec.gen_stress}}

In this section we analyze the stress-energy tensor for the locally anisotropic metric~(\ref{g_generic})
with a radial coordinate dependent exponent 
$\alpha(r_1)=\alpha_0+\alpha_1\,U_{\mathrm{SC}}$~(\ref{alpha_r_1}) with strictly negative
$\alpha_1<0$. In particular we give the details of the computation of the mass-energy density
$\rho_{(\alpha_0,\alpha_1)}$ and the anisotropic pressures $p_{r(\alpha_0,\alpha_1)}$,
$p_{\theta(\alpha_0,\alpha_1)}$ and $p_{\varphi(\alpha_0,\alpha_1)}$, showing that these
quantities have the correct asymptotic leading expressions at the origin and spatial infinity
consistently with the metric ansatz construction assumptions and interpretation
as an interpolation (or deformation) between the cosmological
FLRW metric~(\ref{g_FRW_1}) and the Schwarzschild metric~(\ref{g_SC}).

The stress-energy tensor is related, as usual, to the Einstein tensor by the Einstein
equations
$T^{(\alpha_0,\alpha_1)}_{\mu\nu}=c^4/(8\pi\,G)\,G^{(\alpha_0,\alpha_1)}_{\mu\nu}$
(see equation~(\ref{A.EEQ}) in appendix~\ref{A.defs}).
The Einstein tensor $G_{\mu\nu}^{(\alpha(r_1))}=G_{\mu\nu}^{(\alpha)}+\Delta G_{\mu\nu}$
(excluding the SC mass pole contribution) for a generic exponent $\alpha(r_1)$ dependent
on the radial coordinate is given in equation~(\ref{A.gen.EE_r}) of appendix~\ref{A.generic}
with $G_{\mu\nu}^{(\alpha)}$ given in equation~(\ref{A.gen.EE}) and $\Delta G_{\mu\nu}$ given
in equation~(\ref{A.generic_G_r_I}) of the same appendix. Here we use the index
notation '$(\alpha(r_1))$' when referring to a generic exponent dependent on
the radial coordinate $r_1$ and $'(\alpha_0,\alpha_1)'$ when referring to the specific
exponent $\alpha(r_1)=\alpha_0+\alpha_1\,U_{\mathrm{SC}}$~(\ref{alpha_r_1}).
Either by direct inspection of the Einstein tensor $G_{\mu\nu}^{(\alpha(r_1))}$,
or by noting that the metric~(\ref{g_generic}) is spherical symmetric we conclude that
this symmetry is maintained both by the mass-energy density $\rho_{(\alpha_0,\alpha_1)}$ and
anisotropic pressures $p_{r(\alpha_0,\alpha_1)}$, $p_{\theta(\alpha_0,\alpha_1)}$ and
$p_{\varphi(\alpha_0,\alpha_1)}$. This statement is translated as
that the pressures along the angular directions are identical for all space-time
$p_{\theta(\alpha_0,\alpha_1)}=p_{\varphi(\alpha_0,\alpha_1)}$ while the pressure
along the radial direction may generally be distinct $p_{r(\alpha_0,\alpha_1)}$. We also note
that $\rho_{(\alpha_0,\alpha_1)}$, $p_{r(\alpha_0,\alpha_1)}$, $p_{\theta(\alpha_0,\alpha_1)}$
and $p_{\varphi(\alpha_0,\alpha_1)}$ are scalar quantities, here the pressures
do not correspond to a vector along the spatial directions, instead generally we
have three distinct scalars (only two when spherical symmetry is maintained), hence
frame independent.

In the following we will work in the Cartan frame (Lorentz frame) corresponding to the
local flat frame with the Minkowski metric $\eta_{\hat{\mu}\hat{\nu}}$ for which the
stress-energy tensor is diagonal being explicitly stress and shear free ($T^{(\alpha_0,\alpha_1)}_{\hat{0}\hat{i}}=0$). Here hatted indexes represent the
flat coordinates in the Cartan frame. Assuming a background perfect
fluid with commoving velocity $u_\mu=(c,0,0,0)$ the stress-energy tensor in the
Cartan frame is
\be
T^{(\alpha_0,\alpha_1)}_{\hat{0}\hat{0}}=c^2\rho_{(\alpha_0,\alpha_1)}\ \ ,\ \ T^{(\alpha_0,\alpha_1)}_{\hat{1}\hat{1}}=p_{r(\alpha_0,\alpha_1)}\ \ ,\ \ T^{(\alpha_0,\alpha_1)}_{\hat{2}\hat{2}}=T^{(\alpha_0,\alpha_1)}_{\hat{3}\hat{3}}=p_{\theta(\alpha_0,\alpha_1)}\ ,
\lb{T_anisotropic}
\ee
where as already discussed, due to spherical symmetry, $T^{(\alpha_0,\alpha_1)}_{\hat{2}\hat{2}}=T^{(\alpha_0,\alpha_1)}_{\hat{3}\hat{3}}$.
To explicitly compute $T^{(\alpha_0,\alpha_1)}_{\hat{\mu}\hat{\nu}}$ let us
consider a Cartan tetrad $e^{\hat\mu}=e^{\hat{\mu}}_{\ \mu}dx^\mu$ such that the
coordinate metric is related to the Minkowski metric by $g_{\mu\nu}=e^{\hat{\mu}}_{\ \mu}e^{\hat{\nu}}_{\ \nu}\eta_{\hat{\mu}\hat{\nu}}$. Then, for the locally anisotropic metric~(\ref{g_generic}),
we obtain the following non-null components of the tetrad $e^{\hat{\mu}}_{\ \mu}$
\be
\ba{rclcrclcrcl}
\displaystyle e^{\hat{0}}_{\ 0}&=&\displaystyle \sqrt{1-\frac{2GM}{r_1\,c^2}}&,&
\displaystyle e^{\hat{1}}_{\ 0}&=&\displaystyle -\frac{\dot{a}}{a}\frac{r_1}{c}\left(1-\frac{2GM}{r_1\,c^2}\right)^\frac{\alpha}{2}&,&\\[6mm]
\displaystyle e^{\hat{1}}_{\ 1}&=&\displaystyle \frac{1}{\sqrt{1-\frac{2GM}{r_1\,c^2}}}&,&
\displaystyle e^{\hat{2}}_{\ 2}&=&\displaystyle r_1&,&
\displaystyle e^{\hat{3}}_{\ 3}&=&\displaystyle r_1\sin\theta\ .
\ea
\lb{e_anisotropic}
\ee
The non-null inverse tetrad components $e_{\hat{\mu}}^{\ \mu}$ are straight forwardly obtained
by raising and lowering the flat indexes '$\hat{\mu}$' with the Minkowski metric
$\eta_{\hat{\mu}\hat{\nu}}$ and the coordinate indexes '$\mu$' with the coordinate
metric $g_{\mu\nu}$
\be
\ba{rclcrclcrcl}
\displaystyle e_{\hat{0}}^{\ 0}&=&\displaystyle \frac{1}{\sqrt{1-\frac{2GM}{r_1\,c^2}}}&,&
\displaystyle e_{\hat{0}}^{\ 1}&=&\displaystyle \frac{\dot{a}}{a}\frac{r_1}{c}\left(1-\frac{2GM}{r_1\,c^2}\right)^\frac{\alpha}{2}\ ,\\[6mm]
\displaystyle e_{\hat{1}}^{\ 1}&=&\displaystyle \sqrt{1-\frac{2GM}{r_1\,c^2}}&,&
\displaystyle e_{\hat{2}}^{\ 2}&=&\displaystyle \frac{1}{r_1}&,&
\displaystyle e_{\hat{3}}^{\ 3}&=&\displaystyle \frac{1}{r_1\sin\theta}\ .
\ea
\lb{ei_anisotropic}
\ee
Hence the stress-energy tensor in the Cartan frame is obtained by considering
the contraction of the coordinate stress-tensor with the inverse tetrad
elements $T_{\hat{\mu}\hat{\nu}}=e_{\hat{\mu}}^{\ \mu}e_{\hat{\nu}}^{\ \nu}\,T_{\mu\nu}$.
From the Einstein equation~(\ref{A.EEQ}) we obtain the following relations
\be
\ba{rcl}
T^{(\alpha(r_1))}_{\hat{0}\hat{0}}&=&\displaystyle\frac{c^4}{8\pi\,G}\left(e_{\hat{0}}^{\ 0}e_{\hat{0}}^{\ 0}\,G^{(\alpha(r_1))}_{00}+2e_{\hat{0}}^{\ 0}e_{\hat{0}}^{\ 1}\,G^{(\alpha(r_1))}_{01}+e_{\hat{0}}^{\ 1}e_{\hat{0}}^{\ 1}\,G^{(\alpha(r_1))}_{11}\right)\ ,\\[6mm]
T^{(\alpha(r_1))}_{\hat{0}\hat{1}}&=&\displaystyle\frac{c^4}{8\pi\,G}\left(e_{\hat{0}}^{\ 0}e_{\hat{1}}^{\ 0}\,G^{(\alpha(r_1))}_{01}+e_{\hat{0}}^{\ 1}e_{\hat{1}}^{\ 1}\,G^{(\alpha(r_1))}_{11}\right)=0\ ,\\[6mm]
T^{(\alpha(r_1))}_{\hat{1}\hat{1}}&=&\displaystyle\frac{c^4}{8\pi\,G}\,e_{\hat{1}}^{\ 1}e_{\hat{1}}^{\ 1}\,G^{(\alpha(r_1))}_{11}\ ,\\[6mm]
T^{(\alpha(r_1))}_{\hat{2}\hat{2}}&=&\displaystyle\frac{c^4}{8\pi\,G}\,e_{\hat{2}}^{\ 1}e_{\hat{2}}^{\ 1}\,G^{(\alpha(r_1))}_{22}\ ,\\[6mm]
T^{(\alpha(r_1))}_{\hat{3}\hat{3}}&=&\displaystyle\frac{c^4}{8\pi\,G}\,e_{\hat{3}}^{\ 1}e_{\hat{3}}^{\ 1}\,G^{(\alpha(r_1))}_{33}\ .
\ea
\lb{T_Cartan_transf}
\ee
As already mentioned, the cross component of the stress-energy tensor in the Cartan frame
are null $T_{\hat{0}\hat{i}}=0$ such that in the local flat Minkowski frame the stress-energy
tensor is stress and shear free. The non-null $T_{01}$ is due to the (global) coordinate system
choice and does not represents a measurable physical stress nor shear.

In the following, for compactness of the expressions and direct numerical evaluation,
we will consider the dimensionless radial coordinate rescaled by the Schwarzschild radius
\be
\bar{r}_1=\frac{r_1}{r_{1.\mathrm{SC}}}=\frac{2GM}{c^2}\,r_1\ .
\lb{bar_r}
\ee
Hence from the expressions~(\ref{T_Cartan_transf}) relating the stress-energy
tensor $T_{\hat{\mu}\hat{\nu}}$ in the Cartan frame with the Einstein
tensor $G_{\mu\nu}$~(\ref{A.gen.EE_r}) in the coordinate frame and from the definition
of $T_{\hat{\mu}\hat{\nu}}$ for a (anisotropic) perfect fluid in the Cartan
frame~(\ref{T_anisotropic}) we obtain, for the particular radial coordinate dependent
exponent $\alpha(r_1)=\alpha_0+\alpha_1\,U_{\mathrm{SC}}$~(\ref{alpha_r_1}) the following
solutions for the mass-energy density $\rho_{(\alpha_0,\alpha_1)}$ and the pressures
$p_{r(\alpha_0,\alpha_1)}$ and $p_{\theta(\alpha_0,\alpha_1)}$
\be
\ba{rcl}
\rho_{(\alpha_0,\alpha_1)}&=&\displaystyle\frac{H^2}{8\pi\,G}\,\left(1-\frac{1}{\bar{r}_1}\right)^{\alpha_0-1+\frac{\alpha_1}{\bar{r}_1}}\left(3+\frac{\alpha_0-3}{\bar{r}_1}+\frac{\alpha_1}{\bar{r}_1^2}\right.\\[6mm]
& &\displaystyle\hfill\left.-\frac{\alpha_1}{\bar{r}_1}\left(1-\frac{1}{\bar{r}_1}\right)\log\left(1-\frac{1}{\bar{r}_1}\right)\right)\ ,\\[6mm]
p_{r(\alpha_0,\alpha_1)}&=&\displaystyle\frac{c^2\,H^2}{8\pi\,G}\,\left(1-\frac{1}{\bar{r}_1}\right)^{\frac{\alpha_0}{2}-\frac{1}{2}+\frac{\alpha_1}{2\bar{r}_1}}\left(2(1+q)-\left(\frac{\alpha_0}{\bar{r}_1}+\frac{\alpha_1}{\bar{r}_1^2}\right)\left(1-\frac{1}{\bar{r}_1}\right)^{\frac{\alpha_0}{2}-\frac{1}{2}+\frac{\alpha_1}{2\bar{r}_1}}\right.\\[6mm]
&&\displaystyle\hfill-\left.\left(1-\frac{1}{\bar{r}_1}\right)^{\frac{\alpha_0}{2}+\frac{1}{2}+\frac{\alpha_1}{2\bar{r}_1}}\left(3-\frac{\alpha_1}{\bar{r}_1}\log\left(1-\frac{1}{\bar{r}_1}\right)\right)\right)\ ,\\[6mm]
p_{\theta(\alpha_0,\alpha_1)}&=&p_{\varphi(\alpha_0,\alpha_1)}\\[6mm]
&=&\displaystyle-\frac{c^2\,H^2}{8\pi\,G}\,\left(1-\frac{1}{\bar{r}_1}\right)^{\alpha_0-2+\frac{\alpha_1}{\bar{r}_1}}\times\\[6mm]
&&\displaystyle\hfill\times\left(3+\frac{2(\alpha_0-3)}{\bar{r}_1}+\frac{\alpha_0(\alpha_0-5)+4\alpha_1+6}{2\bar{r}_1^2}+\frac{(2\alpha_0-5)\alpha_1}{2\bar{r}_1^3}+\frac{\alpha_1^2}{2\bar{r}_1^4}\right)\\[6mm]
&&\displaystyle+\frac{c^2\,H^2}{8\pi\,G}\,\frac{\alpha_1}{\bar{r}_1^2}\left(1-\frac{1}{\bar{r}_1}\right)^{\alpha_0-1+\frac{\alpha_1}{\bar{r}_1}}\left(1+\left(\alpha_0+\frac{\alpha_1}{\bar{r}_1}\right)\log\left(1-\frac{1}{\bar{r}_1}\right)\right)\\[6mm]
&&\displaystyle+\frac{c^2\,H^2}{8\pi\,G}\,\frac{\alpha_1}{\bar{r}_1}\left(1-\frac{1}{\bar{r}_1}\right)^{\alpha_0+\frac{\alpha_1}{\bar{r}_1}}\left(2-\frac{\alpha_1}{2\bar{r}_1}\log\left(1-\frac{1}{\bar{r}_1}\right)\right)\log\left(1-\frac{1}{\bar{r}_1}\right)\\[6mm]
&&\displaystyle+\frac{c^2\,H^2}{8\pi\,G}\,(1+q)\left(1-\frac{1}{\bar{r}_1}\right)^{\frac{\alpha_0}{2}-\frac{3}{2}+\frac{\alpha_1}{2\bar{r}_1}}\left(2+\frac{\alpha_0-4}{2\bar{r}_1}+\frac{\alpha_1}{2\bar{r}_1^2}\right)\\[6mm]
&&\displaystyle-\frac{c^2\,H^2}{8\pi\,G}\,\frac{\alpha_1(1+q)}{2\bar{r}_1}\left(1-\frac{1}{\bar{r}_1}\right)^{\frac{\alpha_0}{2}-\frac{1}{2}+\frac{\alpha_1}{2\bar{r}_1}}\log\left(1-\frac{1}{\bar{r}_1}\right)\ .
\ea
\lb{rho_p_gen_r}
\ee
Here we have replaced the derivatives of the scale factor by the Hubble rate $H=\dot{a}/a$~(\ref{H})
and deceleration factor $q=-\ddot{a}/(a\,H^2)$~(\ref{H_q0}) definitions. We note that the expression
for the mass-energy density corresponds to the equation for generic exponent
$\alpha(r_1)$~(\ref{rho_alpha_r}) already considered in the previous section
evaluated for the particular case of the first order exponent~(\ref{alpha_r_1}).

We also recall that, similarly to the analysis carried in the previous section~\ref{sec.alpha_r_0}
(equation~(\ref{rho_tot})), the total mass-energy density
$\rho_{\mathrm{tot}}=\rho_{\mathrm{SC}}+\rho_{(\alpha_0,\alpha_1)}$ is given by the sum of both the
Schwarzschild mass pole contribution $\rho_{\mathrm{SC}}$~(\ref{SC_T})
and the mass-energy density contribution $\rho_{(\alpha_0,\alpha_1)}$~(\ref{rho_p_gen_r})
due to the expanding background deformation. As for the total pressures correspond
to the anisotropic pressures just computed, $p_{r\mathrm{tot}}=p_{r(\alpha_0,\alpha_1)}$
and $p_{\theta\mathrm{tot}}=p_{\theta(\alpha_0,\alpha_1)}$, the Schwarzschild pressure is null
everywhere (including the origin). With respect to the asymptotic limits of these quantities
we obtain at the origin that, for $\alpha_1<0$, both the total mass-energy density
and pressures consistently coincide with the respective Schwarzschild quantities
\be
\ba{rcl}
\rho_{\mathrm{tot}}(0)&=&\displaystyle\lim_{r_1\to 0}\left(\rho_{\mathrm{SC}}+\rho_{(\alpha_0,\alpha_1)}\right)=\rho_{\mathrm{SC}}=M\delta^{(3)}(\vb{r_1})\ ,\\[6mm]
p_{r.\mathrm{tot}}(0)&=&\displaystyle\lim_{r_1\to 0}p_{r(\alpha_0,\alpha_1)}=0\ ,\\[6mm]
p_{\theta.\mathrm{tot}}(0)&=&\displaystyle\lim_{r_1\to 0}p_{\theta(\alpha_0,\alpha_1)}=0\ .
\ea
\ee
Asymptotically, at spatial infinity or at least for large enough $r_1$ such that
the gravitational potential is negligible $U_{\mathrm{SC}}(r_1\sim+\infty)\approx 0$,
independently of the coefficients $\alpha_0$ and $\alpha_1$, we retrieve the
cosmological mass-energy density $\rho_H$ and pressure $p_H$~(\ref{p_rho_FRW_1})
corresponding to the FLRW metric~(\ref{g_FRW_1})
\be
\ba{rcl}
\rho_{\mathrm{tot}}(r_1\sim+\infty)&\approx&\displaystyle\rho_H=\frac{3H^2}{8\pi\,G}\ ,\\[6mm]
p_{r(\alpha_0,\alpha_1)}(r_1\sim+\infty)&\approx&\displaystyle p_H=\frac{(-1+2q)c^2\,H^2}{8\pi\,G}\ ,\\[6mm]
p_{\theta(\alpha_0,\alpha_1)}(r_1\sim+\infty)&\approx&\displaystyle p_H=\frac{(-1+2q)c^2\,H^2}{8\pi\,G}\ .
\ea
\ee
At the Schwarzschild radius $r_1=r_{1.\mathrm{SC}}$, for $\alpha_1<0$ and $\alpha_0+\alpha_1\ge 3$,
the values of these quantities are
\be
\ba{rcl}
\rho_{\mathrm{tot}}(r_{1.\mathrm{SC}})&=&0\ ,\\[6mm]
p_{r(\alpha_0,\alpha_1)}(r_{1.\mathrm{SC}})&=&0\ ,\\[6mm]
p_{\theta(\alpha_0,\alpha_1)}(r_{1.\mathrm{SC}})&=&\left\{\ba{rcl}+\infty&,&\alpha=3\\[3mm]0&,&\alpha>3\ea\right.\ .
\ea
\ee

As for the asymptotic values of the derivatives of the mass-energy density and
pressures we obtain that, both at the origin and at spatial infinity,
are null coinciding, respectively, with the behavior of the SC metric~(\ref{g_SC})
and the FRLW metric~(\ref{g_FRW_1}), $\rho'_{\mathrm{tot}}(0)=p'_{r(\alpha_0,\alpha_1)}(0)=p'_{\theta(\alpha_0,\alpha_1)}(0)=0$ and $\rho'_{\mathrm{tot}}(+\infty)=p'_{r(\alpha_0,\alpha_1)}(+\infty)=p'_{\theta(\alpha_0,\alpha_1)}(+\infty)=0$. At the event horizon these quantities are
\be
\ba{rcl}
\rho'_{\mathrm{tot}}(r_{1.\mathrm{SC}})&=&0\ ,\hspace{3cm}\alpha_0-|\alpha_1|\ge 3\\[5mm]
p'_{r(\alpha_0,\alpha_1)}(r_{1.\mathrm{SC}})&=&
\displaystyle\left\{
\ba{rcl}
\displaystyle\frac{c^2(1+q)\,H^2}{4\pi\,G}&,&\alpha_0-|\alpha_1|=3\ ,\\[3mm]
\displaystyle 0&,&\alpha_0-|\alpha_1|>3\ ,\\[3mm]
\ea\right.\\[12mm]
p'_{\theta(\alpha_0,\alpha_1)}(r_{1.\mathrm{SC}})&=&\displaystyle\left\{
\ba{rcl}
+\infty&,&\alpha_0-|\alpha_1|\in[3,5[\ ,\\[3mm]
\displaystyle\frac{5c^2(1+q)\,H^2}{16\pi\,G}&,&\alpha_0-|\alpha_1|=5\ ,\\[3mm]
0&,&\alpha_0-|\alpha_1|>5\ ,
\ea\right.\
\ea
\lb{limits_rho_p}
\ee
These limits and respective bounds on the coefficient sums $\alpha_0-|\alpha_1|$
are consistent with the space-time being asymptotically flat at the event horizon
for $\alpha_0-|\alpha_1|>5$ as expressed in equation~(\ref{alpha_bounds_0_1})
based in the analysis of the asymptotic behavior of the curvature $R$. Here
the same conclusion is obtained from the asymptotic behavior of the mass-energy
density and pressures, all these quantities and its derivatives are null at the event horizon
for $\alpha_0-|\alpha_1|>5$ matching the respective quantities for flat space-time.
In figure~\ref{fig.rho_p_alpha} are graphically represented the mass-energy density
and pressures for several values of $\alpha_1$ and $\alpha_0-|\alpha_1|$.
\fig{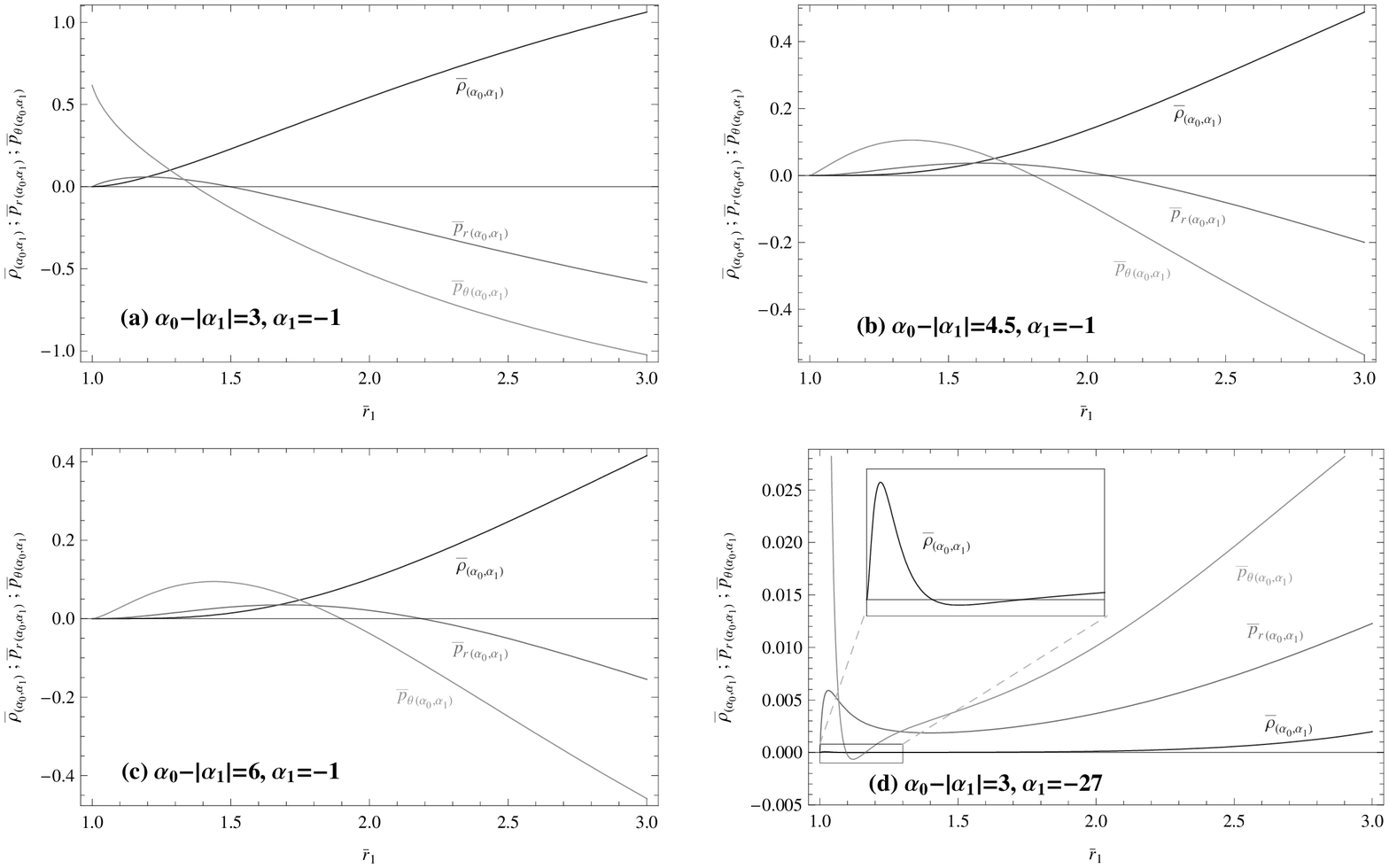}{150mm}{Rescaled mass-energy density $\bar{\rho}_{(\alpha_0,\alpha_1)}=(8\pi G/H^2)\,\rho_{(\alpha_0,\alpha_1)}$ and pressures $\bar{p}_{r(\alpha_0,\alpha_1)}=(8\pi G/(c^2\,H^2))\,\rho_{r(\alpha_0,\alpha_1)}$, $\bar{p}_{\theta(\alpha_0,\alpha_1)}=(8\pi G/(c^2\,H^2))\,\rho_{\theta(\alpha_0,\alpha_1)}$~(\ref{rho_p_gen_r}) as a function of
$\bar{r}_1=r_1/r_{1.\mathrm{SC}}$~(\ref{bar_r}). The assymptotic
values for large $\bar{r}_1$ are the respective quantities for the FRLW metric
$\bar{\rho}_H=(8\pi G/H^2)\,\rho_H=3$ and $\bar{p}_H=(8\pi G/(c^2\,H^2))\,p_H=(2q_0-1)$~(\ref{p_rho_FRW_1}).
Are given the examples for:\hfill\break
\ {\bf(a)} $\alpha_0-|\alpha_1|=3$ with $\alpha_0=4$ and $\alpha_1=-1$, the pressure $\bar{p}_{\theta(\alpha_0,\alpha_1)}$ and its
derivative $\bar{p}'_{\theta(\alpha_0,\alpha_1)}$ diverge at $\bar{r}_1=1$;\hfill\break
\ {\bf(b)} $\alpha_0-|\alpha_1|=4.5$ with $\alpha_0=5.5$ and $\alpha_1=-1$, the pressure derivative $\bar{p}'_{\theta(\alpha_0,\alpha_1)}$ diverges at $\bar{r}_1=1$;\hfill\break
\ {\bf(c)} $\alpha_0-|\alpha_1|=6$ with $\alpha_0=7$ and $\alpha_1=-1$, both mass-energy density, the pressures and their derivatives
are null at $\bar{r}_1=1$;\hfill\break
\ {\bf(d)} $\alpha_0-|\alpha_1|=3$ with $\alpha_0=30$ and $\alpha_1=-27$, characteristics similar to
the case of figure (a), however outside the SC event horizon $\bar{r}_1>1$ exists a region of
negative mass-energy density shown in the blow up. For $\alpha_0=30$ the mass-energy density is
positive definite for $\alpha_1\ge 3-30\Delta_\alpha(30)\approx -23.805$~(\ref{range_A}).}{fig.rho_p_alpha}

In addition, to preserve causality, it is required to demand the energy density to be
positive outside the SC event horizon $r_1>r_{1.\mathrm{SC}}$. This condition is not verified
for all negative values of the parameters $\alpha_1<0$ and $\alpha_0\ge 3+|\alpha_1|$.
By direct inspection of the expression for $\rho_{(\alpha_0,\alpha_1)}$
given in equation~(\ref{rho_p_gen_r}) we note that the sign of the full expression for
$\bar{r}_1>1$ is set by the sign of the last multiplicative factor within brackets.
The zeros of this expression cannot be found analytically, however a direct numerical inspection can
be carried to define the allowed range for the coefficient $\alpha_1$ as a function of $\alpha_0$,
simultaneously considering, at the SC event horizon, either the criteria of the absence of
space-time singularities or space-time being asymptotically flat~(\ref{alpha_bounds_0_1}). Hence
it is further necessary to constraint the coefficient $\alpha_1$ to be either in the range $\alpha_1\in[3-\alpha_0,0[$ or $\alpha_1\in[5-\alpha_0,0[$, respectively. We note that here
we are considering the bounds~(\ref{alpha_bounds_0_1}) keeping
$\alpha_0$ as generic as possible and constraining $\alpha_1$ such that for $\alpha_0>3$
and $\alpha_0>5$ we obtain, respectively, the following allowed ranges for the coefficient $\alpha_1$
\be
\ba{rcl}
\left\{\ba{rcl}\alpha_0&\in&]3,+\infty[\\[6mm]
\alpha_1&\in&{\mathcal{A}}^{(3)}(\alpha_0)\ea\right.&&\mathrm{space-time\ at\ SC\ event\ horizon\ is\ singularity\ free}\\[12mm]
\left\{\ba{rcl}\alpha_0&\in&]5,+\infty]\\[6mm]
\alpha_1&\in&{\mathcal{A}}^{(5)}(\alpha_0)\ea\right.&&\mathrm{space-time\ at\ SC\ event\ horizon\ is\ assymptotically\ flat}
\ea
\lb{alpha_rho_+}
\ee
The sets ${\mathcal{A}}^{(3)}(\alpha_0)$ and ${\mathcal{A}}^{(5)}(\alpha_0)$
are negative ranges depending on the specific value of $\alpha_0$. Specifically are
\be
\ba{rcl}
{\mathcal{A}}^{(3)}(\alpha_0)&=&\left\{\ba{rclclcl}{\mathcal{A}}^{(3)}_1(\alpha_0)&=&]3-\alpha_0,0[&,&\alpha_0&\in&]3,28.347]\ ,\\[6mm]
{\mathcal{A}}^{(3)}_2(\alpha_0)&=&\left]3-\Delta_\alpha(\alpha_0)\times\alpha_0,0\right[&,&\alpha_0&\in&]28.347,+\infty[\ ,\ea\right.\\[12mm]
{\mathcal{A}}^{(5)}(\alpha_0)&=&\left\{\ba{rclclcl}{\mathcal{A}}^{(5)}_1(\alpha_0)&=&]5-\alpha_0,0[&,&\alpha_0&\in&]5,45.226]\ ,\\[6mm]
{\mathcal{A}}^{(5)}_2(\alpha_0)&=&\left]5-\Delta_\alpha(\alpha_0)\times\alpha_0,0\right[&,&\alpha_0&\in&]45.226,+\infty[\ ,\ea\right.\\[12mm]
\ea
\lb{range_A}
\ee
where the values $28.347$ and $45.226$ figuring in the ranges for $\alpha_0$ were computed
numerically and the function $\Delta_\alpha(\alpha_0)$ has no analytic expression
such that, for each particular value of $\alpha_0$, a numerical evaluation must be carried out.
The values of this functions for some values of $\alpha_0$ are listed in table~\ref{table.Delta_alpha}.
\begin{table}[ht]
\begin{center}
\begin{tabular}{r|rrrrrr}
$\alpha_0$&$30$&$50$&$10^2$&$10^3$&$10^5$&$+\infty$\\[6mm]
$\Delta_\alpha(\alpha_0)$&$0.8935$&$0.8887$&$0.8848$&$0.8812$&$0.8808$&$0.8808$
\end{tabular}
\caption{\it \small Values of the functions $\Delta_\alpha(\alpha_0)$~(\ref{range_A}) for several values of $\alpha_0$.\lb{table.Delta_alpha}}
\end{center}
\end{table}

Hence, resuming the final results with respect to the bounds on the
coefficients $\alpha_0$ and $\alpha_1$ for the radial coordinate dependent exponent
$\alpha(r_1)=\alpha_0+\alpha_1\,2GM/(c^2\,r_1)$~(\ref{alpha_r_1}), we have obtained the properties
of space-time described by the locally anisotropic metric~(\ref{g_generic}) for the following
ranges of these coefficients
\be
\ba{l}
\alpha_0>3\ ,\ \alpha_1\in{\mathcal{A}}^{(3)}(\alpha_0)\ :\\[6mm]
\ \ \ \ \mathrm{only\ contribution\ to\ singularity\ at}\ r_1=0\ \mathrm{is\ given\ by\ SC\ mass\ pole}\ ,\\[6mm]
\ \ \ \ \mathrm{mass-energy\ density\ is\ positive\ definite\ for}\ r_1>r_{1.\mathrm{SC}}\ ,\\[6mm]
\ \ \ \ \mathrm{space-time\ is\ singularity\ free\ at}\ r_1=r_{1.\mathrm{SC}}\ ;\\[12mm]
\alpha_0>5\ ,\ \alpha_1\in{\mathcal{A}}^{(5)}(\alpha_0)\ :\\[6mm]
\ \ \ \ \mathrm{only\ contribution\ to\ singularity\ at}\ r_1=0\ \mathrm{is\ given\ by\ SC\ mass\ pole}\ ,\\[6mm]
\ \ \ \ \mathrm{mass-energy\ density\ positive\ definite\ for}\ r_1>r_{1.\mathrm{SC}}\ ,\\[6mm]
\ \ \ \ \mathrm{space-time\ is\ asymptotically\ flat\ at}\ r_1=r_{1.\mathrm{SC}}\ .
\ea
\lb{final_bounds}
\ee

Hence we have finished to build our ansatz. As a final remark we note that when considering
relatively large distances when compared with the SC radius ($r_1\gg r_{1.\mathrm{SC}}$, which
is the case even at planetary scales), for most purposes the parameter $\alpha_1$ works simply as
a regulator of the singularities at the origin and it can be considered as close to null as
wished such that its effects can safely be neglected away from the origin of the coordinate frame.
We will take this approach in the following analysis.

Next we will analyze in detail the modification to the General Relativity Newton law
due to the ansatz for the locally anisotropic metric~(\ref{g_generic}).

\section{The Modified Newton Law\lb{sec.Newton}}
\setcounter{equation}{0}

The main objective in this section is to compute the corrections
to the General Relativity Newton law corresponding to the locally anisotropic metric~(\ref{g_generic})
and analyze the physical effects due to these corrections.
We compute the Newtonian limit for the physical acceleration acting in a test mass
from the geodesic equations considering a massive body with center of mass at the
origin of the coordinate frame independently of considerations concerning
corrections to the observables due to the location of the observer which we will
address somewhere else~\cite{progress}. In the following
we employ spherical non-expanding coordinates $r_1$,
as discussed in section~\ref{sec.review_map}, these coordinates correspond to the
physical spatial lengths.

\subsection{Radial Gravitational Acceleration due to a Central Mass}

In this section we consider the gravitational acceleration of a test mass in the
gravitational field of a central massive object. We assume the limit of non-relativistic
velocities, hence neglect the contributions due to the test mass velocities such that the
only non-null component of the acceleration corresponds to its radial component,
$\ddot{r}_1\approx -c^2\Gamma^1_{\ 00}= -c^2(\,_{(\alpha)}\Gamma^1_{\ 00}+\Delta\Gamma^1_{\ 00}\,)$,
where the connection is given in equations~(\ref{A.gen.connections_1})
and~(\ref{A.gen.connections_r}) of appendix~(\ref{A.generic}).

The corrections to the General Relativity Newton law contain powers of the Hubble rate $H$
(on $H^2$ and $H^4$), hence we consider a decomposition of this modified Newton
law into three distinct factors
\be
\ddot{r}_1=F_{GR}+F_{H^2}+F_{H^4}\ ,
\lb{F_Newton_mod}
\ee
where $F_{GR}$ is the usual expression obtained within General Relativity for flat Minkowski
background~(\ref{F_Newton_GR}), $F_{H^2}$ the second order correction in the Hubble rate ($\sim H^2$)
and $F_{H^4}$ the fourth order correction in the Hubble rate ($\sim H^4$). Specifically these factors are
\be
\ba{rcl}
F_{GR}&=&\displaystyle-\frac{GM}{r_1^2}+\frac{2(GM)^2}{c^2\,r_1^3}\ ,\\[6mm]
F_{H^2}&=&\displaystyle-r_1
\left(1-\frac{2GM}{c^2\,r_1}\right)^\alpha\left(1-\frac{(1-\alpha)GM}{c^2r_1}-(1+q)\left(1-\frac{2GM}{c^2\,r_1}\right)^{\frac{1}{2}-\frac{\alpha}{2}}\right.\\[6mm]
& &\displaystyle\hfill\left.+\frac{r_1}{2}\,\left(1-\frac{2GM}{c^2\,r_1}\right)\log\left(1-\frac{2GM}{c^2\,r_1}\right)\,\alpha'\right)\,H^2\\[6mm]
F_{H^4}&=&\displaystyle-\frac{r_1^3}{2c^2}\,\left(1-\frac{2GM}{c^2\,r_1}\right)^{2\alpha-1}\times\\[6mm]
&&\displaystyle\hfill\times\left(\frac{2\,\alpha\,GM}{c^2\,r_1}+\left(1-\frac{2GM}{c^2\,r_1}\right)\left(2+r_1\log\left(1-\frac{2GM}{c^2\,r_1}\right)\alpha'\right)\right)\,H^4\ .
\ea
\lb{F_Newton_mod_I}
\ee
where we have replace the derivatives of the scale factor by the time dependent Hubble rate
$H=\dot{a}/a$ and deceleration parameter $q=-\ddot{a}/(aH^2)$. Next we consider and discuss
both the perturbative regime and non-perturbative regime with respect to the gravitational field $U_{\mathrm{SC}}$.

\subsubsection{Perturbative Regime\lb{sec.pert}}

Commonly, when deriving the classical Newtonian limit for General Relativity quantities, it is assumed
the weak field approximation~(\ref{weak_field}) and considered an expansion on the gravitational
field $U_{\mathrm{SC}}=2GM/(c^2r_1)$. Here, when applicable, we consider a third order expansion
on the gravitational field $U_{\mathrm{SC}}$ such that the several powers of $(1-U_{\mathrm{SC}})^p$
and the logarithm are approximated by
\be
\ba{rcl}
\displaystyle\left(1-\frac{2GM}{c^2\,r_1}\right)^p&\approx&\displaystyle 1-p\,\frac{2GM}{c^2\,r_1}+\frac{1}{2}\,p\,(p-1)\left(\frac{2GM}{c^2\,r_1}\right)^2\\[6mm]
& &\displaystyle\hfill-\frac{1}{6}\,p\,(p-1)(p-2)\left(\frac{2GM}{c^2\,r_1}\right)^3+O\left(p^4\left(\frac{2GM}{c^2\,r_1}\right)^4\right)\ ,\\[6mm]
\displaystyle\log\left(1-\frac{2GM}{c^2\,r_1}\right)&\approx&\displaystyle 1-\frac{1}{2}\,\left(\frac{2GM}{c^2\,r_1}\right)^2-\frac{1}{3}\left(\frac{2GM}{c^2\,r_1}\right)^3+O\left(\left(\frac{2GM}{c^2\,r_1}\right)^4\right)\ .
\ea
\lb{U_exp}
\ee
Generally the first of these series is convergent independently of the value of the exponent $p$ (as long
as it is finite). The exponent $p=\alpha=\alpha_0+\alpha_1 U_{\mathrm{SC}}$~(\ref{alpha_r_1}) is generally greater than unity so that to attain any desired accuracy for the series expansion it may be necessary
to consider more than the lower order terms, often it is simpler to directly evaluate the exact
expressions. With respect to the first order series expansion we note that is only a valid
approximation when the first order term is less than unity, i.e. for radial distances greater than the Schwarzschild radius times the exponent $p$
\be
r_1>\frac{2\alpha\,GM}{c^2}\ .
\lb{val_1st_order}
\ee
Hence, assuming a small parameter $\alpha_1$ and interpreting this result only with respect to the value
of the parameter $\alpha_0$, the series expansion is no-longer a valid approximation for large values
of this parameter, $\alpha_0>c^2r_1/(2GM)$. In the following discussions we will distinguish between the perturbative regime and non-perturbative regime in the gravitational field which depends both on the
value of the radial coordinate $r_1$ and the value of the parameter $\alpha_0$ in the region of interest. In the perturbative regime we will be at most working
to order $r_1^{-2}$ in the radial coordinate, however it is necessary to consider an expansion to one
higher order to correctly compute the coefficients of the several powers of $r_1$. This is due to the
connection $\Gamma^1_{\ 00}$, within the same order in the Hubble rate powers, having the factors
with $\alpha$ powers multiplied by distinct powers of $r_1$, specifically $r_1^0$ and $r_1^{-1}$.

Hence, considering that the above bound~(\ref{val_1st_order}) is obeyed, we can consider
a series expansion of the modified Newton law~(\ref{F_Newton_mod}) such that the factors
$F_{H^2}$ and $F_{H^4}$~(\ref{F_Newton_mod_I}) are approximate by
\be
\ba{rcl}
F_{H^2}& \approx&\displaystyle-q\,r_1\,H^2+(\alpha_0+1)\,q\,\frac{GM}{c^2}\,H^2+(1+4\alpha_1-\alpha_0^2)(q+1)\frac{(GM)^2}{2\,c^4\,r_1}\,H^2\\[6mm]&&\displaystyle\hfill+O\left(\alpha_0^3\,\frac{(GM)^3}{c^6\,r_1^2}\,H^2\right)\ ,\\[6mm]
F_{H^4}&\approx&\displaystyle -c^2\,\frac{r_1^3}{l_H^4}+\alpha_0\,GM\,\frac{r_1^2}{l_H^4}+O\left(\alpha_0^2\,\frac{2(GM)^2}{c^2}\,\frac{r_1}{l_H^4}\right)\ .
\ea
\lb{F_Newton_mod_I_exp}
\ee
In these series expansions we have explicitly considered the radial coordinate dependent exponent
$\alpha(r_1)=\alpha_0+2\alpha_1\,GM/(c^2\,r_1)$~(\ref{alpha_r_1}). The last expression
is written in terms of the Hubble length $l_H=c/H$ due to these terms being only
relevant for large radial coordinate $r_1\sim l_H$. As for the factor $F_{\mathrm{GR}}$
is exact within the framework of General Relativity (in the non-relativistic velocities
approximation) being already a series expansion on the gravitational field

The lower order terms $-q\,r_1\,H^2$ and
$-c^2r_1^3/l_H^4$ in the expansion of the factor $F_{H^2}$ and $F_{H^4}$ correspond to the
Newton law for the expanding background~(\ref{F_Newton_FRW_1}) described by the FLRW metric~(\ref{g_FRW_1}).
The first of these terms correspond to a repulsive gravitational
interaction (we recall that $q_0<0$) which becomes dominant with respect to the classical
attractive gravitational interaction, $F_{GR}\approx -GM/r_1^2$, for the value of the radial coordinate ${r_1^*}_{(\mathrm{pert})}$
\be
\frac{GM}{({r^*_1}_{(\mathrm{pert})})^2}=|q_0|\,r^*_1\,H_0^2\ \ \Leftrightarrow\ \ {r_1^*}_{(\mathrm{pert})}=\left(\frac{GM}{|q_0|H_0^2}\right)^\frac{1}{3}\ ,
\lb{r_i_1}
\ee
where we derive these expressions assuming today's value of the Hubble rate $H_0$~(\ref{H})
and deceleration parameter $q_0$~(\ref{H_q0}). We note that the value ${r_1^*}_{(\mathrm{pert})}$
obtained perturbatively is a good approximation as long as we assume a small
negative parameter $\alpha_1\approx 0$ (as discussed in the previous section~\ref{sec.ansatz_II}
this is enough to regularize the singularities at the center of mass maintaining the Schwarzschild
mass pole) and a relatively small parameter $\alpha_0$ such that the higher order in the series
expansion~(\ref{F_Newton_mod_I}) are negligible.
\fig{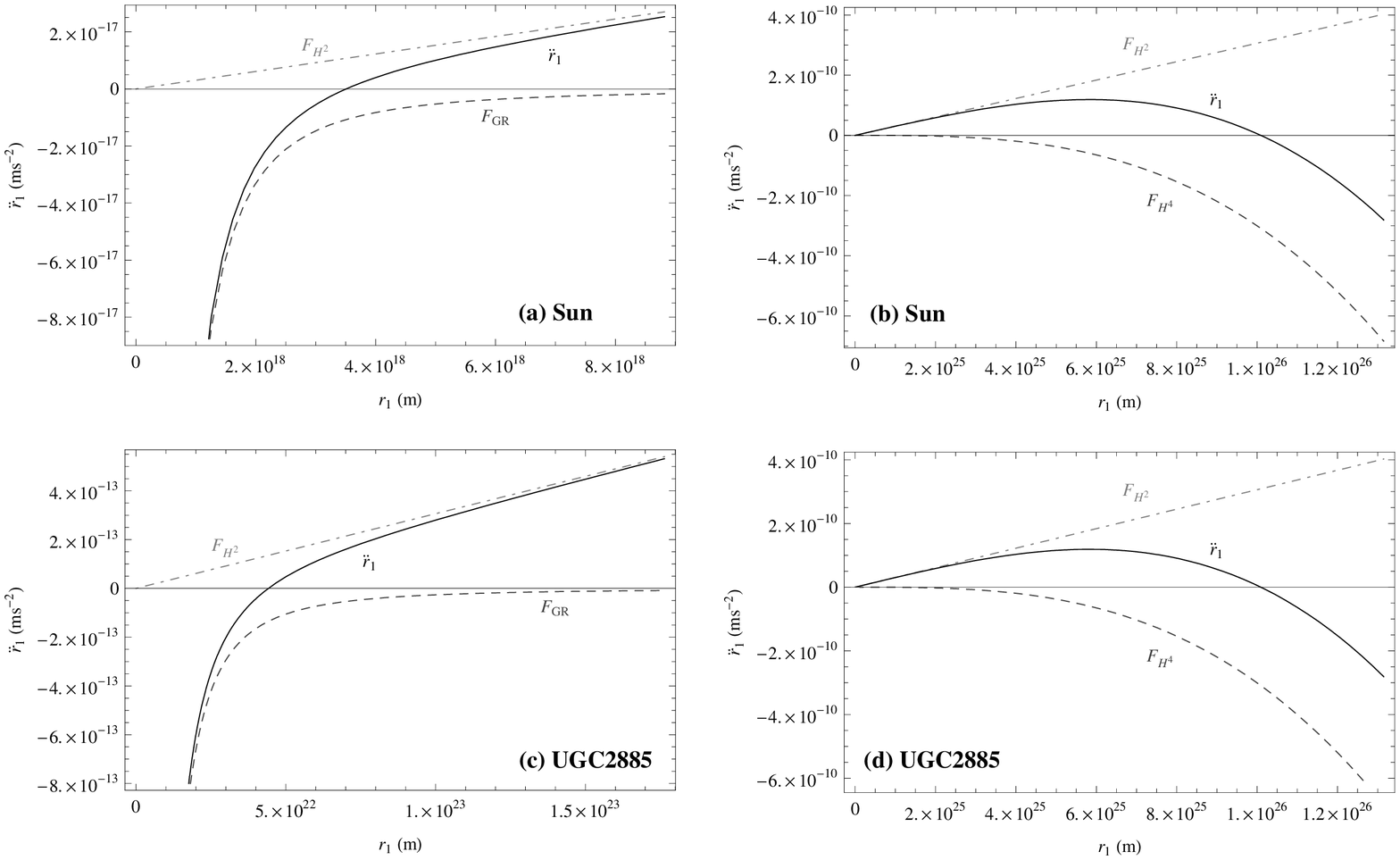}{150mm}{The modified Newton law
$\ddot{r}_1=F_{GR}+F_{H^2}+F_{H^4}$~(\ref{F_Newton_mod}) for the Sun ($M=1.98\times 10^{30}\,Kg$)
and the large galaxy UGC2885 (assuming as approximation a point-like core mass of $M=2\times 10^{12}\,M_{\mathrm{Sun}}=3.96\times 10^{42}\,Kg$)
for $\alpha_0=10$ such that the perturbative regime in the gravitational field is valid. The lower
order terms in the series expansion~(\ref{F_Newton_mod_I_exp}) of~(\ref{F_Newton_mod_I}) are dominant for each of the factors, $F_{GR}\approx -GM/r_1^2$, $F_{H^2}\approx -q_0\,r_1\,H_0^2$ and $F_{H^4}\approx-c^2\,r_1^3/l_H^4$. For relatively small scales the factor $F_{GR}$ is dominant, in the intermediate region the factor $F_{H^2}$
is dominant and for relatively large scales the factor $F_{H^4}$ is dominant such that the effects due to the central mass are only felt at relatively small scales:\hfill\break
\ {\bf(a)} relatively small values of the radial coordinate above the Schwarzschild
radius $r_{1.\mathrm{SC}}=2940\,m$ for the Sun, the modified Newton law is the continuous line such that
the gravitational acceleration changes sign at $r_1=r_1^*=3.51\times 10^{18}\,m$~(\ref{r_i_1}),
the factor $F_{GR}$ is the dashed line, the factor $F_{H^2}$ is the dashed-dotted line and the
factor $F_{H^4}$ is negligible;\hfill\break
\ {\bf(b)} relatively large values of the radial coordinate up to the cosmological horizon
$r_1=l_H=1.31\times 10^{26}\,m$ for the Sun, the modified Newton law is the continuous line such
that the gravitational acceleration changes sign at $r_1=\sqrt{-q}\,l_H=10^{26}\,m$~(\ref{r_FRW_1}),
the factor $F_{GR}$ is negligible, the factor $F_{H^2}$ is the dashed-dotted line and the
factor $F_{H^4}$ is the dashed line;\hfill\break
\ {\bf(c)} relatively small values of the radial coordinate above the Schwarzschild radius
$r_{1.\mathrm{SC}}=5.88\times 10^{15}\,m$ for the large galaxy UGC2885, the modified Newton law
is the continuous line such that the gravitational acceleration changes sign at
$r_1=r_1^*=4.42\times 10^{22}\,m$~(\ref{r_i_1}), the factor $F_{GR}$ is the dashed line, the
factor $F_{H^2}$ is the dashed-dotted line and the factor $F_{H^4}$ is negligible;\hfill\break
\ {\bf(d)} relatively large values of the radial coordinate up to the cosmological horizon
$r_1=l_H=1.31\times 10^{26}\,m$ for the large galaxy UGC2885, the modified Newton law is the
continuous line such that the gravitational acceleration changes sign at $r_1=\sqrt{-q}\,l_H=10^{26}\,m$~(\ref{r_FRW_1}), the factor $F_{GR}$ is negligible,
the factor $F_{H^2}$ is the dashed-dotted line and the factor $F_{H^4}$ is the dashed
line.}{fig.Newton_sun_perturbative}
\clearpage

Hence for $r_1<{r_1^*}_{(\mathrm{pert})}$
the General Relativity Newton acceleration factor $F_{\mathrm{GR}}$ is dominant such that the
gravitational interaction is attractive, at $r_1={r_1^*}_{(\mathrm{pert})}$ both the factors $F_{\mathrm{GR}}$ and $F_{H^2}$ terms cancel such that the gravitational interaction is null
and for $r_1>{r_1^*}_{(\mathrm{pert})}$ the factor $F_{H^2}$ is dominant such that the
gravitational interaction is repulsive. As for large distances, at
$r_1=\sqrt{-q_0}\,l_H$~(\ref{r_FRW_1}), the gravitational acceleration is null and, for
$r_1>\sqrt{-q_0}\,l_H$, the gravitational interaction becomes again attractive
due to the term $-c^2\,r_1^3/l_H^4$ in the factor $F_{H^4}$ becoming dominant as already discussed
in section~\ref{sec.review_map}. It can also be concluded straight forwardly that, although the value
of ${r_1^*}_{(\mathrm{pert})}$~(\ref{r_i_1}) increases with the value
of the mass, for large enough radial coordinate the mass effects are negligible such that,
independently of the value of the mass $M$, the modified Newton law~(\ref{F_Newton_mod})
converges asymptotically to the Newton law for an expanding background~(\ref{F_Newton_FRW_1}) described by
the FLRW metric~(\ref{g_FRW_1}). We recall that this result is consistent with the
assumptions considered when constructing the ansatz for the locally anisotropic metric~(\ref{g_generic})
in section~\ref{sec.ansatz_I}.

Therefore, for small values of the parameter $\alpha_0$, we obtain the lower order perturbative
modified Newton law (the weak field classical limit of the relativistic acceleration) interpolation
between the General Relativity Newton law~(\ref{F_Newton_GR}) obtained from the Schwarzschild metric~(\ref{g_SC}) and the Newton law for an expanding background~(\ref{F_Newton_FRW_1})
obtained from the FLRW metric~(\ref{g_FRW}).
As examples of typical profiles of the perturbative Newton law
for a point-like central mass are plotted in figure~\ref{fig.Newton_sun_perturbative} the
gravitational accelerations felt by a test mass in the gravitational field of the Sun and the large
galaxy UGC2885 for $\alpha_0=10$.

\subsubsection{Non-Perturbative Regime\lb{sec.n_pert}}

For higher values of the parameter $\alpha_0$ the remaining terms on the series expansion (on the
gravitational field) become relevant and contribute significantly to the modified Newton law. Instead of directly consider the validity limit imposed to the first order
term by equation~(\ref{val_1st_order}) obtained from the series expansions of the individual multiplicative
terms~(\ref{U_exp}) of the factors $F_{H^2}$ and $F_{H^4}$~(\ref{F_Newton_mod_I}) we can carry out a 
similar analysis directly in the series expansions for these factors~(\ref{F_Newton_mod_I_exp}).
Hence, for spatial scales for which the factor $F_{H^4}$ is negligible, by directly comparing the
terms on the series of the factor $F_{H^2}$~(\ref{F_Newton_mod_I}), we can conclude that, for some given $r_1$, the series is no-longer accurate when the first order term in $\alpha_0$,
$(\alpha_0+1)q_0\,GM\,H_0^2/c^2$, is of the same order of the zeroth order term in $\alpha_0$,
$-q_0\,r_1\,H_0^2$. When the parameter $\alpha_0$ is above the value that satisfies this equality
the series expansion is no longer valid in the neighborhood of the given value for $r_1$
and the exact expressions for $F_{H^2}$ and $F_{H^4}$~(\ref{F_Newton_mod_I}) must be considered.
In particular, for large values of the parameter $\alpha_0$, the exact value of the radial coordinate
$r_1^*$ for which the gravitational acceleration is null is above the value given in equation~(\ref{r_i_1}), $r_1^*>{r_1^*}_{(\mathrm{pert})}$, and must be computed by
solving the exact expression for the modified Newton law. For the specific case of $r_1$ in the
neighborhood of ${r_1^*}_{(\mathrm{pert})}$ the zeroth order term
$-q_0\,{r_1^*}_{(\mathrm{pert})}\,H_0^2$ coincides with the classical Newton law term
$GM/{r_1^*}_{(\mathrm{pert})}^2$, hence we can directly compare this term with the first order
term $(\alpha_0+1)q_0\,GM\,H^2/c^2$ such that the specific value of the parameter
$\alpha_0=\alpha_0^*$ separating in between these perturbative and non-perturbative
regimes in the gravitational field is
\be
\frac{GM}{({r_1^*}_{(\mathrm{pert})})^2}\approx-(\alpha_0^*+1)\,q\,\frac{GM}{c^2}\,H_0^2\ \ \Leftrightarrow\ \ \alpha_0^*\approx \frac{c^2-|q_0|^\frac{1}{3}(GM\,H_0)^\frac{2}{3}}{|q_0|^\frac{1}{3}(GM\,H_0)^\frac{2}{3}}\approx\frac{c^2}{|q_0|^\frac{1}{3}(GM\,H_0)^\frac{2}{3}}\ .
\lb{alpha_0_star}
\ee
As examples are plotted in figure~\ref{fig.r_non-perturbative} the values of $r_1^*$ computed numerically from the exact expression for the modified Newton law~(\ref{F_Newton_mod}) as a function of the parameter
$\alpha_0$ for the Sun and the large galaxy UGC2885 as well as the respective values $\alpha_0^*$
above which the perturbative regime in the gravitational field is no longer valid.
\fig{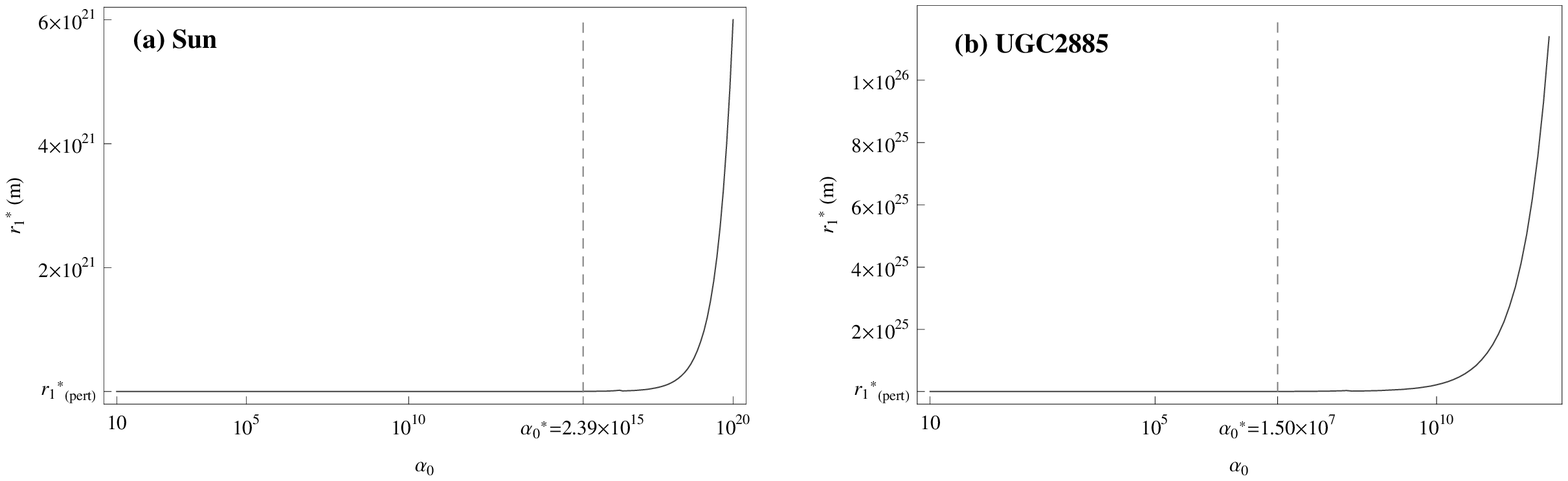}{150mm}{The values of $r_1^*$ for which the modified Newton
acceleration~(\ref{F_Newton_mod}) is null computed numerical from the exact expressions of
$F_{GR}$, $F_{H^2}$ and $F_{H^4}$~(\ref{F_Newton_mod_I}) for the Sun ($M_{\mathrm{Sun}}=1.98\times 10^{30}\,Kg$) and the large galaxy UGC2885 (assuming a point-like core mass of
$M=2\times 10^{12}\,M_{\mathrm{Sun}}=3.96\times 10^{42}\,Kg$):\hfill\break
\ {\bf(a)} the case of the Sun for which ${r_1^*}_{\mathrm(pert)}=3.51\times 10^{18}\,m$ and $\alpha_0^*=2.39\times 10^{15}$;\hfill\break
\ {\bf(b)} the case of the large galaxy UGC2885 for which ${r_1^*}_{\mathrm(pert)}=4.42\times 10^{22}\,m$ and $\alpha_0^*=1.50\times 10^{7}$.}{fig.r_non-perturbative}

We further note that, the higher the value of the parameter $\alpha_0$, the closer the modified Newton
law~(\ref{F_Newton_mod}) is from the General Relativity Newton law for small values of the radial coordinate
(roughly for $r_1<{r_1^*}_{(\mathrm{pert}})$). This result is consistent with the results derived in
section~\ref{sec.alpha_r_sc} when analyzing the curvature close to the Schwarzschild horizon,
for high values of the parameter $\alpha_0$ space-time near this horizon is flat. Furthermore,
for large enough values of this parameter, there is an attractive acceleration contribution
that becomes significant for values of the radial coordinate in between ${r_1^*}_{(\mathrm{pert})}$
and $r_1^*$ increasing the Newtonian acceleration towards the central mass. In
figure~\ref{fig.Newton_alpha} are presented the examples of the modified Newton law~(\ref{F_Newton_mod})
for the Sun and the large Galaxy UGC2885 for values of the parameter $\alpha_0$ above
$\alpha_0^*$~(\ref{alpha_0_star}).
\fig{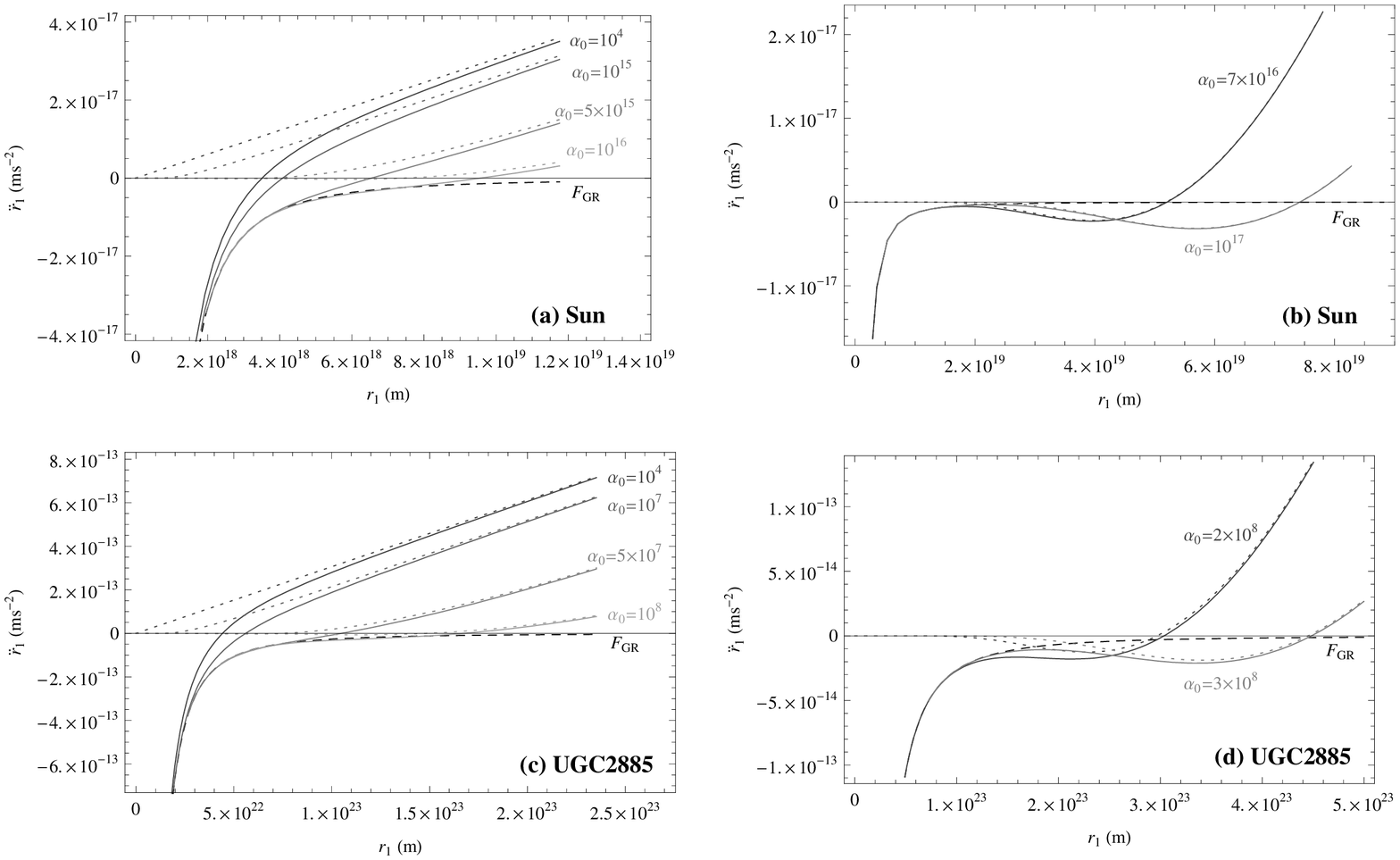}{150mm}{The modified Newton law
$\ddot{r}_1=F_{GR}+F_{H^2}+F_{H^4}$~(\ref{F_Newton_mod}) for values of the parameter $\alpha_0$
above $\alpha_0^*$~(\ref{alpha_0_star}) for which the perturbative regime in the gravitational
field is not valid. Are shown the cases of the Sun
($M_{\mathrm{Sun}}=1.98\times 10^{30}\,Kg$) and the large galaxy UGC2885 (assuming a point-like core
mass of $M=2\times 10^{12}\,M_{\mathrm{Sun}}=3.96\times 10^{42}\,Kg$) with the General Relativity Newton
law $F_{GR}$ represented by the dashed lines, the contribution of the factor $F_{H^2}$ is represented
by the dotted lines, the factor $F_{H^4}$ is negligible and the modified Newton law $\ddot{r}_1$~(\ref{F_Newton_mod}) is represented by the continuous lines:\hfill\break
\ {\bf(a)} for the Sun with $\alpha_0=10^4,10^{15},5\times 10^{15},10^{16}$;\hfill\break
\ {\bf(b)} for the Sun with $\alpha_0=7\times 10^{16},10^{17}$;\hfill\break
\ {\bf(c)} for the large galaxy UGC2885 with $\alpha_0=10^4,10^{7},5\times 10^{7},10^{8}$;\hfill\break
\ {\bf(d)} for the large galaxy UGC2885 with $\alpha_0=2\times 10^{8},3\times 10^{8}$.}{fig.Newton_alpha}

In particular these results imply that, for large values of the parameter $\alpha_0$, while for
planetary scales the effects of expansion may be for most purposes negligible, for galaxy scales
(and above) we may obtain significant deviations from the General Relativity Newton law.
The most straight conclusion is that exists a well defined cutoff corresponding to $r_1=r_1^*$ such that for radial coordinate above this value the gravitational interaction is repulsive due to the expanding
background and no stable orbits exist. In order to further analyze large scale (meaning at least
galactic scale and $r_1<r^*_1$) orbital motion, let us assume the approximation to circular orbits
such that the orbital speed is as usual computed by noting that the gravitational acceleration 
exactly matches the centrifugal acceleration
\be
-\ddot{r}_1=\frac{v_{\mathrm{orb}}^2}{r_1}\ \ \Leftrightarrow\ \ v_{\mathrm{orb}}=\sqrt{-\ddot{r}_1\,r_1}\ .
\lb{v_orb_circ}
\ee
When the acceleration is given by the modified Newton
law~(\ref{F_Newton_mod}) the orbital velocity will significantly deviate from the orbital
velocity for the General Relativity Newton law~(\ref{F_Newton_GR}) for values
of the radial coordinate above ${r_1^*}_{(\mathrm{pert})}$. As an example are
presented in figure~\ref{fig.galaxy_v} the rotation curves for several values of the
parameter $\alpha_0$ for the large galaxy UGC2885 assuming (as a simplification) the core to be
a point-like central mass.
\fig{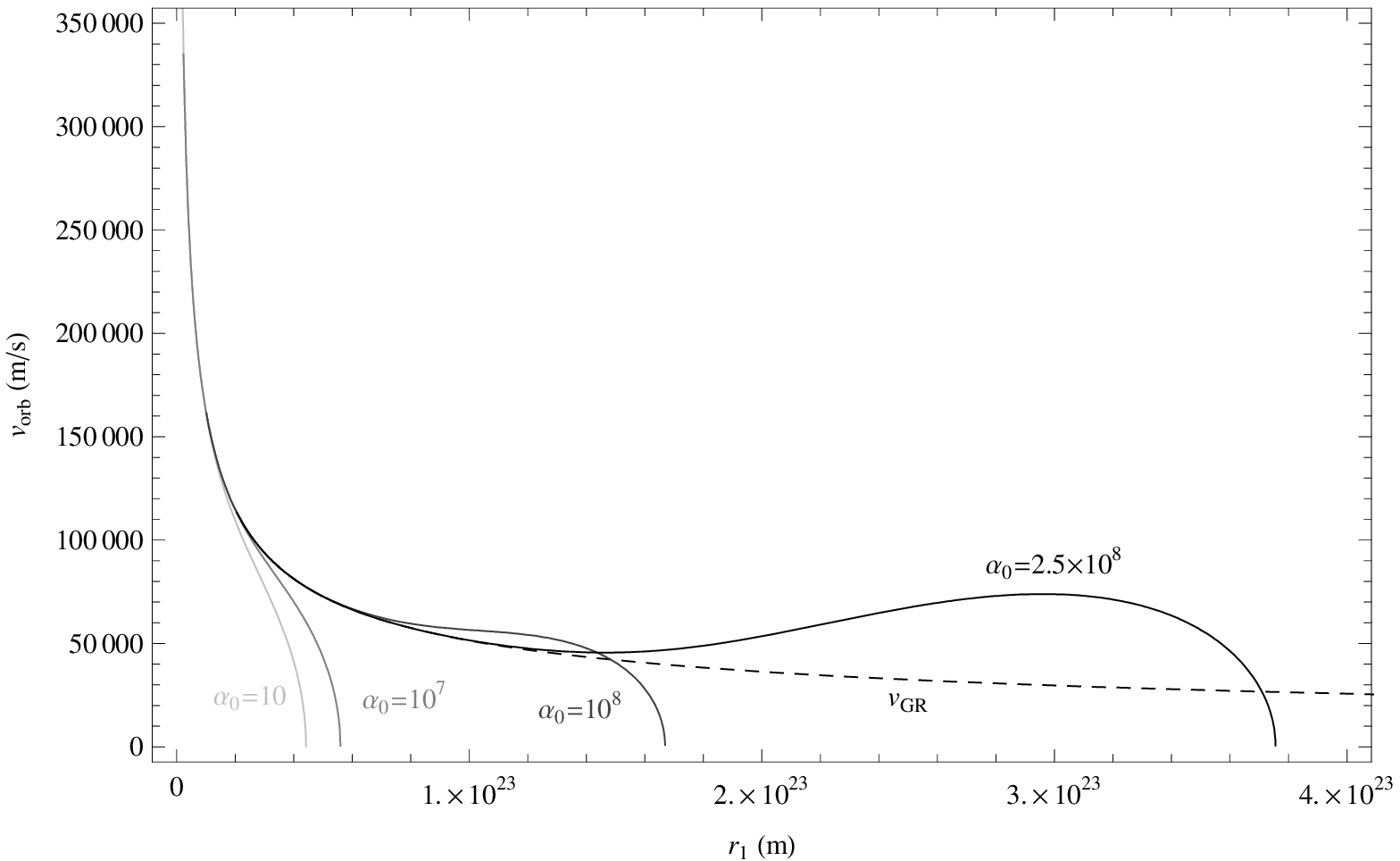}{75mm}{Circular orbital velocities $v_{\mathrm{orb}}=\sqrt{-(F_{GR}+F_{H^2}+F_{H^4})r_1}$~(\ref{F_Newton_mod}) for values of the
parameter $\alpha_0=10,10^7,10^8,2.5\times 10^8$ above $\alpha_0^*=1.5\times 10^7$~(\ref{alpha_0_star})
for the large galaxy UGC2885 (assuming a point-like core mass of
$M=2\times 10^{12}\,M_{\mathrm{Sun}}=3.96\times 10^{42}\,Kg$). The orbital velocity for the
General Relativity Newton law~(\ref{F_Newton_GR}) is represented by the dashed line.}{fig.galaxy_v}

Hence in this section we have shown that the modified Newton law corresponding to the locally
anisotropic metric~(\ref{g_generic}) with exponent $\alpha$ given in equation~(\ref{alpha_r_1})
interpolates between the General Relativity Newton law~(\ref{F_Newton_GR}) for relatively
short radial scales and the Newton law for the expanding background~(\ref{F_Newton_FRW_1})
for relatively large scales. In between there are significant deviations from these
limiting laws, in particular for large values of the parameter $\alpha_0$ there is an increase
of the attractive gravitational acceleration towards the central mass near the value
of the radial coordinate for which the net gravitational acceleration is null.
This result is not completely unwelcome and could contribute, for instance,
for the flattening effect of the rotation velocity curves for galaxies~\cite{galaxies_DM}.
We remark that a large value of this parameter is physically justified by noting that
it maintains the usual General Relativity Newton law at planetary scales such that
space-time is approximately flat near massive bodies.

In order to properly evaluate the effect of the locally anisotropic metric~(\ref{g_generic}) for
planetary motion and analyze if it is possible to obtain any bound on the parameter $\alpha_0$
for the solar system we next analyze the orbit solutions for this metric. We start by taking
the most simple approach by considering circular orbits which allows to estimate the time dependence
of the orbital radius due to the expanding background. Then we proceed to compute the
elliptic orbit solutions to zeroth order on time (static solutions) which allows to obtain estimative
for precession and period corrections due to the expanding background.

\subsection{Circular Orbits Approximation: Time Varying Orbital Radius}

As a first approximation to orbital motion we are considering circular orbits
and compare the effects of the modified Newton law~(\ref{F_Newton_mod})
obtained from the locally anisotropic metric~(\ref{g_generic}). We take as
starting point conservation of angular momentum. For orbital motion the angular momentum
is given by the constant of motion $J=-r_1^2\,d\varphi/d\tau$~(\ref{orbital_constants})
discussed in section~\ref{sec.rev_orbits} when reviewing orbit solutions in the framework
of General Relativity. We note that the expression for
this constant is the same both for classical Keplerian orbits, relativistic orbits
obtained from the Schwarzschild metric~(\ref{g_SC}) and orbits
in an expanding background obtained from the locally anisotropic metric~(\ref{g_generic})
as derived in appendix~\ref{A.orbits}. Hence, recalling that the orbital velocity
for circular orbits is generally given by equation~(\ref{v_orb_circ}) and taking the
usual definition of angular velocity $\dot{\varphi}=\omega=r_1\,v_{\mathrm{orb}}$
with $v_{\mathrm{orb}}=\sqrt{-r_1\,\ddot{r}_1}$~(\ref{v_orb_circ}),
we obtain the following definition of $J_{\mathrm{circ}}$ for a circular orbit of
radius $r_{1.\mathrm{orb}}$
\be
\ba{rcl}
\displaystyle J_{\mathrm{circ}}^2&\approx&\displaystyle \left.-\gamma^2\,r_1^3\,\ddot{r}_1\right|_{r_1=r_{1.\mathrm{orb}}}\\[6mm]
&\approx&\displaystyle \frac{GM\,r_{1.\mathrm{orb}}-\frac{2(GM)^2}{c^2}-F_{H^2}\,r_{1.\mathrm{orb}}^3}{1-\frac{2GM}{c^2\,r_{1.\mathrm{orb}}}-\left(\frac{H\,r_{1.\mathrm{orb}}}{c}\right)^2\left(1-\frac{2GM}{c^2\,r_{1.\mathrm{orb}}}\right)^\alpha}\\[6mm]
&\approx&\displaystyle  GM\,r_{1.\mathrm{orb}}\,\left(1+\left(\frac{H\,r_{1.\mathrm{orb}}}{c}\right)^2\left(1-\frac{2GM}{c^2\,r_{1.\mathrm{orb}}}\right)^\alpha\right)-F_{H^2}\,r_{1.\mathrm{orb}}^3+O(H^4)\ .
\ea
\lb{J_circ}
\ee
Here $\gamma=dt/d\tau$ is the relativistic factor for the generic metric~(\ref{g_generic}) and we are taking the
limit of non-relativistic velocities $\dot{x}^\mu\ll c$ for which
\be
\ba{rcl}
\displaystyle\frac{d^2r_1}{d\tau^2}&\approx&\displaystyle\gamma^2\,\ddot{r}_1\\[6mm]
&=&\displaystyle\frac{\ddot{r}_1}{g_{00}-2g_{01}\,\frac{\dot{r}_1}{c}-g_{11}\,\left(\frac{\dot{r}_1}{c}\right)^2-g_{22}\,\left(\frac{\dot{\varphi}}{c}\right)^2}\\[6mm]
&\approx&\displaystyle\frac{\ddot{r}_1}{1-\frac{2GM}{c^2r_1}-\left(\frac{H\,r_1}{c}\right)^2\left(1-\frac{2GM}{c^2\,r_1}\right)^\alpha}\ .
\ea
\lb{gamma_approx_generic}
\ee
In~(\ref{J_circ}) we are considering the modified Newton law~(\ref{F_Newton_mod}) and neglecting
the term $F_{H^4}$ which is only relevant for cosmological scales ($r_1\sim l_H\sim 10^{26}\,m$)
as discussed in the previous section (here we are considering planetary scales,
at most $r_1\sim 10^{13}\,m$) such that the gravitational acceleration is given by
$\ddot{r}_1\approx F_{GR}+F_{H^2}$ with the factors $F_{GR}$ and $F_{H^2}$
given in equation~(\ref{F_Newton_mod_I}). In the last approximation of equation~(\ref{J_circ})
we have expanded the denominator to lowest order in the gravitational field and $H^2$.

For circular orbits, the main effect obtained due to the corrections of the expanding background
is a time varying radius. Recalling again that $J$ is a constant of motion
(which corresponds to angular momentum) and that the Hubble rate
$H(t)\approx H_0-q_0\,H_0^2\,t$~(\ref{a_exp}) is time dependent
we note that necessarily some other parameter of the equation~(\ref{J_circ}) must be varying
with time to maintain $J$ constant. Assuming that the Newton constant is fixed on time we are
left only with the possibility of a time-varying radius $r_{1.\mathrm{orb}}=r_{1.\mathrm{orb}}(t)$.
Hence differentiating equation~(\ref{J_circ}) and solving the resulting equation
$\dot{J}_{\mathrm{circ}}=0$ for $\dot{r}_{1.\mathrm{orb}}$ we obtain, to lowest order, the time dependence of the orbital radius
\be
\ba{rcl}
\left.\displaystyle\frac{\dot{r}_{1.\mathrm{orb}}}{r_{1.\mathrm{orb}}}\right|_{\dot{G}=0}&=&\displaystyle \frac{2(H_0\,r_{1.\mathrm{orb}})^3}{GM}(1-q_0\,H_0\,t)\,\left(1-\frac{2GM}{c^2\,r_{1.\mathrm{orb}}}\right)^{\frac{\alpha_0}{2}+\frac{1}{2}}\times\\[6mm]
&&\displaystyle\hfill\times\left(1+q_0-\left(1-\frac{(2-\alpha_0)GM}{c^2\,r_{1.\mathrm{orb}}}\right)\left(1-\frac{2GM}{c^2\,r_{1.\mathrm{orb}}}\right)^{\frac{\alpha_0}{2}-\frac{1}{2}}\right)/\\[6mm]
&&\displaystyle/\left(1-\frac{H_0^2\,r_{1.\mathrm{orb}}^3}{GM}\,(1-q_0\,H_0\,t)\,\left(1-\frac{2GM}{c^2\,r_{1.\mathrm{orb}}}\right)^{\frac{\alpha_0}{2}-\frac{3}{2}}\times\right.\\[6mm]
&&\displaystyle\ \ \ \ \times\left((1+q_0)\left(4+\frac{(\alpha_0-15)GM}{c^2\,r_{1.\mathrm{orb}}}-\frac{2(\alpha_0-7)(GM)^2}{c^4\,r_{1.\mathrm{orb}}^2}\right)\right.\\[6mm]
&&\displaystyle\ \ \ \ -\left(4+\frac{(5\alpha_0-14)GM}{c^2\,r_{1.\mathrm{orb}}}+\frac{2(\alpha_0^2-5\alpha_0+6)(GM)^2}{c^4\,r_{1.\mathrm{orb}}^2}\right)\times\\[6mm]
&&\displaystyle\hfill\left.\left.\left(1-\frac{2GM}{c^2\,r_{1.\mathrm{orb}}}\right)^{\frac{\alpha_0}{2}+\frac{1}{2}}\right)\right)\\[6mm]
&\approx&\displaystyle \frac{2(H_0\,r_{1.\mathrm{orb}})^3}{GM}\,\left(1-\frac{2GM}{c^2\,r_{1.\mathrm{orb}}}\right)^{\frac{\alpha_0}{2}+\frac{1}{2}}\times\\[6mm]
&&\displaystyle\hfill\times\left(1+q_0-\left(1-\frac{(2-\alpha_0)GM}{c^2\,r_{1.\mathrm{orb}}}\right)\left(1-\frac{2GM}{c^2\,r_{1.\mathrm{orb}}}\right)^{\frac{\alpha_0}{2}-\frac{1}{2}}\right)+O(H_0^5)\ ,
\ea
\lb{dr_orb_circ}
\ee
where we have neglected the contribution from the parameter $\alpha_1$ and, in the last approximation,
have expanded the denominator to lowest order and kept only the terms on $H_0^3$. We note that this
expression increases with the orbital radius $r_{1.\mathrm{orb}}$ and decreases with the mass $M$.
Hence, for fixed value of the parameter $\alpha_0$, the values for this expression will change
significantly for distinct orbits. We note that this result is consistent with the expected
results for the FLRW metric~(\ref{g_FRW_1}), the expansion effects are larger for larger distances
and taking the interpretation of expansion being due to the net effect of the long range gravitational
interactions we also expect that in the neighborhood of a larger mass $M$ its effects will be
dominant with respect to the effects of more distant objects wile for smaller masses the net effect
of distant objects will become relevant.
With respect to the value of the parameter $\alpha_0$, the time variation of the orbital radius
will be positive with a maximum for the lowest allowed value of the parameter $\alpha_0=3$,
will decrease with increasing $\alpha_0$ until reaching a negative minimum and them increase towards
zero (being still negative, it is only null in the limit $\alpha_0\to+\infty$). We note that both
the values of the parameter $\alpha_0$ for which $\dot{r}_{1.\mathrm{orb}}=0$ and for the negative
minimum of $r_{1.\mathrm{orb}}$ are below $\alpha_0^*$~(\ref{alpha_0_star}) although being close to
this value, hence still for values of $\alpha_0$ for which a series expansion in the gravitational
field can be considered. However due to being relatively close to this value to attain accurate
results perturbatively it would be necessary to consider higher order terms of the series. Hence
we will evaluate numerically the exact expression~(\ref{dr_orb_circ}) in the following discussions.
As an example of the typical values of $\dot{r}_{1.orb}$ as a function of the parameter
$\alpha_0$ are presented in figure~\ref{fig.dr_alpha_circ} the values of $\dot{r}_{1.orb}/r_{1.orb}$
for the Moon orbit with the central mass being Earth and for Venus, Earth and Mars
orbit with the central mass being the Sun.
\fig{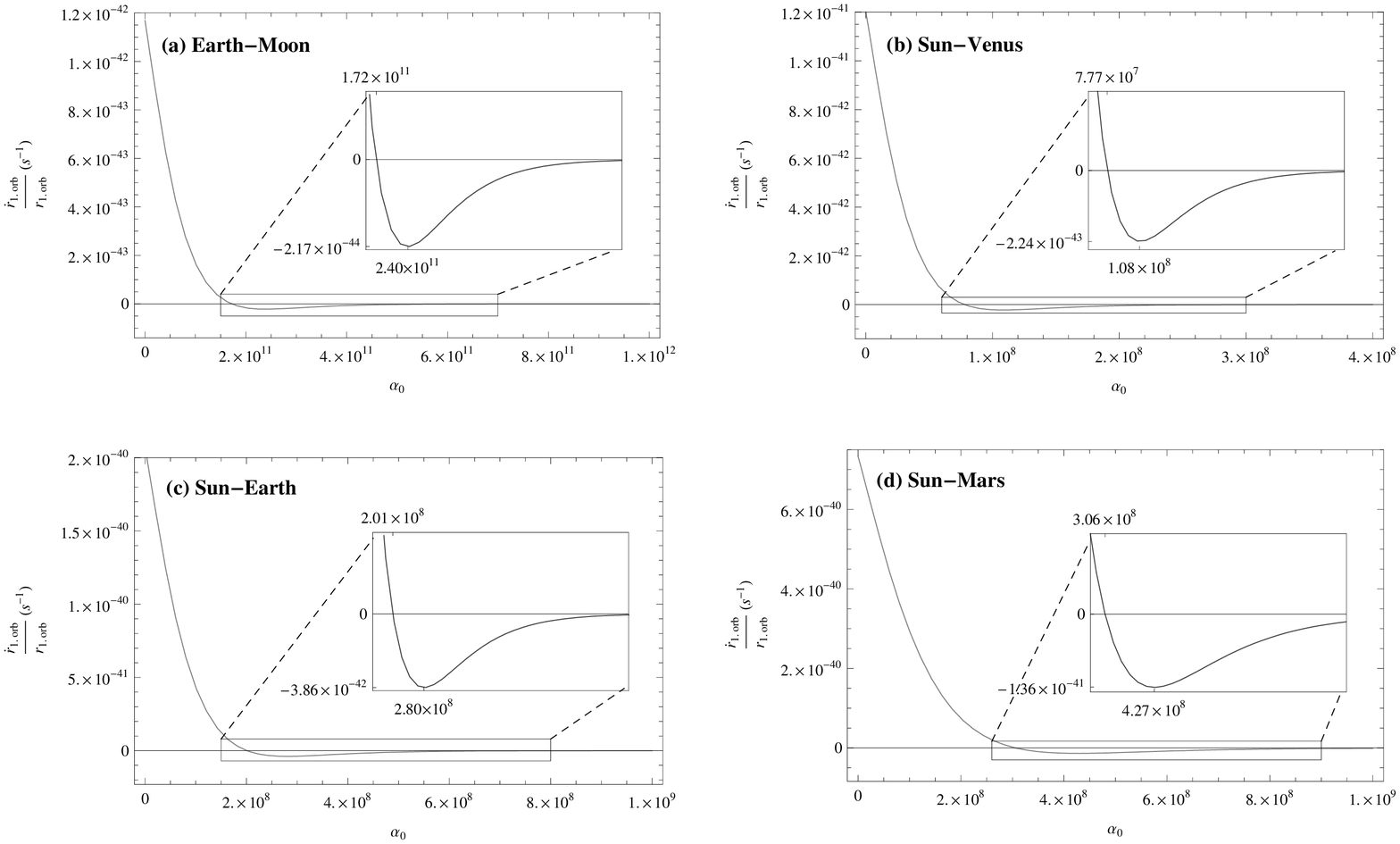}{150mm}{Examples of the profiles of the time variation rate of the orbital
radius $\dot{r}_{1.\mathrm{orb}}/r_{1.\mathrm{orb}}$~(\ref{dr_orb_circ}) as a function of the
parameter $\alpha_0$ assuming the circular orbits approximation:\hfill\break
\ {\bf(a)} for the Earth-Moon orbit, the maximum positive variation is 
$\dot{r}_{1.\mathrm{orb}}/r_{1.\mathrm{orb}}=1.17\times 10^{-42}\,s^{-1}$
corresponding to $\alpha_0=3$, the radius variation is null
$\dot{r}_{1.\mathrm{orb}}/r_{1.\mathrm{orb}}=0\,s^{-1}$ for $\alpha_0=1.72\times 10^{11}$
and the minimum negative variation is $\dot{r}_{1.\mathrm{orb}}/r_{1.\mathrm{orb}}=-2.17\times 10^{-44}\,s^{-1}$ corresponding to $\alpha_0=2.40\times 10^{11}$ ;\hfill\break
\ {\bf(b)} for the Sun-Venus orbit, the maximum positive variation is 
$\dot{r}_{1.\mathrm{orb}}/r_{1.\mathrm{orb}}=1.20\times 10^{-41}\,s^{-1}$
corresponding to $\alpha_0=3$, the radius variation is null
$\dot{r}_{1.\mathrm{orb}}/r_{1.\mathrm{orb}}=0\,s^{-1}$ for $\alpha_0=1.77\times 10^{7}$
and the minimum negative variation is $\dot{r}_{1.\mathrm{orb}}/r_{1.\mathrm{orb}}=-2.24\times 10^{-43}\,s^{-1}$ corresponding to $\alpha_0=1.08\times 10^8$ ;\hfill\break
\ {\bf(c)} for the Sun-Earth orbit, the maximum positive variation is 
$\dot{r}_{1.\mathrm{orb}}/r_{1.\mathrm{orb}}=2.07\times 10^{-40}\,s^{-1}$
corresponding to $\alpha_0=3$, the radius variation is null
$\dot{r}_{1.\mathrm{orb}}/r_{1.\mathrm{orb}}=0\,s^{-1}$ for $\alpha_0=2.01\times 10^{8}$
and the minimum negative variation is $\dot{r}_{1.\mathrm{orb}}/r_{1.\mathrm{orb}}=-3.86\times 10^{-42}\,s^{-1}$ corresponding to $\alpha_0=2.80\times 10^8$ ;\hfill\break
\ {\bf(d)} for the Sun-Mars orbit, the maximum positive variation is 
$\dot{r}_{1.\mathrm{orb}}/r_{1.\mathrm{orb}}=7.34\times 10^{-40}\,s^{-1}$
corresponding to $\alpha_0=3$, the radius variation is null
$\dot{r}_{1.\mathrm{orb}}/r_{1.\mathrm{orb}}=0\,s^{-1}$ for $\alpha_0=3.06\times 10^{8}$
and the minimum negative variation is $\dot{r}_{1.\mathrm{orb}}/r_{1.\mathrm{orb}}=-1.36\times 10^{-41}\,s^{-1}$ corresponding to $\alpha_0=4.27\times 10^8$.}{fig.dr_alpha_circ}
\clearpage

Depending on the value of the parameter $\alpha_0$, the radius variation of
the planetary orbits in the solar system can be, either positive, either negative or some
positive and others negative. For low values of the parameter $\alpha_0\ll 10^8$ the radius
variation are all positive, while for high values of the parameter $\alpha_0\gg 10^8$ the radius
variation are all negative and for values of the parameter $\alpha_0\sim 10^8$ the radius variation
for planets closer to the Sun are positive and for planets farther away from the Sun are negative.
As an example, taking $\alpha_0=10^8$, we have that Mercury and Venus are approaching the Sun while
the remaining planets are drifting away from the Sun. We note that this result is not inconsistent
with global spatial expansion, we recall that for large spatial scales we recover the FLRW
metric~(\ref{g_FRW_1}) and the usual properties of spatial expansion, instead this behavior
should be interpreted as due to the local background deformation in the neighborhood of the
central mass $M$.

In order to compare the values obtained for the expression~(\ref{dr_orb_circ}) with experimental
data we note that the only available experimental results are the values for the time-varying Newton
constant $\dot{G}/G$ (see for instance~\cite{Uzan} for a extended review in this topic).
Similarly to the derivation of equation~(\ref{J_circ}), for the General
Relativity Newton law we have that the constant of motion $J$ for a circular orbit is given by
\be
J^2_{\mathrm{circ}.GR}\approx\left.-\gamma^2\,r_1^3\,\ddot{r}_1\right|_{r_1=r_{1.\mathrm{orb}.GR}}\approx GM\,r_{1.\mathrm{orb}.GR}\,\frac{1-\frac{2GM}{c^2\,r_{1.\mathrm{orb}.GR}}}{1-\frac{2GM}{c^2\,r_{1.\mathrm{orb}.GR}}}=GM\,r_{1.\mathrm{orb}.GR}\ ,
\lb{J_circ_GR}
\ee
where, as before, the non-relativistic velocity limit has been considered, however no expansion on the
gravitational field has been considered such that, up to Special Relativistic corrections, this result
is exact. Hence for a static background no time variation of the orbital radius is possible unless
either the Newton constant $G$ or mass $M$ are also varying on time. Assuming that the mass is not varying
on time, $\dot{M}=0$, and that both the radius and the Newton constant are time dependent $r_{1.\mathrm{orb}.GR}=r_{1.\mathrm{orb}.GR}(t)$ and $G=G(t)$, differentiating the
equation~(\ref{J_circ_GR}) with respect to time, we obtain the following equality
\be
\frac{\dot{G}}{G}=-\frac{\dot{r}_{1.\mathrm{orb}.GR}}{\dot{r}_{1.\mathrm{orb}.GR}}\ .
\ee
Therefore when analyzing experimental data (as well as in numerical calculations) employing
General Relativity in a static background, a varying radius $r_{1.\mathrm{orb}.GR}$ is
equivalent to a varying $G$ which is quoted for several systems in the literature. In particular we
list the results obtained for the moon~\cite{Lunar_01}, Venus~\cite{Venus} and
Mars~\cite{Mars} and compare them with the respective allowed values obtained from
equation~(\ref{dr_orb_circ}) and presented in figure~\ref{fig.dr_alpha_circ}. As far as the
author is aware there are no estimative based on Earth's orbital motion~\cite{Uzan}.
\begin{table}[ht]
\begin{center}
\begin{tabular}{lcc}
Orbits & $\dot{r}_{1.\mathrm{orb}.GR}/r_{1.\mathrm{orb}.GR}=-\dot{G}/G$ & $\dot{r}_{1.\mathrm{orb}}/r_{1.\mathrm{orb}}$\\
      &$(s^{-1})$                       & $(s^{-1})$\\\hline\hline\\[-2mm]
Moon-Earth &$-\dot{G}/G=-(1.9\pm 2.22)\times 10^{-20}$& $\in[-2.17,1.17]\times 10^{-42}$\\[2mm]
Sun-Venus   &$|\dot{G}/G|<4.76\times 10^{-18}$&  $\in[-0.0224,1.2]\times 10^{-41}$\\[2mm]
Sun-Mars   &$-\dot{G}/G=(6.34\pm 31.71)\times 10^{-20}$&  $\in[-0.136,7.34]\times 10^{-40}$
\\[2mm]\hline
\end{tabular}
\caption{\it \small Experimental orbital radius variations and allowed theoretical range obtained from
equation~(\ref{dr_orb_circ}).\lb{table.dr_exp_alpha}}
\end{center}
\end{table}
Although within the error bars the theoretical ranges obtained from equation~(\ref{dr_orb_circ})
are clearly many orders of magnitude below the experimental values. This comparison is not conclusive
due to the error bars being significantly larger than the quoted experimental values and,
depending on the sources and methods employed in the data analysis, the values are quite distinct.
Hence no bounds on the parameter $\alpha_0$ can be drawn from this analysis.

It is also relevant to remark that usually it is assumed that a varying $G$ (equivalent to a varying
orbital radius in GR) is independent of the orbit parameters, this is no longer the case when considering
the locally anisotropic metric~(\ref{g_generic}), equation~(\ref{dr_orb_circ}) does depend
non-linearly both in the central mass value $M$ and the orbital radius $r_{1.\mathrm{orb}}$,
hence it is expected that distinct values are obtained for each orbit (as opposed to GR) which
may justify the dispersion of experimental values in the literature. In table~\ref{table.precession_dT}
we list the allowed ranges for the remaining planets of the solar system. We note that for larger
orbital radius the radius variation significantly increases.

As a final comment we note that from the locally anisotropic metric~(\ref{g_generic}) it is also possible
to consider either a varying $G$ with fixed orbital radius $r_{1.\mathrm{orb}}$ or both
varying $G$ and orbital radius $r_{1.\mathrm{orb}.GR}$. These approaches may be justified in
the context of varying fundamental constant theories~\cite{Uzan} or extended theories of
gravity~\cite{Dirac,Vinti,Dilaton}. Assuming conservation of angular momentum we obtain
the generic expression for the time variation of the radius when $\dot{G}\neq 0$
\be
\ba{rcl}
\displaystyle\frac{\dot{r}_{1.\mathrm{orb}}}{r_{1.\mathrm{orb}}}&=&\displaystyle\left.\frac{\dot{r}_{1.\mathrm{orb}}}{r_{1.\mathrm{orb}}}\right|_{\dot{G}=0}-\frac{\dot{G}}{G}\,\Delta\dot{r}_{1.\mathrm{orb}.\dot{G}}\ ,\\[6mm]
\Delta\dot{r}_{1.\mathrm{orb}.\dot{G}}&=&\displaystyle \left(\left. 1+\left(\frac{H_0\,r_1}{c}\right)^2(1-q_0\,H_0\,t)^2\left(1-\frac{2GM}{c^2\,r_1}\right)^{\frac{\alpha_0}{2}-\frac{1}{2}}\right((\alpha_0+1)(1+q_0)\right.\\[6mm]
&&\displaystyle\left.\left.\left.\hfill-\left((\alpha_0+2)+\frac{2(\alpha_0-2)(\alpha_0+1)GM}{c^2\,r_1}\right)\left(1-\frac{2GM}{c^2\,r_1}\right)^{\frac{\alpha_0}{2}-\frac{1}{2}}\right)\right)/\right(1\\[6mm]
&&\displaystyle+\frac{H_0^2r_1^3}{GM}(1-q_0\,H_0\,t)^2\left((1+q_0)\left(4+\frac{(\alpha_0-7)GM}{c^2\,r_1}\right)\left(1-\frac{2GM}{c^2\,r_1}\right)^{\frac{\alpha_0}{2}-\frac{1}{2}}\right.\\[6mm]
&&\displaystyle\left.\left.-\left(4-\frac{(5\alpha_0-14)GM}{c^2\,r_1}+\frac{2(\alpha_0^2-5\alpha_0+6)(GM)}{c^4\,r_1^2}\right)\left(1-\frac{2GM}{c^2\,r_1}\right)^{\alpha_0-1}\right)\right)\\[6mm]
&\approx&\displaystyle \left. 1+\left(\frac{H_0\,r_1}{c}\right)^2\left(1-\frac{2GM}{c^2\,r_1}\right)^{\frac{\alpha_0}{2}-\frac{1}{2}}\right((\alpha_0+1)(1+q_0)\\[6mm]
&&\displaystyle\hfill-\left.\left((\alpha_0+2)+\frac{2(\alpha_0-2)(\alpha_0+1)GM}{c^2\,r_1}\right)\left(1-\frac{2GM}{c^2\,r_1}\right)^{\frac{\alpha_0}{2}-\frac{1}{2}}\right)+O\left(H_0^5\right)\ ,
\ea
\ee
where $\left.\dot{r}_{1.\mathrm{orb}}/r_{1.\mathrm{orb}}\right|_{\dot{G}=0}$ is given in
equation~(\ref{dr_orb_circ}). It is possible to maintain the orbital radius fixed on time by
fine-tuning the time variation of $G$ such that this equation is null, $\dot{r}_{1.\mathrm{orb}}=0$.
Otherwise if the time variation of $G$ is given by some extended theory of gravity
its effects simply add to the orbital radius variation. We remark however that in the
present framework, for which we consider only the expanding background, our previous analysis
is consistent and does not require a varying $G$.

\subsection{Perturbative Static Elliptical Orbit Solutions}

With the objective of estimating the orbital precession and orbital period
corrections to an elliptical orbit due to the expanding background we
will proceed to derive orbital solutions for the locally anisotropic metric~(\ref{g_generic}).
It is hard, if not impossible, to obtain a analytical solution considering the differential
equations for a time varying Hubble rate $H$. The main difficulty is that energy conservation
is no-longer given by a constant of motion, instead we have a non-linear second order differential
equation on the function $t(\varphi)$ coupled to the differential equation
for $u(\varphi)=1/r_1(\varphi)$. Hence, for technical simplification purposes,
we are taking the static orbit approach by considering a fixed Hubble rate $H=H_0$.
Also we note that a estimative for the orbital radius time dependence has already been
computed in the previous section considering the approximation to circular orbits.

The differential equation describing an static orbit of a test particle in the gravitational
field of a central mass $M$ for the locally anisotropic metric~(\ref{g_generic}), to order $H_0^2$,
is given in equation~(\ref{A.Eq_u}) of Appendix~\ref{A.orbits} for the inverse radial
coordinate function $u(\varphi)=1/r_1(\varphi)$ and fixed angular coordinate $\theta=\pi/2$.
We note that for planetary scales the corrections to order $H_0^4$ are negligible.
So far the author failed to obtain a treatable analytic solution to this differential equation
considering the exact expressions for the corrections due to the expanding background.
Here, when refering to corrections we mean with respect to the General Relativity
orbit equation~(\ref{eqd_u}) discussed in section~\ref{sec.rev_orbits}.
For orbits in the solar system
the function $u$ has small values ($0.5\times 10^{-12}<u<0.5\times 10^{-10}\,m^{-1}$, where $r_{1.\mathrm{orb}}$ is the orbit semi-major axis) such that we can consider a series expansion on $u$
of the corrections due to the expanding background. However we note that this expansion is equivalent
to an expansion on the weak gravitational field and, as already discussed in section~\ref{sec.pert},
this perturbative regime is valid only for values of the exponent parameter $\alpha_0$
(we are assuming a negligible negative valor for the parameter $\alpha_1$) up to~(\ref{val_1st_order})
\be
\alpha_{0.\mathrm{max.pert}}\approx \frac{c^2\,r_{1.\mathrm{orb}}}{2GM}\ .
\lb{alpha_max_pert}
\ee
Above this value it is either necessary to consider higher order terms on the series expansion
or to consider the exact expressions. Nevertheless we remark that, for a fixed value of the
radial coordinate $r_1$, and larger values of $\alpha_0>\alpha_{0.\mathrm{max.pert}}$ the
corrections given by the exact expression due to the expanding background will decrease significantly
in absolute value becoming, for very large values of the parameter $\alpha_0\gg \alpha_{0.\mathrm{max.pert}}$, negligible. Considering a rough numerical estimative
we conclude that the maximum deviation for the corrections for higher values of the
parameter $\alpha_0>\alpha_{0.\mathrm{max.pert}}$ is less than 20\% of the value obtained
in the perturbative regime for $\alpha_0=\alpha_{0.\mathrm{max.pert}}$.
As an example are plotted in figure~\ref{fig.orbits_perturbative} the values
of the exact and perturbative correction terms on $H_0^2$ as presented in the differential
equation~(\ref{A.Eq_u}) for the case of Earth's orbit.
\fig{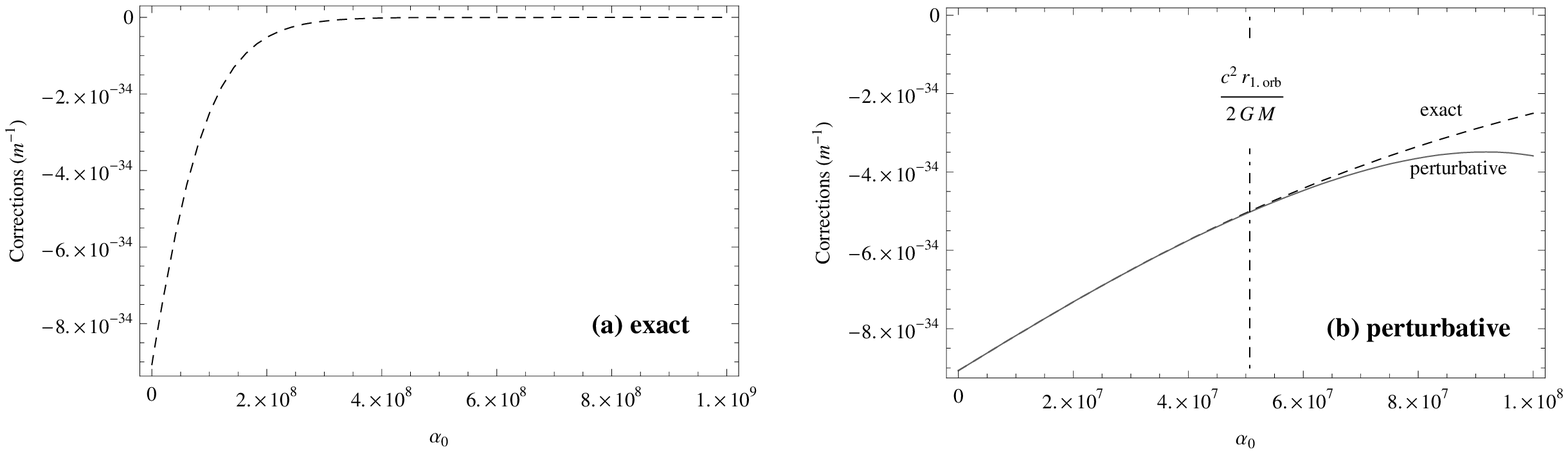}{150mm}{Plot of the exact (dashed line) and perturbative
(continuous line) expressions for the corrections to the General Relativity orbit differential
equation due to the expanding background given in equation~(\ref{A.Eq_u}) as a function of the
parameter $\alpha_0$ for Earth's orbit. The perturbative regime is valid up to $\alpha_{0.\mathrm{max.pert}}=c^2r_{1.\mathrm{orb}}/(2GM)\approx 0.5\times 10^7$~(\ref{alpha_max_pert}):\hfill\break
\ {\bf(a)} plot of the exact expressions up to $\alpha_0=10^9$, the corrections asymptoticaly vanish
for large $\alpha_0\gg \alpha_{0.\mathrm{max.pert}}$;\hfill\break
\ {\bf(b)} plot of the exact and perturbative expressions up to $\alpha_0=10^8$, the perturbative
and exact expressions approximately match up to $\alpha_{0.\mathrm{max.pert}}$.}{fig.orbits_perturbative}

Hence, for technical simplification, in order to approximately estimate
the orbit precession and period corrections due to the expanding background and compare our results
with the same effects due to standard General Relativity we are considering a perturbative series
expansion on $u$ to the orbit differential equation~(\ref{A.Eq_u}) of Appendix~\ref{A.orbits}.
We note that the General Relativity corrections to the orbit differential equation are
second order in $u$~(\ref{eqd_u}). To match at least the same order of these corrections
we are considering a series expansion up to second order such that we obtain
\be
\ba{rcl}
u''(\varphi)+A\,u(\varphi)&\approx&\displaystyle \frac{GM}{J^2}\,B+\frac{3GM}{c^2}\,C\,u^2\\[6mm]
&&\displaystyle-\left(\frac{H_0}{J}\right)^2\,\frac{1}{u^3}+\alpha_0\left(\frac{H_0}{J}\right)^2\,\frac{GM}{c^2}\,\frac{1}{u^2}+O\left(u^3\right)\ ,
\ea
\lb{orbits_exp}
\ee
where the correction coefficients are
\be
\ba{rcl}
A&=&1+\delta_A\ ,\ B\ =\ 1+\delta_B\ ,\ C\ =\ 1+\delta_C\ ,\\[6mm]
\delta_A&=&\displaystyle-2(\alpha_0-1)\alpha_0\left(\frac{GM\,H_0}{c^3}\right)^2\left(1+(\alpha_0^2-5\alpha_0+6)\frac{(GM)^2}{3c^2\,J^2}\right)\ ,\\[6mm]
\delta_B&=&\displaystyle-\alpha_0\,\left(\frac{H_0}{c^2}\right)^2\left(J^2+(\alpha_0^2-3\alpha_0+2)\frac{2(GM)^2}{3c^2}\right)\ ,\\[6mm]
\delta_C&=&\displaystyle-\frac{2}{3}\alpha_0\left(\alpha_0^2-3\alpha_0+2\right)\,\left(\frac{GM\,H_0}{c^3}\right)^2\left(1+(\alpha_0^2-7\alpha_0+12)\frac{(GM)^2}{5c^2\,J^2}\right)\ .
\ea
\lb{A_B_C}
\ee
As an example of the magnitude of the several correction terms for Earth's orbit we have that
the constant term in the right-hand side of the differential equation is of order
$\sim 7\times 10^{-12}\,m^{-1}$, the General Relativistic correction is of order
$\sim 2\times 10^{-19}\,m^{-1}$ and the corrections on $H_0^2$ are of order $\sim 9\times 10^{-34}\,m^{-1}$
for $\alpha_0=3$ and $\sim -3\times 10^{-41}\,m^{-1}$ for $\alpha_0=10^9$.

To solve the differential equation~(\ref{orbits_exp}) we take the same approach discussed in
section~\ref{sec.rev_orbits}. We start by solving the differential equation considering only the
dominant term in the right-hand side of~(\ref{orbits_exp}) obtaining
\be
u_{0.H^2}''+A\,u_{0.H^2}=\frac{GM}{J^2}\,B\ \ \Rightarrow\ \ u_{0.H^2}=\frac{1+e\,\cos(\sqrt{A}\,\varphi)}{d}\ ,
\lb{u_0_H}
\ee
where $d$ is defined in terms of the ellipse semi-major axis $r_{1.\mathrm{orb}}$ and
the eccentricity $e$, being related to the constant of motion $J$ and the correction coefficients $A$
and $B$ by the following equalities
\be
d=r_{1.\mathrm{orb}}(1-e^2)=\frac{J^2}{GM}\,\frac{A}{B}\ \ \Leftrightarrow\ \ J^2=r_{1.\mathrm{orb}}\,GM\,(1-e^2)\,\frac{B}{A}\ .
\lb{J_d_H}
\ee
We note that, besides this correction to the relation between the constant of motion
$J$ and the parameter $d$ given by the ratio of the correction coefficients $B/A$~(\ref{A_B_C}), there
is a small precession effect proportional to $H_0^2$ due to the factor $\sqrt{A}$ in the cosine argument
of the solution~(\ref{u_0_H}). We will deal with this effect in detail later.

Next let us compute the corrections to the solution $u_{0.H^2}$ by considering the remaining terms in
right-hand side of equation~(\ref{orbits_exp}) evaluated for the function $u_{0.H^2}$ such that the
full solution is
\be
u=u_{0.H^2}+u_{\mathrm{GR}.H^2}+u_{H^2}\ .
\lb{u_H2}
\ee
Here the functions $u_{\mathrm{GR}.H^2}$ and $u_{H^2}$ correspond respectively to the (modified) General
Relativity and expanding background corrections approximated to order $H_0^2$ being the solutions of the
following differential equations
\be
\ba{rcl}
\displaystyle u_{\mathrm{GR}.H^2}''+A\,u_{\mathrm{GR}.H^2}&=&\displaystyle\frac{3GM}{c^2}\,C\,u_{0.H^2}^2\\[6mm]
&=&\displaystyle\frac{3GM}{c^2}\,C\,\frac{(1+e\cos(\sqrt{A}\,\varphi))^2}{d^2}\ ,\\[6mm]
\displaystyle u_{H^2}''+A\,u_{H^2}&=&\displaystyle -\left(\frac{H_0}{J}\right)^2\,\frac{1}{u_{0.H^2}^3}+\alpha_0\left(\frac{H_0}{J}\right)^2\,\frac{GM}{c^2}\,\frac{1}{u_{0.H^2}^2}\\[6mm]
&=&\displaystyle -\left(\frac{H_0}{J}\right)^2\,\frac{1}{(1+e\cos(\sqrt{A}\,\varphi))^3}\\[6mm]
&&\displaystyle+\alpha_0\left(\frac{H_0}{J}\right)^2\,\frac{GM}{c^2}\,\frac{1}{(1+e\cos(\sqrt{A}\,\varphi))^2}\ ,
\ea
\ee
such that we obtain
\be
\ba{rcl}
u_{\mathrm{GR}.H^2}&=&\displaystyle\frac{C}{A}\,\frac{\alpha_{\mathrm{GR}}}{d}\left(\left(1+\frac{e^2}{2}\right)-\frac{e^2}{6}\,\cos(2\sqrt{A}\varphi)+\sqrt{A}\,e\,\varphi\sin(\sqrt{A}\,\varphi)\right)\ ,\\[6mm]
u_{H^2}(\varphi)&=&\displaystyle \frac{d^3\,H_0^2}{A\,J^2(1-e^2)}\left(\frac{\alpha_0\,GM}{c^2\,d}+\frac{(-4+e^2)+3e^2\cos(2\sqrt{A}\,\varphi)}{4(1-e^2)(1+e\,\cos(\sqrt{A}\,\varphi))}\right.\\[6mm]
&&\displaystyle\left. -\left(\frac{3}{2(1-e^2)}-\frac{\alpha_0\,GM}{c^2\,d}\right)\frac{2e\,\arctan\left(\sqrt{\frac{1-e}{1+e}}\,\tan\left(\frac{\sqrt{A}\,\varphi}{2}\right)\right)\sin(\sqrt{A}\,\varphi)}{\sqrt{1-e^2}}\right)\ .
\ea
\lb{u_GR_H2}
\ee
Where $\alpha_{GR}=3GM/(c^2\,d)$~(\ref{alpha_GR}) and we have set the integration constants such that
no $\cos(\sqrt{A}\,\varphi)$ neither $\sin(\sqrt{A}\,\varphi)$ are present (see the discussion of
equation~(\ref{integration_constants}) for the justification of this choice). The modified General
Relativity solution $u_{\mathrm{GR}.H^2}$ has the same structure of the usual solution
$u_{\mathrm{GR}}$~(\ref{uGR}), the first term is a constant that can be neglected, the second term has
a period that is a multiple of the one of solution $u_{0.H^2}$~(\ref{u_0_H}) contributing
a small shift to the orbital period and the last term contributes to the orbital precession.
As for the solution $u_{H^2}$ has a similar structure, the first term is a constant that can be neglected, the second term contributes a small shift to the orbital period and
the last term contributes to the orbital precession. This last result is justified by noting that
the analytic continuation of the inverse of a function corresponds to the argument of the function
(in this way $\arctan(\tan \varphi)=\varphi$ increases monotonically with $\varphi$), here due to the correction being small compared to the dominant term we will expand the functions to lower order
as we did in equation~(\ref{cos_exp}). Therefore we are considering the following series expansions
for the oscillatory terms of $u_{0.H^2}$~(\ref{u_0_H}), the last term of~$u_{\mathrm{GR}.H^2}$~(\ref{u_GR_H2}) and the last term of $u_{H^2}$~(\ref{u_GR_H2})
\be
\ba{l}
\displaystyle\ \ \cos(\sqrt{A}\,\varphi)\ \approx\ 1-A^2\,\frac{\varphi^2}{2}+O(\varphi^4)\ \ ,\ \ 
f\,\sin(\sqrt{A}\,\varphi)\ \approx\ 2\sqrt{A}\,\frac{\varphi^2}{2}+O(\varphi^4)\ ,\\[6mm]
\displaystyle\ \ \arctan\left(\sqrt{\frac{1-e}{1+e}}\tan\left(\frac{\sqrt{A}}{2}\,\varphi\right)\right)\ \approx\ A\,\sqrt{\frac{1-e}{1+e}}\,\frac{\varphi^2}{2}+O(\varphi^4)\ .
\ea
\lb{tan_exp}
\ee
Hence, neglecting the constant terms in the solutions $u_{\mathrm{GR}.H^2}$ and $u_{H^2}$~(\ref{u_GR_H2}),
gathering the several terms and respective coefficients of the series expansions~(\ref{tan_exp}) and approximating these by a cosine function as we did in equation~(\ref{cos_exp}) we can rewrite the full
solution $u$~(\ref{u_H2}) as
\be
\ba{rcl}
u&\approx&\displaystyle\frac{1}{d}\left(1+e\,\cos\left(\left(1-\frac{\Delta\varphi_{\mathrm{GR}}}{2\pi}-\frac{\Delta\varphi_{H^2}}{2\pi}\right)\,\varphi\right)\right)+u_{\mathrm{osc.GR}}+u_{\mathrm{osc}.H^2}\ ,\\[6mm]
\displaystyle\frac{\Delta\varphi_{H^2}}{2\pi}&=&\displaystyle-\frac{\delta_A}{2}+\alpha_{\mathrm{GR}}\,\delta_C+\frac{d^3\,H_0^2}{(1-e)(1+e)^\frac{3}{2}}\,\left(\frac{\alpha_0}{c^2\,d}-\frac{3}{2(1-e^2)\,GM}\right)+O(H_0^4)\ ,\\[6mm]
u_{\mathrm{osc}.H^2}&=&\displaystyle \frac{(\delta_C-\delta_A)\alpha_{\mathrm{GR}}}{6}\,e^2\cos(2\varphi)+(h\,d)^2\frac{-4+e^2+3e^2\cos(2\varphi)}{4(1-e^2)^2\,GM\,\left(1+e\cos\varphi\right)}+O(H^4)
\ea
\lb{precession_H2}
\ee
where $\Delta\varphi_{\mathrm{GR}}/(2\pi)=\alpha_{\mathrm{GR}}$ is the precession
per turn of the orbit due to General Relativity corrections given in equation~(\ref{precession_GR})
and $\Delta\varphi_{H^2}/(2\pi)$ is the precession per turn due to the expanding background.
The factor $u_{\mathrm{osc.GR}}$ is the oscillatory factor of the correction to the orbit solution
due to General Relativity given in equation~(\ref{u_GR_exp}) and $u_{\mathrm{osc}.H^2}$ 
is the oscillatory factor correction to the orbit solution due to the expanding background.
For both these factors we considered the approximation $\sqrt{A}\,\varphi\approx\varphi$ in the
argument of the cosines. This approximation is justified by noting that
$\delta_A$~(\ref{A_B_C}) is a small perturbation ($A=1+\delta_A$) such that
over one turn of the orbit its contribution to the period correction is negligible.
In deriving these expressions we have used the following equality for the constant
of motion $J^2=GM\,d\,A/B$~(\ref{J_d_H}) and considered an expansion on $\delta_A$, $\delta_B$
and $\delta_C$~(\ref{A_B_C}), keeping only the lower order terms in $H_0^2$.

In order to compute the period correction to the orbits due to the expanding background we
use the same method of equation~(\ref{dT_GR}) discussed in section~\ref{sec.rev_orbits}.
Then the period is
\be
T=T_0+\Delta T_{\mathrm{GR}}+\Delta T_{H^2}\ \ ,\ \ \Delta T_{H^2}=-\frac{2}{|J|}\int_0^{2\pi}d\varphi\,\frac{u_{\mathrm{osc}.H^2}}{u_0^2}\left(1-\frac{GM}{c^2}u_0\right)\ ,
\lb{dT_H2}
\ee
where $u_0$ corresponds to the classical Keplerian orbits solution~(\ref{uo}), $\Delta T_{\mathrm{GR}}$
is the period correction due to General Relativity given in equation~(\ref{dT_GR}) and $\Delta T_{H^2}$
is the period correction due to the expanding background with $u_{\mathrm{osc}.H^2}$ given in
equation~(\ref{precession_H2}).

\begin{table}[ht]
\begin{center}
{\tiny
\begin{tabular}{lcclll}
Planet& GR precession~(\ref{precession_GR}) & $\Delta T_{GR}$~(\ref{dT_GR}) & $H^2$ precession~(\ref{precession_H2}) & $\Delta T_{H^2}$~(\ref{dT_H2})& $\dot{r}_{1.\mathrm{orb}}/r_{1.\mathrm{orb}}~(\ref{dr_orb_circ})$\\
      &$(arcsec/century)$                       & $(s/century)$                 &$(arcsec/century)$ &$(s/century)$ &$(s^{-1}/century)$\\\hline\hline\\[-2mm]
Mercury &$10.35$& $5.66\times 10^{-2}$ &$\in[-6.04,-0.122]\times 10^{-14}$ &$\in[1.27,1.30]\times 10^{-16}$ & $\in[-0.0706,3.80]\times 10^{-32}$       \\
Venus   &$5.30$&  $8.01\times 10^{-8}$ &$\in[-3.63,-0.0906]\times 10^{-13}$ &$=1.93\times 10^{-15}$ & $\in[-0.0460,2.48]\times 10^{-31}$         \\
Earth   &$3.84$&  $3.64\times 10^{-6}$ &$\in[-9.59,-0.239]\times 10^{-13}$ &$=8.29\times 10^{-15}$ & $\in[-0.122,6.54]\times 10^{-31}$         \\
Mars    &$2.54$&  $4.48\times 10^{-3}$ &$\in[-3.44,-0.0809]\times 10^{-12}$ &$=5.62\times 10^{-14}$ & $\in[-0.043,2.31]\times 10^{-30}$         \\
Jupiter &$0.74$&  $6.12\times 10^{-4}$ &$\in[-1.36,-0.0331]\times 10^{-10}$ &$=1.39\times 10^{-11}$ & $\in[-0.171,9.22]\times 10^{-29}$         \\
Saturn  &$0.40$&  $1.48\times 10^{-3}$ &$\in[-8.48,-0.206]\times 10^{-10}$ &$=2.18\times 10^{-10}$ & $\in[-0.107,5.76]\times 10^{-28}$         \\
Uranos  &$0.20$&  $8.97\times 10^{-4}$ &$\in[-6.81,-0.167]\times 10^{-9}$ &$=4.96\times 10^{-9}$ & $\in[-0.0861,4.63]\times 10^{-27}$         \\
Neptune &$0.13$&  $4.18\times 10^{-6}$ &$\in[-2.60,-0.0650]\times 10^{-8}$ &$=3.70\times 10^{-8}$ & $\in[-0.0330,1.77]\times 10^{-26}$         \\
Pluto   &$0.10$&  $1.28$               &$\in[-6.67,-0.125]\times 10^{-8}$ &$\in[1.44,1.52]\times 10^{-7}$ & $\in[-0.0749,4.03]\times 10^{-26}$        
\\\hline
\end{tabular}}
\caption{\it \small Orbital precession and period corrections due to General Relativity (not including the
effects of the Sun quadrupole $J_2$) as given by equations~(\ref{precession_GR}) and~(\ref{dT_GR})
and due to the expanding background as given by equations~(\ref{precession_H2}) and~(\ref{dT_H2}).
It was considered the range for the parameter
$\alpha_0\in[3,\alpha_{0.\mathrm{max.pert}}]$~(\ref{alpha_max_pert}) for which the perturbative
expansion~(\ref{orbits_exp}) is valid,
when no interval for the values of the correction is given is due to the respective upper and lower
values differing by less than two precision digits. The allowed ranges for orbital radius
time variation $\dot{r}_{1.\mathrm{orb}}/r_{1.\mathrm{orb}}$ as given by~(\ref{dr_orb_circ}) are
also quoted for the circular orbits approximation and the full range of the
parameter $\alpha_0\in[3,+\infty[$. \lb{table.precession_dT}}
\end{center}
\end{table}

In table~\ref{table.precession_dT} are listed the estimative for orbital precession and period
corrections both for the General Relativity corrections and for the expanding background corrections
in the solar system. As can readily be concluded the effects due to the expanding background are
several orders of magnitude lower than the respective GR effects (from 12 orders of magnitude for
mercury to 7 orders of magnitude for Pluto) being, for most purposes, negligible. We also note that
the quoted values for these corrections are well below the experimentally detected deviations from
the theoretical predictions (usually obtained through numerical simulations) for the orbital motion
in the solar system~\cite{DM_orbits}. Hence, no bounds for the parameter $\alpha_0$ can be drawn from this
analysis.

\section{Conclusions\lb{sec.conclusions}}
\setcounter{equation}{0}

\subsection{Resume of Results}

In this work we have built an ansatz for a locally anisotropic metric~(\ref{g_generic}) describing
matter in an globally expanding background, our universe. This metric interpolates
between the Schwarzschild metric~(\ref{g_SC}) near massive bodies and the FLRW metric~(\ref{g_FRW_1})
at spatial infinity where spatial isotropy is retrieved. By considering a two parameter radial
coordinate dependent exponent on the shift function, $\alpha=\alpha_{0}+\alpha_1\,2GM/(c^2\,r_1)$~(\ref{alpha_r_1}),
we have maintain space-time complete, free of singularities except for the Schwarzschild mass pole
at the origin and the total mass, within a shell of finite radius, finite. The negative parameter
$\alpha_1<0$ plays the role of a regulator
removing the essential singularities at the center of mass. Its upper bounds have been analyzed
in order to maintain the mass-energy density positive, however we note that it can, for most purposes, be
considered as close to zero as desired being negligible outside the event horizon $r_1>2GM/(c^2\,r_1)$.
As for the parameter $\alpha_0> 3$ fine-tunes the transition between small and large scale
physics. For $\alpha_0>5$ space-time is asymptotically flat near the event horizon and for relatively
larger values we have shown that, although for relatively small spatial scales
space-time can be contracting, for large spatial scales it will always be expanding consistently with
the global expansion assumption. In particular this characteristic may contribute to dark matter
effects at galactic scales maintaining both the usual physics of planetary systems (small scales)
and global expansion (large scale). Hence we may have manage to solve the long standing puzzle of
consistently describing matter in a expanding background first approached by McVittie.

We have analyzed in some detail the orbital motion in the solar system considering the
locally anisotropic metric~(\ref{g_generic}) and have concluded that the corrections due
to the expanding background are negligible by many orders of magnitude with respect to
the General Relativity corrections as well as to the detected experimental deviations such that
no bounds for the parameter $\alpha_0$ can be obtained based in experimental observations on the
solar system. This result is welcome in the sense that the effects of expansion are negligible for small scale systems such that the standard General Relativity in Minkowski
background describes these systems to a very good accuracy. Also we note that the anomalous Pioneer acceleration~\cite{Pioneer} towards the Sun of
$a_p\approx 8\times 10^{-10}\,ms^{-2}$ cannot possible be explained by the expanding background effects.
In the range of the Pioneer distance from the Sun $r_1\approx [2.85,6.60]\times 10^{12}\,m$ the maximum
background correction to the General Relativity Newton law towards the Sun corresponding
to a value of the exponent parameter $\alpha_0\sim 10^{10}$ is of order
$\ddot{r}_1\sim 10^{-25}\,ms^{-2}$, hence $15$ orders of magnitude below the measured acceleration.

As a final remark it is relevant to note that we have at most considered
a test mass and a central mass when deriving the analytical results in this work.
When considering many body interactions the net effect of the corrections due to the expanding
background may become slightly more significant, mostly for distant astrophysical body interactions
(the expansion effects are proportional to the distance). In order to implement such
corrections in numerical simulations it is necessary to derive the modified Newton law in
isotropic coordinates for inclusion of the corrections in the PPN formalism~\cite{PPN} which is the
most widely employed description of many body gravitational systems.

\subsection{Outlook: Possible Contributions to Dark Matter Effects}

The corrections to the General Relativity Newton law~(\ref{F_Newton_mod}) due to the locally anisotropic
metric~(\ref{g_generic}) discussed in section~\ref{sec.Newton} increase the orbital velocities
for galactic scales. This effect is similar to the ones attributed either to the dark matter hypothesis~\cite{DM}
or modified theories of gravity~\cite{MOND,STV}. In either of these cases the modified
Newton law at galaxy scales may explain the observed deviation from General
Relativity predictions such as the flattening of the galaxies rotation curves~\cite{galaxies_DM}
and the deviation from the predicted gravitational lensing by astrophysical
objects~\cite{lensing_DM,modulation_DM}.

As discussed in section~\ref{sec.Newton} the
locally anisotropic metric~(\ref{g_generic}), for large values of the parameter $\alpha_0$,
also predicts a significant deviation of the orbital velocities for galaxy size scales
as exemplified in figure~\ref{fig.galaxy_v} for a central point-mass with the value of the
core mass of the large galaxy UGC2885. As can readily be verified, when considering
the full mass of the galaxy to be point-like the radial distance at which this effect
is verified is too large to significantly change the orbital velocity within the
radius of the galaxy. However by comparison with the same effect for the Sun
(see the modified Newton acceleration in figure~\ref{fig.Newton_alpha} for comparison)
we conclude that when considering several massive objects with a large parameter
$\alpha_0$ within the galaxy we may expect that the modifications to the orbital
velocity will be felt for lower values of the radial distance, hence contributing
to the dark matter effects. With respect to large scale observations we note
that the local anisotropy detected in the background radiation~\cite{review}
may help to set bounds on the parameter $\alpha_0$ allowing to properly evaluate
to which extend can the local anisotropy contribute to the astrophysical dark matter
effects. We will give an account of these issues somewhere else~\cite{progress}.

As a relevant final remark we note that the exponent $\alpha$ has been introduced phenomenological when 
building the metric ansatz~(\ref{g_generic}) without any more fundamental reasoning concerning its meaning.
In particular if it represents some sort of \textit{unknown} gravitational
interaction associated to some sort of matter or if it is some kind of fundamental constant
identical for all masses. If it is to be interpreted as an effective description of dark matter
we may expect that for each body the parameter $\alpha_0$ has a distinct value. For each
galaxy (or set of bodies within each galaxy) its value can be inferred from the deviation of the usual General Relativity Newton law
and fitted accordingly. Although our construction can be interpreted as that the locally anisotropic
metric~(\ref{g_generic}) represents the net effect of long range gravitational interaction in local
masses due to the expansion of the universe, the specific nature of the interaction that it may describe
or its meaning in terms of the known physical interactions (or fundamental principles) is not obvious
to the author and is here left as an unsolved question. We note however that a dilaton-like scalar field~\cite{Dilaton,strings,modgrav}
with local solutions of the form $\Phi=\alpha\,\log(1-U_{\mathrm{SC}})$ such that
$(1-U_{\mathrm{SC}})^\alpha=e^{\alpha\,\log (1-U_{\mathrm{SC}})}=e^\Phi$ could justify this metric.
The author is unaware of a specific theory that predicts these particular solutions.

\ \\
{\Large\bf Appendixes}

\appendix

\section{Conventions and Definitions\lb{A.defs}}
\setcounter{equation}{0}

In this appendix we list the General Relativity and Special Relativity conventions
employed in this work. For units we use the International System maintaining all the physical dimensionfull
constants explicitly in the equations. The Einstein convention is considered such that,
unless explicitly stated, sums are implied when repeated indexes are present with the following
convention for greek and roman indexes
\be
\mu=0,i\ \ ,\ \ i=1,2,3\ .
\ee
For the definition of Cartesian coordinate parameterization
we consider $x^0=c\,t$, $x^1=x$, $x^2=y$ and $x^3=z$ in an infinite interval.
As for the definitions of the spherical coordinate parameterization and its mapping to
Cartesian coordinates we consider
\be
x^0=c\, t\ ,\ x^1=r=\sqrt{x^2+y^2+z^2}\ ,\ x^2=\theta=\arctan\frac{\sqrt{x^2+y^2}}{z}\ ,\ x^3=\varphi=\arctan\frac{y}{x},
\ee
with $\theta\in[-\pi/2,\pi/2[$, $\varphi\in[0,2\pi[$ and $r\in[0,+\infty[$.
The inverse mapping between spherical and Cartesian coordinates is
\be
x=r\,\sin\theta\cos\varphi\ \ ,\ \ y=r\,\sin\theta\sin\varphi\ \ ,\ \ z=r\,\cos\theta\ .
\ee
Bold face quantities
$\vb{r}$ ($\vb{x}$) represent spatial vectors for spherical coordinates (Cartesian coordinates)
and, when not specified, a dotted function $\dot{f}$ represents simple derivation with
respect to $t$ and a primed function $f'$ with respect to $r$.

For a generic metric $g_{\mu\nu}$ we have the following definition of the infinitesimal
length square
\be
ds^2=c^2\,\eta_{00}d\tau^2=g_{\mu\nu} dx^\mu dx^\nu\ ,
\lb{A.ds}
\ee
being $\tau$ the proper time corresponding to the free falling frame and we adopt
the metric signature $(+,-,-,-)$. The connections are defined as
\be
\Gamma^\mu_{\ \nu\delta}=\frac{1}{2}\, g^{\mu\rho}\left(g_{\rho\nu,\delta}+g_{\rho\delta,\nu}-g_{\nu\delta,\rho}\right)\ ,
\lb{A.Gamma}
\ee
where the index '$,\delta$' denotes simple derivation with respect to $x^\delta$ and
the covariant derivative for a vector is defined as
\be
q^\mu_{\ ;\nu}=\frac{Dq^\mu}{dx^\nu}=q^\mu_{\ ,\nu}+\Gamma^{\mu}_{\ \nu\delta}q^\delta\ .
\lb{A.D}
\ee
We note that, considering non-natural units ($c\neq 1$), for derivatives with respect to the
coordinate time $t=x^0/c$ we obtain the expression $q^\mu_{\ ;t}=q^\mu_{\ ,t}+c\,\Gamma^{\mu}_{\ 0\delta}q^\delta$.

The Riemann tensor is defined as
\be
R^\mu_{\ \nu\lambda\rho}=\Gamma^\mu_{\ \nu\rho,\lambda}-\Gamma^\mu_{\ \nu\lambda,\rho}+\Gamma^\mu_{\ \delta\lambda}\Gamma^\delta_{\ \nu\rho}-\Gamma^\mu_{\ \delta\rho}\Gamma^\delta_{\ \nu\lambda}\ ,
\lb{A.R}
\ee
and, unless otherwise stated, we express the Einstein equations in the following form
\be
\ba{rcl}
\displaystyle G_{\mu\nu}&=&\displaystyle \frac{8\pi\,G}{c^4}\,T_{\mu\nu}\ ,\\[5mm]
\displaystyle G_{\mu\nu}&=&\displaystyle R_{\mu\nu}-\frac{1}{2}g_{\mu\nu}R\ ,\\[5mm]
\displaystyle T_{\mu\nu}&=&\displaystyle \left(\frac{p}{c^2}+\rho\right)u_\mu u_\nu-g_{\mu\nu}\,p\ ,
\ea
\lb{A.EEQ}
\ee
where $G_{\mu\nu}$ is the Einstein tensor, $T_{\mu\nu}$ is the stress-energy tensor,
$R_{\mu\nu}=R^\alpha_{\ \mu\alpha\nu}$ is the Ricci tensor and $R=g^{\delta\rho}R^\alpha_{\ \delta\alpha\rho}$ is the Ricci scalar. In the last line we consider the stress-energy tensor
for a perfect isotropic fluid where $p$ is the pressure, $\rho$ the mass-density and
$u_\mu=(c,v_i)$ the fluid velocities. In the commoving frame of the fluid the four-velocity is
given by $u_\mu=(c,0)$ and we obtain the tensor components
\be
\displaystyle T^{\mathrm{com}}_{00}=\rho\,c^2+p\left(1-g_{00}\right)\ ,\ T^{\mathrm{com}}_{0i}=-p\,g_{0i}\ ,\ 
T^{\mathrm{com}}_{ij}=-p\,g_{ij}\ ,
\lb{A.T_com}
\ee
The sign conventions in these expressions depend on the metric signature choice.
We note that under a coordinate transformation $x^\mu=x^\mu(\bar{x}^\nu)$ the contravariant
velocity components generally change to $\bar{v}_\nu=\partial x^\mu/\partial \bar{x}^\nu v_\mu$,
hence as long as we consider transformations for which $x^0=\bar{x}^0$ we remain in the commoving
frame of the fluid and equations~(\ref{A.T_com}) are still valid.

For light-like trajectories we compute the geodesics directly from the distance element~(\ref{A.ds}),
$ds^2=0$. As for time-like trajectories of massive objects we consider the usual geodesic obtained
from the Euler-Lagrange equations that minimize the length $L=\sqrt{ds^2}$. The solution for the relativistic factor $dt/d\tau=\gamma$ is obtained either from integrating the geodesic equation
for $x^0$ or directly from~(\ref{A.ds})
\be
\ba{rcl}
\displaystyle\eta_{00}c^2&=&\displaystyle c^2\,g_{00}\left(\frac{dt}{d\tau}\right)^2+c\,g_{0i}\frac{dx^i}{dt}\left(\frac{dt}{d\tau}\right)^2+g_{ij}\frac{dx^i}{dt}\frac{dx^j}{dt}\left(\frac{dt}{d\tau}\right)^2\ ,\\[6mm]
\displaystyle\gamma&=&\displaystyle \frac{dt}{d\tau}=\sqrt{\frac{\eta_{00}}{g_{00}+\frac{2g_{0i}\dot{x}^i}{c}+\frac{g_{ij}\dot{x}^i\dot{x}^j}{c^2}}}\ .
\ea
\lb{A.gamma}
\ee
Then the geodesic equations for the remaining space coordinates are
\be
\ba{rcl}
\displaystyle \frac{d^2x^i}{d\tau^2}&=&\displaystyle-\,\Gamma^i_{\ \mu\nu}\,\frac{dx^\mu}{d\tau}\frac{dx^\nu}{d\tau} ,\\[6mm]
\displaystyle \ddot{x}^i&=&\displaystyle-c^2\,\Gamma^i_{\ 00}-2c\,\Gamma^{i}_{\ 0j}\dot{x}^j-\Gamma^{i}_{\ jk}\dot{x}^j\dot{x}^k-\gamma^{-1}\dot{\gamma}\,\dot{x}^i\ .
\ea
\lb{A.geo}
\ee

For projections of generic vectors from the four-dimensional space-time manifold to the spatial
hyper-surface at a given fixed time we take the following definitions~\cite{Israel} for
the intrinsic metric $^{(3)}g_{ij}$ and three-dimensional vector components $^{(3)}w^i$
\be
\ba{rcl}
^{(3)}g_{ij}&=&-\,^{(4)}g_{ij}\ ,\\[6mm]
^{(3)}w^i&=&^{(3)}g^{ij}\,^{(4)}w_i=\,^{(3)}g^{ij}\left(^{(4)}g_{ij}\,^{(4)}w^i+\,^{(4)}g_{i0}\,^{(4)}w^0\right)\ .
\ea
\lb{Israel_proj}
\ee
In this equation $^{(4)}g_{\mu\nu}$ stands for the usual four-dimensional metric components
and $^{(4)}w^\mu$ stand for the four-dimensional vector to be projected. In particular
this projection is applicable to the space-time coordinates $^{(4)}x^\mu$.

In this work, both in order to simplify the technical details and for easier physical
interpretation of the derived quantities we consider several coordinate systems corresponding,
for spherical coordinates, to the following radial coordinate definitions
labeled by an underscore index as:

\begin{enumerate}
\item[$r$] -- a radial coordinate with integration measure $\sqrt{-g}=a^3\,r^2\,\sin\theta$,
and a time dependent area for the 2-sphere, $A(t)=4 \pi\, a^2\, r^2$, being $a=a(t)$
a generic cosmological solution for the universe scale factor. In this work we refer
to it as the \textit{\bf non-expanding coordinates} due to the geometrical spatial lengths
expanding over time while coordinates lengths are \textit{fixed}.

\item[$r_1$] -- a radial coordinate with integration measure $\sqrt{-g}=r_1^2\,\sin\theta$,
hence with a time-independent area for the 2-sphere, $A=4\pi\, r_1^2$. These are the usual Schwarzschild
coordinates and we will refer to it as \textit{\bf expanding coordinates} due to the coordinates
lengths being coincident with the physical geometrical lengths, hence expanding over time. $r_1$ is related
to $r$ as
\be
r_1=a\,r\ .
\lb{A.r1.r}
\ee

\item[$r_2$] -- the usual {\bf expanding isotropic coordinates} employed for the Schwarzschild metric
for which the speed of light is the same along all space directions. $r_2$ is related to $r_1$ as
\be
r_1=r_2\left(1+\frac{GM}{2r_2\,c^2}\right)^2\ \mathrm{for}\ r_1\ge\frac{2GM}{c^2}\ .
\lb{A.r2.r1}
\ee

\item[$r_3$] -- {\bf non-expanding isotropic coordinates} for which the metric is explicitly dependent on the scale factor $a$. $r_3$ is related to $r_2$ as
\be
r_2=a\,r_3\ .
\lb{A.r3.r2}
\ee
\end{enumerate}

In table~\ref{table.units} are summarized the IS units for the quantities
and the values of the fundamental constants employed in this work.
\begin{table}[ht]
\begin{center}
\begin{tabular}{clll}
symbol&units&value&name\\[2mm]\hline\hline\\[-2mm]
$c$&$m\,s^{-1}$&$2.998\times 10^8$&speed\ of\ light\\[2mm]
$G$&$Kg^{-1}\,m^3\,s^{-2}$&$6.670\times 10^{-11}$&Newton\ Constant\\[2mm]
$M$&$Kg$& &mass\\[2mm]
$x^\mu$&$m$&&space-time\ coordinates\\[2mm]
$g_{\mu\nu}$&$1$&&metric\ tensor\\[2mm]
$\Gamma^{\mu}_{\ \nu\delta}$&$m^{-1}$&&metric\ connections\\[2mm]
$R$&$m^{-2}$&&Ricci\ scalar\ (coordinate curvature)
\\\hline
\end{tabular}
\caption{\it \small IS units for the fundamental constants, metric and other derived quantities.\lb{table.units}}
\end{center}
\end{table}

\section{Cosmological Metric\lb{A.FRW}}
\setcounter{equation}{0}

In this appendix we summarize technical details for the FLRW metric representing
a homogeneous and isotropic, globally flat, expanding universe.

\subsection{Non-Expanding Coordinates\lb{A.cosm_nexpanding}}

In the commoving frame for non-expanding coordinates this metric is given by
equation~(\ref{g_FRW}) and the non-null connections are, for Cartesian coordinates,
\be
\Gamma^0_{\ ij}=\delta_{ij}\,\frac{a\,\dot{a}}{c}\ ,\ \Gamma^i_{\ 0j}=\Gamma^i_{\ j0}=\delta^i_{\ j}\,\frac{1}{c}\,\frac{\dot{a}}{a}\ .
\lb{A.FRW.conn_cart}
\ee
The Ricci scalar (scalar curvature) for this metric is
\be
R_{\mathrm{FLRW}}=-6\,\left(\frac{\dot{a}}{a\,c}\right)^2-6\,\frac{\ddot{a}}{a\,c^2}\ ,
\lb{A.FRW_R}
\ee
and the non-null Einstein tensor components are
\be
G_{00}=3\,\left(\frac{\dot{a}}{a\,c}\right)^2\ \ ,\ \ G_{ij}=-\delta_{ij}\,\frac{1}{c^2}\,\left(\dot{a}^2+2a\,\ddot{a}\right)\ .
\lb{A.FRW.EEQ}
\ee

\subsection{Expanding Coordinates\lb{A.cosm_expanding}}

For expanding coordinates this metric is given by equation~(\ref{g_FRW_1}), for
Cartesian coordinates the non-null connections are
\be
\ba{rclcrcl}
\Gamma^0_{\ 00}&=&\displaystyle\delta_{ij}x_1^ix_1^j\,\left(\frac{\dot{a}}{a\,c}\right)^3&,&\Gamma^0_{\ 0i}&=&\displaystyle\Gamma^0_{\ i0}\,=\,-\delta_{ij}x_1^j\left(\frac{\dot{a}}{a\,c}\right)^2\ ,\\[6mm]
\Gamma^0_{\ ij}&=&\displaystyle\delta_{ij}\,\frac{\dot{a}}{a\,c}&,&\Gamma^i_{\ 00}&=&\displaystyle x^i\,\delta_{jk}x_1^jx_1^k\,\left(\frac{\dot{a}}{a\,c}\right)^4-x^i\,\frac{\ddot{a}}{a\,c^2}\ ,\\[6mm]
\Gamma^i_{\ 0j}&=&\displaystyle\Gamma^i_{\ j0}\,=\,-x^i\,\delta_{jk}x^k\,\left(\frac{\dot{a}}{a\,c}\right)^3&,& \Gamma^i_{\ jk}&=&\displaystyle\delta_{jk}\,x^i\,\left(\frac{\dot{a}}{a\,c}\right)^2\ ,
\lb{A.FRW_1.conn_cart}
\ea
\ee
and, for spherical coordinates are
\be
\ba{rclcrclcrcl}
\Gamma^0_{\ 00}&=&\displaystyle r_1^2\,\left(\frac{\dot{a}}{a\,c}\right)^3&,&\Gamma^0_{\ 01}&=&\displaystyle\Gamma^0_{\ 10}\,=\,-r_1\,\left(\frac{\dot{a}}{a\,c}\right)^2\ ,\\[6mm]
\Gamma^0_{\ 11}&=&\displaystyle\frac{\dot{a}}{a\,c}&,&\Gamma^0_{022}&=&\displaystyle r_1^2\Gamma^0_{\ 11}\\[6mm]
\Gamma^0_{\ 22}&=&r_1^2\sin^2\theta\Gamma^0_{\ 33}&,&\Gamma^1_{\ 00}&=&\displaystyle-\frac{r_1\,\ddot{a}}{a\,c^2}+r_1^3\,\left(\frac{\dot{a}}{a\,c}\right)^4\ ,\\[6mm]
\Gamma^1_{\ 01}&=&\Gamma^1_{\ 10}\,=\,-\Gamma^0_{\ 00}&,&\Gamma^1_{\ 11}&=&-\Gamma^0_{\ 01}\\[6mm]
\Gamma^1_{\ 22}&=&\displaystyle-r_1+r_1^3\,\left(\frac{\dot{a}}{a\,c}\right)^2&,&
\Gamma^1_{\ 33}&=&\sin^2\theta\Gamma^1_{\ 22}\\[6mm]
\Gamma^2_{\ 12}&=&\displaystyle\Gamma^2_{\ 21}\,=\,\frac{1}{r}&,&\Gamma^2_{\ 33}&=&-\cos\theta\sin\theta\ ,\\[6mm]
\Gamma^3_{\ 13}&=&\Gamma^3_{\ 31}\,=\,\Gamma^2_{\ 12}\ &,&\Gamma^3_{\ 23}&=&\displaystyle\Gamma^3_{\ 32}\,=\,\frac{\cos\theta}{\sin\theta}\ .
\ea
\lb{A.FRW_1.conn_pol}
\ee
The Ricci scalar (scalar curvature) is given by the same expression than for
expanding coordinates~(\ref{A.FRW_R}) and the non-null Einstein tensor components
for spherical coordinates are
\be
\ba{rcl}
G_{00}&=&\displaystyle 3\,\left(\frac{\dot{a}}{a\,c}\right)^2-r_1^2\,\left(\frac{\dot{a}}{a\,c}\right)^2\,\left(\left(\frac{\dot{a}}{a\,c}\right)^2+2\frac{\ddot{a}}{a\,c^2}\right)\ ,\\[5mm]
G_{01}&=&\displaystyle r_1\,\frac{\dot{a}}{a\,c}\,\left(\left(\frac{\dot{a}}{a\,c}\right)^2+2\frac{\ddot{a}}{a\,c^2}\right)\ ,\\[5mm]
G_{ij}&=&\displaystyle g_{ij}\left(\left(\frac{\dot{a}}{a\,c}\right)^2+2\frac{\ddot{a}}{a\,c^2}\right)\ .
\ea
\lb{A.FRW_1.EEQ}
\ee

\subsection{Short-Scale Time Evolution of the Expansion Factor\lb{A.cosm_H}}

When the time $t_0$ considered for the series expansion of the universe scale factor $a$ given
in equation~(\ref{a_exp}) does not coincide with the time $t_{\mathrm{exp}}$ corresponding to
the experimental measurement of the Hubble rate $H_{\mathrm{exp}}=\left.\dot{a}/a\right|_{t=t_{\mathrm{exp}}}$
and the deceleration parameter $-q_{\mathrm{exp}}/H_{\mathrm{exp}}^2=\left.-\ddot{a}/a\right|_{t=t_{\mathrm{exp}}}$
there is a small correction for these quantities when evaluated at $t_0$, $H_0=\left.\dot{a}/a\right|_{t=t_0}$
and  $-q_0/H_0^2=\left.-\ddot{a}/a\right|_{t=t_0}$. Similarly to the expansion~(\ref{a_exp})
let us take a series expansions to third order on the product of the constant Hubble rate by
time $(H_0t)^3$ such that in the neighborhood of $t_{\mathrm{exp}}$
\be
\ba{rcl}
a&=&\displaystyle a_{\mathrm{exp}}\left(1+\frac{\dot{a}_{\mathrm{exp}}}{a_{\mathrm{exp}}}(t-t_{\mathrm{exp}})+\frac{1}{2}\frac{\ddot{a}_{\mathrm{exp}}}{a_{\mathrm{exp}}}(t-t_{\mathrm{exp}})^2+\frac{1}{6}\frac{\dot{\ddot{a}}_{\mathrm{exp}}}{a_{\mathrm{exp}}}(t-t_{\mathrm{exp}})^3\right)\\[6mm]
& &\displaystyle\hfill+O\left((t-t_{\mathrm{exp}})^4\right)\\[6mm]
&\approx&\displaystyle a_{\mathrm{exp}}\left(1+H_{\mathrm{exp}}(t-t_{\mathrm{exp}})-\frac{1}{2}q_{\mathrm{exp}}H^2_{\mathrm{exp}}(t-t_{\mathrm{exp}})^2-\frac{1}{6}s_{\mathrm{exp}}q_{\mathrm{exp}}H^3_{\mathrm{exp}}(t-t_{\mathrm{exp}})^3\right)\\[6mm]
& &\displaystyle \hfill+O\left(H_0^4(t-t_{\mathrm{exp}})^4\right)\ ,
\ea
\lb{A.exp_a_exp}
\ee
where in the last line we replace the derivatives of the scale factor by the Hubble rate
$H$~(\ref{H}), deceleration parameter $q$~(\ref{H_q0}) and variation
of the deceleration parameter $s$~(\ref{H_s0}) evaluated at $t_{\mathrm{exp}}$.
Hence expanding the Hubble rate $H$ and deceleration parameter $q$ in the neighborhood
of $t_{\mathrm{exp}}$ we obtain, for their value at the reference time $t=t_0$
the following first order (on time) relations
\be
\ba{rcl}
H_0&=&\displaystyle\left.\frac{\dot{a}}{a}\right|_{t=t_0}=H_{\mathrm{exp}}\left(1-(q_{\mathrm{exp}}+1)\,H_{\mathrm{exp}}\,(t_0-t_{\mathrm{exp}})\right)\ ,\\[6mm]
q_0&=&\displaystyle-\left.\frac{\ddot{a}}{a}\frac{1}{H^2}\right|_{t=t_0}=q_{\mathrm{exp}}\left(1+(1+2q_{\mathrm{exp}}-s_{\mathrm{exp}})\,H_{\mathrm{exp}}\,(t_0-t_{\mathrm{exp}})\right)\ .
\ea
\lb{A.H_0_exp}
\ee
Without loss of generality the above series expansions can also be employed with a negative value for $t_{\mathrm{exp}}$. For the specific case of expanding coordinates $x^i_1$ the reference time
is set to zero, $t_0=0$.

\section{McVittie Metric\lb{A.McV}}
\setcounter{equation}{0}

The McVittie metric for a point-like central mass $M$ at the origin of the coordinate frame
in an expanding background was originally derived for non-expanding isotropic spherical
coordinates corresponding to $r_3$ as defined in equation~(\ref{A.r3.r2})
\be
ds^2=c^2\left(\frac{1-\frac{GM}{2\,a\,r_3\,c^2}}{1+\frac{GM}{2\,a\,r_3\,c^2}}\right)^2dt^2-a^2\left(1+\frac{GM}{2\,a\,r_3\,c^2}\right)^4\left(dr_3^2+r_3^2d\theta^2+r_3^2\sin^2\theta\,d\varphi^2\right)\ .
\lb{A.McV.g_r3}
\ee

In the following our aim is to express this metric in the several coordinate choices corresponding
to non-expanding coordinates $r$, expanding coordinates $r_1$ and expanding isotropic coordinates $r_2$
as defined in appendix~\ref{A.defs}. Hence considering the coordinate transformation~(\ref{A.r3.r2})
we obtain
\be
\ba{rcl}
r_3&=&\displaystyle \frac{r_2}{a}\ ,\\[5mm]
ds^2&=&\displaystyle c^2\left[\left(\frac{1-\frac{GM}{2\,r_2\,c^2}}{1+\frac{GM}{2\,r_2\,c^2}}\right)^2-\frac{r_2^2}{c^2}\,\frac{\dot{a}}{a}\,\left(1+\frac{GM}{2\,r_2\,c^2}\right)^4\right]dt^2+\frac{2r_2}{c}\,\frac{\dot{a}}{a}\,\left(1+\frac{GM}{2\,r_2\,c^2}\right)^4\,dt\,dr_2\\[6mm]
& &\displaystyle -\left(1+\frac{GM}{2\,r_2\,c^2}\right)^4\left(dr_2^2+r_2^2d\theta^2+r_2^2\sin^2\theta\,d\varphi^2\right).
\ea
\lb{A.McV.g_r2}
\ee
The transformation~(\ref{A.r2.r1}) is, generally, not bijective. Then, instead of proceeding
directly from the above metric, we can directly infer the expression for the metric in the
coordinates $r_1$
\be
\ba{rcl}
r_2&=&\displaystyle r_1\left(1+\frac{GM}{2\,r_1\,c^2}\right)^2\ ,\\[5mm]
ds^2&=&\displaystyle c^2\left(1-\frac{2\,GM}{r_1\,c^2}-\frac{r_1^2}{c^2}\,\left(\frac{\dot{a}}{a}\right)^2\right)dt^2+\frac{\dot{a}}{a}\,\frac{r_1}{\sqrt{1-\frac{2\,GM}{r_1\,c^2}}}\,dt\,dr_1\\[6mm]
& &\displaystyle -\frac{dr_1^2}{1-\frac{2\,GM}{r_1\,c^2}}-r_1^2\left(d\theta^2+\sin^2\theta d\varphi^2\right)\ .
\ea
\lb{A.McV.g_r1}
\ee
Finally considering the coordinate transformation~(\ref{A.r1.r}) we obtain
\be
\ba{rcl}
r_1&=&\displaystyle a\,r\ ,\\[5mm]
ds^2&=&\displaystyle c^2\left(1-\frac{2\,GM}{a\,r\,c^2}-\frac{\dot{a}^2\,r^2}{c^2}\,\left(1-\frac{2}{\sqrt{1-\frac{2\,GM}{a\,r\,c^2}}}+\frac{1}{1-\frac{2\,GM}{a\,r\,c^2}}\right)\right)dt^2\\[6mm]
& & \displaystyle+2\,a\,\dot{a}\,r\left(\frac{1}{\sqrt{1-\frac{2\,GM}{a\,r\,c^2}}}-\frac{1}{1-\frac{2\,GM}{a\,r\,c^2}}\right)\,dt\,dr\\[6mm]
& &\displaystyle -\frac{a^2\,dr^2}{1-\frac{2\,GM}{a\,r\,c^2}}-a^2\,r^2\left(d\theta^2+\sin^2\theta d\varphi^2\right)\ .
\ea
\lb{A.McV.g_r}
\ee

We note that the transformation to isotropic coordinates is valid for large radial coordinate
as expressed in~(\ref{A.r2.r1}), this may be the reason why McVittie originally did not notice that
the Schwarzschild radius is singular.

\section{Locally Anisotropic Metric\lb{A.generic}}
\setcounter{equation}{0}

Except for the curvature invariant ${\mathcal{R}}_\alpha$,
in the following expressions we are omitting the Schwarzschild mass pole contribution
which, when required, is considered in the main text.

\subsection{Constant Exponent $\alpha$}

The non-null connections $_{(\alpha)}\Gamma^\beta_{\ \mu\nu}$ for the metric~(\ref{g_generic}) 
with a constant exponent $\alpha$ in the shift function~(\ref{N_1_alpha_1}) expressed in
expanding spherical coordinates $r_1$ are, for $\beta=0$,
\be
\ba{rcl}
_{(\alpha)}\Gamma^0_{\ 00}&=&\displaystyle-\frac{GM}{c^2\,r_1}\,\left(\frac{\dot{a}}{a\,c}\right)\,\left(1-\frac{2GM}{c^2\,r_1}\right)^{\frac{\alpha}{2}-\frac{1}{2}}\\[6mm]
& &\displaystyle+\frac{r_1^2}{c^3}\,\left(\frac{\dot{a}}{a}\right)^3\,\left(1-\frac{2GM}{c^2\,r_1}\right)^{\frac{3\alpha}{2}-\frac{3}{2}}\left(1+\frac{(\alpha-2)\,GM}{c^2\,r_1}\right)\ ,\\[6mm]
_{(\alpha)}\Gamma^0_{\ 01}&=&_{(\alpha)}\Gamma^0_{10}\\[5mm]
&=&\displaystyle\frac{GM}{c^2\,r_1^2}\left(1-\frac{2GM}{c^2\,r_1}\right)^{-1}\\[6mm]
& &\displaystyle -r_1\,\left(\frac{\dot{a}}{a\,c}\right)^2\,\left(1-\frac{2GM}{c^2\,r_1}\right)^{\alpha-2}\,\left(1+\frac{(\alpha-2)\,GM}{c^2\,r_1}\right)\ ,\\[6mm]
_{(\alpha)}\Gamma^0_{\ 11}&=&\displaystyle\left(\frac{\dot{a}}{a\,c}\right)\,\left(1-\frac{2GM}{c^2\,r_1}\right)^{\frac{\alpha}{2}-\frac{5}{2}}\,\left(1+\frac{(\alpha-2)\,GM}{c^2\,r_1}\right)\ ,\\[6mm]
_{(\alpha)}\Gamma^0_{\ 22}&=&\displaystyle\,r_1^2\,\left(\frac{\dot{a}}{a\,c}\right)\,\left(1-\frac{2GM}{c^2\,r_1}\right)^{\frac{\alpha}{2}-\frac{1}{2}}\ ,\\[6mm]
_{(\alpha)}\Gamma^0_{\ 33}&=&\sin^2\theta\,_{(\alpha)}\Gamma^0_{\ 22}\ ,
\ea
\lb{A.gen.connections_0}
\ee
for $\beta=1$,
\be
\ba{rcl}
_{(\alpha)}\Gamma^1_{\ 00}&=&\displaystyle\frac{GM}{c^2\,r_1^2}\left(1-\frac{2GM}{c^2\,r_1}\right)-r_1\,\left(\frac{\ddot{a}}{a\,c^2}\right)\,\left(1-\frac{2GM}{c^2\,r_1}\right)^{\frac{\alpha}{2}+\frac{1}{2}}\\[6mm]
& &\displaystyle+r_1\,\left(\frac{\dot{a}}{a\,c}\right)^2\,\left(1-\frac{2GM}{c^2\,r_1}\right)^{\frac{\alpha}{2}+\frac{1}{2}}\left(1-\left(1-\frac{2GM}{c^2\,r_1}\right)^{\frac{\alpha}{2}-\frac{1}{2}}\,\left(1+\frac{(\alpha-1)\,GM}{c^2\,r_1}\right)\right)\\[6mm]
& &\displaystyle+r_1^3\,\left(\frac{\dot{a}}{a\,c}\right)^4\,\left(1-\frac{2GM}{c^2\,r_1}\right)^{2\alpha-1}\,\left(1+\frac{(\alpha-2)\,GM}{c^2\,r_1}\right)\ ,\\[6mm]
_{(\alpha)}\Gamma^1_{\ 01}&=&_{(\alpha)}\Gamma^1_{\ 10}\\[5mm]
&=&\displaystyle\frac{GM}{c^2\,r_1}\,\left(\frac{\dot{a}}{a\,c}\right)\,\left(1-\frac{2GM}{c^2\,r_1}\right)^{\frac{\alpha}{2}-\frac{1}{2}}\\[6mm]
& &\displaystyle-r_1^2\,\left(\frac{\dot{a}}{a\,c^3}\right)^3\,\left(1-\frac{2GM}{c^2\,r_1}\right)^{\frac{3\alpha}{2}-\frac{3}{2}}\left(1+\frac{(\alpha-2)\,GM}{c^2\,r_1}\right)\ ,\\[6mm]
_{(\alpha)}\Gamma^1_{\ 11}&=&\displaystyle-\frac{GM}{c^2\,r_1^2}\,\left(1-\frac{2GM}{c^2\,r_1}\right)^{-1}\\[6mm]
& &\displaystyle+r_1\,\left(\frac{\dot{a}}{a\,c}\right)^2\left(1-\frac{2GM}{c^2\,r_1}\right)^{\alpha-2}\left(1+\frac{(\alpha-2)GM}{c^2\,r_1}\right)\ ,\\[6mm]
_{(\alpha)}\Gamma^1_{\ 22}&=&\displaystyle -r_1\,\left(1-\frac{2GM}{c^2\,r_1}\right)+r_1^3\,\left(\frac{\dot{a}}{a\,c}\right)^2\,\left(1-\frac{2GM}{c^2\,r_1}\right)^\alpha\ ,\\[6mm]
_{(\alpha)}\Gamma^1_{\ 33}&=&\sin^2\theta\,_{(\alpha)}\Gamma^1_{\ 22}\ ,
\ea
\lb{A.gen.connections_1}
\ee
and the remaining connections coincide with the ones for spherical coordinates in flat space-time
\be
\ba{rclcrcl}
_{(\alpha)}\Gamma^2_{\ 12}&=&\displaystyle_{(\alpha)}\Gamma^2_{\ 21}=\frac{1}{r_1}& ,&_{(\alpha)}\Gamma^2_{\ 33}&=&\displaystyle-\cos\theta\sin\theta\ ,\\[6mm]
_{(\alpha)}\Gamma^3_{\ 13}&=&\displaystyle_{(\alpha)}\Gamma^3_{\ 31}=\frac{1}{r_1}& ,&_{(\alpha)}\Gamma^3_{\ 23}&=&\displaystyle_{(\alpha)}\Gamma^3_{\ 32}=\frac{\cos\theta}{\sin\theta}\ .
\ea
\lb{A.gen.connections_2}
\ee
In these expressions we have ordered the terms in increasing powers of the Hubble rate
($\dot{a}/a=H$ and $\ddot{a}/a=-q_0H^2$).

The Ricci scalar (curvature) is
\be
R_\alpha=-6\,R_{I}\,\left(\frac{\dot{a}}{a\,c}\right)^2-6\,R_{II}\,\left(\frac{\ddot{a}}{a\,c^2}\right)\ ,
\lb{A.generic_R}
\ee
where the coefficients $R_I$ and $R_{II}$ are functions of $r_1$ only
\be
\ba{rcl}
R_{I}&=&\displaystyle\frac{2\alpha\,GM}{c^2\,r_1}\left(1-\frac{2GM}{c^2\,r_1}\right)^{\alpha-2}\left(1-\frac{(7-\alpha)\,GM}{3c^2\,r_1}\right)\\[6mm]
&&\displaystyle-\left(1-\frac{2GM}{c^2\,r_1}\right)^{\frac{\alpha}{2}-\frac{3}{2}}\left(1-\frac{(6-\alpha)\,GM}{3c^2\,r_1}\right)+2\left(1-\frac{2GM}{c^2\,r_1}\right)^{\alpha}\\[6mm]
R_{II}&=&\displaystyle\left(1-\frac{2GM}{c^2\,r_1}\right)^{\frac{\alpha}{2}-\frac{3}{2}}\left(1-\frac{(6-\alpha)\,GM}{3\,c^2\,r_1}\right)\ ,
\ea
\lb{A.generic_R_I}
\ee
where we organize the several terms in growing powers of $(1-2GM/(c^2\,r_1))$.
The non-null components of the Einstein tensor are
\be
\ba{rcl}
G^{(\alpha)}_{00}&=&\displaystyle 3\,\left(\frac{\dot{a}}{a\,c}\right)^2\,\left(1-\frac{2GM}{c^2\,r_1}\right)^\alpha\,\left(1+\frac{2(3-\alpha)\,GM}{3c^2\,r_1}\right)\\[6mm]
& &\displaystyle+r_1^2\left(\frac{\dot{a}}{a\,c}\right)^4\,\left(1-\frac{2GM}{c^2\,r_1}\right)^{\frac{3\alpha}{2}-\frac{1}{2}}\times\\[6mm]
& &\displaystyle\hfill\times\left(2-3\left(1-\frac{2GM}{c^2\,r_1}\right)^{\frac{\alpha}{2}-\frac{1}{2}}\left(1+\frac{2(\alpha-3)\,GM}{3c^2\,r_1}\right)\right)\\[6mm]
& &\displaystyle -2r_1^2\,\left(\frac{\dot{a}}{a\,c}\right)^2\,\left(\frac{\ddot{a}}{a\,c^2}\right)\,\left(1-\frac{2GM}{c^2\,r_1}\right)^{\frac{3\alpha}{2}-\frac{1}{2}}\ ,\\[6mm]
G^{(\alpha)}_{01}&=&\displaystyle \frac{g_{01}}{g_{11}}\,G^{(\alpha)}_{11}\,=\,-\frac{\dot{a}\,r_1}{a\,c}\left(1-\frac{2GM}{c^2\,r_1}\right)^{\frac{\alpha}{2}+\frac{1}{2}}\,G^{(\alpha)}_{11}\ ,\\[6mm]

G^{(\alpha)}_{11}&=&\displaystyle -\left(\frac{\dot{a}}{a\,c}\right)^2\left(1-\frac{2GM}{c^2\,r_1}\right)^{\frac{\alpha}{2}-\frac{3}{2}}\times\\[6mm]
&&\displaystyle\hfill\times\left(2-3\left(1-\frac{2GM}{c^2\,r_1}\right)^{\frac{\alpha}{2}-\frac{1}{2}}\left(1+\frac{2(\alpha-3)\,GM}{3c^2\,r_1}\right)\right)\\[6mm]
& &\displaystyle -2\,\left(\frac{\ddot{a}}{a\,c^2}\right)\left(1-\frac{2GM}{c^2\,r_1}\right)^{\frac{\alpha}{2}-\frac{3}{2}}\ ,\\[6mm]
G^{(\alpha)}_{22}&=&\displaystyle r_1^2\,\left(\frac{\dot{a}}{a\,c}\right)^2\left(1-\frac{2GM}{c^2\,r_1}\right)^{\frac{\alpha}{2}-\frac{3}{2}}\,\left(2+\frac{(\alpha-4)GM}{c^2\,r_1}-\left(1-\frac{2GM}{c^2\,r_1}\right)^{\frac{\alpha}{2}-\frac{1}{2}}\right.\times\\[6mm]
& &\displaystyle\hfill\left.\times\left(3+\frac{4(\alpha-3)\,GM}{c^2\,r_1}+\frac{2(\alpha^2-5\alpha+6)\,(GM)^2}{c^4\,r_1^2}\right)\right)\\[6mm]
& &\displaystyle -2r_1^2\,\left(\frac{\ddot{a}}{a\,c^2}\right)\left(1-\frac{2GM}{c^2\,r_1}\right)^{\frac{\alpha}{2}-\frac{3}{2}}\,\left(1+\frac{(\alpha-4)\,GM}{2c^2\,r_1}\right)\ ,\\[6mm]
G^{(\alpha)}_{33}&=&\displaystyle\sin^2\theta\,G^{(\alpha)}_{22}\ . 
\ea
\lb{A.gen.EE}
\ee

The curvature invariant is
\be
\ba{rcl}
{\mathcal{R}}_\alpha&=&R_{\mu\nu\delta\rho}R^{\mu\nu\delta\rho}\\[6mm]
&=&\displaystyle{\mathcal{R}}_{I}+{\mathcal{R}}_{II}\left(\frac{\dot{a}}{a\,c}\right)^2+{\mathcal{R}}_{III}\left(\frac{\ddot{a}}{a\,c^2}\right)+{\mathcal{R}}_{IV}\left(\frac{\dot{a}}{a\,c}\right)^4+{\mathcal{R}}_{V}\frac{\dot{a}^2\,\ddot{a}}{a^3\,c^4}+{\mathcal{R}}_{VI}\left(\frac{\ddot{a}}{a\,c^2}\right)^2\ ,
\ea
\lb{A.generic_RR}
\ee
where the several coefficients ${\mathcal{R}}_I$ to ${\mathcal{R}}_{VI}$
are
\be
\ba{rcl}
\displaystyle {\mathcal{R}}_{I}&=&\displaystyle\frac{48(GM)^2}{c^4\,r_1^6}\ ,\\[6mm]
\displaystyle {\mathcal{R}}_{II}&=&\displaystyle-\frac{16\alpha\,(GM)^2}{c^4\,r_1^4}\,\left(1-\frac{2GM}{c^2\,r_1}\right)^{\frac{\alpha}{2}-\frac{3}{2}}\left(1-\frac{2(\alpha-1)GM}{c^2\,r_1}\left(1-\frac{2GM}{c^2\,r_1}\right)^{\frac{\alpha}{2}-\frac{1}{2}}\right)\ ,\\[6mm]
\displaystyle {\mathcal{R}}_{III}&=&\displaystyle\frac{16\alpha\,(GM)^2}{c^4\,r_1^4}\,\left(1-\frac{2GM}{c^2\,r_1}\right)^{\frac{\alpha}{2}-\frac{3}{2}}\ ,\\[6mm]
{\mathcal{R}}_{IV}&=&\displaystyle+\frac{16\alpha^2\,(GM)^2}{c^4r_1^2}\left(1-\frac{2GM}{c^2\,r_1}\right)^{2\alpha-4}\times\\[6mm]
&&\displaystyle\hfill\times\left(3+\frac{2(\alpha-7)\,GM}{c^2\,r_1}+\frac{(\alpha^2-6\alpha+17)(GM)^2}{c^4\,r_1^2}\right)\\[6mm]
&&\displaystyle+\frac{16\alpha\,GM}{c^2\,r_1}\left(3-\frac{7GM}{c^2\,r_1}\right)\left(1-\frac{2GM}{c^2\,r_1}\right)^{2\alpha-2}+24\left(1-\frac{2GM}{c^2\,r_1}\right)^{2\alpha}\\[6mm]
&&\displaystyle+\frac{4\alpha^2\,(GM)^2}{c^4\,r_1^2}\left(1-\frac{2GM}{c^2\,r_1}\right)^{\alpha-3}+\frac{8\alpha\,GM}{c^2\,r_1}\left(1-\frac{2GM}{c^2\,r_1}\right)^{\alpha-2}\\[6mm]
&&\displaystyle+12\left(1-\frac{2GM}{c^2\,r_1}\right)^{\alpha-1}-\frac{16\alpha^2\,(GM)^2}{c^4\,r_1^2}\left(1-\frac{2GM}{c^2\,r_1}\right)^{\frac{3\alpha}{2}-\frac{9}{2}}\times\\[6mm]
&&\displaystyle\hfill\times\left(2+\frac{(\alpha-9)\,GM}{c^2\,r_1}-\frac{2(\alpha-5)\,(GM)^2}{c^4\,r_1^2}\right)\\[6mm]
&&\displaystyle-\frac{8\alpha\,GM}{c^2\,r_1}\left(1-\frac{2GM}{c^2\,r_1}\right)^{\frac{3\alpha}{2}-\frac{5}{2}}\,\left(5-\frac{12GM}{c^2\,r_1}\right)-24\left(1-\frac{2GM}{c^2\,r_1}\right)^{\frac{3\alpha}{2}-\frac{1}{2}}\ ,
\ea
\lb{A.generic_RR_I}
\ee
and
\be
\ba{rcl}
\displaystyle 
{\mathcal{R}}_{V}&=&\displaystyle-\frac{8\alpha^2\,(GM)^2}{c^4\,r_1^2}\,\left(1-\frac{2GM}{c^2\,r_1}\right)^{\alpha-3}-\frac{16\alpha\,GM}{c^2\,r_1}\,\left(1-\frac{2GM}{c^2\,r_1}\right)^{\alpha-2}\\[6mm]
&&\displaystyle-24\left(1-\frac{2GM}{c^2\,r_1}\right)^{\alpha-1}+\frac{16\alpha^2\,(GM)^2}{c^4\,r_1^2}\,\left(1-\frac{2GM}{c^2\,r_1}\right)^{\frac{3\alpha}{2}-\frac{7}{2}}\left(2+\frac{(\alpha-5)GM}{c^2\,r_1}\right)\\[6mm]
&&\displaystyle+\frac{8\alpha\,GM}{c^2\,r_1}\left(1-\frac{2GM}{c^2\,r_1}\right)^{\frac{3\alpha}{2}-\frac{5}{2}}\,\left(5-\frac{12GM}{c^2\,r_1}\right)+24\left(1-\frac{2GM}{c^2\,r_1}\right)^{\frac{3\alpha}{2}-\frac{1}{2}}\ ,\\[6mm]
{\mathcal{R}}_{VI}&=&\displaystyle \frac{4\alpha^2(GM)^2}{c^4\,r_1^2}\left(1-\frac{2GM}{c^2\,r_1}\right)^{\alpha-3}\\[6mm]
&&\displaystyle
+\frac{8\alpha\,GM}{c^2\,r_1}\left(1-\frac{2GM}{c^2\,r_1}\right)^{\alpha-2}+12\left(1-\frac{2GM}{c^2\,r_1}\right)^{\alpha-1}\ ,
\ea
\lb{A.generic_RR_II}
\ee
where we organized the several terms in growing powers of the factor $(1-2GM/(c^2\,r_1))$.

\subsection{Space Dependent Exponent $\alpha(r_1)$}

When the exponent $\alpha$ in the shift function~(\ref{N_1_alpha_1}) of the metric
ansatz~(\ref{g_generic}) is dependent on the radial coordinate $r_1$, $\alpha=\alpha(r_1)$, as
introduced in equation~(\ref{alpha_r_1}), the connections are
\be
_{(\alpha(r_1))}\Gamma^{\beta}_{\ \mu\nu}=_{(\alpha)}\Gamma^{\beta}_{\ \mu\nu}+\Delta \Gamma^{\beta}_{\ \mu\nu}\ ,
\lb{A.gen_connections_r}
\ee
where $_{(\alpha)}\Gamma^{\beta}_{\ \mu\nu}$ are the connections for constant exponent $\alpha$
given in the above equations~(\ref{A.gen.connections_0}),~(\ref{A.gen.connections_1}) and~(\ref{A.gen.connections_2}) and the non-null corrections $\Delta \Gamma^{\beta}_{\ \mu\nu}$
are expressed in terms of the derivative ${\alpha}'=\partial\alpha(r_1)/\partial r_1$ as
\be
\ba{rcl}
\Delta\Gamma^0_{\ 00}&=&\displaystyle+\frac{r_1^3}{2}\,\left(\frac{\dot{a}}{a\,c}\right)^3\,\left(1-\frac{2GM}{c^2\,r_1}\right)^{\frac{3\alpha}{2}-\frac{1}{2}}\,\log\left(1-\frac{2GM}{c^2\,r_1}\right)\,{\alpha}'
\ ,\\[6mm]
\Delta\Gamma^0_{\ 01}&=&\Delta\Gamma^0_{10}\\[5mm]
&=&\displaystyle -\frac{r_1^2}{2}\,\left(\frac{\dot{a}}{a\,c}\right)^2\,\left(1-\frac{2GM}{c^2\,r_1}\right)^{\alpha-1}\,\log\left(1-\frac{2GM}{c^2\,r_1}\right)\,{\alpha}'\ ,\\[6mm]
\Delta\Gamma^0_{\ 11}&=&\displaystyle\frac{r_1}{2}\,\left(\frac{\dot{a}}{a\,c}\right)\,\left(1-\frac{2GM}{c^2\,r_1}\right)^{\frac{\alpha}{2}-\frac{3}{2}}\,\log\left(1-\frac{2GM}{c^2\,r_1}\right)\,{\alpha}'\ ,\\[6mm]
\Delta\Gamma^1_{\ 00}&=&\displaystyle-\frac{r_1^2}{2}\left(\frac{\dot{a}}{a\,c}\right)^2\left(1-\frac{2GM}{c^2\,r_1}\right)^{\alpha+1}\,\log\left(1-\frac{2GM}{c^2\,r_1}\right)\,{\alpha}'\\[6mm]
& &\displaystyle+\frac{r_1^4}{2}\left(\frac{\dot{a}}{a\,c}\right)^4\left(1-\frac{2GM}{c^2\,r_1}\right)^{2\alpha}\,\log\left(1-\frac{2GM}{c^2\,r_1}\right)\,{\alpha}'\ ,\\[6mm]
\Delta\Gamma^1_{\ 01}&=& \Delta\Gamma^1_{\ 10}\\[5mm]
&=&\displaystyle -\frac{r_1^3}{2}\,\left(\frac{\dot{a}}{a\,c}\right)^3\,\left(1-\frac{2GM}{c^2\,r_1}\right)^{\frac{3\alpha}{2}-\frac{1}{2}}\,\log\left(1-\frac{2GM}{c^2\,r_1}\right)\,{\alpha}'\ ,\\[6mm]
\Delta\Gamma^1_{\ 11}&=&\displaystyle\frac{r_1^2}{2}\,\left(\frac{\dot{a}}{a\,c}\right)^3\,\left(1-\frac{2GM}{c^2\,r_1}\right)^{\alpha-1}\,\log\left(1-\frac{2GM}{c^2\,r_1}\right)\,{\alpha}'\ ,\\[6mm]
\ea
\lb{A.gen.connections_r}
\ee
The curvature is
\be
R_{\alpha(r_1)}=-6(R_{I}+\Delta R_{I})\left(\frac{\dot{a}}{a\,c}\right)^2-6(R_{II}+\Delta R_{II})\left(\frac{\ddot{a}}{a\,c^2}\right)\ ,
\lb{A.gen_R_r}
\ee
where the coefficients $R_I$ and $R_{II}$ are given in equation~(\ref{A.generic_R_I}) and the
corrections due to the radial coordinate dependence of the exponent $\alpha$ are expressed in terms
of the derivatives ${\alpha}'=\partial\alpha/\partial r_1$ and
${\alpha}''=\partial^2\alpha/\partial r_1^2$ as
\be
\ba{rcl}
\Delta R_{I}&=&\displaystyle \frac{2GM}{3c^2}\left(1-\frac{2GM}{c^2\,r_1}\right)^{\alpha-1}\left(1+\alpha\log\left(1-\frac{2GM}{c^2r_1}\right)\right){\alpha}'\\[6mm]
& &\displaystyle
-r_1\,\left(\frac{1}{6}\left(1-\frac{2GM}{c^2\,r_1}\right)^{\frac{\alpha}{2}-\frac{1}{2}}-\frac{4}{3}\left(1-\frac{2GM}{c^2\,r_1}\right)^{\alpha}\right)\log\left(1-\frac{2GM}{c^2r_1}\right){\alpha}'\\[6mm]
&&\displaystyle+\frac{r_1^2}{6}\left(1-\frac{2GM}{c^2\,r_1}\right)^{\alpha}\,\log\left(1-\frac{2GM}{c^2\,r_1}\right)\left(\left({\alpha}'\right)^2+\left({\alpha}''\right)^2\right)\ ,\\[6mm]
\Delta R_{II}&=&\displaystyle \frac{r_1}{6}\left(1-\frac{2GM}{c^2\,r_1}\right)^{\frac{\alpha}{2}-\frac{1}{2}}\log\left(1-\frac{2GM}{c^2\,r_1}\right){\alpha}'\ .
\ea
\lb{A.generic_R_r_I}
\ee

The Einstein tensor is
\be
G^{(\alpha(r_1))}_{\mu\nu}=G^{(\alpha)}_{\mu\nu}+\Delta G_{\mu\nu}\ ,
\lb{A.gen.EE_r}
\ee
where $G^{(\alpha)}_{\mu\nu}$ is the Einstein tensor for constant $\alpha$
given in~(\ref{A.gen.EE}) and the corrections $\Delta G_{\mu\nu}$ due to the 
dependence on the radial coordinate of the exponent $\alpha$ are
\be
\ba{rcl}
\Delta G_{00}&=&\displaystyle r_1\,\left(\frac{\dot{a}}{a\,c}\right)^2\left(1-\frac{2GM}{c^2\,r_1}\right)^{\alpha+1}\log\left(1-\frac{2GM}{c^2r_1}\right){\alpha}'\\[6mm]
&&\displaystyle -r_1^3\,\left(\frac{\dot{a}}{a\,c}\right)^4\left(1-\frac{2GM}{c^2\,r_1}\right)^{2\alpha}\log\left(1-\frac{2GM}{c^2r_1}\right){\alpha}'\ ,\\[6mm]
\Delta G_{01}&=&\displaystyle \frac{g_{01}}{g_{11}}\,\Delta G_{11}\,=\,-\frac{\dot{a}\,r_1}{a\,c}\left(1-\frac{2GM}{c^2\,r_1}\right)^{\frac{\alpha}{2}+\frac{1}{2}}\,\Delta G_{11}\ ,\\[6mm]
\Delta G_{11}&=&\displaystyle-r_1\,\left(\frac{\dot{a}}{a\,c}\right)^2\left(1-\frac{2GM}{c^2\,r_1}\right)^{\alpha-1}\log\left(1-\frac{2GM}{c^2\,r_1}\right){\alpha}'\ ,\\[6mm]
\Delta G_{22}&=&\displaystyle-\frac{r_1^3}{2}\,\left(\frac{\dot{a}}{a\,c}\right)^2\left(\frac{4GM}{c^2\,r_1}\left(1-\frac{2GM}{c^2\,r_1}\right)^{\alpha-1}\left(1+\alpha\log\left(1-\frac{2GM}{c^2\,r_1}\right)\right){\alpha}'\right.\\[6mm]
&&\displaystyle-\left(1-\frac{2GM}{c^2\,r_1}\right)^{\frac{\alpha}{2}-\frac{1}{2}}\log\left(1-\frac{2GM}{c^2\,r_1}\right){\alpha}'+r_1\,\left(1-\frac{2GM}{c^2\,r_1}\right)^{\alpha}\times\\[6mm]
&&\displaystyle\hfill\times\left.\log\left(1-\frac{2GM}{c^2\,r_1}\right)\left(\frac{6}{r_1}{\alpha}'+\log\left(1-\frac{2GM}{c^2\,r_1}\right)\left({\alpha}'\right)^2+{\alpha}''\right)\right)\\[6mm]
&&\displaystyle -\frac{r_1^3}{2}\left(\frac{\ddot{a}}{a\,c^2}\right)\left(1-\frac{2GM}{c^2\,r_1}\right)^{\frac{\alpha}{2}-\frac{1}{2}}\log\left(1-\frac{2GM}{c^2\,r_1}\right){\alpha}'\ ,\\[6mm]
\Delta G_{33}&=&\displaystyle \sin^2\theta\,\Delta G_{22}\ .
\ea
\lb{A.generic_G_r_I}
\ee

The curvature invariant is
\be
\ba{rcl}
{\mathcal{R}}_{\alpha(r_1)}&=&\displaystyle{\mathcal{R}}_{I}+({\mathcal{R}}_{II}+\Delta{\mathcal{R}}_{II})\left(\frac{\dot{a}}{a\,c}\right)^2+({\mathcal{R}}_{III}+\Delta{\mathcal{R}}_{III})\left(\frac{\ddot{a}}{a\,c^2}\right)\\[6mm]
& &\displaystyle+({\mathcal{R}}_{IV}+\Delta{\mathcal{R}}_{IV})\left(\frac{\dot{a}}{a\,c}\right)^4+({\mathcal{R}}_{V}+\Delta{\mathcal{R}}_{V})\frac{\dot{a}^2\,\ddot{a}}{a^3\,c^4}+({\mathcal{R}}_{VI}+\Delta{\mathcal{R}}_{VI})\left(\frac{\ddot{a}}{a\,c^2}\right)^2\ ,
\ea
\lb{A.generic_RR_r}
\ee
where the coefficients ${\mathcal{R}}_I$ to ${\mathcal{R}}_{VI}$ correspond to the coefficients
of the curvature invariant~(\ref{A.generic_RR}) for constant $\alpha$ given in equation~(\ref{A.generic_RR_I}) and~(\ref{A.generic_RR_II}). The corrections to these coefficients
$\Delta{\mathcal{R}}_{I}$ due to the radial coordinate dependence of the exponent $\alpha$ are
\be
\ba{rcl}
\Delta{\mathcal{R}}_{II}&=&\displaystyle \frac{8GM}{c^2\,r_1^2}\,\left(\frac{4GM}{c^2\,r_1}\left(1-\frac{2GM}{c^2\,r_1}\right)^{\alpha-1}\left(1+\alpha\log\left(1-\frac{2GM}{c^2\,r_1}\right)\right){\alpha}'\right.\\[6mm]
&&\displaystyle -\left(1-\frac{2GM}{c^2\,r_1}\right)^{\frac{\alpha}{2}-\frac{1}{2}}\log\left(1-\frac{2GM}{c^2\,r_1}\right){\alpha}'+r_1\,\left(1-\frac{2GM}{c^2\,r_1}\right)^{\alpha}\times\\[6mm]
&&\displaystyle\hfill\times\left.\log\left(1-\frac{2GM}{c^2\,r_1}\right)\left(\frac{2}{r_1}{\alpha}'+\log\left(1-\frac{2GM}{c^2\,r_1}\right)\left({\alpha}'\right)^2+{\alpha}''\right)\right)\ ,\\[6mm]
\Delta{\mathcal{R}}_{III}&=&\displaystyle\frac{8GM}{c^2\,r_1^2}\left(1-\frac{2GM}{c^2\,r_1}\right)^{\frac{\alpha}{2}-\frac{1}{2}}\log\left(1-\frac{2GM}{c^2\,r_1}\right){\alpha}'
\ea
\lb{A.generic_RR_r_I}
\ee
and
\be
\ba{rcl}
\Delta{\mathcal{R}}_{IV}&=&\displaystyle 4{\alpha}'\,r_1\,\Delta{\mathcal{R}}_{IV-1}+\left({\alpha}'\right)^2\,r_1^2\,\Delta{\mathcal{R}}_{IV-2}+4{\alpha}''\,r_1^2\,\log\left(1-\frac{2GM}{c^2\,r_1}\right)\Delta{\mathcal{R}}_{IV-3}\\[6mm]
&&\displaystyle+2{\alpha}'\,r_1^3\log\left(1-\frac{2GM}{c^2\,r_1}\right)\left({\alpha}''+\log\left(1-\frac{2GM}{c^2\,r_1}\right)\left({\alpha}'\right)^2\right)\Delta{\mathcal{R}}_{IV-4}\\[6mm]
&&\displaystyle+\left(1-\frac{2GM}{c^2\,r_1}\right)^{2\alpha}\left(r_1^2\,\log\left(1-\frac{2GM}{c^2\,r_1}\right)\right)^2\times\\[6mm]
&&\displaystyle\hfill\times\left({\alpha}''+\log\left(1-\frac{2GM}{c^2\,r_1}\right)\left({\alpha}'\right)^2\right)^2\ ,\\[6mm]
\Delta{\mathcal{R}}_{V}&=&\displaystyle -8\left(\frac{\alpha\,GM}{c^2}\left(1-\frac{2GM}{c^2\,r_1}\right)^{\alpha-2}+\left(1-\frac{2GM}{c^2\,r_1}\right)^{\alpha-1}\right)\log\left(1-\frac{2GM}{c^2\,r_1}\right){\alpha}'\\[6mm]
&&\displaystyle-\frac{8\alpha\,GM}{c^2}\left(1-\frac{2GM}{c^2\,r_1}\right)^{\frac{3\alpha}{2}-\frac{5}{2}}\left(\frac{2GM}{c^2\,r_1}+\left(5+\frac{(3\alpha-11)GM}{c^2\,r_1}\right)\times\right.\\[6mm]
&&\displaystyle\hfill\left.\times\log\left(1-\frac{2GM}{c^2\,r_1}\right)\right){\alpha}'+\frac{16GM}{c^2}\left(1-\frac{2GM}{c^2\,r_1}\right)^{\frac{3\alpha}{2}-\frac{3}{2}}{\alpha}'\\[6mm]
&&\displaystyle +2r_1\left(1-\frac{2GM}{c^2\,r_1}\right)^{\frac{3\alpha}{2}-\frac{1}{2}}\left(14+r_1^2\log\left(1-\frac{2GM}{c^2\,r_1}\right){\alpha}''\right)\log\left(1-\frac{2GM}{c^2\,r_1}\right){\alpha}'\\[6mm]

&&\displaystyle+\frac{8GM}{c^2}\left(1-\frac{2GM}{c^2\,r_1}\right)^{\frac{3\alpha}{2}-\frac{1}{2}}r_1\log\left(1-\frac{2GM}{c^2\,r_1}\right)\left({\alpha}'\right)^2\\[6mm]
&&\displaystyle-2\left(\left(1-\frac{2GM}{c^2\,r_1}\right)^{\alpha-1}-\frac{6\alpha\,GM}{c^2\,r_1}\left(1-\frac{2GM}{c^2\,r_1}\right)^{\frac{3\alpha}{2}-\frac{3}{2}}\right.\\[6mm]
&&\displaystyle\hfill\left.-6\left(1-\frac{2GM}{c^2\,r_1}\right)^{\frac{3\alpha}{2}-\frac{1}{2}}\right)\left(r_1\log\left(1-\frac{2GM}{c^2\,r_1}\right){\alpha}'\right)^2\\[6mm]
&&\displaystyle +4\left(1-\frac{2GM}{c^2\,r_1}\right)^{\frac{3\alpha}{2}-\frac{3}{2}}\left(1+\frac{(\alpha-2)GM}{c^2\,r_1}\right)r_1^2\log\left(1-\frac{2GM}{c^2\,r_1}\right){\alpha}''\\[6mm]
&&\displaystyle +2\left(1-\frac{2GM}{c^2\,r_1}\right)^{\frac{3\alpha}{2}-\frac{1}{2}}\left(r_1\log\left(1-\frac{2GM}{c^2\,r_1}\right){\alpha}'\right)^3\ ,\\[6mm]

\Delta{\mathcal{R}}_{VI}&=&\displaystyle \left(\frac{4\alpha\,GM}{c^2\,r_1}\left(1-\frac{2GM}{c^2\,r_1}\right)^{\alpha-2}+\right.\\[6mm]
&&\displaystyle\hfill\left.\left(1-\frac{2GM}{c^2\,r_1}\right)^{\alpha-1}\left(4+r_1\log\left(1-\frac{2GM}{c^2\,r_1}\right){\alpha}'\right)\right)r_1\,\log\left(1-\frac{2GM}{c^2\,r_1}\right){\alpha}'\ ,
\ea
\lb{A.generic_RR_r_II}
\ee
where the coefficients $\Delta{\mathcal{R}}_{IV-1}$ to $\Delta{\mathcal{R}}_{IV-4}$ in
the definition of $\Delta{\mathcal{R}}_{IV}$ are
\be
\ba{rcl}
\Delta{\mathcal{R}}_{IV-1}&=&\displaystyle\frac{8\alpha\,(GM)^2}{c^4\,r_1^2}\left(1-\frac{2GM}{c^2\,r_1}\right)^{2\alpha-3}\left(1+\frac{(\alpha-3)(GM)}{c^2\,r_1}\right)\\[6mm]
&&\displaystyle+\frac{4\,GM}{c^2\,r_1}\left(1-\frac{2GM}{c^2\,r_1}\right)^{2\alpha-1}-\frac{4\,(GM)^2}{c^4\,r_1^2}\left(1-\frac{2GM}{c^2\,r_1}\right)^{\frac{3\alpha}{2}-\frac{5}{2}}\\[6mm]
&&\displaystyle-\frac{4\,GM}{c^2\,r_1}\left(1-\frac{2GM}{c^2\,r_1}\right)^{\frac{3\alpha}{2}-\frac{3}{2}}+\log\left(1-\frac{2GM}{c^2\,r_1}\right)\times\\[6mm]
&&\displaystyle\times\left(\frac{8\alpha^2\,(GM)^2}{c^4\,r_1^2}\left(2+\frac{(\alpha-5)\,GM}{c^2\,r_1}\right)\left(1-\frac{2GM}{c^2\,r_1}\right)^{2\alpha-3}\right.\\[6mm]
&&\displaystyle+\frac{8\alpha\,GM}{c^2\,r_1}\left(2-\frac{5GM}{c^2\,r_1}\right)\left(1-\frac{2GM}{c^2\,r_1}\right)^{2\alpha-2}+8\left(1-\frac{2GM}{c^2\,r_1}\right)^{2\alpha}\\[6mm]
&&\displaystyle+\left(1+\frac{(\alpha-2)\,GM}{c^2\,r_1}\right)\left(1-\frac{2GM}{c^2\,r_1}\right)^{\alpha-2}-7\left(1-\frac{2GM}{c^2\,r_1}\right)^{\frac{3\alpha}{2}-\frac{1}{2}}\\[6mm]
&&\displaystyle\left.-\frac{2\alpha\,GM}{c^2\,r_1}\left(5+\frac{(3\alpha-11)\,GM}{c^2\,r_1}\right)\left(1-\frac{2GM}{c^2\,r_1}\right)^{\frac{3\alpha}{2}-\frac{5}{2}}\right)\ ,\\[6mm]
\Delta{\mathcal{R}}_{IV-2}&=&\displaystyle\frac{16(GM)^2}{c^2\,r_1^2}\left(1-\frac{2GM}{c^2\,r_1}\right)^{2\alpha-2}\left(1+2\alpha\log\left(1-\frac{2GM}{c^2\,r_1}\right)\right)\\[6mm]
&&\displaystyle -\frac{8GM}{c^2\,r_1}\left(1-\frac{2GM}{c^2\,r_1}\right)^{\frac{3\alpha}{2}-\frac{3}{2}}\times\\[6mm]
&&\displaystyle\hfill\times\left(1+\frac{3\alpha}{2}\log\left(1-\frac{2GM}{c^2\,r_1}\right)\right)\log\left(1-\frac{2GM}{c^2\,r_1}\right)\\[6mm]
&&\displaystyle +\log\left(1-\frac{2GM}{c^2\,r_1}\right)\left(\frac{32GM}{c^2\,r_1}\left(1-\frac{2GM}{c^2\,r_1}\right)^{2\alpha-1}+\log\left(1-\frac{2GM}{c^2\,r_1}\right)\times\right.\\[6mm]
&&\displaystyle\times\left(\frac{8\alpha\,GM}{c^2\,r_1}\left(5+\frac{(3\alpha-11)GM}{c^2\,r_1}\right)\left(1-\frac{2GM}{c^2\,r_1}\right)^{2\alpha-2}\right.\\[6mm]
&&\displaystyle\left.\left.+24\left(1-\frac{2GM}{c^2\,r_1}\right)^{2\alpha}+\left(1-\frac{2GM}{c^2\,r_1}\right)^{\alpha-1}-12\left(1-\frac{2GM}{c^2\,r_1}\right)^{\frac{3\alpha}{2}-\frac{1}{2}}\right)\right)\ ,\\[6mm]
\Delta{\mathcal{R}}_{IV-3}&=&\displaystyle +\frac{2\alpha\,GM}{c^2\,r_1}\left(1+\frac{(\alpha-3)GM}{c^2\,r_1}\right)\left(1-\frac{2GM}{c^2\,r_1}\right)^{2\alpha-2}+\left(1-\frac{2GM}{c^2\,r_1}\right)^{2\alpha}\\[6mm]
&&\displaystyle-\frac{\alpha\,GM}{c^2\,r_1}\left(1-\frac{2GM}{c^2\,r_1}\right)^{\frac{3\alpha}{2}-\frac{3}{2}}-\left(1-\frac{2GM}{c^2\,r_1}\right)^{\frac{3\alpha}{2}-\frac{1}{2}}\ ,\\[6mm]
\Delta{\mathcal{R}}_{IV-4}&=&\displaystyle \frac{4GM}{c^2\,r_1}\left(1-\frac{2GM}{c^2\,r_1}\right)^{2\alpha-1}\left(1+\alpha\log\left(1-\frac{2GM}{c^2\,r_1}\right)\right)\\[6mm]
&&\displaystyle+\left(4\left(1-\frac{2GM}{c^2\,r_1}\right)^{2\alpha}-\left(1-\frac{2GM}{c^2\,r_1}\right)^{\frac{3\alpha}{2}-\frac{1}{2}}\right)\log\left(1-\frac{2GM}{c^2\,r_1}\right)\ .
\ea
\ee

\subsection{Planetary Orbits\lb{A.orbits}}

In this section we derive the perturbative differential equation describing orbits
of a test mass in the gravitational field of a central mass $M$ for the locally anisotropic metric~(\ref{g_generic}) with a radial coordinate dependent exponent
$\alpha(r_1)=\alpha_0+\alpha_1\,U_{\mathrm{SC}}$~(\ref{alpha_r_1}).

We take the same approach of section~\ref{sec.rev_orbits} starting from equation~(\ref{L_orbits}).
For the specific case of the locally anisotropic metric~(\ref{g_generic}) and considering
an orbit lying in the plane of constant coordinate $\theta=\pi/2$ such that $d\theta=0$
and $\sin\theta=1$, we obtain the following equation
\be
\ba{rcl}
c^2&=&\displaystyle\left(1-U_{\mathrm{SC}}-\left(\frac{H\,r_1}{c}\right)^2\left(1-U_{\mathrm{SC}}\right)^\alpha\right)\,(c\,\dot{t})^2\\[6mm]
&&\displaystyle+2\left(\frac{H\,r_1}{c}\right)\left(1-U_{\mathrm{SC}}\right)^{\frac{\alpha}{2}-\frac{1}{2}}\,c\,\dot{t}\,\dot{r}_1-\frac{\dot{r}_1^2}{1-U_{\mathrm{SC}}}-r_1^2\dot{\varphi}^2\ ,
\ea
\lb{A.L_orbit}
\ee
where we have replaced the derivatives of the scale factor by the Hubble rate $H=\dot{a}/a$~(\ref{H}),
have written the equations in terms of the Schwarzschild gravitational potential
$U_{\mathrm{SC}}$~(\ref{U_SC}) and the dotted quantities represent derivatives with respect to
the proper time $\tau$. We will further take the approximation of static orbits considering
that the Hubble rate is a constant corresponding to the measurement of this rate at the reference
time $t_0$, $H_0=H(t_0)$~(\ref{H}). In this way the Lagrangian is independent of the time coordinate
and a conserved constant of motion corresponding to energy exists given by the functional
variation of the Lagrangian with respect to $c\,\dot{t}$
\be
\ba{rcl}
\displaystyle\frac{1}{m}\,\frac{\delta{\mathcal{L}}}{\delta(c\,\dot{t})}&=&\displaystyle\frac{2E_H}{m\,c}\\[6mm]
&=&\displaystyle 2\left(1-U_{\mathrm{SC}}-\left(\frac{H_0\,r_1}{c}\right)^2\left(1-U_{\mathrm{SC}}\right)^\alpha\right)\,(c\,\dot{t})+2\left(\frac{H_0\,r_1}{c}\right)\left(1-U_{\mathrm{SC}}\right)^{\frac{\alpha}{2}-\frac{1}{2}}\,\dot{r}_1\ .
\ea
\ee
This equation can be solved for $c\,\dot{t}$
\be
c\,\dot{t}=\frac{\frac{E_H}{m\,c}-\left(\frac{H_0\,r_1}{c}\right)\left(1-U_{\mathrm{SC}}\right)^{\frac{\alpha}{2}-\frac{1}{2}}\,\dot{r}_1}{1-U_{\mathrm{SC}}-\left(\frac{H_0\,r_1}{c}\right)^2\left(1-U_{\mathrm{SC}}\right)^\alpha}\ ,
\ee
such that replacing this solution in the Lagrangian~(\ref{A.L_orbit}) we obtain
\be
c^2=\frac{\left(\frac{E_H}{m\,c}\right)^2-\left(\frac{H_0\,r_1}{c}\right)^2\left(1-U_{\mathrm{SC}}\right)^{\alpha-1}\,\dot{r}_1^2}{1-U_{\mathrm{SC}}-\left(\frac{H_0\,r_1}{c}\right)^2\left(1-U_{\mathrm{SC}}\right)^\alpha}-\frac{\dot{r}_1^2}{1-U_{\mathrm{SC}}}-r_1^2\dot{\varphi}^2\ .
\ee
Multiplying by the factor $1-U_{\mathrm{SC}}-\left(\frac{H_0\,r_1}{c}\right)^2\left(1-U_{\mathrm{SC}}\right)^\alpha$
and gathering the constant terms in the left-hand side of the equation we obtain
\be
\ba{rcl}
\displaystyle c^2\left(1-\left(\frac{E_H}{m\,c^2}\right)^2\right)&=&\displaystyle \frac{2GM}{r_1}-\dot{r}_1^2-r_1^2\dot{\varphi}^2\left(1-\frac{2GM}{c^2\,r_1}\right)\\[6mm]
&&\displaystyle+r_1^2\dot{\varphi}^2\left(\frac{r_1\,H_0}{c}\right)^2\left(1-\frac{2GM}{c^2\,r_1}\right)^\alpha+\left(r_1\,H_0\right)^2\left(1-\frac{2GM}{c^2\,r_1}\right)^\alpha\ .
\ea
\lb{A.L_orbit_E}
\ee
The terms in the first line of this equation match the usual General Relativity terms, while the
terms in the second line are the lower order corrections (in $H_0^2$) due to the expanding background.

In addition the Lagrangian is independent of the coordinate $\varphi$ such that
a constant of motion corresponding to angular momentum exists given by the variational
derivation of the Lagrangian with respect to $\dot{\varphi}$
\be
\frac{1}{m}\,\frac{\delta{\mathcal{L}}}{\delta\dot{\varphi}}=2J=-2r_1^2\dot{\varphi}\ .
\ee
We note that this expression matches the same quantity for Keplerian and General Relativity
orbits~(\ref{orbital_constants}).
Further re-expressing the time derivatives $\dot{r}_1$ as derivatives with respect to $\varphi$,
$dr_1/d\tau=dr_1/d\varphi\times d\varphi/d\tau$ and considering the change of variables $u=1/r$
as given in equation~(\ref{orb_var_transf}), replacing the respective derivatives in equation~(\ref{A.L_orbit_E}), differentiating it and factoring out an overall factor of $2u'J^2$ we
obtain the lower order equation in $H_0^2$ for the function $u(\varphi)$ which describes an
orbiting test mass in the gravitational field of a point-like central mass $M$
\be
\ba{rcl}
u''+u&=&\displaystyle\frac{GM}{J^2}+\frac{3GM}{c^2}\,u^2\\[6mm]
&&\displaystyle -\frac{GM}{c^2}\left(\frac{H_0}{c}\right)^2\,\left(1-\frac{2GM}{c^2}\,u\right)^{-1+\alpha_0+\alpha_1\,\frac{2GM}{c^2}\,u}\times\\[6mm]
&&\hfill\displaystyle\times\left(\alpha_0+\alpha_1\,\frac{2GM}{c^2}\,u-\alpha_1\,\left(1-\frac{2GM}{c^2}\,u\right)\,\log\left(1-\frac{2GM}{c^2}\,u\right)\right)\\[6mm]
&&\displaystyle -\left(\frac{H_0}{J}\right)^2\,\frac{1}{u^3}\,\left(1-\frac{2GM}{c^2}\,u\right)^{-1+\alpha_0+\alpha_1\,\frac{2GM}{c^2}\,u}\,\left(1-\frac{2GM}{c^2}\,u+\frac{GM}{c^2}\,u\,\times\right.\\[6mm]
&&\displaystyle\hfill\times\left.\left(\alpha_0+\alpha_1\,\frac{2GM}{c^2}\,u-\alpha_1\,\left(1-\frac{2GM}{c^2}\,u\right)\,\log\left(1-\frac{2GM}{c^2}\,u\right)\right)\right)\ .\\[6mm]
\ea
\lb{A.Eq_u}
\ee
The terms in the first line match the usual terms in the respective General Relativity
differential equation~(\ref{eqd_u}) while the terms in the second and third lines are the
corrections due to the expanding background.

\begin{table}[ht]
\begin{center}
{\tiny
\begin{tabular}{l|ccccccccc}
Planet&Mercury &Venus   &Earth   &Mars    &Jupiter &Saturn  &Uranos  &Neptune &Pluto   \\[2mm]
$r_{1.\mathrm{orb}}\,(\times 10^9\,m)$ &$57.91$&$108.21$&$149.60$&$227.92$&$778.57$&$1433.53$&$2872.46$&$4495.06$&$5906.38$\\
$e$ &$0.2056$ &$0.0067$ &$0.0167$ &$0.0935$ &$0.0489$ &$0.0565$ &$0.0457$ &$0.0113$ &$0.2488$
\end{tabular}}
\caption{\it \small Planetary orbits parameters considered: the semi-major axis
$r_{1.\mathrm{orb}}$ and the eccentricity $e$~\cite{NASA}.\lb{table.planet_data}}
\end{center}
\end{table}
In table~\ref{table.planet_data} are listed the orbital parameters for the
planets considered in section~\ref{sec.Newton} when estimating the
corrections due to General Relativity and due to the expanding background
listed in table~\ref{table.precession_dT}.

\ \\
{\Large\bf Acknowledgements}

This work was supported by grant SFRH/BPD/34566/2007 from FCT-MCTES.

\end{document}